\documentstyle[11pt,epsf]{article}
\begin{document}

\newcommand{\A}{\u{a}}
\newcommand{\h}{\^{\i}}

\begin{center}
{\bf Los Alamos Electronic ArXives} 

{\bf http://xxx.lanl.gov/physics/0004072\\}
\end{center}

\bigskip
\bigskip

\begin{center}
{\huge \bf ELEMENTARY QUANTUM MECHANICS}\end{center}


\bigskip

\begin{center}
{\large \bf HARET C. ROSU}\end{center}
\begin{center} e-mail: rosu@ifug3.ugto.mx\\
fax: 0052-47187611\\
phone: 0052-47183089  \end{center}

\bigskip
\bigskip

\vskip 2ex
\centerline{
\epsfxsize=280pt
\epsfbox{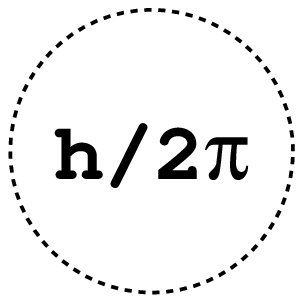}}
\vskip 4ex

\vspace{5.5cm}

\begin{center} Copyright \copyright 2000 by the author.
All commercial rights are reserved.
\end{center}
\vspace{0.2cm}
\centerline{\bf April 2000}

\vspace{2cm}


\begin{center} Abstract \end{center}

\bigskip 

\noindent
This is the first graduate course on elementary quantum mechanics in Internet
written for the benefit of undergraduate and graduate students. 
It is a translation (with corrections) 
of the Romanian version of the course, which I did at the 
suggestion of several students from different countries.
The topics 
included refer to the postulates of quantum mechanics, one-dimensional barriers 
and wells, angular momentum and spin, WKB method, harmonic oscillator, 
hydrogen atom, quantum scattering, and partial waves.\\

\bigskip



\noindent


\newpage

\centerline{{\huge CONTENTS}}

\vspace{0.5cm}

\noindent

{\bf 0. Forward} \hfill ... 4
\\

{\bf 1. Quantum postulates} \hfill ... 5
\\

{\bf 2. One-dimensional rectangular barriers and wells} \hfill ... 23
\\

{\bf 3. Angular momentum and spin} \hfill ... 45
\\

{\bf 4. The WKB method} \hfill ... 75
\\

{\bf 5. The harmonic oscillator} \hfill ... 89
\\

{\bf 6. The hydrogen atom} \hfill ... 111
\\

{\bf 7. Quantum scattering} \hfill ... 133
\\

{\bf 8. Partial waves}\hfill ... 147
\\

There are about 25 illustrative problems.

\bigskip
\bigskip

\centerline{Spacetime nonrelativistic atomic units} 
$$a_H=\hbar ^2/m_ee^2=0.529 \cdot 10^{-8}{\rm cm}$$ 
$$t_{H}=\hbar ^3/m_ee^4=0.242\cdot 10^{-16}{\rm sec}$$ 

\bigskip
\centerline{Planck relativistic units of space and time} 
$$l_P=\hbar/m_Pc=1.616 \cdot 10^{-33}{\rm cm}$$ 
$$t_{P}=\hbar/m_Pc^2=5.390\cdot 10^{-44}{\rm sec}$$

\newpage

{\sl

\section*{{\huge 0. FORWARD}}  

The energy quanta occured in 1900 in the work of Max Planck (Nobel prize, 1918)
on the black body electromagnetic radiation. Planck's ``quanta of light" have
been used
by Einstein (Nobel prize, 1921) to explain the photoelectric effect, but
the first ``quantization" of a quantity having units of action (the angular
momentum) belongs to Niels Bohr (Nobel Prize, 1922). This
opened the road to the universalization of quanta, since the action is the
basic functional to describe any type of motion. However, only in the
1920's the formalism of quantum mechanics has been developed in a systematic
manner. The remarkable works of that decade contributed in a decisive way
to the rising of quantum mechanics at the level of fundamental theory of the
universe, with successful 
technological applications. Moreover, it is quite probable that many
of the cosmological misteries may be disentangled by means of various
quantization procedures of the gravitational field, advancing our 
understanding of the origins of the universe. On the other hand, in recent years,
there is a strong surge of activity in the information aspect of
quantum mechanics. This aspect, which was generally
ignored in the past,
aims at a very attractive ``quantum computer" technology.

At the philosophical level, the famous paradoxes of quantum mechanics,
which are perfect examples of the difficulties of `quantum' thinking, are
actively pursued ever since they have been first posed. Perhaps the most
famous of them is the EPR paradox (Einstein, Podolsky, Rosen, 1935) on the
existence of {\em elements of physical reality}, or in EPR words:
``If, without in any way disturbing a system, we can predict with certainty
(i.e., with probability equal to unity) the value of a physical quantity, then
there exists an element of physical reality corresponding to this physical
quantity."
Another famous paradox is that of Schr\"odinger's cat which is related to
the fundamental quantum property of entanglement and the way we
understand and detect it.
What one should emphasize is that all these delicate points are the sourse
of many interesting and innovative experiments (such as the so-called 
``teleportation" of quantum states) pushing up the technology.

Here, I present eight elementary topics in nonrelativistic quantum mechanics
from a course in Spanish (``castellano")
on quantum mechanics that I taught in
the Instituto de F\'{\i}sica, Universidad de Guanajuato (IFUG), Le\'on, Mexico,
during the semesters of 1998. 

\hfill Haret C. Rosu}

\newpage

\newpage

\section*{\huge 1. THE QUANTUM POSTULATES} 

The following six postulates can be considered as the basis for 
theory and experiment in quantum mechanics in its most used form, which is
known as the Copenhagen interpretation.
\begin{enumerate}
\item[{\bf P1.}-]
To any physical quantity L, which is well defined at the classical level,
one can associate a hermitic operator $\hat{L}$.
\end{enumerate}
\begin{enumerate}
\item[{\bf P2.}-]
To any stationary physical state in which a quantum system can be found one 
can associate a (normalized) wavefunction.
$\psi$ ($\parallel\psi\parallel _{{\cal L}^2}^2=1$).
\end{enumerate}
\begin{enumerate}
\item[{\bf P3.}-]
In (appropriate) experiments, the physical quantity L can take only the 
eigenvalues of  
$\hat{L}$. Therefore the eigenvalues should be real, a condition which is 
fulfilled only by hermitic operators.
\end{enumerate}
\begin{enumerate}
\item[{\bf P4.}-]
What one measures is always the mean value $\overline{L}$
of the physical quantity (i.e., operator) $\hat{L}$
in a state $\psi _{n}$, which, theoretically speaking, is 
the corresponding diagonal matrix element    

 $\langle \psi _{n}\mid\hat{L}\mid \psi _{n}\rangle=\overline{L}$.

\end{enumerate}
\begin{enumerate}
\item[{\bf P5.}-]
The matrix elements of the operators corresponding to the cartesian coordinate 
and momentum,
$\widehat{x_{i}}$ and $\widehat{p_{k}}$, when calculated with the 
wavefunctions $f$ and $g$ satisfy the Hamilton equations of motion of classical
mechanics in the form:\\
$$\frac{d}{dt}\langle f\mid\widehat{p_{i}}\mid{g}\rangle=
-\langle f\mid\frac{\partial\widehat{H}}
{\partial\widehat{x_{i}}}\mid{g}\rangle$$
$$\frac{d}{dt}\langle f\mid\widehat{x_{i}}\mid{g}\rangle
=\langle f\mid\frac{\partial\widehat{H}}
{\partial\widehat{p_{i}}}\mid{g}\rangle~,$$

where $\widehat{H}$ is the hamiltonian operator, whereas the derivatives with
respect to operators are defined as at point 3 of this chapter.
\end{enumerate}
\begin{enumerate}
\item[{\bf P6.}-]
The operators $\widehat{p_{i}}$ and $\widehat{x_{k}}$ have the following   
commutators:
\end{enumerate}
%

$$
\qquad \; [\widehat{p_{i}},\widehat{x_{k}}]  =  -i\hbar\delta _{ik},
$$
$$
[\widehat{p_{i}},\widehat{p_{k}}]  =  0, 
$$
$$
[\widehat{x_{i}},\widehat{x_{k}}]  =  0
$$
%
\begin{center}
$\hbar=h/2\pi=1.0546\times10^{-27}$ erg.sec.
\end{center}
%
\begin{enumerate}
\item[1.-]
\underline{The correspondence between classical and quantum quantities}

This can be done by substituting $x_{i}$, $p_{k}$ with $\widehat{x_{i}}$
$\widehat{p_{k}}$. The function L is supposed to be analytic (i.e., it can be
developed in Taylor series). If the L function does not contain mixed products 
$x_{k}p_{k}$, 
the operator $\hat{L}$ is directly hermitic.\\
Exemple:
\begin{center}

$T=(\sum_{i}^3p_{i}^2)/2m$ $\longrightarrow$
$\widehat{T}=(\sum_{i}^3\widehat{p}^2)/2m$.\\
\end{center}

If L contains mixed products $x_{i}p_{i}$ and higher powers of them, 
$\hat{L}$ is not hermitic, 
and in this case 
L is substituted by $\hat\Lambda$, the hermitic part of $\hat{L}$ 
($\hat\Lambda$ is an autoadjunct operator).\\
Exemple:
\begin{center}

$w(x_{i},p_{i})=\sum_{i}p_{i}x_{i}$ $\longrightarrow$             
$\widehat{w}=1/2\sum_{i}^3(\widehat{p_{i}}\widehat{x_{i}}+\widehat
{x_{i}}\widehat{p_{i}})$.\\
\end{center}

In addition, one can see that we have no time operator. In quantum mechanics,
time is only a
parameter that can be introduced in many ways. This is so because time does
not depend on the canonical variables, merely the latter depend on time. 

\end{enumerate}
\begin{enumerate}
\item[2.-]
\underline{Probability in the discrete part of the spectrum}

If $\psi_{n}$ is an eigenfunction of the operator $\hat{L}$, then:\\

$\overline{L}=<n\mid\hat{L}\mid{n}>=<n\mid\lambda_{n}\mid{n}>=
\lambda_{n}<n\mid{n}>=\delta_{nn}\lambda_{n}=\lambda_{n}$.\\

Moreover, one can prove that $\overline{L}^k=(\lambda_{n})^k$.

If the function $\phi$ is not an eigenfunction of $\hat{L}$, one can make use of
the expansion in the complete system of eigenfunctions of
$\hat{L}$ to get:
\begin{center}

 $\hat{L}\psi_{n}=\lambda_{n}\psi_{n}$,\hspace{10mm} 
 $\phi=\sum_{n}a_{n}\psi_{n}$\\
\end{center}
and combining these two relationships one gets:\\

\begin{center}
 $\hat{L}\phi=\sum_{n}\lambda_{n}a_{n}\psi_{n}$.\\
\end{center}
In this way, one is able to calculate the matrix elements 
of the operator $\hat{L}$:\\
\begin{center}
$\langle \phi\mid\hat{L}\mid{\phi}\rangle=
\sum_{n,m}a_{m}^{\ast}a_{n}\lambda_{n}\langle m\mid{n}\rangle
=\sum_{m}\mid{a_{m}}\mid^2\lambda_{m}$,\\
\end{center}
telling us that the result of the experiment is $\lambda_{m}$ with a probability 
$\mid{a_{m}}\mid^2$.\\
If the spectrum is discrete, according to {\bf P4} this means that 
$\mid{a_{m}}\mid^2$, that is the coefficients of the expansion in a complete set
of eigenfunctions, 
determine the probabilitities to observe the eigenvalue $\lambda_{n}$.\\
If the spectrum is continuous, using the following definition

\begin{center}
$\phi(\tau)=\int{a}(\lambda)\psi{(\tau,\lambda)}d\lambda$,\\
\end{center}
one can calculate the matrix elements in the continuous part of the spectrum
\begin{center}
$\langle\phi\mid{\hat{L}}\mid{\phi}\rangle$\\
\end{center}
\begin{center}
$=\int{d}\tau\int{a}^\ast(\lambda)\psi^\ast(\tau,\lambda)d\lambda\int\mu{a}
(\mu)\psi(\tau,\mu)d\mu$\\
\end{center}
\begin{center}
$=\int\int{a}^{\ast} a(\mu)\mu\int\psi^\ast(\tau,\lambda)\psi(tau,\mu)d\lambda
 {d}\mu{d}\tau$\\
\end{center}
\begin{center}
$=\int\int{a}^\ast(\lambda){a}(\mu)\mu\delta(\lambda-\mu){d}\lambda{d}\mu$\\
\end{center}
\begin{center}
$=\int{a}^\ast(\lambda)a(\lambda)\lambda{d}\lambda$\\
\end{center}
\begin{center}
$=\int\mid{a}(\lambda)\mid^2\lambda{d}\lambda$.\\
\end{center}
In the continuous case, $\mid{a}(\lambda)\mid^2$ should be understood as the 
probability density for observing the eigenvalue $\lambda$ belonging to the 
continuous spectrum. 
Moreover, the following holds\\
\begin{center}
 $\overline{L}=\langle \phi\mid\hat{L}\mid\phi\rangle$.
\end{center}

One usually says that $\langle \mu\mid \Phi\rangle$ is the representation of
$\mid \Phi\rangle$ in the representation $\mu$, where $\mid \mu\rangle$ is an
 eigenvector of $\hat{M}$.
\end{enumerate} 
\newpage
\begin{enumerate}
\item[3.-]
\underline{Definition of the derivate with respect to an operator}\\
\begin{center}
$\frac{\partial{F(\hat{L})}}{\partial\hat{L}}={\rm lim}_{\epsilon\rightarrow
\infty}
\frac{F(\hat{L}+\epsilon\hat{I})-F(\hat{L})}{\epsilon}.$
\end{center}
\end{enumerate}
\begin{enumerate}
\item[4.-]\underline{The operators of cartesian momenta}

Which is the explicit form of $\widehat{p_{1}}$, $\widehat{p_{2}}$ and 
$\widehat{p_{3}}$, if the arguments of the wavefunctions are the cartesian
coordinates $x_{i}$ ?\\
Let us consider the following commutator:\\
\begin{center}
$[\widehat{p_{i}}, \widehat{x_{i}}^2]= \widehat{p_{i}}\widehat{x_{i}}^2-\widehat{x_{i}}^2
\widehat{p_{i}}$\\
\end{center}
\begin{center}
$= \widehat{p_{i}}\widehat{x_{i}}\widehat{x_{i}}-\widehat{x_{i}}\widehat{p_{i}}
\widehat{x_{i}}+\widehat{x_{i}}\widehat{p_{i}}\widehat{x_{i}}-\widehat{x_{i}}
\widehat{x_{i}}
\widehat{p_{i}}$\\
\end{center}
\begin{center}
$=(\widehat{p_{i}}\widehat{x_{i}}-\widehat{x_{i}}\widehat{p_{i}})\widehat{x_{i}}
+\widehat{x_{i}}(\widehat{p_{i}}\widehat{x_{i}}-\widehat{x_{i}}\widehat{p_{i}})$\\
\end{center}
\begin{center}
$=[\widehat{p_{i}},
\widehat{x_{i}}]\widehat{x_{i}}+\widehat{x_{i}}[\widehat{p_{i}}, \widehat{x_{i}}]$\\
\end{center}
\begin{center}
$=-i\hbar\widehat{x_{i}}-i\hbar\widehat{x_{i}}=-2i\hbar\widehat{x_{i}}$.\\
\end{center}
In general, the following holds:\\

\begin{center}
$\widehat{p_{i}}\widehat{x_{i}}^n-\widehat{x_{i}}^n\widehat{p_{i}}=
-ni\hbar\widehat{x_{i}}^{n-1}.$\\
\end{center}
Then, for all analytic functions we have:\\
\begin{center}
$\widehat{p_{i}}\psi(x)-\psi(x)\widehat{p_{i}}=-i\hbar\frac{\partial\psi}
{\partial{x_{i}}}$.\\
\end{center}
Now, let $\widehat{p_{i}}\phi=f(x_{1},x_{2},x_{3})$ be the manner in which
$\widehat{p_{i}}$ acts on $\phi(x_{1},x_{2},x_{3})=1$.
Then:

 $\widehat{p_{i}}\psi=-i\hbar\frac{\partial\psi}{\partial{x_{1}}}+f_{1}\psi$ 
and similar relationships hold for $x_{2}$ and $x_{3}$.\\
From the commutator $[\widehat{p_{i}},\widehat{p_{k}}]=0$ it is easy to get 
$\nabla\times\vec{f}=0$
and therefore $f_{i}=\nabla_{i}F$.\\
The most general form of $\widehat{p_{i}}$ is
$\widehat{p_{i}}=-i\hbar\frac{\partial}{\partial{x_{i}}}+\frac{\partial{F}}
{\partial{x_{i}}}$, where $F$ is an arbitrary function.
The function $F$ can be eliminated by the unitary transformaton
$\widehat{U}^\dagger=\exp(\frac{i}{\hbar}F)$.\\

\begin{center}
$\widehat{p_{i}}=\widehat{U}^\dagger(-i\hbar\frac{\partial}{\partial{x_{i}}}+
\frac{\partial{F}}{\partial{x_{i}}})\widehat{U}$\\
\end{center}
\begin{center}

$=\exp^{\frac{i}{\hbar}F}(-i\hbar
\frac{\partial}{\partial{x_{i}}}+\frac{\partial{F}}{\partial{x_{i}}})
\exp^{\frac{-i}{\hbar}F}$\\
\end{center}
\begin{center}
$=-i\hbar\frac{\partial}{\partial{x_{i}}}$\\
\end{center}
leading to  
\hspace{10mm} $\widehat{p_{i}}=-i\hbar\frac{\partial}{\partial{x_{i}}}$ 
$\longrightarrow$ $\widehat{p}=-i\hbar\nabla$.\\

\end{enumerate}
\begin{enumerate}
\item[5.-]
\underline{Calculation of the normalization constant}

Any wavefunction $\psi(x)$ $\in$ ${\cal L}^2$ of variable $x$ can be written 
in the form:\\
\begin{center}
$\psi(x)=\int\delta(x-\xi)\psi(\xi)d\xi$\\
\end{center}
that can be considered as the expansion of $\psi$ in eigenfunction of the 
operator position (cartesian coordinate) 
$\hat{x}\delta(x-\xi)=\xi(x-\xi)$.
Thus, $\mid\psi(x)\mid^2$ is the probability density of the coordinate in the
state $\psi(x)$. From here one gets the interpretation of the norm\\
\begin{center}
$\parallel\psi(x)\parallel^2=\int\mid\psi(x)\mid^2 dx=1$.\\
\end{center}
Intuitively, this relationship tells us that the system described by  
$\psi(x)$ should be encountered at a certain point on the real axis, 
although we can know only approximately the location.\\
The eigenfunctions of the momentum operator are:\\
$-i\hbar\frac{\partial\psi}{\partial{x_{i}}}=p_{i}\psi$, and by integrating one 
gets
$\psi(x_{i})=A\exp^{\frac{i}{\hbar}p_{i}x_{i}}$. $x$ and $p$ 
have continuous spectra and therefore the normalization
is performed by means of the Dirac delta function.\\
Which is the explicit way of getting the normalization constant ?\\
This is a matter of the following Fourier transforms:\\
$f(k)=\int{g(x)}\exp^{-ikx}dx$,\hspace{3mm}$g(x)=\frac{1}{2\pi}\int{f(k)}\exp
^{ikx}dk.$\\
It can also be obtained with the following procedure.
Consider the unnormalized wavefunction of the free particle\\
$\phi_{p}(x)=A\exp^{\frac{ipx}{\hbar}}$ and the formula 
\begin{center}
$\delta(x-x^{'})=\frac{1}{2\pi}\int_{-\infty}^{\infty}\exp^{ik(x-x^{'})}dx~.$\\
\end{center}
One can see that \\

\begin{center}
$\int_{-\infty}^{\infty}\phi_{p^{'}}^{\ast}(x)\phi_{p}(x)dx$\\
\end{center}
\begin{center}
$=\int_{-\infty}^{\infty}A^{\ast}\exp^{\frac{-ip^{'}x}{\hbar}}A\exp^{\frac{ipx}
{\hbar}}dx$\\
\end{center}
\begin{center}
$=\int_{-\infty}^{\infty}\mid{A}\mid^2\exp^{\frac{ix(p-p^{'})}{\hbar}}dx$\\
\end{center}
\begin{center}
$=\mid{A}\mid^2\hbar\int_{-\infty}^{\infty}\exp^{\frac{ix(p-p^{'})}{\hbar}}
d\frac{x}{\hbar}$\\
\end{center}
\begin{center}
$=2\pi\hbar\mid{A}\mid^2\delta(p-p^{'})$\\
\end{center}
and therefore the normalization constant is:
\begin{center}              
$A=\frac{1}{\sqrt{2\pi\hbar}}$.\\
\end{center}
Moreover, the eigenfunctions of the momentum form a complete system (in the 
sense of the continuous case) for all functions of the ${\cal L}^2$ class.\\

\begin{center}
$\psi(x)=\frac{1}{\sqrt{2\pi\hbar}}\int{a(p)}\exp^{\frac{ipx}{\hbar}}dp$\\
\end{center}
\begin{center}
$a(p)=\frac{1}{\sqrt{2\pi\hbar}}\int\psi(x)\exp^{\frac{-ipx}{\hbar}}dx$.\\
\end{center}
These formulae provide the connection between the x and p representations.

\end{enumerate}
\begin{enumerate}
\item[6.-]
\underline{The momentum (p) representation}\\ 

The explicit form of the operators $\hat{p_{i}}$ and
$\hat{x_{k}}$ can be obtained either from the commutation relationships or through
the usage of the kernels \\
\begin{center}

$x(p,\beta)=U^{\dagger}xU=\frac{1}{2\pi\hbar}\int\exp^{\frac{-ipx}{\hbar}}x
\exp^{\frac{i\beta{x}}{\hbar}}dx$\\
\end{center}
\begin{center}
$=\frac{1}{2\pi{\hbar}}\int\exp^{\frac{-ipx}{\hbar}}(-i\hbar\frac{\partial}
{\partial\beta}\exp^{\frac{i\beta{x}}{\hbar}})$.\\
\end{center}
The integral is of the form:
$M(\lambda,\lambda^{'})=\int{U^{\dagger}}(\lambda,x)\widehat{M}U(\lambda^{'},x)
dx$, and using $\hat{x}f=\int{x}(x,\xi)f(\xi)d\xi$,
the action of $\hat{x}$ on $a(p)$ $\in$ ${\cal L}^2$ is:\\
\begin{center}
$\hat{x}a(p)=\int{x}(p,\beta)a(\beta)d\beta$\\
\end{center}
\begin{center}
$=\int(\frac{1}{2\pi\hbar}\int
\exp^{\frac{-ipx}{\hbar}}(-i\hbar\frac{\partial}{\partial\beta}\exp^{\frac
{i\beta{x}}{\hbar}})dx)a(\beta)d\beta$\\
\end{center}
\begin{center}
$=\frac{-i}{2\pi}\int\int\exp^{\frac{-ipx}{\hbar}}\frac{\partial}{\partial\beta}
\exp^{\frac{i\beta{x}}{\hbar}}a(\beta)dxd\beta$\\
\end{center}
\begin{center}
$=\frac{-i\hbar}{2\pi}\int\int\exp^{\frac{-ipx}{\hbar}}\frac
{\partial}{\partial\beta}\exp^{\frac{i\beta{x}}{\hbar}}a(\beta)d\frac{x}{\hbar}
d\beta$\\
\end{center}
\begin{center}
$=\frac{-i\hbar}{2\pi}\int\int\exp^{\frac{ix(\beta-p)}{\hbar}}\frac{\partial}
{\partial\beta}a(\beta)d\frac{x}{\hbar}d\beta$\\
\end{center}
\begin{center}
$=-i\hbar\int\frac{\partial{a(p)}}{\partial\beta}\delta(\beta-p)d\beta
=-i\hbar\frac{\partial{a(p)}}
{\partial{p}}$,\\

\end{center}
where \hspace{15mm}$\delta(\beta-p)=\frac{1}{2\pi}\int\exp^{\frac{ix(\beta-p)}
{\hbar}}d\frac{x}{\hbar}$.\\

The momentum operator in the p representation is defined by the kernel:\\

\begin{center}
$p(p,\beta)=\widehat{U}^{\dagger}p\widehat{U}$\\
\end{center}
\begin{center}
$=\frac{1}{2\pi\hbar}\int\exp^{\frac{-ipx}{\hbar}}
(-i\hbar\frac{\partial}{\partial{x}})\exp^{\frac{i\beta{x}}{\hbar}}dx$\\
\end{center}
\begin{center}
$=\frac{1}{2\pi\hbar}\int\exp^{\frac{-ipx}{\hbar}}\beta\exp^{\frac{i\beta{x}}{\hbar}}dx
=\beta\lambda(p-\beta)$ \\
\end{center}
leading to $\hat{p}a(p)=pa(p)$.\\

It is worth noting that $\hat{x}$ and $\hat{p}$, although hermitic 
operators for all
f(x) $\in$ ${\cal L}^2$, are not hermitic for their own eigenfunctions.\\
If $\hat{p}a(p)=p_{o}a(p)$ and $\hat{x}=\hat{x}^\dagger$ $\hat{p}=
\hat{p}^\dagger$, then\\
\begin{center}
$<a\mid\hat{p}\hat{x}\mid{a}>-<a\mid\hat{x}\hat{p}\mid{a}>=-i\hbar<a\mid{a}>$\\
\end{center}
\begin{center}
$p_{o}[<a\mid\hat{x}\mid{a}>-<a\mid\hat{x}\mid{a}>]=-i\hbar<a\mid{a}>$\\
\end{center}
\begin{center}
$p_{o}[<a\mid\hat{x}\mid{a}>-<a\mid\hat{x}\mid{a}>]=0$
\end{center}
The left hand side is zero, whereas the right hand side is indefinite, 
which is a contradiction.
\end{enumerate}
  
\begin{enumerate}
\item[7.-]
\underline{Schr\"{o}dinger and Heisenberg representations} \\

The equations of motion given by {\bf P5} have different interpretations 
because in the expression  
$\frac{d}{dt}\langle f\mid\hat{L}\mid{f}\rangle$ one can consider the 
temporal dependence
as belonging either to the wavefunctions or operators, or both to
wavefunctions and operators. We shall consider herein only the first two 
cases.\\
\begin{itemize}
\item
For an operator depending on time $\widehat{O}=\widehat{O(t)}$ we have:\\

\begin{center}
$\hat{p_{i}}=-\frac{\partial\widehat{H}}{\partial\hat{x_{i}}}$,\hspace{5mm}
$\hat{x_{i}}=\frac{\partial\widehat{H}}{\partial\hat{p_{i}}}$\\
\end{center}
\begin{center}
$[\hat{p},f]=\hat{p}f-f\hat{p}=-i\hbar\frac{\partial{f}}{\partial\hat{x_{i}}}$\\
\end{center}
\begin{center}
$[\hat{x},f]=\hat{x}f-f\hat{x}=-i\hbar\frac{\partial{f}}{\partial\hat{p_{i}}}$\\
\end{center}
and the Heisenberg equations of motion are easily obtained:\\

\begin{center}
$\hat{p_{i}}=\frac{-i}{\hbar}[\hat{p},\widehat{H}]$,\hspace{5mm}
$\hat{x_{i}}=\frac{-i}{\hbar}[\hat{x},\widehat{H}]$.
\end{center}

\end{itemize}
\begin{itemize}
\item
If the wavefunctions are time dependent one can still use 
$\hat{p_{i}}=\frac{-i}{\hbar}[\hat{p_{i}},\widehat{H}]$, because being a 
consequence of the commutation relations it does not depend on representation\\

\begin{center}
$\frac{d}{dt}<f\mid\hat{p_{i}}\mid{g}>=\frac{-i}{\hbar}<f\mid[\hat{p},{\widehat{H}}]
\mid{g}>$.\\
\end{center}

If now $\hat{p_{i}}$ and ${\widehat{H}}$ do not depend on time, taking into
account the hermiticity, one gets:\\

\begin{center}
$(\frac{\partial{f}}{\partial{t}},\hat{p_{i}}g)+(\hat{p_{i}}f,\frac
{\partial{g}}{\partial{t}})$\\
\end{center}
\begin{center}
$=\frac{-i}{\hbar}(f,\hat{p_{i}}\hat{H}g)+\frac
{i}{\hbar}(f,\hat{H}\hat{p_{i}}g)$\\
\end{center}
\begin{center}
$=\frac{-i}{\hbar}(\hat{p}f,\hat{H}g)+\frac{i}{\hbar}(\hat{H}f,\hat{p_{i}}g)$\\
\end{center}

\begin{center}
$(\frac{\partial{f}}{\partial{t}}+\frac{i}{\hbar}\hat{H}f,\hat{p_{i}}g)+
(\hat{p_{i}}f,\frac{\partial{g}}{\partial{t}}-\frac{i}{\hbar}\hat{H}g)=0$\\
\end{center}
The latter relationship holds for any pair of functions 
$f(x)$ and
$g(x)$ at the initial moment if each of them satisfies
the equation\\
\begin{center}
$i\hbar\frac{\partial\psi}{\partial{t}}=H\psi$.\\
\end{center}
This is the Schr\"{o}dinger equation. It describes the system by means
of time-independent operators and makes up the so-called 
Schr\"{o}dinger representation.
\end{itemize}
In both representations the temporal evolution of the system
is characterized by the operator
$\widehat{H}$, which can be obtained from Hamilton's function of classical
mechanics.\\
Exemple: $\widehat{H}$ for a particle in a potential 
$U(x_{1},x_{2},x_{3})$ we have:\\

$\widehat{H}=\frac{\hat{p^2}}{2m}+U(x_{1},x_{2},x_{3})$, which in the
x representation is:\\
\begin{center}
$\widehat{H}=-\frac{\hbar^{2}}{2m}\nabla ^2_{x}+U(x_{1},x_{2},x_{3})$.
\end{center}         
\end{enumerate} 
\begin{enumerate}
\item[8.-] \underline{The connection between the S and H representations}

{\bf P5} is correct in both Schr\"{o}dinger's representation and
Heisenberg's.
This is why, the mean value of any observable coincides in the two 
representations. Thus, there is a unitary transformation that can be used
for passing from one to the other.
Such a transformation is of the form 
$\hat{s}^\dagger=\exp^{\frac{-i\hat{H}t}{\hbar}}$. In order to pass to the 
Schr\"{o}dinger representation one should use the  
Heisenberg transform $\psi=\hat{s^{\dagger}}f$ with $f$ and $\hat{L}$, 
whereas to pass to 
Heisenberg's representation the 
Schr\"{o}dinger transform $\hat{\Lambda}=\hat{s^{\dagger}}\hat{L}\hat{s}$ 
with $\psi$ 
and $\hat{\Lambda}$ is of usage.
One can obtain the Schr\"{o}dinger equation as follows: since in the  
transformation $\psi=\hat{s^{\dagger}}f$ the function $f$ does not depend
on time, we shall derivate the 
transformation with respect to time to get:\\

$\frac{\partial{\psi}}{\partial{t}}=\frac{\partial{s^{\dagger}}}{\partial{t}}f=
\frac{\partial}{\partial{t}}(\exp^{\frac{-i\widehat{H}t}{\hbar}})f=\frac{-i}
{\hbar}\widehat{H}\exp^{\frac{-i\widehat{H}t}{\hbar}}f=\frac{-i}{\hbar}\widehat{H}\hat{s^{\dagger}}f=\frac{-i}{\hbar}\widehat{H}\psi$.\\

Therefore:\\
\begin{center}
$i\hbar\frac{\partial\psi}{\partial{t}}=\widehat{H}\psi$.\\
\end{center}
Next we get the Heisenberg equations: putting the
Schr\"{o}dinger transform in the form 
$\hat{s}\hat{\Lambda}\hat
{s^{\dagger}}=\hat{L}$ and performing the derivatives with respect
to time one gets Heisenberg's equation\\

\begin{center}
$\frac{\partial\hat{L}}{\partial{t}}=\frac{\partial\hat{s}}{\partial{t}}
\hat{\Lambda}\hat{s^{\dagger}}+\hat{s}\hat{\Lambda}\frac{\partial\hat
{s^{\dagger}}}{\partial{t}}=\frac{i}{\hbar}\widehat{H}\exp^{\frac{i\widehat{H}t}{\hbar}}\hat\Lambda\hat{s^{\dagger}}-\frac{i}{\hbar}\hat{s}\hat\lambda
\exp^{\frac{-i\hat{H}t}{\hbar}}\widehat{H}$\\
\end{center}
\begin{center}
$=\frac{i}{\hbar}(\widehat{H}\hat{s}\hat{\Lambda}
\hat{s^{\dagger}}-\hat{s}\hat{\Lambda}\hat{s^{\dagger}}\widehat{H})=\frac{i}{\hbar}(\widehat{H}\hat{L}-\hat{L}\widehat{H})=\frac{i}{\hbar}[\widehat{H},\hat{L}]$.\\
\end{center}
Thus, we have:\\
\begin{center}
$\frac{\partial\hat{L}}{\partial{t}}=\frac{i}{\hbar}[\widehat{H},\hat{L}]$.\\
\end{center}
Moreover, Heisenberg's equation can be written in the form:\\
\begin{center}
$\frac{\partial\hat{L}}{\partial{t}}=\frac{i}{\hbar}\hat{s}[\widehat{H},\hat
{\Lambda}]\hat{s^{\dagger}}$.\\
\end{center}
$\hat{L}$ is known as an integral of motion, which, if $\frac{d}{dt}
<\psi\mid\hat{L}\mid\psi>=0$, is characterized by the following 
commutators:\\
\begin{center}
$[\widehat{H},\hat{L}]=0$,\hspace{6mm} $[\widehat{H},\hat\Lambda]=0$.

\end{center}
\end{enumerate} 
\begin{enumerate}
\item[9.-] \underline{Stationary states}

The states of a quantum system described by the eigenfunctions
of $\widehat{H}$ are called stationary states and the corresponding set of
eigenvalues is known as the energy spectrum of the system. 
In such cases, the Schroedinger equation is:\\
\begin{center}
$i\hbar\frac{\partial\psi_{n}}{\partial{t}}=E_{n}\psi_{n}=\widehat{H}\psi_{n}$.\\
\end{center}
The solutions are of the form: 
\hspace{11mm}$\psi_{n}(x,t)=\exp^{\frac{-iE_{n}t}{\hbar}}\phi_{n}(x)$.\\
\begin{itemize}
\item
The probability is the following:\\
\begin{center}
$\delta(x)=\mid\psi_{n}(x,t)\mid^2=\mid\exp^{\frac{-iE_{n}t}{\hbar}}\phi_{n}(x)
\mid^2$\\
\end{center}
\begin{center}
$=\exp^{\frac{iE_{n}t}{\hbar}}\exp^{\frac{-iE_{n}t}{\hbar}}\mid\phi_{n}(x)
\mid^2=\mid\phi_{n}(x)\mid^2$.\\
\end{center}
Thus, the probability is constant in time.
\end{itemize}
\begin{itemize}
\item
In the stationary states, the mean value of any commutator
of the form $[\widehat{H},\hat{A}]$ is zero, where $\hat{A}$ is an arbitrary
operator:\\
\begin{center}
$<n\mid\widehat{H}\hat{A}-\hat{A}\widehat{H}\mid{n}>=<n\mid\widehat{H}\hat{A}\mid{n}>-
<n\mid\hat{A}\widehat{H}\mid{n}>$\\
\end{center}
\begin{center}
$=<n\mid{E_{n}}\hat{A}\mid{n}>-<n\mid\hat{A}E_{n}\mid{n}>$\\
\end{center}
\begin{center}
$=E_{n}<n\mid\hat{A}\mid{n}>-E_{n}<n\mid\hat{A}\mid{n}>=0$.\\
\end{center}
\end{itemize}
\begin{itemize}
\item The virial theorem in quantum mechanics 
- if $\widehat{H}$ is a hamiltonian operator 
of a particle in the field $U(r)$, using\\ 
$\hat{A}=1/2\sum_{i=1}^3(\hat{p_{i}}\hat{x_{i}}-\hat{x_{i}}
\hat{p_{i}})$ one gets:\\
\begin{center}
$<\psi\mid[\hat{A},\widehat{H}]\mid\psi>=0=<\psi\mid\hat{A}\widehat{H}-\widehat{H}\hat{A}
\mid\psi>$\\
\end{center}
\begin{center}
$=\sum_{i=1}^3<\psi\mid\hat{p_{i}}\hat{x_{i}}\widehat{H}-\widehat{H}\hat{p_{i}}\hat{x_{i}}\mid\psi>$\\
\end{center}
\begin{center}
$=\sum_{i=1}^3<\psi\mid[\widehat{H},\hat{x_{i}}]\hat{p_{i}}+
\hat{x_{i}}[\widehat{H},\hat{p_{i}}]\mid\psi>$.\\
\end{center}
Using several times the commutators and $\hat{p_{i}}
=-i\hbar\nabla _{i}$,
$\hat{H}=\widehat{T}+U(r)$, one can get:\\
\begin{center}
$<\psi\mid[\hat{A},\widehat{H}]\mid\psi>=0$\\
\end{center}
\begin{center}
$=-i\hbar(2<\psi\mid\widehat{T}\mid\psi>-<\psi\mid\vec{r}\cdot\nabla{U(r)}\mid
\psi>)$.\\
\end{center}
This is the virial theorem. If the potential is $U(r)=U_{o}r^{n}$, 
then a form of the virial theorem similar to that
in classical mechanics can be obtained with the only difference 
that it refers to mean values\\
\begin{center}
$\overline{T}=\frac{n}{2}\overline{U}$.
\end{center}
\end{itemize}
\begin{itemize}
\item
For a Hamiltonian $\widehat{H}=-\frac{\hbar^2}{2m}\nabla ^2+U(r)$ and
$[\vec{r},H]=\frac{-i\hbar}{m}\vec{p}$, calculating the matrix elements
one finds:\\
\begin{center}
$(E_{k}-E_{n})<n\mid\vec{r}\mid{k}>=\frac{i\hbar}{m}<n\mid\hat{p}\mid{k}>$.
\end{center}
\end{itemize}  
\end{enumerate} 
\begin{enumerate}
\item[10.-]\underline{ 
The nonrelativistic probability current density}

The following integral:\\
\begin{center}
$\int\mid{\psi_{n}}(x)\mid^2dx=1$,\\
\end{center}
is the normalization of an eigenfunction of the discrete spectrum
in the coordinate representation. It appears as a condition 
on the microscopic motion in a finite region of space.\\ 
For  the wavefunctions of the continuous spectrum $\psi_{\lambda}(x)$ one cannot
give a direct probabilistic interpretation.\\
Let us consider a given wavefunction $\phi$ $\in$ ${\cal L}^2$, 
that we write as a linear combination of eigenfunctions of the continuum:\\
\begin{center}
$\phi=\int{a(\lambda)}\psi_{\lambda}(x)dx.$\\
\end{center}
One says that $\phi$ corresponds to an infinite motion.\\
In many cases, the function $a(\lambda)$ is not zero only in a small 
neighborhood of a point $\lambda=\lambda_{o}$. In such a case,
$\phi$ is known as a wavepacket.\\
We shall calculate now the rate of change of the probability of finding the
system in the volume $\Omega$.\\
\begin{center}
$P=\int_{\Omega}\mid\psi(x,t)\mid^2dx=\int_{\Omega}\psi^{\ast}(x,t)
\psi(x,t)dx$.\\
\end{center}
Derivating the integral with respect to time leads to\\
\begin{center}
$\frac{dP}{dt}=\int_{\Omega}(\psi\frac{\partial{\psi^{\ast}}}{\partial{t}}+
\psi^{\ast}\frac{\partial{\psi}}{\partial{t}})dx$.\\
\end{center}
Using now the Schr\"{o}dinger equation in the integral of the right hand side, 
one gets:\\
\begin{center}
$\frac{dP}{dt}=\frac{i}{\hbar}\int_{\Omega}(\psi\hat{H}\psi^{\ast}-\psi^{\ast}
\hat{H}\psi)dx$.\\
\end{center}
Using the identity $f\nabla ^2{g}-g\nabla ^2{f}=div[(f) grad{(g)}-(g) 
grad{(f)}]$
and also the Schr\"{o}dinger equation in the form:\\
\begin{center}
$\hat{H}\psi=\frac{\hbar^2}{2m}\nabla ^2{\psi}$\\
\end{center}
and subtituting in the integral, one gets:\\

\begin{center}
$\frac{dP}{dt}=\frac{i}{\hbar}\int_{\Omega}[\psi(-\frac{\hbar^2}{2m}\nabla{\psi
^{\ast}})-\psi^{\ast}(\frac{-\hbar^2}{2m}\nabla{\psi})]dx$\\
\end{center}
\begin{center}
$=-\int_{\Omega}\frac{i\hbar}{2m}(\psi\nabla{\psi^{\ast}}-\psi^{\ast}\nabla
\psi)dx$\\
\end{center}
\begin{center}
$=-\int_{\Omega}div\frac{i\hbar}{2m}(\psi\nabla{\psi^{\ast}}-\psi^{\ast}
\nabla{\psi})dx$.\\
\end{center}
By means of the divergence theorem, the volume integral can be transformed
in a surface one leading to:\\
\begin{center}
$\frac{dP}{dt}=-\oint\frac{i\hbar}{2m}(\psi\nabla{\psi^{\ast}}-\psi^{\ast}
\nabla{\psi})dx$.\\
\end{center}
The quantity $\vec J(\psi)=\frac{i\hbar}{2m}(\psi\nabla{\psi^{\ast}}-
\psi^{\ast}\nabla{\psi})$ is known as the probability density current, 
for which one can easily get the following continuity equation\\
\begin{center}
$\frac{d\rho}{dt}+div(\vec J)=0$.\\
\end{center}
\begin{itemize}
\item
If $\psi(x)=AR(x)$, where $R(x)$ is a real function,
then: $\vec J(\psi)=0$.\\
\end{itemize}
\begin{itemize}
\item
For momentum eigenfunctions $\psi(x)=\frac{1}{(2\pi{\hbar})^3/2}
\exp^{\frac{i\vec{p}\vec{x}}{\hbar}}$, one gets:\\

\begin{center}
$J(\psi)=\frac{i\hbar}{2m}(\frac{1}{(2\pi{\hbar})^3/2}\exp^{\frac
{i\vec{p}\vec{x}}{\hbar}}(\frac{i\vec{p}}{\hbar(2\pi{\hbar})^3/2}\exp
^{\frac{-i\vec{p}\vec{x}}{\hbar}})$\\
\end{center}
\begin{center}
$-(\frac{1}{(2\pi{\hbar})^3/2}\exp^{\frac{-i\vec
{p}\vec{x}}{\hbar}}\frac{i\vec{p}}{\hbar(2\pi{\hbar})^3/2}
\exp^{\frac{i\hbar
\vec{p}\vec{x}}{\hbar}}))$\\
\end{center}
\begin{center}
$=\frac{i\hbar}{2m}(-\frac{2i\vec{p}}{\hbar(2\pi{\hbar})^3})=\frac{\vec{p}}
{m(2\pi{\hbar})^3}$,\\
\end{center}
which shows that the probability density current does not depend on the 
coordinate.
\end{itemize}
\end{enumerate} 
\begin{enumerate}
\item[11.-]\underline{Operator of spatial transport}

If $\widehat{H}$ is invariant at translations of arbitrary vector 
$\vec{a}$,\\
\begin{center}
$\widehat{H}(\vec{r}+\vec{a})=\widehat{H}\vec{(r)}$~,\\
\end{center}
then there is an operator $\widehat{T}(\vec{a})$ which is unitary 
$\widehat{T}^{\dagger}(\vec
{a})\widehat{H}(\vec{r})\widehat{T}(\vec{a})=\widehat{H}(\vec{r}+\vec{a})$.\\
Commutativity of translations
\begin{center}
 $\widehat{T}(\vec{a})\widehat{T}(\vec{b})=
\widehat{T}(\vec{b})\widehat{T}(\vec{a})=\widehat{T}(\vec{a}+\vec{b})$,
\end{center}
implies that $\widehat{T}$ is of the form $\widehat{T}=\exp^{i\hat{k}a}$,
where $\hat{k}=\frac{\hat{p}}{\hbar}$.\\
In the infinitesimal case:
\begin{center}
$\widehat{T}(\delta\vec{a})\widehat{H}\widehat{T}(\delta\vec{a})\approx(\hat{I}+i\hat{k}
\delta\vec{a})\widehat{H}(\hat{I}-i\hat{k}\delta\vec{a})$,
\end{center}
\begin{center}
$\widehat{H}(\vec{r})+i[\hat{K},\widehat{H}]\delta\vec{a}=\widehat{H}(\vec{r})+(\nabla\widehat{H})\delta\vec{a}$.\\
\end{center}
Moreover, $[\hat{p},\widehat{H}]=0$, where $\hat{p}$ is an integral of the 
motion.
The sistem of wavefunctions of the form $\psi(\vec{p},\vec{r})=\frac
{1}{(2\pi\hbar)^3/2}\exp^{\frac{i\vec{p}\vec{r}}{\hbar}}$ and the unitary
transformation leads to 
$\exp^{\frac{i\vec{p}\vec{a}}{\hbar}}\psi(\vec{r})=\psi(\vec{r}+\vec{a})$.
The operator of spatial transport $\widehat{T}^\dagger
=\exp^{\frac{-i\vec{p}\vec{a}}
{\hbar}}$ is the analog of
 $\hat{s}^\dagger=\exp^{\frac{-i\hat{H}t}{\hbar}}$, which is the operator
of time `transport' (shift).
\end{enumerate}   
\begin{enumerate}
\item[12.-]\underline{Exemple: The `crystal' (lattice) Hamiltonian}

If $\widehat{H}$ is invariant for a discrete translation  
(for exemple, in a crystal lattice) 
$\widehat{H}(\vec{r}+\vec{a})=\widehat{H}(\vec{r})$, where
$\vec{a}=\sum_{i}\vec{a_{i}}n_{i}$, $n_{i}$ $\in$ $N$ and $a_{i}$ are  
baricentric vectors, then:
\begin{center}
$\widehat{H}(\vec{r})\psi(\vec{r})=E\psi(\vec{r})$,
\end{center}
\begin{center}
$\widehat{H}(\vec{r}+\vec{a})\psi(\vec{r}+\vec{a})=E\psi(\vec{r}+\vec{a})=\hat{H}
(\vec{r})\psi(\vec{r}+\vec{a})$.
\end{center}
Consequently, $\psi(\vec{r})$ and $\psi(\vec{r}+\vec{a})$ are wavefunctions 
for the same eigenvalue of $\widehat{H}$.
The relationship between $\psi(\vec{r})$ and $\psi(\vec{r}+\vec{a})$ 
can be saught for in the form 
$\psi(\vec{r}+\vec{a})=\hat{c}(\vec{a})\psi(\vec{r})$, where
$\hat{c}(\vec{a})$ is a gxg matrix (g is the order of degeneration of level E). 
Two column matrices, $\hat{c}(\vec{a})$ and 
 $\hat{c}(\vec{b})$ commute and therefore they
are diagonalizable simultaneously.\\
Moreover, for the diagonal elements, 
$c_{ii}(\vec{a})c_{ii}(\vec{b})=c_{ii}(\vec{a}+\vec{b})$ holds for 
i=1,2,....,g, having solutions of the type
$c_{ii}(a)=\exp^{ik_{i}a}$. Thus, 
$\psi_{k}(\vec{r})=U_{k}(\vec{r})\exp^{i\vec{k}\vec{a}}$, where $\vec{k}$
is a real arbitrary vector and the function
$U_{k}(\vec{r})$ is periodic of 
period $\vec{a}$, $U_{k}(\vec{r}+\vec{a})=U_{k}(\vec{r})$.\\
The assertion that the eigenfunctions of a periodic $\hat{H}$ of the lattice 
type $\hat{H}(\vec{r}+\vec{a})=\hat{H}(\vec{r})$ can be written 
$\psi_{k}(\vec{r})=U_{k}(\vec{r})\exp{i\vec{k}\vec{a}}$, where
$U_{k}(\vec{r}+\vec{a})=U_{k}(\vec{r})$ is known as Bloch's theorem.
In the continuous case, $U_{k}$ should be constant, because the  
constant is the only function periodic for any $\vec{a}$.
The vector $\vec{p}=\hbar\vec{k}$ is called quasimomentum  
(by analogy with the continuous case). The vector $\vec{k}$ is not determined 
univoquely, 
because one can add any vector $\vec{g}$ for which 
$ga=2\pi{n}$, where $n$ $\in$ $N$.\\
The vector $\vec{g}$ can be written 
$\vec{g}=\sum_{i=1}^{3}\vec{b_{i}}m_{i}$, where
$m_{i}$ are integers and $b_{i}$ are given by\\
\begin{center}
$\vec{b_{i}}=2\pi\frac{\hat{a_{j}}\times
\vec{a_{k}}}{\vec{a_{i}}(\vec{a_{j}}\times\vec
{a_{k}})}$, \\
\end{center}
for $i\neq{j}\neq{k}$. $\vec{b_{i}}$ are the baricentric vectors of the 
lattice.\\

\end{enumerate}
\underline{Recommended references}


\noindent 1. E. Farhi, J. Goldstone, S. Gutmann, ``How probability arises in
quantum mechanics",
Annals of Physics {\bf 192}, 368-382 (1989)

\noindent 2. N.K. Tyagi in Am. J. Phys. {\bf 31}, 624 (1963) gives a very short
proof of the Heisenberg uncertainty principle, which asserts that the 
simultaneous measurement of two noncommuting hermitic operators
results in an uncertainty given by the value of their commutator.

\noindent 
3. H.N. N\'u\~nez-Y\'epez et al., ``Simple quantum systems in the momentum 
representation", physics/0001030 (Europ. J. Phys., 2000).

\noindent 4. J.C. Garrison, ``Quantum mechanics of periodic systems",
Am. J. Phys. {\bf 67}, 196 (1999).

\noindent 5. F. Gieres, ``Dirac's formalism and mathematical surprises in 
quantum mechanics", quant-ph/9907069 (in English); quant-ph/9907070
(in French).

\noindent
{\bf 1N. Notes}

\noindent
1. For ``the creation of quantum mechanics...", Werner Heisenberg has been
awarded the
Nobel prize in 1932 (delivered in 1933). The paper
``Z\"ur Quantenmechanik. II", [``On quantum mechanics.II", 
Zf. f. Physik {\bf 35}, 557-615
(1926) (received by the Editor on 16 November 
1925) by M. Born, W. Heisenberg and
P. Jordan, is known as the ``work of the three people", being considered as
the work that really opened the vast horizons of quantum mechanics.

\noindent
2. For ``the statistical interpretation of the wavefunction"
Max Born was awarded the Nobel prize in 1954.

\bigskip

\section*{{\huge 1P. Problems}} 
{\bf Problema 1.1}:
Let us consider two operators, A and B, which commutes by hypothesis.
In this case, one can derive the following relationship:\\

$e^{A}e^{B}=e^{(A+B)}e^{(1/2[A,B])}$.\hspace{10mm}

\bigskip

{\bf Solution}

\noindent
Defining an operator F(t), as a function of real variable t, of the form:
$F(t)=e^{(At)}e^{(Bt)}$,\\
then:
$\frac{dF}{dt}=Ae^{At}e^{Bt}+e^{At}Be^{Bt}=(A+e^{At}Be^{-At})
F(t)$.\\
Applying now the formula $[A,F(B)]=[A,B]F^{'}(B)$, we have \\
$[e^{At},B]=t[A.B]e^{At}$, and therefore:
$e^{At}B=Be^{At}+t[A,B]e^{At}~.$\\
Multiplying both sides of the latter equation by $\exp^{-At}$
and substituting in the first equation, we get:\\

$\frac{dF}{dt}=(A+B+t[A,B])F(t)$.\\

The operators A , B and [A,B] commutes by hypothesis. Thus, we can 
integrate the differential equation as if  
$A+B$ and $[A,B]$ would be scalar numbers.\\
We shall have:\\

$F(t)=F(0)e^{(A+B)t+1/2[A,B]t^2}$.\\

Putting $t=0$, one can see that $F(0)=1$ and therefore :\\

$F(t)=e^{(A+B)t+1/2[A,B]t^2}$.\\

Putting now $t=1$, we get the final result.\\

\bigskip

\noindent
{\bf Problem 1.2}:
Calculate the commutator $[X,D_{x}]$.

\bigskip 

{\bf Solution}

\noindent
The calculation is performed by applying the commutator 
to an arbitrary function $\psi(\vec{r})$:\\
$[X,D_{x}]\psi(\vec{r})=(x\frac{\partial}{\partial{x}}-
\frac{\partial}{\partial{x}}x)\psi(\vec{r})=
x\frac{\partial}{\partial{x}}\psi(\vec{r})-
\frac{\partial}{\partial{x}}[x\psi(\vec{r})]\\
=x\frac{\partial}{\partial{x}}\psi(\vec{r})-
\psi(\vec{r})-x\frac{\partial}{\partial{x}}\psi(\vec{r})=-\psi(\vec{r})$.\\
Since this relationship is satisfied for any $\psi(\vec{r})$, 
one can conclude that $[X,D_{x}]=-1$.         

\bigskip
 
\noindent
{\bf Problem 1.3}:
Check that the trace of a matrix is invariant of changes of discrete 
orthonormalized bases.

\bigskip

{\bf Solution}

\noindent
The sum of the diagonal elements
of a matrix representation of an operator A in an arbitrary basis
does not depend on the basis.\\
This important property can be obtained by passing from an 
orthonormalized discrete basis {$\mid{u_{i}}>$} to another orthonormalized
discrete basis {$\mid{t_{k}}>$}. We have:\\
$\sum_{i}<u_{i}\mid{A}\mid{u_{i}}>=\sum_{i}<u_{i}
\mid \left(\sum_{k}\mid{t_{k}}><t_{k}\mid \right)A\mid{u_{i}}>$\\

\noindent
(where we have used the completeness relationship for the states
$t_{k}$). The right hand side is:\\

$\sum_{i,j}<u_{i}\mid{t_{k}}><t_{k}\mid{A}\mid{u_{i}}>=\sum_{i,j}
<t_{k}\mid{A}\mid{u_{i}}><u_{i}\mid{t_{k}}>$,\\

\noindent
(the change of the order in the product of two scalar numbers is allowed). 
Thus, we can replace
$\sum_{i}\mid{u_{i}}><u_{i}\mid$ with unity 
(i.e., the completeness relationship for the states $\mid{u_{i}}>$), 
in order to get finally:
$$\sum_{i}<u_{i}\mid{A}\mid{u_{i}}>=\sum_{k}<t_{k}\mid{A}\mid{t_{k}}>~.$$
Thus, we have proved the invariance property for matriceal traces. 

\bigskip

\noindent
{\bf Problem 1.4}:
If for the hermitic operator $N$ there are the hermitic operators $L$ and
$M$ such that : $[M,N]=0$, $[L,N]=0$, $[M,L]\neq 0$, then the eigenfunctions
of $N$ are degenerate. 

\bigskip

{\bf Solution}

\noindent
Let $\psi(x;\mu , \nu)$ be the common eigenfunctions of $M$ and $N$ 
(since they commute they are simultaneous observables). 
Let $\psi(x;\lambda , \nu)$ be the common eigenfunctions
of $L$ and $N$ 
(again, since they commute they are simultaneous observables). 
The Greek parameters denote the eigenvalues of the corresponding operators. 
Let us consider for simplicity sake that $N$ has a discrete spectrum. Then:
$$
f(x)=\sum _{\nu}a_{\nu}\psi(x;\mu , \nu)=\sum _{\nu}b_{\nu}
\psi(x;\lambda , \nu)~.
$$
We calculate now the matrix element $<f|ML|f>$:
$$
<f|ML|f>=\int\sum_{\nu}\mu _{\nu}a_{\nu}\psi ^{*}(x;\mu,\nu)\sum_{\nu ^{'}}
\lambda _{\nu ^{'}}b_{\nu ^{'}}\psi(x;\lambda , \nu ^{'})dx~.
$$
If all the eigenfunctions of $N$ are nondegenerate
then $<f|ML|f>=\sum _{\nu}\mu _{\nu}a_{\nu}\lambda _{\nu}b_{\nu}$.
But the same result can be obtained if one calculates
$<f|LM|f>$ and the commutator would be zero. Thus, at least some of the 
eigenfunctions of $N$ should be degenerate.


\newpage
\newcommand{\aple}{\mbox{${}_{\textstyle\sim}^{\textstyle<}$}}
\newcommand{\apge}{\mbox{${}_{\textstyle\sim}^{\textstyle>}$}}
\newcommand{\slsh}[1]{\mbox{$\displaystyle {#1}\!\!\!{/}$}}
\newcommand{\lpr}{\mbox{$ \displaystyle O_L $}}
\newcommand{\rpr}{\mbox{$ \displaystyle O_R $}}
\newcommand{\GeV}{\mbox{$\rm  \, GeV $}}
\section*{\huge 2. ONE DIMENSIONAL RECTANGULAR BARRIERS AND WELLS} 


\subsection*{Regions of constant potential}

\qquad In the case of a rectangular potential, $V(x)$ is a constant
function $V(x)=V$ in a certain region of the one-dimensional space.
In such a region, the Schr\"odinger eq. can be written:
\begin{equation}
\frac{d^2}{dx^2} \psi(x) + \frac{2m}{\hbar^2} (E-V)\psi(x) = 0
\end{equation}
 
One can distinguish several cases:

{\bf (i) $E>V$}

Let us introduce the positive constant $k$, defined by
\begin{equation}
k = \frac{\sqrt{2m(E-V)}}{\hbar}
\end{equation}

\noindent
Then, the solution of eq. (1) can be written:
\begin{equation}
\psi(x) = Ae^{ikx} + A'e^{-ikx}
\end{equation}

\noindent
where $A$ and $A'$ are complex constants.


{\bf (ii) $E<V$}

This condition corresponds to segments of the real axis which would be
prohibited to any particle from the viewpoint of classical mechanics.
In this case, one introduces the positive constant
$q$ defined by:

\begin{equation}
 q = \frac{\sqrt{2m(V-E)}}{\hbar} 
\end{equation}
and the solution of (1) can be written:
\begin{equation}
\psi(x) = Be^{q x} + B'e^{-q x}~,
\end{equation}
where $B$ and $B'$ are complex constants.

{\bf (iii) $E = V$}

\noindent
In this special case, $\psi(x)$ is a linear function of $x$.

\noindent
\subsection*{The behaviour of $\psi(x)$ at a discontinuity of the 
potential}

\qquad One might think that at the point
$x=x_1$, where the potential $V(x)$ is discontinuous, the wavefunction
$\psi(x)$ behaves in a more strange way, maybe discontinuously for example. 
This is not so: $\psi(x)$ and
$\frac{d\psi}{dx}$ are continuous, and only the second derivative
is discontinuous at $x=x_1$.

\noindent
\subsection*{General look to the calculations}

\qquad The procedure to determine the stationary states in rectangular 
potentials is the following: in all regions in which
$V(x)$ is constant we write $\psi(x)$ in any of the two forms
(3) or (5) depending on application; next, we join smoothly
these functions according to the continuity conditions for $\psi(x)$ 
and $\frac{d\psi}{dx}$ at the points where $V(x)$ is discontinuous.

\noindent
\section*{Examination of several simple cases}
\qquad Let us make explicite calculations for some simple stationary states
according to the proposed method.

\subsection*{The step potential}

\vskip 2ex
\centerline{
\epsfxsize=280pt
\epsfbox{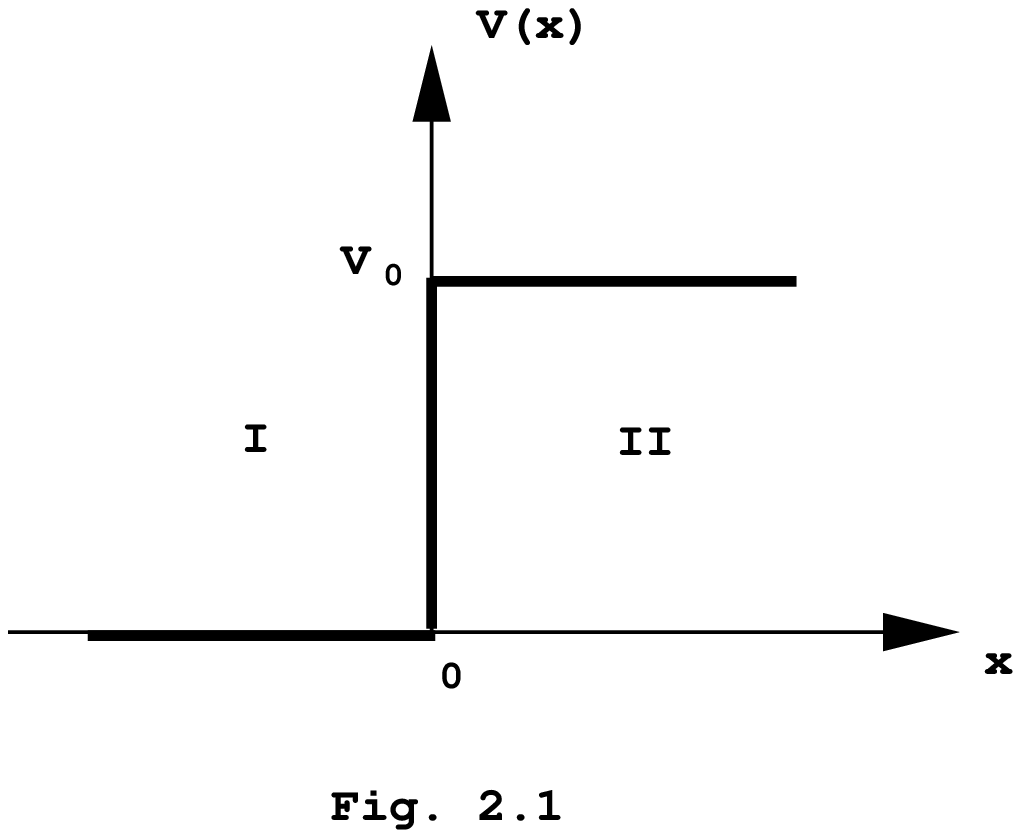}}
\vskip 4ex

{\bf a.} $E>V_0$ case; {\it partial reflexion }

Let us put eq. (2) in the form:

\begin{eqnarray}
k_1 &=& \frac{\sqrt{2mE}}{\hbar}  \\
\nonumber\\
k_2 &=& \frac{\sqrt{2m(E-V_0)}}{\hbar}
\end{eqnarray}

The solution of eq. (1) has the form of eq. (3) in the regions $I (x<0)$ 
and $II (x>0)$:
\begin{eqnarray}
\psi_I &=& A_1e^{ik_1x} + A_1'e^{-ik_1x}  \nonumber
\nonumber\\
\psi_{II} &=& A_2e^{ik_2x} + A_2'e^{-ik_2x} \nonumber
\end{eqnarray}
In region I eq. (1) takes the form
\begin{eqnarray}
\psi''(x) + \frac{2mE}{\hbar^2}\psi(x) = \psi''(x) + k^2\psi(x) = 0 \nonumber
\end{eqnarray}

\noindent
and in the region II:
\begin{eqnarray}
\psi''(x) - \frac{2m}{\hbar^2} [V_0-E]\phi(x) = \psi''(x) - q^2\psi(x) = 0 \nonumber
\end{eqnarray}
If we limit ourselves to the case of an incident 
particle `coming'
from $x=-\infty$, we have to choose $A_2'=0$ and
we can determine the ratios $A_1'/A_1$ and $A_2/A_1$. The joining conditions
give then:
\begin{itemize}
\item
$\psi_I = \psi_{II}$,\qquad {\mbox at} $x=0:$
\begin{equation}
A_1+A_1' = A_2
\end{equation}
\item
$\psi'_I = \psi'_{II}$,\qquad {\mbox at} $x=0:$
\begin{equation}
A_1ik_1 - A_1'ik_1 = A_2ik_2
\end{equation}
\end{itemize}

\noindent
Substituting $A_1$ and $A_1'$ from (8) in (9):
\begin{eqnarray}
A_1' &=& \frac{A_2(k_1 - k_2)}{2k_1} \\
\nonumber\\
A_1 &=& \frac{A_2(k_1 + k_2)}{2k_1}
\end{eqnarray}
\noindent
From the two expressions of the constant $A_2$ in (10) and (11) one gets
\begin{equation}
\frac{A_1'}{A_1} = \frac{k_1 - k_2}{k_1 + k_2}
\end{equation}
and from (11) it follows:
\begin{equation}
\frac{A_2}{A_1} = \frac{2k_1}{k_1+k_2}~.
\end{equation}

\noindent
$\psi(x)$ is a superposition of two waves. The first
(the $A_1$ part) corresponds to an incident wave of 
momentum $p = \hbar k_1$, propagating from the left to the right. The second
(the $A_1'$ part) corresponds to a reflected particle
of momentum $-\hbar k_1$ propagating in opposite direction. Since we have 
already chosen
$A_2' = 0$, it follows that $\psi_{II}(x)$ contains a single wave, which is 
associated to a transmitted particle. (We will show later how it is possible 
by employing the concept of probability current to define
the transmission coefficient T as well as the reflection
coefficient R for the step potential). These
coefficients give the probability that a particle coming from $x
=-\infty$ can pass through or get back from the step at $x=0$. 
Thus, we obtain:
\begin{equation}
R = | \frac{A_1'}{A_1}|^2~,
\end{equation}

\noindent
whereas for $T$:
\begin{equation}
T = \frac{k_2}{k_1}| \frac{A_2}{A_1}|^2~.
\end{equation}

Taking into account (12) and (13) one is led to:
\begin{eqnarray}
R &=& 1- \frac{4 k_1 k_2}{(k_1 + k_2)^2} \\
\nonumber\\
T &=& \frac{4 k_1 k_2}{(k_1 + k_2)^2}~.
\end{eqnarray}
  
It is easy to check that $R+T=1$. 
It is thus sure that the particle will be either transmitted or reflected. 
Contrary to the predictions of classical mechanics, the incident particle
has a nonzero probability of not going back.

It is also easy to check using (6), (7) and (17), 
that if $E \gg V_0$ then $T \simeq 1$: when the energy of the particle is 
sufficently big in comparison with the height of the step, everything happens as
if the step does not exist for the particle.

\bigskip

Consider the following natural form of the solution in region I:
\begin{eqnarray}
\psi_I = A_1e^{ik_1x} + Ae^{-ik_1x}  \nonumber
\end{eqnarray}

\begin{equation}
j = -\frac{i\hbar}{2m}(\phi^* \bigtriangledown \phi - 
\phi \bigtriangledown \phi^*)
\end{equation}

with $A_1 e^{ik_1x}$ and its conjugate $A_1^* e^{-ik_1x}$:
\begin{eqnarray}
j &=& -\frac{i\hbar}{2m}[(A_1^* e^{-ik_1x})
(A_1 i k_1 e^{ik_1x})-(A_1 e^{ik_1x})(-A_1^* i k_1 e^{-ik_1x})] \nonumber \\
\nonumber \\
j &=& \frac{\hbar k_1}{m}|A_1|^2~. \nonumber
\end{eqnarray}

Now with $A e^{-ik_1x}$ and its conjugate $A^{*} e^{ik_1x}$ 
one is led to:

\noindent
\begin{eqnarray}
j = -\frac{\hbar k_1}{m}|A|^2~. \nonumber
\end{eqnarray}

In the following we wish to check the proportion of reflected current
with respect to the incident current (or more exactly, we want to check
the relative probability that the particle is returned back):
\begin{eqnarray}
R = \frac{|j(\phi_-)|}{|j(\phi_+)|} = 
\frac{| -\frac{\hbar k_1}{m}|A|^2|}{| \frac{\hbar k_1}{m}|A_1|^2|} = 
|\frac{A}{A_1}|^2~.
\end{eqnarray}

Similarly, the proportion of transmission with respect to incidence
(that is the probability that the particle is transmitted) is,
taking now into account the solution in the region II:
\begin{eqnarray}
T = \frac{|\frac{\hbar k_2}{m}|A_2|^2|}{| \frac{\hbar k_1}{m}|A_1|^2|} =
 \frac{k_2}{k_1}|\frac{A_2}{A_1}|^2~.
\end{eqnarray}

{\bf b}. $E<V_0$ case; {\it total reflection}

In this case we have:
\begin{eqnarray}
k_1 &=& \frac{\sqrt{2mE}}{\hbar}  \\ 
\nonumber\\
q_2 &=& \frac{\sqrt{2m(V_0-E)}}{\hbar}
\end{eqnarray}
In the region $I (x<0)$, the solution of eq. (1) [written as
$\psi(x)'' + k_1^2\psi(x) = 0$] has the form given in eq. (3):

\begin{equation}
\psi_I = A_1e^{ik_1x} + A_1'e^{-ik_1x}~, 
\end{equation}
 
\noindent
whereas in the region $II (x>0)$, the same eq. (1) [now written as 
$\psi(x)'' -  q_2^2\psi(x) = 0$] has the form of eq. (5):
\begin{equation}
\psi_{II} = B_2e^{q_2x} + B_2'e^{-q_2x}~.
\end{equation}

\noindent
In order that the solution be kept finite 
when $x \rightarrow + \infty$, it is necessary that:
\begin{equation}
B_2 = 0~.
\end{equation}
The joining condition at $x=0$ give now:

\begin{itemize}
\item
$\psi _I = \psi_{II}$,\qquad {\mbox at} $x=0:$
\begin{equation}
A_1 + A_1' = B_2'
\end{equation}
\item
$\psi'_I = \psi'_{II}$,\qquad {\mbox at} $x=0:$
\begin{equation}
A_1 ik_1 - A_1' ik_1 = - B_2' q_2~.
\end{equation}
\end{itemize}

\noindent
Substituting $A_1$ and $A_1'$ from (26) in (27) we get:
\begin{eqnarray}
A_1' &=& \frac{B_2'(i k_1 + q_2)}{2i k_1} \\
\nonumber\\
A_1 &=& \frac{B_2'(i k_1 -  q_2)}{2i k1}~.
\end{eqnarray}
\noindent
Equating the expressions for the constant $B_2'$ from (28) and (29) leads to:
\begin{equation}
\frac{A_1'}{A_1} = \frac{i k_1 + q_2}{i k_1 - q_2} = \frac{k_1 - iq_2}
{k_1 + iq_2}, 
\end{equation}
so that from (29) we have:
\begin{equation}
\frac{B_2'}{A_1} = \frac{2i k_1}{i k_1 - q_2} =\frac{2 k_1}{k_1 - iq_2}~. 
\end{equation}

\noindent
Therefore, the reflection coefficient $R$ is:
\begin{equation}
R = | \frac{A_1'}{A_1}|^2 = | \frac{k_1 - i q_2}{k_1 + i q_2}|^2 = 
\frac{k_1^2 + q_2^2}{k_1^2 + q_2^2} = 1~.   
\end{equation}
\noindent
As in classical mechanics, the microparticle is always reflected 
(total reflexion). However, there is an important difference, namely, 
because of the existence of the so-called evanescent wave 
$e^{-q_2x}$, the particle has a nonzero
probability to find itself in a spatial region which is classicaly forbidden.  
This probability decays exponentially
with $x$ and turns to be negligible when $x$ overcome 
$1/q_{2}$ corresponding to the evanescent wave. Notice also that
$A_1'/A_1$ is a complex quantity. A phase difference occurs as a consequence of
the reflexion, which physically is due to the fact that the particle is 
slowed down when entering the region $x>0$. 
There is no analog phenomenon for this in classical mechanics (but there is 
of course such an analog in optical physics).

\subsection*{Rectangular barrier}

\vskip 2ex
\centerline{
\epsfxsize=180pt
\epsfbox{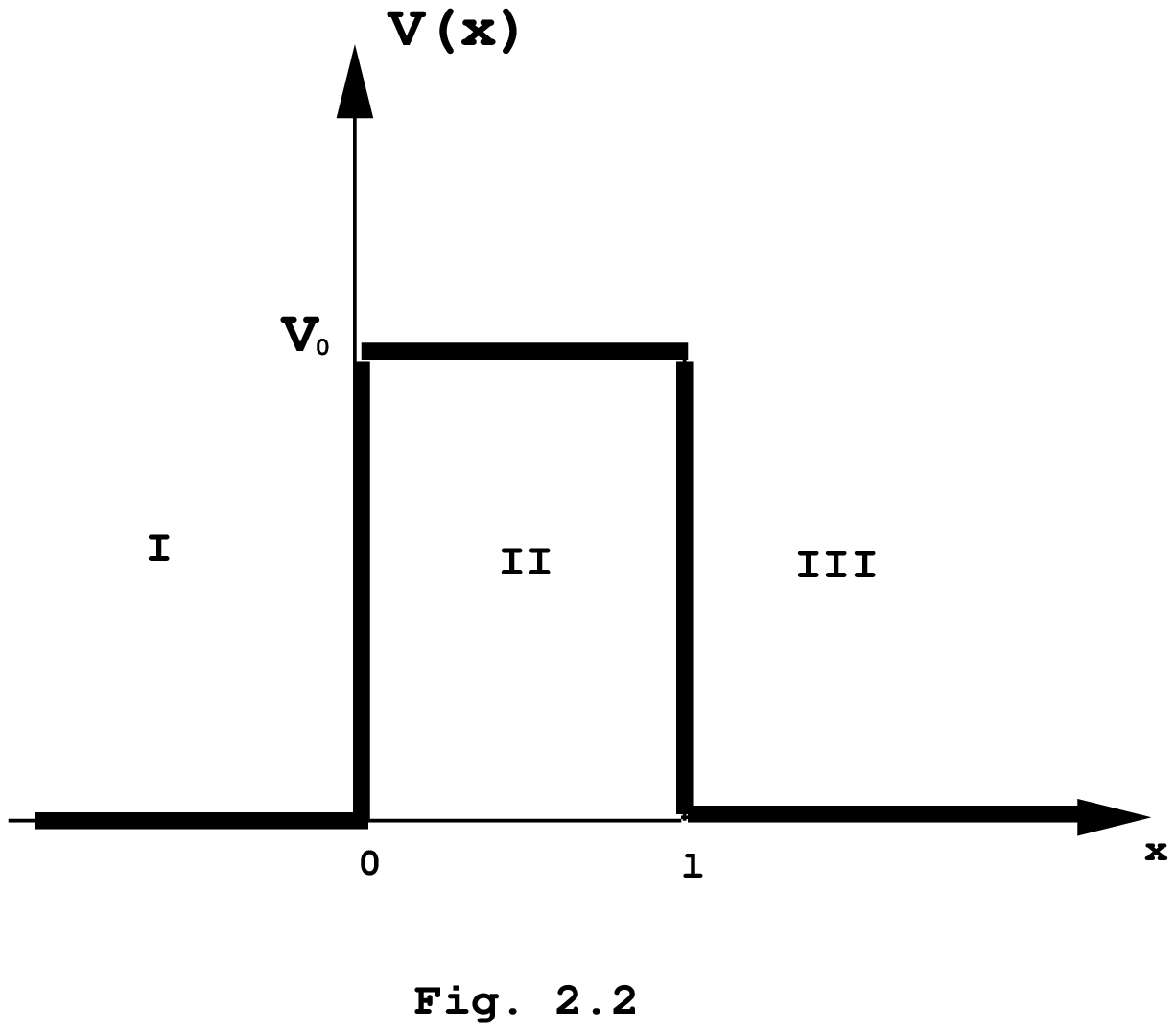}}
\vskip 4ex

{\bf a. $E>V_0$} case; {\it resonances}

Here we put  eq. (2) in the form:
\begin{eqnarray}
k_1 &=& \frac{\sqrt{2mE}}{\hbar}  \\
\nonumber\\
k_2 &=& \frac{\sqrt{2m(E-V_0)}}{\hbar}~.
\end{eqnarray}

The solution of eq.~(1) is as in eq. (3) in the regions
$I (x<0)$, $II (0<x<a$) and $III (x>a):$
\begin{eqnarray}
\psi_I &=& A_1e^{ik_1x} + A_1'e^{-ik_1x} \nonumber
\nonumber\\
\psi_{II} &=& A_2e^{ik_2x} + A_2'e^{-ik_2x}\nonumber
\nonumber\\
\psi_{III} &=& A_3e^{ik_1x} + A_3'e^{-ik_1x}~.\nonumber   
\end{eqnarray}

If we limit ourselves to the case of an incident particle coming 
from $x=-\infty$, we have to choose $A_3'=0$. 

\begin{itemize}
\item
$\psi_I = \psi_{II}$,\qquad {\mbox at} $x=0:$
\begin{equation}
A_1 + A_1' = A_2 + A_2'
\end{equation}
\item
$\psi'_I = \psi'_{II}$,\qquad {\mbox at} $x=0:$
\begin{equation}
A_1ik_1 - A_1'ik_1 = A_2ik_2 - A_2'ik_2
\end{equation}
\item
$\psi_{II} = \psi_{III}$,\qquad {\mbox at} $x=a:$
\begin{equation}
A_2e^{ik_2a} + A_2'e^{-ik_2a} = A_3e^{ik_1a} 
\end{equation}
\item
$\psi'_{II} = \psi'_{III}$,\qquad {\mbox at} $x=a:$ 
\begin{equation}
A_2ik_2e^{ik_2a} - A_2'ik_2e^{-ik_2a} = A_3ik_1e^{ik_1a}~. 
\end{equation}
\end{itemize}

\noindent
The joining conditions at $x=a$ give $A_2$ and $A_2'$ as functions
of $A_3$, whereas those at $x=0$ give $A_1$ and $A_1'$ as functions
of $A_2$ and $A_2'$ (thus, as functions of $A_3$). 
This procedure is shown in detail in the following.

\noindent 
Substituting $A_2'$ from eq. (37) in (38) leads to:
\begin{equation}
A_2 = \frac{A_3e^{ik_1a}(k_2+k_1)}{2k_2e^{ik_2a}}~.
\end{equation}

\noindent
Substituting $A_2$ from eq. (37) in (38) leads to:
\begin{equation}
A_2' = \frac{A_3e^{ik_1a}(k_2-k_1)}{2k_2e^{-ik_2a}}~.
\end{equation}
\noindent
Substituting $A_1$ from eq. (35) in (36) leads to:
\begin{equation}
A_1' = \frac{A_2(k_2-k_1)-A_2'(k_2+k_1)}{-2k_1}~.
\end{equation}

\noindent
Substituting $A_1'$ from eq. (35) in (36) gives:
\begin{equation}
A_1 = \frac{A_2(k_2+k_1)-A_2'(k_2-k_1)}{2k_1}~.
\end{equation}

\noindent
Now, substituting the eqs. (39) and (40) in (41), we have:
\begin{equation}
A_1' = i \frac{(k_2^2 - k_1^2)}{2 k_1 k_2} (\sin k_2a) e^{ik_1a}A_3~.
\end{equation}

\noindent
Finally, substituting the eqs. (39) and (40) in (42) we get:
\begin{equation}
A_1 = [\cos k_2a - i\frac{k_1^2 + k_2^2}{2 k_1 k_2} \sin k_2a] e^{ik_1a}A_3~.
\end{equation}
$A_1'/A_1$ and $A_3/A_1$ [these ratios can be obtained 
by equating (43) and (44), 
and by separating, respectively, in eq.~(44)] allow the calculation of the 
reflexion coefficient $R$ as well as of the transmission one $T$. For this type 
of barrier, they are given by the following formulas:
\begin{equation}
R = |A_1'/A_1|^2 = \frac{(k_1^2 - k_2^2)^2\sin^2 k_2a}{4k_1^2k_2^2 + 
(k_1^2-k_2^2)^2 \sin^2 k_2a},
\end{equation}
\begin{equation}
T=|A_3/A_1|^2=\frac{4k_1^2k_2^2}{4k_1^2 k_2^2 + (k_1^2 - k_2^2)^2 \sin^2 k_2a}.
\end{equation}

\noindent 
It is easy to see that they check $R + T = 1$.


{\bf b.} $E<V_0$ case; {\it the tunnel effect}

\qquad Now, let us take the eqs. (2) and (4):
\begin{eqnarray}
k_1 &=& \frac{\sqrt{2mE}}{\hbar}  \\
\nonumber\\
q_2 &=& \frac{\sqrt{2m(V_0 - E)}} {\hbar}~.
\end{eqnarray}

The solution of eq.~(1) has the form given in eq.~(3) in the regions 
$I (x<0)$ and $III 
(x>a)$, while in the region $II (0<x<a$) has the form of eq.~(5):
\begin{eqnarray}
\psi_I &=& A_1e^{ik_1x} + A_1'e^{-ik_1x}\nonumber
\nonumber\\
\psi_{II} &=& B_2e^{q_2x} + B_2'e^{-q_2x}\nonumber
\nonumber\\
\psi_{III} &=& A_3e^{ik_1x} + A_3'e^{-ik_1x}~.\nonumber   
\end{eqnarray}

The joining conditions at $x=0$ and $x=a$ allow the calculation of the 
transmission coefficient of the barrier. 
As a matter of fact, it is not necessary to repeat the calculation:
merely, it is sufficient to replace $k_2$ by $-i q_2$
in the equation obtained in the first case of this section.


\subsection*{Bound states in rectangular well}
{\bf a. Well of finite depth}

\vskip 2ex
\centerline{
\epsfxsize=280pt
\epsfbox{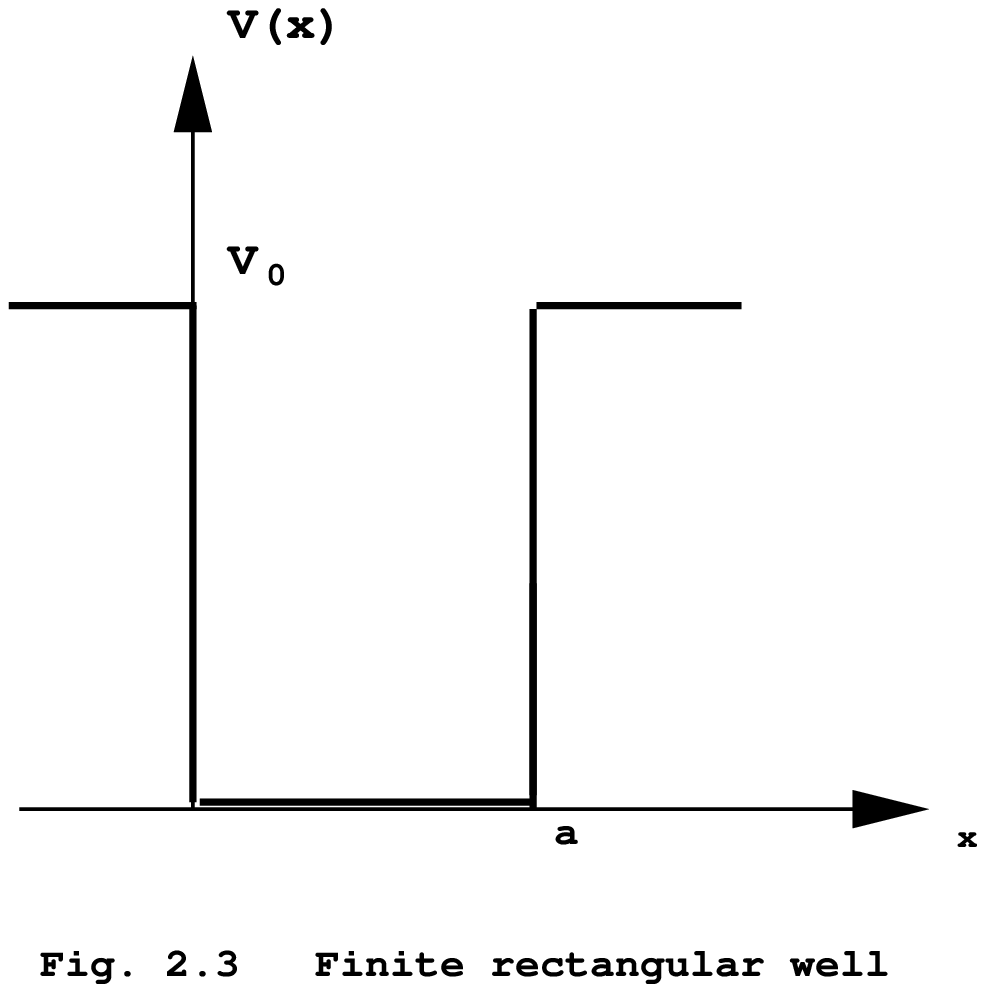}}
\vskip 4ex

We first study the case $0<E<V_0$ 
($E>V_0$ is similar to the calculation in the previous section).

For the exterior regions I, $(x<0)$ and III, $(x>a)$ we employ eq. (4):

\begin{equation}
q = \frac{\sqrt{2m(V_0-E)}}{\hbar}~.
\end{equation}

For the central region  II $(0<x<a)$ we use eq.~(2):

\begin{equation}
k = \frac{\sqrt{2m(E)}}{\hbar}~.
\end{equation}

The solution of eq.~(1) has the form of eq.~(5) in the exterior regions
and of eq.~(3) in the central region:
\begin{eqnarray}
\psi_I &=& B_1e^{q x} + B_1'e^{-q x}\nonumber
\nonumber\\
\psi_{II} &=& A_2e^{ikx} + A_2'e^{-ikx}\nonumber
\nonumber\\
\psi_{III} &=& B_3e^{q x} + B_3'e^{-q x} \nonumber   
\end{eqnarray}

In the region $(0<x<a)$ eq. (1) has the form:
\begin{equation}
\psi(x)'' + \frac{2mE}{\hbar^2}\psi(x) = \psi(x)'' + k^2\psi(x) = 0
\end{equation}

\noindent
while in the exterior regions:
\begin{equation}
\psi(x)'' - \frac{2m}{\hbar^2} [V_0-E]\phi(x) = \psi(x)'' - q^2\psi(x) = 0~.
\end{equation}

Because $\psi$ should be finite in the region I, we impose:
\begin{equation}
B_1'=0~.
\end{equation}
The joining conditions give:

$\psi_I = \psi_{II}$,\qquad at $x=0:$
\begin{equation}
B_1 = A_2 + A'_2
\end{equation}

$\psi'_I = \psi'_{II}$,\qquad at $x=0:$
\begin{equation}
B_1 q = A_2ik - A'_2ik
\end{equation}

$\psi_{II} = \psi_{III}$,\qquad at $x=a:$
\begin{equation}
A_2e^{ika} + A'_2e^{-ika} = B_3e^{q a} + B'_3e^{-q a}
\end{equation}

$\psi'_{II} = \psi'_{III}$,\qquad at $x=a:$
\begin{equation}
A_2ike^{ika} - A'_2ike^{-ika} = B_3q e^{q a} - B'_3q e^{-q a}
\end{equation}

Substituting the constants $A_2$ and $A'_2$ from eq. (54) in eq. (55) 
we get
\begin{eqnarray}
A'_2 &=& \frac{B_1(q-ik)}{-2ik}\nonumber
\nonumber\\
A_2 &=& \frac{B_1(q+ik)}{2ik}~,
\end{eqnarray}
respectively.

Substituting the constant $A_2$ and the constant $A'_2$ from eq. (56) in 
eq. (57) we get
\begin{eqnarray}
B'_3e^{-q a}(ik + q) + B_3e^{q a}(ik-q) + A'_2e^{-ika}(-2ik) &=& 0\nonumber
\nonumber\\
2ikA_2e^{ika} + B'_3e^{-q a}(-ik+q) + B_3E^{q a}(-ik-q) &=& 0~,
\end{eqnarray}
respectively.

\noindent
Equating $B'_3$ from eqs. (59) and taking into account the eqs (58) leads to
\begin{equation}
\frac{B_3}{B_1} = \frac{e^{-q a}}{4ikq}[e^{ika}(q+ik)^2 - e^{-ika}(q - ik)^2]~.
\end{equation}

Since $\psi(x)$ should be finite in region III as well, we require 
$B_3=0$. Thus
\begin{equation}
[\frac{q - ik}{q + ik}]^2 = \frac{e^{ika}}{e^{-ika}} = e^{2ika}~.
\end{equation}

Because $q$ and $k$ depend on $E$, eq. (1) can be satisfied 
for some particular values of $E$. 
The condition that $\psi(x)$ should be finite in all spatial regions 
imposes the quantization of the energy. Two cases are possible:

{\bf (i)} if
\begin{equation}
\frac{q - ik}{q + ik} = - e^{ika}~,
\end{equation}

\noindent
equating in both sides the real and the imaginary parts, respectively, we have
\begin{equation}
\tan(\frac{ka}{2}) =\frac{q}{k}~. 
\end{equation}
Putting
\begin{equation}
k_0 = \sqrt{\frac{2mV_0}{\hbar}} = \sqrt{k^2 + q^2}
\end{equation}
one gets
\begin{equation}
\frac{1}{\cos^2(\frac{ka}{2})} = 1 + \tan^2(\frac{ka}{2}) = \frac{k^2 + q^2}{k^2} = (\frac{k_0}{k})^2
\end{equation}

Eq.~(63) is therefore equivalent to the system of eqs.

\begin{eqnarray}
|\cos(\frac{ka}{2})| &=& \frac{k}{k_0}
\nonumber\\
\tan(\frac{ka}{2}) &>& 0   
\end{eqnarray}

The energy levels are determined by the intersection of a straight line of slope 
$\frac{1}{k_0}$ with the first set of dashed cosinusoides
in fig. 2.4. Thus, we get a certain number of energy levels
whose wavefunctions are even. This fact becomes clearer
if we substitute (62) in (58) and (60). It is easy to check
that $B'_3 = B_1$ and $A_2 = A'_2$ leading to $\psi(-x) =\psi(x)$.

{\bf (ii)} if
\begin{equation}
\frac{q - ik}{q + ik} = e^{ika}~,
\end{equation}
a similar calculation gives
\begin{eqnarray}
|\sin(\frac{ka}{2})| &=& \frac{k}{k_0}
\nonumber\\
\tan(\frac{ka}{2}) &<& 0~.   
\end{eqnarray}
 
The energy levels are in this case determined by the intersection of the same
straight line with the second set of dashed cosinusoides 
in fig. 2.4. The obtained levels are interlaced with those 
found in the case (i). One can easily show that the corresponding wavefunctions 
are odd.

\vskip 2ex
\centerline{
\epsfxsize=280pt
\epsfbox{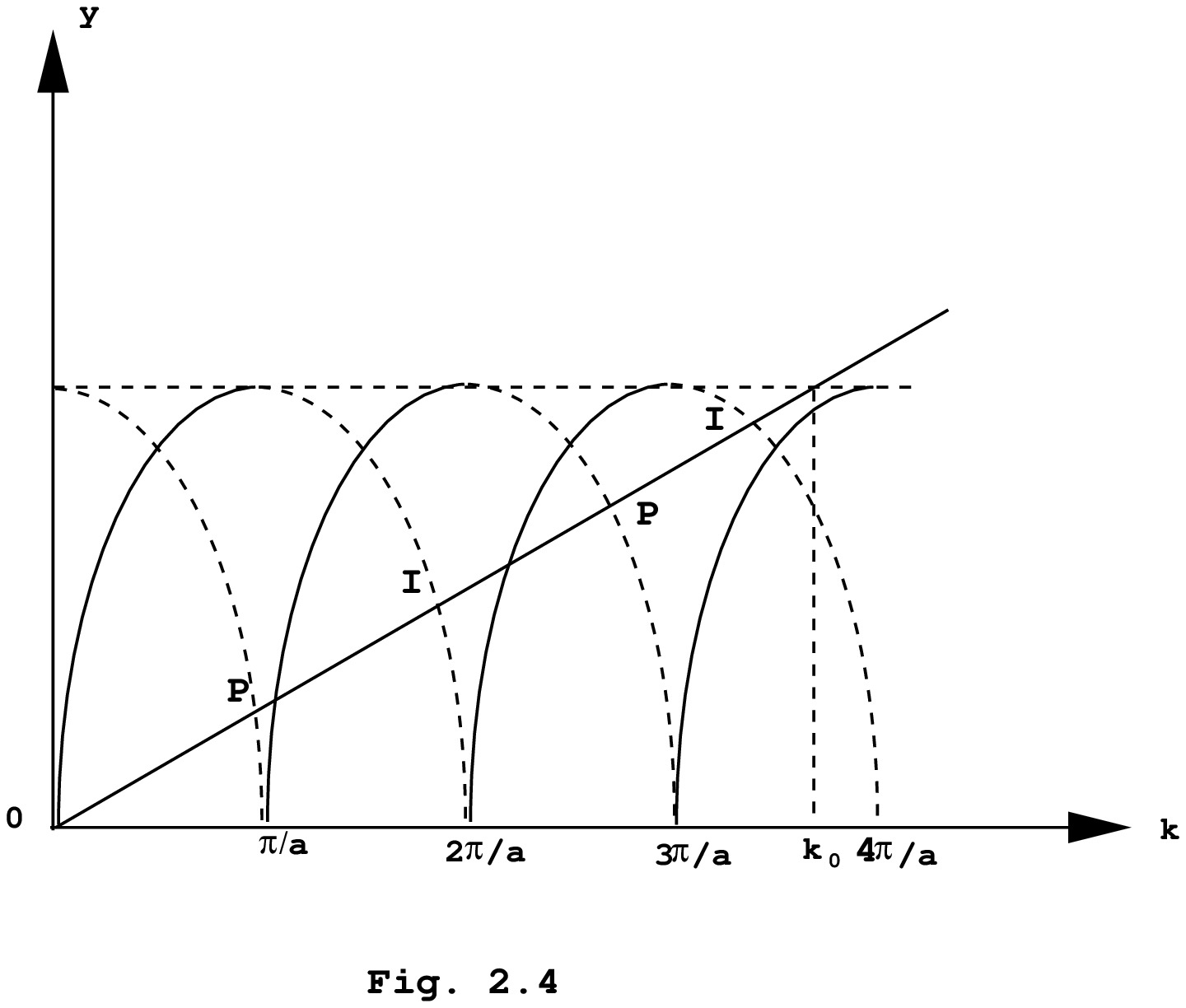}}
\vskip 4ex

{\bf b. Well of infinite depth}

\noindent
In this case it is convenient to put $V(x)$ equal to zero for $0<x<a$ 
and equal to infinity for the rest of the real axis. 
Putting
\begin{equation}
k = \sqrt{\frac{2mE}{\hbar^2}}~,
\end{equation}
$\psi(x)$ should be zero outside the interval $[0,a]$ and continuous
at $x=0$ and $x=a$.

\noindent
For $0 \leq x \leq a$:
\begin{equation}
\psi(x) = Ae^{ikx} + A'e^{-ikx}~.
\end{equation}
Since $\psi(0)=0$, one can infer that $A' = -A$, leading to:
\begin{equation}
\psi(x) = 2iA\sin(kx)~.
\end{equation}
Moreover, $\psi(a)=0$ and therefore
\begin{equation}
k = \frac{n\pi}{a}~,
\end{equation}
where $n$ is an arbitrary positive integer. If we normalize the
function (71), taking into account (72), then we obtain the stationary
wavefunctions
\begin{equation}
\psi_n(x) = \sqrt{\frac{2}{a}}\sin(\frac{n\pi x}{a})
\end{equation}
with the energies
\begin{equation}
E_n = \frac{n^2\pi^2\hbar^2}{2ma^2}~.
\end{equation}
The quantization of the energy levels is extremely simple in this case.
The stationary energies are proportional with the natural numbers
squared. 

\bigskip

\section*{{\huge 2P. Problems}}

\vspace*{4mm}

\subsection*{Problem 2.1: The attractive $\delta$ potential}

Suppose we have a potential of the form:

\begin{eqnarray}
V(x) = -V_0 \delta(x);\qquad  V_0 > 0; \qquad x \in \Re. \nonumber 
\end{eqnarray}
The corresponding wavefunction $\psi(x)$ is assumed continuous.
 
%
a) Obtain the bound states ($E<0$), if they exist, localized in this type of
potential.

b) Calculate the dispersion of a plane wave falling on the $\delta$ 
potential and obtain the  {\it reflexion coefficient}
\begin{eqnarray}
R = \frac{|\psi _{refl}|^2}{|\psi _{inc}|^2}|_{x=0}~, \nonumber
\end{eqnarray}
where $\psi _{refl}$, $\psi _{inc}$ are the reflected and incoming waves,
respectively.

\noindent
{\it Suggestion}: To determine the behavior of $\psi(x)$ in x=0, 
it is better to proceed by integrating the Schr\"odinger equation in the 
interval 
($-\varepsilon ,+\varepsilon$), and then to apply the limit 
$\varepsilon$ $\rightarrow$ $0$.

{\bf Solution.} a) The Schr\"odinger eq. is:
\begin{equation}
\frac{d^2}{dx^2} \psi(x) + \frac{2m}{\hbar^2} (E+V_0 \delta(x))\psi(x) = 0~.
\end{equation}
Far from the origin we have a differential eq. of the form
\begin{equation}
\frac{d^2}{dx^2} \psi (x) = - \frac{2mE}{\hbar^2}\psi(x).
\end{equation}
Consequently, the wavefunctions are of the form 
\begin{equation}
\psi (x) = Ae^{-q x} + Be^{q x} \qquad {\rm for} \qquad x>0 \qquad 
{\rm and} \qquad x<0,
\end{equation}
where $q = \sqrt{-2mE/ \hbar^2}$ $ \in\Re.$ Since $|\psi|^2$ should be 
${\cal L}^2$ integrable , we cannot accept that a part of it grows exponentially. 
Moreover, the wavefunction should be continuous at the origin. 
With these conditions, we have
\begin{eqnarray}
\psi(x) &=& Ae^{q x}; \qquad (x<0), \nonumber
\nonumber\\
\psi(x) &=& Ae^{-q x}; \qquad (x>0).
\end{eqnarray} 
%
Integrating the Schr\"odinger eq. between $-\varepsilon$ and 
$+\varepsilon$, we get
\begin{equation}
-\frac{\hbar^2}{2m}[\psi'(\varepsilon)-\psi'(-\varepsilon)] - 
V_0\psi(0) = E\int^{+\varepsilon} _{-\varepsilon} \psi(x)dx 
\approx 2\varepsilon E\psi(0)
\end{equation}
Introducing now the result (78) and taking into account the 
limit $\varepsilon \rightarrow 0$, we have
\begin{equation}
-\frac{\hbar^2}{2m}(-q A-q A) - V_0 A = 0~,
\end{equation}
or $E = -m(V_0^2 / 2\hbar^2)$ [$-\frac{V_{0}^2}{4}$ in units 
of $\frac{\hbar ^2}{2m}$]. Clearly, there is a single discrete energy.
The normalization constant is found to be 
$A = \sqrt{mV_0/ \hbar^2}$. The wavefunction of the bound state 
will be $\psi _{o}=Ae^{V_0|x|/2}$, where $V_0$ is in $\frac{\hbar ^2}{2m}$ units.
\begin{eqnarray}
\nonumber
\end{eqnarray} 
b) Take now the wavefunction of a plane wave
\begin{equation}
\psi(x) = A e^{ikx}, \qquad k^2 = \frac{2mE}{\hbar^2}~. 
\end{equation}
It moves from the left to the right and is reflected by the potential. 
If $B$ and $C$ are the amplitudes of the reflected and transmitted 
waves, respectively, then we have
\begin{eqnarray}
\psi(x) &=& Ae^{ikx} + Be^{-ikx}; \qquad (x<0), \nonumber
\nonumber\\
\psi(x) &=& Ce^{ikx}; \qquad \qquad \qquad (x>0).
\end{eqnarray} 

The joining conditions and the relationship  
$\psi'(\varepsilon)-\psi'(-\varepsilon) = - f\psi(0)$ cu $f = 2mV_0 / \hbar^2$ 
lead to
\begin{eqnarray}
A + B &=& C \qquad  \qquad  \qquad B = -\frac{f}{f+2ik}A,     \nonumber
\nonumber\\
ik(C - A + B) &=& -fC \qquad  \qquad  C = \frac{2ik}{f+2ik}A.
\end{eqnarray} 
The reflection coefficient will be
\begin{eqnarray}
R = \frac{|\psi_{refl}|^2}{|\psi_{inc}|^2}|_{x=0} = \frac{|B|^2}{|A|^2} = 
\frac{m^2V_0^2}{m^2V_0^2 + \hbar^4k^2}.
\end{eqnarray}
If the potential is very strong  
($V_0 \rightarrow \infty$), one can see that $R \rightarrow 1$, 
i.e., the wave is totally reflected.

{\it The transmission coefficient}, on the other hand, will be
\begin{eqnarray}
T = \frac{|\psi_{trans}|^2}{|\psi_{inc}|^2}|_{x=0} = \frac{|C|^2}{|A|^2} 
= \frac{\hbar^4 k^2}{m^2V_0^2 + \hbar^4k^2}.
\end{eqnarray} 
Again, if the potential is very strong
($V_0 \rightarrow \infty$) then $T \rightarrow 0$,i.e., the transmitted wave 
fades rapidly on the other side of the potential.

In addition, $R + T = 1$ as expected, which is a check of the calculation.

\subsection*{Problem 2.2:
Particle in a 1D potential well of finite depth}

Solve the 1D Schr\"odinger eq.  
for a finite depth potential well given by 
\[
V(x) = \left\{
\begin{array}{ll}
-V_0&\mbox{dac\A\ $|x| \leq a$}\\
0&\mbox{dac\A\ $|x|>a$~.}
\end{array}
\right.
\]

Consider only the bound spectrum ($E<0$).

\vskip 2ex
\centerline{
\epsfxsize=280pt
\epsfbox{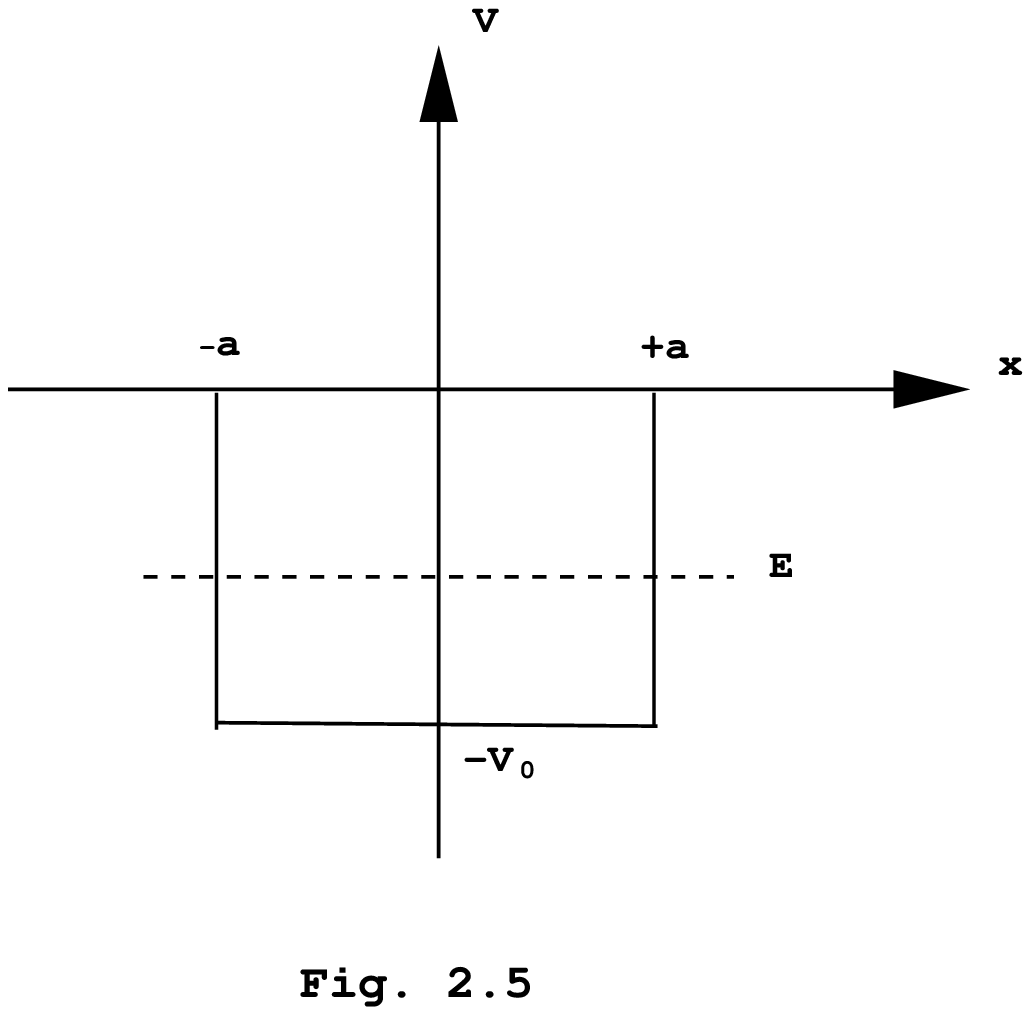}}
\vskip 4ex


\noindent
{\bf Solution.} 

a) The wavefunction for $|x|<a$ and $|x|>a$.

The corresponding Schr\"odinger eq. is
\begin{equation}
-\frac{\hbar^2}{2m}\psi ^{''}(x) + V(x)\psi(x) = E\psi(x)~.
\end{equation}
Defining
\begin{equation}
q^2 = -\frac{2mE}{\hbar^2}, \qquad  k^2 = \frac{2m(E+V_0)}{\hbar^2}~,
\end{equation}
%

\noindent
we get:
\begin{eqnarray}
{\rm 1)~ for ~~~x<-a:}\qquad \psi ^{''}_{1}(x) - q^2 \psi _1 &=& 0,\ 
\psi _1 = A_1e^{q x} + B_1e^{-q x};\nonumber
\nonumber\\
{\rm 2)~ for ~-a\leq x\leq a:}~\psi ^{''}_{2}(x) + k^2 \psi _2 &=& 0, 
\ \psi _2 = A_2 \cos(kx) + B_2 \sin(kx); \nonumber
\nonumber\\
{\rm 3)~ for ~~~x>a:~~}\qquad \psi ^{''}_{3}(x) - q^2 \psi _3 &=& 0,\ 
\psi _3 = B_3 e^{q x} + B_3 e^{-q x}. \nonumber   
\end{eqnarray}

b) Formulation of the boundary conditions. 

\noindent
The normalization of the bound states requires  
solutions going to zero at infinity. This means  
$B_1=A_3=0$. Moreover, $\psi(x)$ should be continuously differentiable. 
All particular solutions are fixed in such a way that
$\psi$ and $\psi'$ are continuous for that value of  
$x$ corresponding to the boundary between the interior and the outside regions.  
The second derivative $\psi''$ displays the discontinuity
the `box' potential imposes. 
Thus we are led to:
\begin{eqnarray}
\psi_1(-a) &=& \psi_2(-a),\qquad  \psi_2(a) = \psi_3(a), \nonumber
\nonumber\\
\psi'_1(-a) &=& \psi'_2(-a),\qquad  \psi'_2(a) = \psi'_3(a).
\end{eqnarray} 

c) The eigenvalue equations. 

From (88) we get four linear and homogeneous eqs
for the coefficients $A_1$, $A_2$, $B_2$ and $B_3$:
\begin{eqnarray}
A_1 e^{-qa} &=& A_2\cos(ka) - B_2\sin(ka), \nonumber
\nonumber\\
qA_1 e^{-qa} &=& A_2k\sin(ka) + B_2k\cos(ka),  \nonumber
\nonumber\\
B_3 e^{-qa} &=& A_2\cos(ka)  + B_2\sin(ka), \nonumber
\nonumber\\
-qB_3 e^{-qa} &=& -A_2k\sin(ka) + B_2k\cos(ka).
\end{eqnarray} 


\noindent
Adding and subtracting, one gets a system of eqs. which is easier
to solve:
\begin{eqnarray}
 (A_1+B_3) e^{-qa} &=& 2A_2\cos(ka) \nonumber
\nonumber\\
q(A_1+B_3) e^{-qa} &=& 2A_2k\sin(ka) \nonumber
\nonumber\\
(A_1-B_3)  e^{-qa} &=& -2B_2\sin(ka) \nonumber
\nonumber\\
q(A_1-B_3) e^{-qa} &=&  2B_2k\cos(ka).
\end{eqnarray} 
Assuming $A_1+B_3 \neq 0$ and $A_2 \neq 0$, the first two eqs give 
\begin{equation}
q = k\tan(ka)~,
\end{equation}
which inserted in the last two eqs gives
\begin{equation}
A_1 = B_3; \qquad B_2 = 0.
\end{equation}
The result is the symmetric solution
$\psi(x) = \psi(-x)$, also called of {\it positive parity}.

A similar calculation for $A_1 - B_3 \neq 0$ and $B_2 \neq 0$
leads to
\begin{equation}
q = -k\cot(ka) \qquad y \qquad A_1 = -B_3; \qquad A_2 = 0. 
\end{equation}
The obtained wavefunction is antisymmetric, 
corresponding to a {\it negative} parity

d) Quantitative solution of the eigenvalue problem.

The equation connecting  
$q$ and $k$, already obtained previously, gives the condition to get the 
eigenvalues. Using the notation
\begin{equation}
\xi = ka, \qquad \eta = qa,
\end{equation}
from the definition (87) we get
\begin{equation}
\xi^2 + \eta^2 = \frac{2mV_0a^2}{\hbar^2} = r^2.
\end{equation}
On the other hand, using (91) and (93) we get the equations 
\begin{eqnarray}
\eta = \xi \tan(\xi), \qquad \eta = -\xi\cot(\xi). \nonumber
\end{eqnarray}
Thus, the sought energy eigenvalues can be obtained from the intersections
of these two curves with the circle defined by
(95) in the plane $\xi$-$\eta$ (see fig. 2.6).

\vskip 2ex
\centerline{
\epsfxsize=280pt
\epsfbox{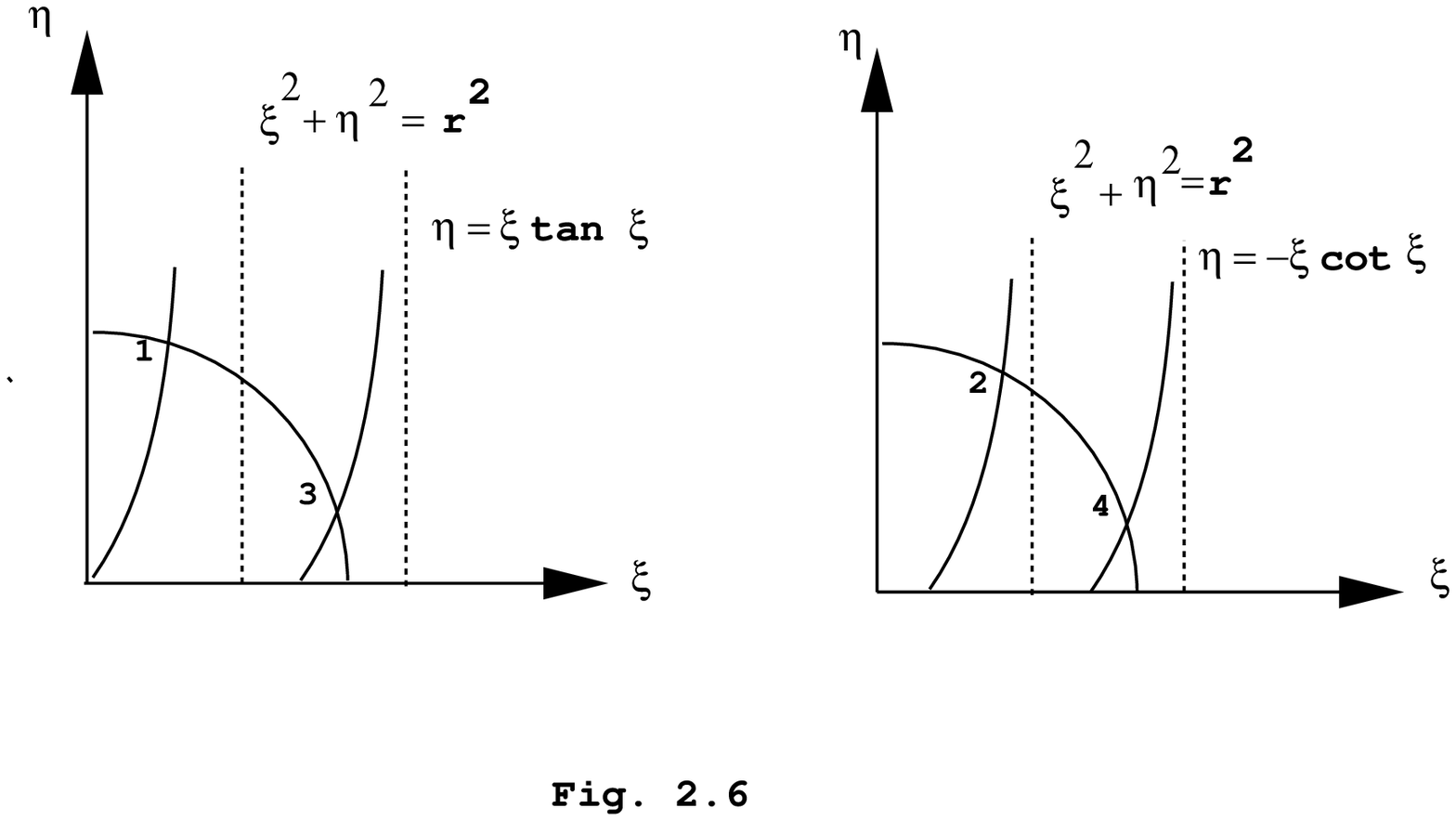}}
\vskip 4ex


There is at least one solution for arbitrary values of the parameter
$V_0$, in the positive parity case, because the tangent function 
passes through the origin. For the negative parity, 
the radius of the circle should be greater than a certain lower bound for the 
two curves to intersect. 
Thus, the potential should have a certain depth related to a given spatial scale  
$a$ and a given mass scale $m$,
to allow for negative parity solutions. The number of energy levels
grows with $V_0$, $a$, and $m$. For the case in which
$mVa^2 \rightarrow \infty$, the intersections are obtained from
\begin{eqnarray}
\tan(ka) &=& \infty \qquad \longrightarrow \qquad  ka=\frac{2n-1}{2}\pi, 
\nonumber
\nonumber\\
-\cot(ka) &=& \infty \qquad \longrightarrow \qquad ka = n \pi, 
\end{eqnarray} 
where $n=1,2,3, \, \ldots $;
by combining the previous relations
\begin{equation}
k(2a) = n \pi.
\end{equation}
For the energy spectrum this fact means that
\begin{equation}
E_n = \frac{\hbar^2}{2m}(\frac{n \pi}{2a})^2 - V_0.
\end{equation}
Widening the well and/or the mass of the particle $m$, 
the diference between two neighbour eigenvalues will decrease. 
The lowest level ($n=1$) is not localized at $-V_0$, but slightly upper.
This `small' difference is called {\it zero point energy}.

e) The forms of the wavefunctions are shown in fig. 2.7. 

\vskip 2ex
\centerline{
\epsfxsize=280pt
\epsfbox{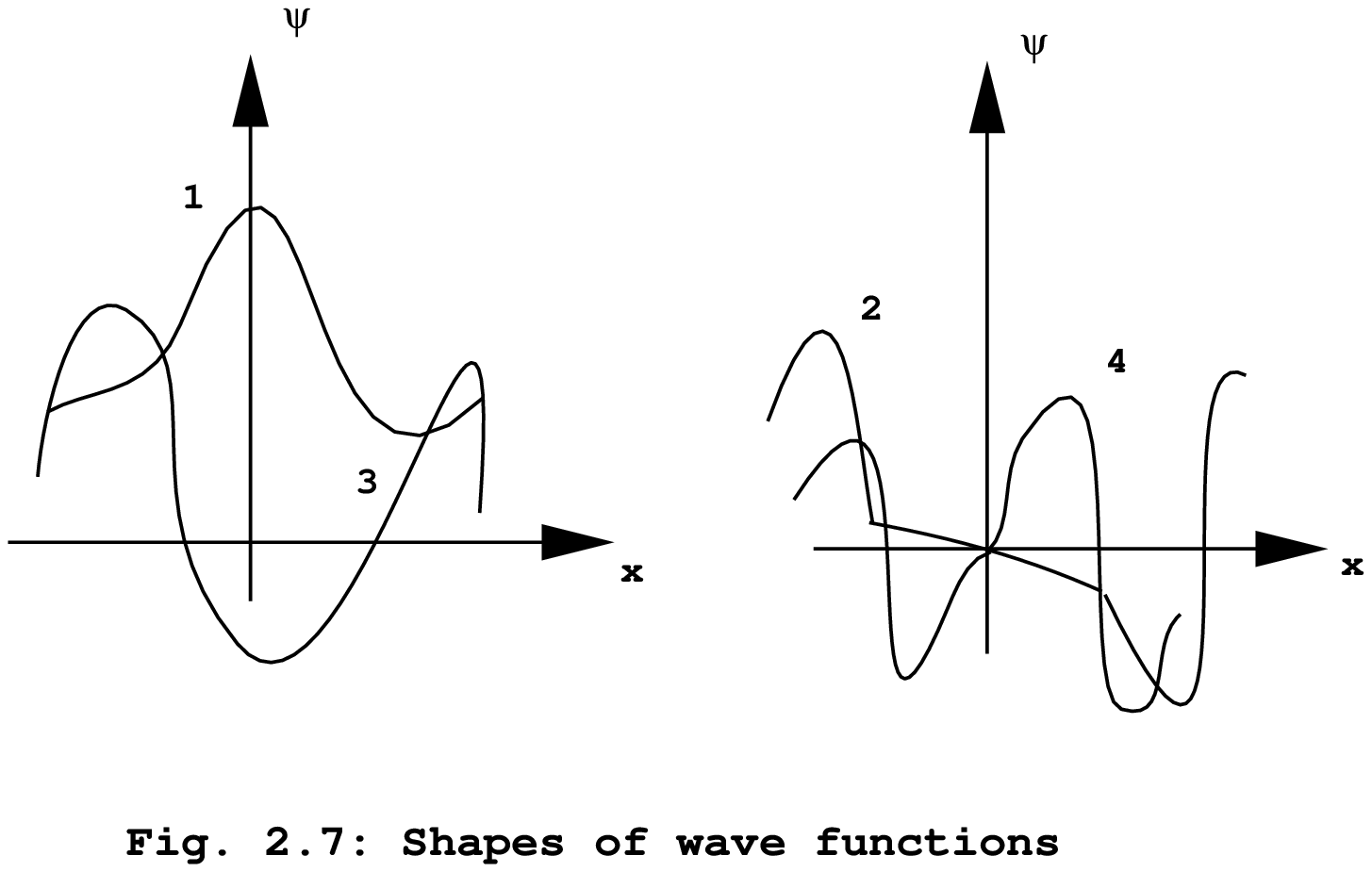}}
\vskip 4ex



\subsection*{Problem 2.3:
Particle in 1D rectangular well of infinite depth}

Solve the 1D Schr\"odinger eq. for a particle 
in a potential well of infinite depth as given by:
\[
V(x) = \left\{
\begin{array}{ll}
0&\mbox{for $x'<x<x'+2a$}\\
\infty&\mbox{for $x'\geq x~~{\rm o}~~x\geq x'+2a$.}
\end{array}
\right.
\]
The solution in its general form is
\begin{equation}
\psi(x) = A\sin(kx) + B\cos(kx)~,
\end{equation}
where
\begin{equation}
k = \sqrt{\frac{2mE}{\hbar^2}}~.
\end{equation}
Since $\psi$ should fulfill $\psi(x') = \psi(x'+2a) = 0$, 
we get:
\begin{eqnarray}
A~\sin(kx')~~~ +~~~ B~\cos(kx') = 0 \\
A\sin[k(x'+2a)] + B\cos[k(x'+2a)] = 0~.
\end{eqnarray} 
Multiplying (101) by $\sin[k(x'+2a)]$ and (102) by $\sin(kx')$ and next
subtracting the latter result from the first we get:
\begin{equation}
B[~~\cos(kx') \sin[k(x'+2a)] - \cos[k(x'+2a)]\sin(kx')~~] = 0~,
\end{equation}
and by means of a trigonometric identity:
\begin{equation}
B \sin(2ak) = 0
\end{equation}
Multiplying (101) by $\cos[k(x'+2a)]$ and 
subtracting (102) multiplied by $\cos(kx')$ leads to:
\begin{equation}
A[~~\sin(kx') \cos[k(x'+2a)] - \sin[k(x'+2a)]\cos(kx')~~] = 0~,
\end{equation}
and by means of the same trigonometric identity:
\begin{equation}
A \sin[k(-2ak)] = A \sin[k(2ak)] =  0~.
\end{equation}

Since we do not take into account the trivial solution $\psi=0$, 
using  
(104) and (106) one has $\sin(2ak)=0$ that takes place only if
$2ak = n \pi$, with $n$ an integer. Accordingly, $k=n \pi/2a$ and since
$k^2=2mE/\hbar^2$ then it comes out that the eigenvalues are given by
the following expression:
\begin{equation}
E = \frac{\hbar^2\pi^2n^2}{8a^2m}~.
\end{equation}
The energy is quantized because only for each
$k_n = n\pi/2a$ one gets a well-defined energy $E_n=[n^2/2m][\pi\hbar/2a]^2$.

The general form of the solution is:
\begin{equation}
\psi_n = A\sin(\frac{n\pi x}{2a}) + B\cos(\frac{n\pi x}{2a}),
\end{equation}
and it can be normalized
\begin{equation}
1 = \int_{x'} ^{x'+2a} \psi \psi^* dx = a(A^2+B^2),
\end{equation}
wherefrom:
\begin{equation}
A = \pm \sqrt{1/a - B^2}~.
\end{equation}
Substituting this value of $A$ in (101) one gets:
\begin{equation}
B = \mp\frac{1}{\sqrt{a}}\sin(\frac{n\pi x'}{2a})~,
\end{equation}
and plugging $B$ in (110) we get
\begin{equation}
A = \pm\frac{1}{\sqrt{a}}\cos(\frac{n\pi x'}{2a})~.
\end{equation}
Using the upper signs for $A$ and $B$, 
by substituting their values in
(108) we obtain:
\begin{equation}
\psi_n =\frac{1}{\sqrt{a}}\sin(\frac{n\pi}{2a})(x-x')~.
\end{equation}
Using the lower signs for $A$ and $B$, one gets 
\begin{equation}
\psi_n =-\frac{1}{\sqrt{a}}\sin(\frac{n\pi}{2a})(x-x').
\end{equation}

\newpage
\begin{center}
{\huge 3. ANGULAR MOMENTUM AND SPIN }
\end{center}

\section*{Introduction}
It is known from {\it Classical Mechanics} that the {\it angular momentum} 
$\bf{l}$ for macroscopic particles is given by
\setcounter{equation}{0}
\begin{equation}
{\bf l=r} \times {\bf p},
\label{1}
\end{equation}
where $\bf r$ and $\bf p$ are the radius vector and the linear 
momentum, respectively.

However, in {\it Quantum Mechanics}, one can find 
operators of angular momentum type  
(OOAMT), which are not compulsory expressed only in terms of the  
coordinate $\hat{x}_j$ and the momentum $\hat{p}_k$
and acting only on the eigenfunctions in the x representation. 
Consequently, it is very important to settle first of all general commutation
relations for the OOAMT components. 

In {\it Quantum Mechanics}
$\bf l$ is expressed by the operator 
\begin{equation}
{\bf l}=-i\hbar {\bf r} \times \nabla,
\label{2}
\end{equation}
whose components are operators satisfying the following commutation rules 

\begin{equation}
[l_x,l_y]=il_z, \qquad  [l_y,l_z]=il_x, \qquad  [l_z,l_x]=il_y.
\label{3}
\end{equation}
Moreover, each of the components commutes with the square of the angular
momentum, i.e.

\begin{equation}
l^2=l^2_x+l^2_y+l^2_z, \qquad [l_i,l^2]=0, \qquad i=1,2,3.
\label{4}
\end{equation}
These relations, besides being correct for the angular momentum, 
are fulfilled for the important OOAMT class of spin operators, which miss 
exact analogs in {\it classical mechanics}.

\noindent
These commutation relations are fundamental
for getting the spectra of the aforementioned operators as well as for their
differential representations.

\section*{The angular momentum}
For an arbitrary point of a fixed space (FS), one can introduce a function
$\psi(x,y,z)$, for which let's consider two cartesian systems
$\Sigma$ and $\Sigma '$, where $\Sigma '$ is obtained by the rotation of the
$z$ axis of $\Sigma$.

In the  general case, a FS refers to a coordinate system, which is different of 
$\Sigma$ and ${\Sigma}'$.

Now, let's compare the values of $\psi$ at two points of the FS
with the same coordinates (x,y,z) in $\Sigma$ and ${\Sigma}'$, which is 
equivalent to the vectorial rotation
\begin{equation}
\psi(x',y',z') = R \psi(x,y,z)
\label{5}
\end{equation}
where $R$ is a rotation matrix in {\sc R}$^3$
\begin{equation}
\left(\begin{array}{c}
x' \\ y' \\ z' 
\end{array}\right)
=
\left(\begin{array}{ccc}
\cos \phi & -\sin \phi & 0 \\
\sin \phi &  \cos \phi & 0 \\
0         &  0         & z 
\end{array}\right)
\left(\begin{array}{c}
x \\ y \\ z
\end{array}\right).
\label{6}
\end{equation}
Then
\begin{equation}
R \psi (x,y,z) =
\psi(x \cos \phi -y \sin \phi ,
     x \sin \phi  +y \cos \phi , z).
\label{7}
\end{equation}

On the other hand, it is important to recall
that the wavefunctions are frame independent
and that the transformation at rotations
within the FS is achieved by means of unitary operators. Thus,
to determine the form of the unitary operator $U^\dagger(\phi)$ that passes
$\psi$ to $\psi '$, one usually considers an infinitesimal rotation $d\phi$,
keeping only the linear terms in $d\phi$ when one expands $\psi '$
in Taylor series in the neighborhood of $x$
\begin{eqnarray}\vspace*{-20pt}
\psi(x',y',z') & \approx &  \psi(x+yd\phi, xd\phi+y, z), \nonumber\\
	       & \approx & \psi(x,y,z) 
+ d\phi\left(y \frac{\partial \psi}{\partial x} 
      - x \frac{\partial \psi}{\partial y}\right), \nonumber\\
               & \approx & (1-id\phi l_z)\psi(x,y,z),
\label{8} 
\end{eqnarray}

\noindent
where we have used the notation\footnote{The proof of (8) 
is displayed as problem 3.1}
\begin{equation}
l_z = \hbar^{-1}(\hat{x}\hat{p}_y -\hat{y}\hat{p}_x ).
\label{9}
\end{equation}
As one will see later, this corresponds to the projection operator 
onto $z$ of the angular momentum according to the definition (2) 
unless the factor $\hbar ^{-1}$. In this way, the rotations of finite 
angle $\phi$ can be represented as exponentials of the form
\begin{equation}
\psi(x', y', z) = e^{il_z \phi}\psi(x,y,z),
\label{10}
\end{equation}
where
\begin{equation}
\hat{U}^\dagger(\phi)=e^{il_z\phi}.
\label{11}
\end{equation}
In order to reassert the concept of rotation, we will consider it in a more 
general approach with the help of the vectorial operator
$\hat{\vec{A}}$ acting on $\psi$, 
assuming that $\hat{A}_x$, $\hat{A}_y$, $\hat{A}_z$ have the same form in
$\Sigma$ and $\Sigma '$, that is, the mean values of
$\hat{\vec{A}}$ as calculated in $\Sigma$ and $\Sigma '$ should be equal
when they are seen from the FS
\begin{eqnarray}
&&\int \psi^*(\vec{r}')
(\hat{A}_x\hat{\imath}' + \hat{A}_y\hat{\jmath}' + \hat{A}_z\hat{k}')
\psi^*(\vec{r}')\,d\vec{r} \nonumber\\
&& \qquad =\int \psi^*(\vec{r})
(\hat{A}_x\hat{\imath} + \hat{A}_y\hat{\jmath} + \hat{A}_z\hat{k})
\psi^*(\vec{r})\,d\vec{r},
\label{12}
\end{eqnarray}
where

\begin{equation}
\hat{\imath}' = \hat{\imath}\cos\phi + \hat{\jmath}\sin\phi, \qquad
\hat{\jmath}' = \hat{\imath}\sin\phi + \hat{\jmath}\cos\phi, \qquad
\hat{k}' = \hat{k}.
\label{13}
\end{equation}

Thus, by combining (10), (12) and (13) we get
\begin{eqnarray}
e^{il_z\phi} \hat{A}_x e^{-il_z\phi}&=& \hat{A}_x\cos\phi -\hat{A}_y\sin\phi,
\nonumber\\
e^{il_z\phi} \hat{A}_y e^{-il_z\phi}&=& \hat{A}_x\sin\phi -\hat{A}_y\cos\phi,
\nonumber\\
e^{il_z\phi} \hat{A}_z e^{-il_z\phi}&=& \hat{A}_z.
\label{14}
\end{eqnarray}

Again, considering infinitesimal rotations and expanding the left hand sides in
(14), one can determine the commutation relations 
of $\hat{A}_x$, $\hat{A}_y$ and $\hat{A}_z$ with $\hat{l}_z$
\begin{equation}
[l_z,A_x]=iA_y, \qquad  [l_z,A_y]=-iA_x, \qquad  [l_z,A_z]= 0,
\label{15}
\end{equation}
and similarly for $l_x$ and $l_y$. 
\\

The basic conditions to obtain these commutation relations are

\begin{itemize}

\item[$\star$]
The eigenfunctions transform as in (7) when $\Sigma \rightarrow \Sigma '$.

\item[$\star$]
The components $\hat{A}_x$, $\hat{A}_y$, $\hat{A}_z$ have the same form in
$\Sigma$ and $\Sigma '$.

\item[$\star$] 
The kets corresponding to the mean values of $\hat{A}$ in $\Sigma$ and 
$\Sigma '$ coincide (are the same) for a FS observer.

\end{itemize}

One can also use another representation in which $\psi(x,y,z)$
does not change when $\Sigma \rightarrow \Sigma '$ and the vectorial operators  
transform as ordinary vectors. In order to pass to such a  
representation when we rotate by $\phi$
around $z$ one makes use of the operator $\hat{U}(\phi)$, 
that is

\begin{equation}
e^{il_z\phi} \psi'(x,y,z) = \psi(x,y,z),
\label{16}
\end{equation}
and therefore
\begin{equation}
e^{-il_z\phi} \hat{\vec{A}} e^{il_z\phi} = \hat{\vec{A}}.
\label{17}
\end{equation}
Using the relationships (14) we obtain
\begin{eqnarray}
\hat{A}_x' & = & \hat{A}_x\cos\phi + \hat{A}_y\sin\phi 
             =   e^{-il_z\phi} \hat{A}_x e^{il_z\phi}, \nonumber\\ 
\hat{A}_y' & = & -\hat{A}_x\sin\phi + \hat{A}_y\cos\phi 
             =  e^{-il_z\phi} \hat{A}_y e^{il_z\phi}, \nonumber\\ 
\hat{A}_z' & = & e^{-il_z\phi} \hat{A}_z e^{il_z\phi}.
\label{18}
\end{eqnarray}

Since the transformations of the new representation are performed by means of
unitary operators, the commutation relations do not change.

\subsection*{Remarks}

\begin{itemize}

\item[$\star$]
The operator $\hat{A}^2$ is invariant at rotations, that is
\begin{equation}
e^{-il_z\phi} \hat{A}^2 e^{il_z\phi} = \hat{A}'^2 = \hat{A}^2~.
\label{19}
\end{equation}

\item[$\star$]
It follows that 
\begin{equation}
[\hat{l}_i, \hat{A}^2] = 0~.
\label{20}
\end{equation}

\item[$\star$]
If the Hamiltonian operator is of the form 
\begin{equation}
\hat{H} = \frac{1}{2m}\hat{p}^2 + U(|\vec{r}|),
\label{21}
\end{equation}
then it remains invariant under rotations in any axis 
passing through the coordinate origin 
\begin{equation}
[\hat{l}_i, \hat{H}] = 0~,
\label{22}
\end{equation}
where $\hat{l}_i$ are integrals of the motion.
\end{itemize}

\subsection*{Definition}
If $\hat{A}_i$ are the components of a vectorial operator acting 
on a wavefunction depending only on the coordinates 
and if there are
operators $\hat{l}_i$ that satisfy the following commutation relations
\begin{equation}
[\hat{l}_i, \hat{A}_j] = i\varepsilon_{ijk}\hat{A}_k, \qquad
[\hat{l}_i, \hat{l}_j] = i\varepsilon_{ijk}\hat{l}_k~,
\label{23}
\end{equation}

\noindent
then $\hat{l}_i$ are known as the components of the angular momentum
operator and we can infer from (20) and (23) that 
\begin{equation}
[\hat{l}_i, \hat{l}^2]=0.
\label{24}
\end{equation}
 
Consequently the three operatorial components associated to the
components of a classical angular momentum 
satisfy commutation relations of the type (23), (24).
Moreover, one can prove that these relations lead to specific geometric 
properties of the rotations in a 3D euclidean space. 
This takes place if we adopt a more general point of view 
by defining an angular momentum operator  
$\bf J$ (we shall not use the hat symbol for simplicity of writing)
as any set of three observables $J_x$, $J_y$ \c{s}i
$J_z$ which fulfill the commutation relations
\begin{equation}
[J_i, J_j] = i\varepsilon_{ijk}J_k.
\label{25}
\end{equation}

Moreover, let us introduce the operator 
\begin{equation}
{\bf J}^2 = J^2_x + J^2_y + J^2_z,
\label{26}
\end{equation}
the scalar square of the angular momentum $\bf J$. This operator is 
hermitic because $J_x$, $J_y$ and $J_z$ are hermitic
and it is easy to show that $\bf J^2$ commutes with the three components
of $\bf J$
\begin{equation}
[{\bf J}^2, J_{i}]=0.
\label{27}
\end{equation}

Since $\bf J^2$ commutes with each of the components it follows that there is
a complete system of eigenfunctions, i.e.
\begin{equation}
{\bf J^2}\psi_{\gamma \mu} = f(\gamma ^2)\psi_{\gamma \mu}, \qquad
    J_i\psi_{\gamma \mu} = g(\mu)    \psi_{\gamma \mu},
\label{28}
\end{equation}
where, as it will be shown in the following, the eigenfunctions depend
on two subindices, which will be determined together with the form of the 
functions $f(\gamma)$ and $g(\mu)$. 
The operators $J_i$ and $J_k$ $(i \neq k)$ do not commute, i.e. they do not have 
common eigenfunctions. For physical and mathematical reasons, we are interested
to determine the common eigenfunctions of 
${\bf J^2}$ and $J_{z}$, that is, we shall take $i=z$ in (28).

Instead of using the components $J_x$ and $J_y$ of the angular momentum
$\bf J$,
it is more convenient to work with the following 
linear combinations
\begin{equation}
J_+ = J_x + iJ_y, \qquad J_{-} = J_x - iJ_y.
\label{29}
\end{equation}
Contrary to the operators $a$ and $a^\dagger$
of the harmonic oscillator (see chapter 5), these operators are not hermitic,
they are only adjunct to each other.
The following properties are easy to prove
\begin{equation}
[J_z,   J_{\pm}] = \pm J_{\pm}, \qquad 
[J_{+}, J_{-}]   =     2J_z, 
\label{30}
\end{equation}
\begin{equation}
[J^2, J_{+}]   =  [J^2, J_{-}] = [J^2, J_{z}] = 0.
\label{31}
\end{equation}
\begin{equation}
J_z(J_{\pm}\psi_{\gamma\mu}) = \{J_{\pm}J_z + [J_z, J_{\pm}]\}\psi_{\gamma\mu}
=(\mu \pm 1) (J_{\pm} \psi_{\gamma\mu}).
\label{32}
\end{equation}

Therefore $J_{\pm}\psi_{\gamma\mu}$ are eigenfunctions of $J_z$
corresponding to the eigenvalues $\mu \pm 1$, that is these functions are
identical up to the constant factors
$\alpha_\mu$ and $\beta_\mu$ (to be determined)
\begin{eqnarray}
J_{+} \psi_{\gamma\mu - 1} &=& \alpha_\mu \psi_{\gamma\mu}, \nonumber\\
J_{-}\psi_{\gamma\mu}         &=& \beta_\mu \psi_{\gamma\mu-1}.
\label{33}
\end{eqnarray}
On the other hand
\begin{equation}
\alpha^*_{\mu} = (J_{+}\psi_{\gamma \mu -1} , \psi_{\gamma \mu}) 
               = (\psi_{\gamma \mu -1} J_{-} \psi_{\gamma \mu})= \beta_\mu~.
\label{34}
\end{equation}
Therefore, taking a phase of the type $e^{ia}$ (where $a$ is real) for the 
function
$\psi_{\gamma \mu}$ one can put $\alpha_\mu$ real and equal to $\beta _\mu$, 
which means
\begin{equation}
J_{+}\psi_{\gamma , \mu - 1} = \alpha \mu \psi_{\gamma \mu},
J_{-}\psi_{\gamma \mu    } = \alpha \mu \psi_{\gamma , \mu - 1},
\label{35}
\end{equation}
and therefore
\begin{eqnarray}
\gamma &=& (\psi_{\gamma \mu}, [J_x^2 + J_y^2 + J_z^2] \psi_{\gamma \mu})
= \mu^2 + a + b, \nonumber\\
a & = & (\psi_{\gamma \mu}, J_{x}^{2} \psi_{\gamma \mu}) = 
       (J_x\psi_{\gamma \mu}, J_x \psi_{\gamma \mu}) \geq 0, \nonumber\\
b & = &  (\psi_{\gamma \mu}, J_{y}^{2} \psi_{\gamma \mu}) = 
        (J_y\psi_{\gamma \mu}, J_y \psi_{\gamma \mu}) \geq 0.
\label{36}
\end{eqnarray}

The normalization constant cannot be negative. This implies
\begin{equation}
\gamma \geq \mu^2,
\label{37}
\end{equation}
for a fixed $\gamma$; thus, $\mu$ has both superior and inferior limits
(it takes values in a finite interval). 

Let $\Lambda$ and $\lambda$ be these limits, respectively, for a given
$\gamma$
\begin{equation}
J_{+}\psi_{\gamma \Lambda} = 0, \qquad J_{-}\psi_{\gamma \lambda} = 0.
\label{38}
\end{equation}

Using the following operatorial identities
\begin{eqnarray}
J_{-}J_{+} &=& {\bf J^2} - J^2_z + J_z = {\bf J^2} - J_z(J_z-1),\nonumber\\
J_{+}J_{-} &=& {\bf J^2} - J^2_z + J_z = {\bf J^2} - J_z(J_z+1),
\label{39}
\end{eqnarray}
acting on $\psi _ {\gamma \Lambda}$ as well as on
$\psi_{\gamma \lambda}$ one gets
\begin{eqnarray}
\gamma - \Lambda^2 - \Lambda &=& 0, \nonumber \\
\gamma - \lambda^2 + \lambda &=& 0, \nonumber \\
(\lambda - \lambda + 1) (\lambda + \lambda) &=& 0.
\label{40}
\end{eqnarray}

In addition, 
\begin{equation}
\Lambda \geq \lambda \rightarrow \Lambda = -\lambda = J \rightarrow
\gamma = J(J+1).
\label{41}
\end{equation} 

For a given $\gamma$ the projection $\mu$ of the momentum takes
$2J+1$ values that differ by unity, from $J$ to $-J$. 
Therefore, the
difference $\Lambda -\lambda = 2J$ should be an integer and consequently the
eigenvalues of $J_z$ that are labelled by $m$ are integer
\begin{equation}
m=k, \qquad k=0, \pm 1, \pm 2, \, \ldots \, ,
\label{42}
\end{equation}
or half-integer
\begin{equation}
m=k + {1 \over 2}, \qquad k=0, \pm 1, \pm 2, \, \ldots \, .
\label{43}
\end{equation}

A state having a given $\gamma = J(J+1)$ presents a degeneration of order
$g=2J+1$
with regard to the eigenvalues $m$ (this is so because
$J_i,~J_k$ commute with $J^2$ but do not commute between themselves.
 
By a ``state of angular momentum $J$''  one usually understands
a state of $\gamma = J(J+1)$ having the maximum projection of its momentum,
i.e. $J$. Quite used notations for angular momentum states are
$\psi_{jm}$ and the Dirac ket one $|jm\rangle$.

Let us now obtain the matrix elements of $J_x,~J_y$ 
in the representation in which
$J^2$ and $J_z$ are diagonal. In this case, one obtains from (35) and (39)
the following relations
\begin{eqnarray}
J_{-}J_{+} \psi_{jm-1} = \alpha_mJ_{-}\psi_{jm} = \alpha_m\psi_{jm-1},
\nonumber\\
J(J+1)-(m-1)^2-(m-1) = \alpha_m^2,\nonumber\\
\alpha_m = \sqrt{(J+m)(J-m+1)}.
\label{44}
\end{eqnarray}

Combining (44) and (35) leads to
\begin{equation}
J_+\psi_{jm-1} = \sqrt{(J+m)(J-m+1)}\psi_{jm}~.
\label{45}
\end{equation}
It follows that the matrix element of $J_{+}$ is
\begin{equation}
\langle jm | J_+ | jm-1 \rangle = \sqrt{(J+m)(J-m+1)} \delta_{nm},
\label{46}
\end{equation}
and analogously
\begin{equation}
\langle jn | J_{-} | jm \rangle = -\sqrt{(J+m)(J-m+1)} \delta_{nm-1}~.
\label{47}
\end{equation}
Finally, from the definitions (29) for $J_{+},\ J_{-}$ one easily gets  
\begin{eqnarray}
\langle jm | J_x | jm-1 \rangle &=& {1 \over 2}\sqrt{(J+m)(J-m+1)} ,
\nonumber\\
\langle jm | J_y | jm-1 \rangle &=& {-i \over 2}\sqrt{(J+m)(J-m+1)}~.
\label{48}
\end{eqnarray}

\subsection*{Partial conclusions}
\begin{itemize}
\item[$\alpha$] \underline{{\it Properties of the eigenvalues of   
$\bf J$ and $J_z$}}\\
If $j(j+1)\hbar^2$ and $m\hbar$ are eigenvalues of $\bf J$ and $J_z$
associated to the eigenvectors $|kjm\rangle$, then $j$ and $m$ satisfy the
inequality 
\[
-j \leq m \leq j.
\]

\item[$\beta$] \underline{{\it Properties of the vector
$J_{-}|kjm\rangle$}}\\
Let $|kjm\rangle$ be an eigenvector of $\bf J^2$ and $J_{z}$ with the 
eigenvalues $j(j+1)\hbar^2$ and $m\hbar$
\begin{itemize}
\item{(i)}
If $m=-j$, then $J_{-}|kj-j\rangle=0$.
\item{(ii)}
If $m>-j$, then $J_{-}|kjm \rangle$ is a nonzero eigenvector of 
$J^2$ and $J_z$ with the eigenvalues $j(j+1)\hbar^2$ and $(m-1)\hbar$.
\end{itemize}

\item[$\gamma$] \underline{{\it Properties of the vector 
$J_+|kjm\rangle$}}\\
Let $|kjm\rangle$ be a (ket) eigenvector of $\bf J^2$ and $J_z$ for the 
eigenvalues $j(j+1)\hbar ^2$ and $m\hbar$
\begin{itemize}
\item[$\star$]
If $m=j$, then $J_+|kjm\rangle =0.$
\item[$\star$]
If $m<j$, then $J_+|kjm\rangle$ is a nonzero eigenvector of $\bf J^2$ 
and $J_z$ with the eigenvalues $j(j+1)\hbar^2$ and $(m+1)\hbar$
\end{itemize}

\item[$\delta$] \underline{{\it Consequences of the previous properties}}
\begin{eqnarray}
J_z|kjm\rangle &=& m\hbar|kjm\rangle,\nonumber\\
J_+|kjm\rangle &=& m\hbar\sqrt{j(j+1) - m(m+1)}|kjm+1\rangle, \nonumber\\
J_-|kjm\rangle &=& m\hbar\sqrt{j(j+1) - m(m-1)}|kjm+1\rangle. \nonumber
\end{eqnarray}
\end{itemize}

\section*{Applications of the orbital angular momentum}
Until now we have considered those properties of the angular momentum 
that could be derived only from the commutation relations. 
Let us go back to the orbital 
momentum $\bf l$ of a particle without intrinsic rotation  
and let us examine how one can apply the theory of the previous section in the 
important particular case
\begin{equation}
[\hat{l}_i, \hat{p}_j] = i\varepsilon_{ijk}\hat{p}_k.
\label{49}
\end{equation} 
First, $\hat{l}_z$ and $\hat{p}_j$ have a common system of eigenfunctions.
On the other hand, the Hamiltonian of a free particle
\[
\hat{H} = \left(\frac{\hat{\vec{p}}}{\sqrt{2m}}\right)^2,
\]
being the square of a vectorial operator has a complete system of 
eigenfunctions with $\hat{L^2}$ and $\hat{l}_z$. In addition, this implies 
that the free particle can be found in a state of well-defined $E$, $l$, and $m$.

\subsection*{Eigenvalues and eigenfunctions of $\bf l^2$ and 
$\bf l_{z}$}
It is more convenient to work in spherical coordinates because, as we will see,
various angular momentum operators act only on the angle variables
$\theta,\ \phi$ and not on $r$.
Thus, instead of describing $r$ by its cartesian components
$x,\ y,\ z$ we determine the arbitrary point $M$ of vector radius
$\bf r$ by the spherical 3D coordinates
\begin{equation}
x=r\cos\phi\sin\theta, \qquad  
y=r\sin\phi\sin\theta, \qquad
z=r\cos\theta,
\label{50}
\end{equation}
where
\[
r \geq 0, \qquad 
0 \leq \theta \leq \pi, \qquad 
0 \leq \phi   \leq 2\pi.
\]

Let $\Phi(r, \theta,\phi)$ and $\Phi'(r, \theta,\phi)$ be the wavefunctions
of a particle in $\Sigma$ and $\Sigma'$, respectively, in which
the infinitesimal rotation is given by ~$\delta\alpha$ around the ~$z$ axis
\begin{eqnarray}
\Phi'(r, \theta,\phi) &=& \Phi(r, \theta,\phi+\delta\alpha),\nonumber\\
&=& \Phi(r, \theta,\phi) + \delta\alpha\frac{\partial \Phi}{\partial\phi},
\label{51}
\end{eqnarray}
or
\begin{equation}
\Phi'(r, \theta,\phi) = (1+i\hat{l}_z\delta\alpha)\Phi(r, \theta,\phi).
\label{52}
\end{equation}

It follows that
\begin{equation}
\frac{\partial \Phi}{\partial \phi} = i\hat{l_z}\Phi, \qquad 
\hat{l}_z = -i{\partial \over \partial \phi}.
\label{53}
\end{equation}

For an inifinitesimal rotation in $x$
\begin{eqnarray}
\Phi'(r, \theta,\phi) &=& \Phi+\delta\alpha
\left(\frac{\partial \Phi}{\partial \theta}
      \frac{\partial \theta}{\partial \alpha} +
      \frac{\partial \Phi}{\partial \theta}
      \frac{\partial \phi}{\partial \alpha}
\right), \nonumber\\
&=& (1+i\hat{l}_x\delta\alpha)\Phi(r, \theta,\phi),
\label{54}
\end{eqnarray}
but in this rotation 
\begin{equation}
z' = z+y\delta\alpha; \qquad
z' = z+y\delta\alpha; \qquad x' = x
\label{55}
\end{equation}
and from (50) one gets
\begin{eqnarray}
r\cos(\theta + d\theta) &=& 
r\cos\theta + r\sin\theta\sin\phi\delta\alpha, \nonumber\\
r\sin\phi\sin(\theta + d\theta) &=& 
r\sin\theta\sin\phi + r\sin\theta\sin\phi -r\cos\theta\delta\alpha,
\label{56}
\end{eqnarray}
i.e.
\begin{equation}
\sin\theta d\theta = \sin\theta\sin\phi \, \delta\alpha \rightarrow
{d\theta \over d\alpha} = -\sin\phi,
\label{57}
\end{equation}
and
\begin{eqnarray}
\cos\theta\sin\phi \, d\theta + \sin\theta\cos\phi \, d\phi 
&=& -\cos\theta \, \delta\alpha,\nonumber\\
\cos\phi\sin\theta{d\phi \over d\alpha} &=& -\cos\theta - 
\cos\theta\sin\phi{d\theta \over d\alpha}~.
\label{58}
\end{eqnarray}
Substituting (57) in (56) leads to 
\begin{equation}
\frac{d\phi}{d\alpha} = -\cot\theta\cos\phi~.
\label{59}
\end{equation}
With (56) and (58)  substituted in (51) and comparing the right hand sides of
(51) one gets
\begin{equation}
\hat{l}_x = i\left(
\sin\phi {\partial \over \partial \theta} + 
\cot \theta\cos\phi {\partial \over \partial \phi}
\right).
\label{60}
\end{equation}

For the rotation in $y$, the result is similar
\begin{equation}
\hat{l}_y = i\left(
-\cos\phi {\partial \over \partial \theta} + 
\cot \theta\sin\phi {\partial \over \partial \phi}
\right).
\label{61}
\end{equation}

Using $\hat{l}_x, \ \hat{l}_y$ one can also obtain   
$\hat{l}_\pm, \ \hat{l}^2$
\begin{eqnarray}
\hat{l}_\pm &=& \exp\left[\pm i\phi
\left(
\pm{\partial \over \partial\theta} + i\cot\theta{\partial \over \partial\phi}
\right)\right], \nonumber\\
\hat{l}^2 &=& \hat{l}_{-}\hat{l}_{+} + \hat{l}^2 + \hat{l}_z, \nonumber\\
&=&-\left[
{1 \over \sin^2\theta}
{\partial^2 \over \partial \phi^2} +
{1 \over \sin\theta}
{\partial \over \partial \theta}
\bigg(
\sin\theta{\partial \over \partial\theta} 
\bigg)\right].
\label{62}
\end{eqnarray}
From (62) one can see that $\hat{l}^2$ is identical up to
a constant to the angular part of the Laplace operator at a fixed radius
\begin{equation}
\nabla^2 f = {1 \over r^2}{\partial \over \partial r}
\left( r^2{\partial f \over \partial r} \right) + 
{1 \over r^2}
\left[
{1 \over \sin\theta}
{\partial \over \partial\theta} 
\left(
\sin\theta{\partial f \over \partial\theta}
\right) + 
{1 \over \sin^2\theta}{\partial^2 \over \partial \phi^2}
\right].
\label{63}
\end{equation}

\subsection*{The eigenfunctions of $l_z$}
\begin{eqnarray}
\hat{l}_z\Phi_m = m\Phi = -i{\partial \Phi_m \over \partial \phi},
\nonumber\\
\Phi_m = {1 \over \sqrt{2\pi}}e^{im\phi}.
\label{64}
\end{eqnarray}

\subsection*{Hermiticity conditions of $\hat{l}_z$}
\begin{equation}
\int_0^{2\pi} f^*\hat{l}_zg\,d\phi = 
\left( \int_0^{2\pi} g^*\hat{l}_zf\,d\phi \right)^* +
f^*g(2\pi) - f^*g(0).
\label{65}
\end{equation}

It follows that $\hat{l}_z$ is hermitic in the class of functions for which
\begin{equation}
f^*g(2\pi) = f^*g(0).
\label{66}
\end{equation}

The eigenfunctions $\Phi_m$ of $\hat{l}_z$ belong to the integrable
class 
${\cal L}^2(0, 2\pi)$ and they fulfill (66), as it happens for any
function that can be expanded in $\Phi_m(\phi)$
\begin{eqnarray}
F(\phi) &=& \sum^k a_ke^{ik\phi}, \qquad k = 0, \pm 1, \pm 2, \, \ldots \, ,
\nonumber\\
G(\phi) &=& \sum^k b_ke^{ik\phi},\qquad  
k = \pm 1/2,\pm 3/2, \pm 5/2 \, \ldots \, ,
\label{67}
\end{eqnarray}
with $k$ only integers or half-integers, but not for combinations of 
$F(\phi)$ and $G(\phi)$.
The correct choice of $m$ is based on the common eigenfunctions of  
$\hat{l}_z$ and $\hat{l}^2$.

\subsection*{Spherical harmonics}
In the $\{\bf \vec{r}\}$ representation, the eigenfunctions associated
to the eigenvalues
$l(l+1)\hbar^2$ of $\bf l^2$ and $m\hbar$ 
of $l_z$ are solutions of the partial differential equations
\begin{eqnarray}
-\left({\partial^2 \over \partial \theta^2} + {1 \over \tan\theta}
{\partial \over \partial \theta} + {1 \over \sin^2\theta} 
{\partial^2 \over \partial \phi^2}\right)\psi(r, \theta, \phi) &=&
l(l+1)\hbar ^2\psi(r, \theta, \phi), \nonumber\\
-i{\partial \over \partial \phi}\psi(r, \theta, \phi) &=& 
m\hbar\psi(r, \theta, \phi). 
\label{68}
\end{eqnarray}

Taking into account that the general results presented above can be applied
to the orbital momentum,
we infer that $l$ can be an integer or half-integer and that, for fixed
$l$, $m$ can only take the values 
\[
-l, -l+1, \, \dots \, ,l-1, l.
\]

In (68), $r$ is not present in the differential operator, so that
it can be considered as a parameter. Thus, considering only
the dependence on $\theta,\ \phi$ of $\psi$, one uses the notation
$Y_{lm}(\theta, \phi)$ for these common eigenfunctions of 
$\bf l^2$ and $l_z$, corresponding to the eigenvalues 
$l(l+1)\hbar^2, m\hbar$. They are known as spherical harmonics
\begin{eqnarray}
{\bf l^2}Y_{lm}(\theta, \phi) &=& l(l+1)\hbar^2Y_{lm}(\theta,\phi), \nonumber\\
 l_z Y_{lm}(\theta, \phi)     &=& m\hbar Y_{lm}(\theta,\phi).
\label{69}
\end{eqnarray}

For more rigorousness, one should introduce one more index in order to 
distinguish among the various solutions of (69) corresponding to the same 
$(l,m)$ pairs for particles with spin. 
If the spin is not taken into account, these equations have a unique
solution (up to a constant factor) for each allowed pair of $(l,m)$; this is so
because the subindices $l,m$ are sufficient in this context.
The solutions $Y_{lm}(\theta,\, \phi)$ have been found by the method
of the separation of variables in spherical variables 
(see also the chapter {\em The hydrogen atom})
%
\begin{equation}
\psi_{lm}(r, \theta,\phi) = f(r)\psi_{lm}(\theta,\phi),
\label{70}
\end{equation}
where $f(r)$ is a function of $r$, which looks as an integration constant
from the viewpoint of the partial differential equations in (68). 
The fact that $f(r)$ is arbitrary proves that
$\bf L^2$ and $l_z$ do not form a complete set of observables\footnote{By 
definition, the hermitic operator 
A is an observable if the orthogonal system of eigenvectors
form a base in the space of states.} 
in the space 
$\varepsilon_r$\footnote{Each quantum state of a particle is characterized
by a vectorial state belonging to an abstract vectorial space
$\varepsilon_r$.} of functions of $\vec{r}$ ($r, \theta, \phi$).

In order to normalize $\psi_{lm}(r, \theta, \phi)$, it is convenient
to normalize $Y_{lm}(\theta, \phi) $ and $f(r)$ separately 
%
\begin{eqnarray}
\int_0^{2\pi}d\phi\int_0^{\pi}\sin\theta|\psi_{lm}(\theta, \phi)|^2d\theta 
&=& 1, 
\label{71} \\
\int_0^\infty r^2|f(r)|^2dr &=& 1.
\label{72}
\end{eqnarray}

\subsection*{The values of the pair $(l,m)$}

\noindent
($\alpha$):  {\it $l,m$ should be integers}\\
Using $l_z = {\hbar \over i}{\partial \over \partial \phi}$, we can write
(69) as follows 
\begin{equation}
{\hbar \over i}{\partial \over \partial \phi}Y_{lm}(\theta, \phi) = m\hbar 
Y_{lm}(\theta, \phi).
\label{73}
\end{equation}
Thus,  
\begin{equation}
Y_{lm}(\theta, \phi) = F_{lm}(\theta, \phi)e^{im\phi}.
\label{74}
\end{equation}

If $0\leq \phi < 2\pi$, then we should tackle the condition of 
covering all space according to the requirement of dealing with a function
continuous in any angular zone, i.e. 
c\A\
\begin{equation}
Y_{lm}(\theta, \phi=0) = Y_{lm}(\theta, \phi=2\pi),
\label{75}
\end{equation}
implying 
\begin{equation}
e^{im\pi} = 1.
\label{76}
\end{equation}

As has been seen, $m$ is either an integer or a half-integer; for the 
application to the orbital momentum, $m$ should be an integer. 
($e^{2im\pi}$
would be $-1$ if $m$ is a half-integer).

\noindent
($\beta$): For a given value of $l$, all the corresponding $Y_{lm}$
can be obtained by algebraic means using
\begin{equation}
l_+Y_{ll}(\theta, \phi)=0,
\label{77}
\end{equation}
which combined with eq.~(62) for $l_+$ leads to
\begin{equation}
\left({d \over d\theta} - l\cot\theta \right) F_{ll}(\theta) = 0.
\label{78}
\end{equation}
This equation can be immediately integrated if we notice the relationship
\begin{equation}
\cot\theta d\theta = \frac{d(\sin\theta)}{\sin\theta}~.
\label{79}
\end{equation}
Its general solution is

\begin{equation}
F_{ll}=c_l(\sin\theta)^l,
\label{80}
\end{equation}
where $c_l$ is a normalization constant.

It follows that for any positive or zero value of $l$, there is a function 
$Y_{ll}(\theta, \phi)$, which up to a constant factor is
\begin{equation}
Y^{ll}(\theta, \phi) = c_l(\sin\theta)^le^{il\phi}.
\label{81}
\end{equation}

Using repeatedly the action of $l_{-}$, one can build the whole set of functions
$Y_{ll-1}(\theta, \phi), \, \dots \, , Y_{l0}(\theta, \phi),$  
$\dots\, , Y_{l-l}(\theta, \phi)$. Next, we look at the way in which these 
functions can be put into correspondence with the eigenvalue pair
$l(l+1)\hbar, m\hbar $ (where $l$ is an arbitrary positive integer 
such that $l\leq m \leq l$ ). Using
(78), we can make the conclusion that any other eigenfunction 
$Y_{lm}(\theta, \phi)$ can be unambigously obtained from $Y_{ll}$. 

\subsection*{Properties of spherical harmonics}

\noindent
$\alpha\ $ {\it Iterative relationships}\\
From the general results of this chapter, we have 
\begin{equation}
l_\pm Y_{lm}(\theta,\phi)=\hbar\sqrt{l(l+1)-m(m\pm 1)}Y_{lm\pm1}(\theta, \phi).
\label{82}
\end{equation}
Using (62) for 
$l_\pm $ and the fact that $Y_{lm}(\theta,\phi)$
is the product of a $\theta$-dependent function and $e^{\pm i\phi}$, one gets
\begin{equation}
e^{\pm i\phi}\left(\frac{\partial}{\partial \theta} - m\cot\theta \right)
Y_{lm}(\theta,\phi) = \sqrt{l(l+1)-m(m\pm 1)}Y_{lm\pm 1}(\theta, \phi)
\label{83}
\end{equation}

\noindent
$\beta \ $ {\it Orthonormalization and completeness relationships}\\
Equation (68) determines the spherical harmonics only up to 
a constant factor. 
We shall now choose this factor such that to have 
the orthonormalization of  
$Y_{lm}(\theta, \phi)$ (as functions of the angular variables $\theta,\ \phi$)
\begin{equation}
\int^{2\pi}_0d\phi\int^\pi_0\sin\theta\,d\theta Y^*_{lm}(\theta, \phi)
Y_{lm}(\theta, \phi) = \delta_{l'l}\delta_{m'm}.
\label{84}
\end{equation}
In addition, any continuous function of 
$\theta,\ \phi$ can be expressed by means of the spherical harmonics as follows
\begin{equation}
f(\theta, \phi) = \sum^\infty_{l=0}\sum^l_{m= -l}c_{lm}Y_{lm}(\theta,\phi),
\label{85}
\end{equation}
where
\begin{equation}
c_{lm}=\int^{2\pi}_0d\phi\int^\pi_0\sin\theta\,d\theta\, Y^*_{lm}(\theta, \phi)
f(\theta, \phi).
\label{86}
\end{equation}

The results (85), (86) are consequences of defining
the spherical harmonics as an orthonormalized and complete
base in the space $\varepsilon_{\Omega}$ of functions of $\theta,\ \phi$. 
The completeness relationship is
\begin{eqnarray}
 \sum^\infty_{l=0}\sum^l_{m=l}Y_{lm}(\theta,\phi)Y^*_{lm}(\theta',\phi)
&=&\delta(\cos\theta-\cos\theta' )\delta(\phi, \phi),\nonumber\\
&=&\frac{1}{\sin\theta}\delta(\theta-\theta')\delta(\phi, \phi).
\label{87} 
\end{eqnarray}

\noindent
The `function' $\delta(\cos\theta-\cos\theta' )$ occurs 
because the integral over the variable
$\theta$ is performed by using the differential element 
$\sin\theta\,d\theta = -d(\cos\theta)$.

\subsection*{Parity operator $\cal P$ for spherical harmonics}
The behavior of ${\cal P}$ in 3D is rather close to that in 1D. 
When it is applied to a function of cartesian coordinates
$\psi(x,y,z)$ changes the sign of each of the coordinates
\begin{equation}
{\cal P}\psi(x,y,z) = \psi(-x,-y,-z).
\label{88}
\end{equation}
$\cal P$ has the properties of a hermitic operator, being also a unitary
operator, as well as a projector since
${\cal P}^2$ is an identity operator 
\begin{eqnarray}
\langle \psi(\bf{r}) | {\cal P} | \psi(\bf{r}) \rangle =
\langle \psi(\bf{r}) | \psi(\bf{-r}) \rangle  = 
\langle \psi(-\bf{r}) | \psi(\bf{r'}) \rangle , 
\nonumber\\
{\cal P}^2 |\bf{r}\rangle = {\cal P}({\cal P}|\bf{r}\rangle )  = 
\cal{P}|-\bf{r}\rangle  = |\bf{r}\rangle.
\label{89}
\end{eqnarray}
Therefore
\begin{equation}
{\cal P}^2 = \hat{1}, 
\label{90}
\end{equation}
for which the eigenvalues are $P=\pm1$. 
The eigenfunctions are called even if $P = 1$ and odd if
$P=-1$. In nonrelativistic quantum mechanics, the operator $\hat{H}$
for a conservative system is invariant with regard to discrete
unitary transformations, i.e.
\begin{equation}
{\cal P}\hat{H}{\cal P} = {\cal P}^{-1}\hat{H}{\cal P} = \hat{H}.
\label{91}
\end{equation}
Thus, $\hat{H}$ commutes with $\cal P$ and the parity of the state
is a constant of the motion. In addition, $\cal P$ commutes with the 
operators $\hat{l}$ and $\hat{l}_\pm$
\begin{equation}
[{\cal P}, \hat{l}_i] = 0, \qquad [{\cal P}, \hat{l}_\pm]=0.
\label{92}
\end{equation}
%
Because of all these properties, one can have the important class of wave 
functions which are common eigenfunctions of the triplet 
${\cal P},\ \hat{l}^2$ and $\hat{l}_z$. It follows from (92) that 
the parities of the states which difer only in $\hat{l}_z$ coincide. 
In this way, one can identify the parity 
of a particle of definite orbital angular momentum $\hat{l}$.

In spherical coordinates, we shall consider the following change of variables
\begin{equation}
r\rightarrow r, \qquad \theta \rightarrow \pi-\theta \qquad
\phi \rightarrow \pi+\phi.
\label{93}
\end{equation}

\noindent
Thus, using a standard base in the space of wavefunctions 
of a particle without `intrinsic rotation', the radial part of the base functions 
$\psi_{klm}(\vec{r})$ is not changed by the parity operator. 
Only the spherical harmonics will change.
From the trigonometric standpoint, the transformations (93) are as follows
\begin{equation}
\sin(\pi-\theta) \rightarrow \sin\theta, \qquad \cos(\pi-\theta) \rightarrow
-\cos\theta \qquad e^{im(\pi +\phi} \rightarrow (-1)^me^{im\phi}
\label{94}
\end{equation}
leading to the following transformation of the function $Y_{ll}(\theta, \phi)$ 
\begin{equation}
Y_{ll}(\phi - \theta, \pi + \phi) = (-1)^l Y_{ll}(\theta, \phi)~.
\label{95}
\end{equation}
From (95) it follows that the parity of $Y_{ll}$ is $(-1)^l$.
On the other hand, $l_{-}$ (as well as $l_{+}$ is invariant to the 
transformations 
\begin{equation}
{\partial \over \partial(\pi-\theta)} \rightarrow -{\partial 
\over \partial\theta},
\quad {\partial \over \partial(\pi+\phi)} 
\rightarrow {\partial \over \partial\phi}\quad
e^{i(\pi+\phi)}\rightarrow -e^{i\phi}\quad
{\rm cot}(\pi-\theta)\rightarrow -{\rm cot}\theta~.
\label{96}
\end{equation}
In other words, $l_{\pm}$ are even. Therefore, we infer that the parity of
any spherical harmonics is
$(-1)^l$, that is it is invariant under azimuthal changes  
\begin{equation}
Y_{lm}(\phi - \theta, \pi + \phi) = (-1)^l Y_{lm}(\theta, \phi).
\label{97}
\end{equation}
In conclusion, the spherical harmonics are functions of well-defined parity,
which is independent of $m$, even if $l$ is even and odd if $l$ is odd.

\section*{The spin operator}
Some particles have not only orbital angular momentum with regard to external
axes but also a  
{\it proper momentum \/}, which is known as {\it spin\/} denoted here by
$\hat{S}$. This operator is not related to normal rotation 
with respect to `real' axes in space, although it fulfills commutation relations 
of the same type as those of the orbital angular momentum, i.e.
\begin{equation}
[\hat{S}_i,\hat{S}_j]= i\varepsilon_{ijk}\hat{S}_k~,
\label{98}
\end{equation}
together with the following properties

\begin{itemize}

\item[(1).]
For the spin operator all the formulas of the orbital angular momentum
from (23) till (48) are satisfied. 

\item[(2).] 
The spectrum of the spin projections is a sequence of either integer or
half-integer numbers differing by unity.

\item[(3).]
The eigenvalues of $\hat{S}^2$ are the following 
\begin{equation}
\hat{S}^2\psi _{s}=S(S+1)\psi_s.
\label{99}
\end{equation}

\item[(4).]
For a given $S$, the components $S_z$ can take only $2S+1$ values, 
from $-S$ to $+S$.
 
\item[(5).]
Besides the usual dependence on $\vec{r}$ and/or
$\vec{p}$,
the eigenfunctions of the particles with spin depend also on a discrete variable,
(characteristic for the spin)~$\sigma$ denoting the projection of the spin on
the $z$ axis.

\item[(6).]
The wavefunctions $\psi(\vec{r}, \sigma)$ of a particle with spin can be 
expanded in eigenfunctions of given spin projection $S_z$, i.e.
\begin{equation}
\psi(\vec{r}, \sigma) = \sum_{\sigma = -S}^S \psi_\sigma(\vec{r})\chi(\sigma),
\label{100}
\end{equation}
where $\psi_\sigma(\vec{r})$ is the orbital part and $\chi(\sigma)$ is the 
spinorial part.

\item[(7).]
The spin functions (the spinors) $\chi(\sigma_i)$ are orhtogonal for any
pair $\sigma_i \ne \sigma_k$. The functions
$\psi_\sigma(\vec{r})\chi(\sigma)$ in the sum of (100) are the components
of a wavefunction of a particle with spin.

\item[(8).]
The function $\psi_\sigma(\vec{r})$ is called the orbital part of the spinor, or 
shortly orbital.

\item[(9)]
The normalization of the spinors is done as follows
\begin{equation}
\sum_{\sigma = -S}^S ||\psi_\sigma(\vec{r})|| = 1.
\label{101}
\end{equation}

\end{itemize}

The commutation relations allow to determine the explicit form of the spin
operators (spin matrices) acting in the space of the eigenfunctions of definite
spin projections.

Many `elementary' particles, such as the electron, the neutron, the proton, 
etc.\ have a spin of $1/2$ (in units of $\hbar$) and therefore the projection
of their spin takes only two values, 
($S_z = \pm 1/2$ (in $\hbar$ units), respectively. They belong to the 
fermion class because of their statistics when they form many-body systems.

On the other hand, the matrices $S_x,\ S_y,\ S_z$ in the space of 
$\hat{S}^2,\ \hat{S}_z$ are
\begin{eqnarray}
S_x = {1 \over 2}\left(\begin{array}{cc}
 0 & 1  \\
 1 & 0   
\end{array}\right), \qquad
&&S_y = {1 \over 2}\left(\begin{array}{cc}
 0 & -i  \\
 i &  0   
\end{array}\right), \nonumber\\
\nonumber\\
S_z = {1 \over 2}\left(\begin{array}{cc}
 1 &  0  \\
 0 & -1  
\end{array}\right), \qquad
&&S^2 = {3 \over 4}\left(\begin{array}{cc}
 1 &  0  \\
 0 &  1  
\end{array}\right).
\label{102}
\end{eqnarray}

\subsection*{Definition of the Pauli matrices}
The matrices
\begin{equation}
\sigma_i = 2S_i
\label{103}
\end{equation}
are called the Pauli matrices. They are hermitic and have the same 
characteristic  eq.
\begin{equation}
\lambda^2 - 1 = 0.
\label{104}
\end{equation}
Therefore, the eigenvalues of $\sigma_x,\ \sigma_y$ and  $\sigma_z$ 
are
\begin{equation}
\lambda = \pm 1.
\label{105}
\end{equation}
%
The algebra of these matrices is the following 
\begin{equation}
\sigma_i^2 = \hat{I}, \qquad  \sigma_k\sigma_j = -\sigma_j\sigma_k
= i\sigma_z,\qquad \sigma_j\sigma_k = i\sum_l \varepsilon_{jkl}\sigma_l. +
\delta_{jk}I~.
\label{106}
\end{equation}

\noindent
In the case in which the spin system has spherical symmetry 
\begin{equation}
\psi_1(r, +\textstyle{1\over 2}), \qquad
\psi_1(r, -\textstyle{1\over 2})~,
\label{107}
\end{equation}
are different solutions because of the different projections $S_z$.
The value of the probability of one or another projection is determined
by the square moduli $||\psi_{1}||^2$ or $||\psi_{2}||^2$, respectively, such
that
\begin{equation}
||\psi_1||^2 + ||\psi_2||^2 = 1.
\label{109}
\end{equation}
Since the eigenfunctions of $S_z$ have two components, then
\begin{equation}
\chi_1= \left(\begin{array}{c}  1  \\ 0  
\end{array}\right), \qquad
\chi_2= \left(\begin{array}{c}  0  \\ 1  
\end{array}\right),
\label{109}
\end{equation}
so that the eigenfunction of a spin one-half particle can be written
as a column matrix
\begin{equation}
\psi = \psi_1\chi_1 + \psi_2\chi_2 = 
\left(\begin{array}{c}  \psi_1 \\ \psi_2 \end{array}\right).
\label{110}
\end{equation}

\noindent
In the following, the orbitals will be replaced by numbers because we are
interested only in the spin part.

\section*{Transformations of spinors to rotations}
Let $\psi$ be the wavefunction of a spin system in $\Sigma$. 
We want to determine the probability of the spin projection 
in a arbitrary direction in 3D space, which one can always chose as the 
$z'$ of $\Sigma'$.
As we have already seen in the case of the angular momentum
there are two viewpoints in trying to solve this problem 

\begin{itemize}

\item[($\alpha$)]
$\psi$ does not change when $\Sigma \rightarrow \Sigma'$ and the operator
$\hat{\Lambda}$ transforms as a vector. We have to find the eigenfunctions
of the projections  
$S'_z$ and to expand $\psi$ in these eigenfunctions. 
The square moduli of the coefficients give the result
\begin{eqnarray}
\hat{S}_x' = \hat{S}_x\cos\phi + \hat{S}_y\sin\phi &=& 
e^{-il\phi}\hat{S}_x e^{il\phi},\nonumber\\
\hat{S}_y' = -\hat{S}_x\sin\phi + \hat{S}_y\cos\phi &=& 
e^{-il\phi}\hat{S}_y e^{il\phi},\nonumber\\
\hat{S}_z' = -\hat{S}_z = e^{il\phi}\hat{S}_z,
\label{111}
\end{eqnarray}
for infinitesimal rotations. Then, from the commutation relations for spin
one can find
\begin{equation}
\hat{l} =\hat{S}_z,
\label{112}
\end{equation}
where $\hat{l}$ is the infinitesimal generator.

\item[($\beta$)]
The second representation is:\\
\noindent
$\hat{S}$ does not change when $\Sigma \rightarrow \Sigma'$ and 
the components of $\psi$ does change.
The transformation to this representation can be performed through a unitary 
transformation of the form
\begin{eqnarray}
\hat{V}^\dagger\hat{S}'\hat{V} &=& \hat{\Lambda}, \nonumber\\
\left(\begin{array}{c}  \psi_1'  \\  \psi_2'  \end{array}\right) &=&
\hat{V}^\dagger
\left(\begin{array}{c}  \psi_1  \\  \psi_2  \end{array}\right)~. 
\label{113}
\end{eqnarray}
Using (111) and (113) one gets  
\begin{eqnarray}
\hat{V}^\dagger e^{-i\hat{S}_z\phi}\hat{S} e^{i\hat{S}_z\phi}
\hat{V}&=&\hat{S},\nonumber\\
\hat{V}^\dagger &=& e^{i\hat{S}_z\phi},
\label{114}
\end{eqnarray}
and from (114) we are led to
\begin{equation}
\left(\begin{array}{c}  \psi_1'  \\  \psi_2'  \end{array}\right) =
e^{i\hat{S}_z\phi}
\left(\begin{array}{c}  \psi_1  \\  \psi_2  \end{array}\right)~. 
\label{115}
\end{equation}
Using the explicit form of $\hat{S}_z$ and the properties of the Pauli matrices
one can find the explicit form of $\hat{V}^\dagger_z$, such that
\begin{equation}
\hat{V}^\dagger_z(\phi) = \left(\begin{array}{cc}  
e^{{i \over 2}\phi}  & 0 \\  
0 &  e^{{-i \over 2}\phi}  \end{array}\right).
\label{116}
\end{equation}

\end{itemize}

\section*{A result of Euler}
One can reach any reference frame $\Sigma'$ of arbitrary orientation with regard
to
$\Sigma$ through only three rotations; the first of angle $\phi$
around $z$, the next of angle $\theta$   
around $x'$ and the last of angle $\psi_a$
around $z'$, i.e. This important result belongs to Euler.
The parameters $(\varphi, \theta, \psi_a)$ are called Euler's angles. Thus
\begin{equation}
\hat{V}^\dagger(\varphi, \theta, \psi_a) = 
\hat{V}^\dagger_{z'}(\psi_a)\hat{V}^\dagger_{x'}(\theta)
\hat{V}^\dagger_{z}(\varphi).
\label{117}
\end{equation}

The matrices $\hat{V}^\dagger_z$ are of the form (116), whereas 
$\hat{V}^\dagger_x$ is of the form
\begin{equation}
\hat{V}^\dagger_x(\varphi) = \left(\begin{array}{cc}  
\cos{\theta \over 2}  & i\sin{\theta \over 2} \\  
i\sin{\theta \over 2} &  \cos{\theta \over 2}  
\end{array}\right),
\label{118}
\end{equation}
so that  
\begin{equation}
\hat{V}^\dagger(\varphi, \theta, \psi_a) = 
\left(\begin{array}{cc}  
e^{i{\varphi + \psi_a \over 2}}\cos{\theta \over 2}   & 
ie^{i{\psi_a - \varphi \over 2}}\sin{\theta \over 2}  \\  
ie^{i{\varphi - \psi_a \over 2}}\sin{\theta \over 2}  &
e^{-i{\varphi + \psi_a \over 2}}\cos{\theta \over 2} 
\end{array}\right).
\label{119}
\end{equation}

It comes out in this way that by the rotation of $\Sigma$, the components
of the spinorial function transforms as follows
\begin{eqnarray}
\psi'_1 &=& \psi_1 e^{i{\varphi + \psi_a \over 2}}\cos{\theta \over 2} +
       i\psi_2 e^{i{\psi_a - \varphi \over 2}}\sin{\theta \over 2},\nonumber\\ 
\psi'_2 &=&i\psi_1 e^{i{\varphi - \psi_a \over 2}}\sin{\theta \over 2} +
\psi_2 e^{-i{\varphi + \psi_a \over 2}}\cos{\theta \over 2}.
\label{120}
\end{eqnarray}
From (120) one can infer that there is a one-to-one
mapping between any rotation in $E_3$ and a linear transformation of
$E_2$, the two-dimensional Euclidean space. This mapping is related to the 
two components of the spinorial wavefunction. 
The rotation in
$E_3$ does not imply a rotation in $E_2$, which means that
\begin{equation}
\langle \Phi' | \psi' \rangle = 
\langle \Phi | \psi \rangle = 
\Phi^*_1\psi_1 + \Phi^*_2\psi_2.
\label{121}
\end{equation}

From (119) one finds that (121) does not hold; nevertheless there is an 
invariance in the 
transformations (119) in the space $E_2$ of spinorial wavefunctions
\begin{equation}
\{\Phi | \psi \}= 
\psi_1\Phi_2 - \psi_2\Phi_1.
\label{122}
\end{equation} 

The linear transformations that preserve invariant bilinear forms invariant are 
called binary transformations.

A physical quantity with two components for which a rotation of the coordinate
system is a binary transformation is know as  
{\it a spin of first order} or shortly {\it spin}.

\subsection*{ The spinors of a system of two fermions}
The eigenfunctions of $_i\hat{s}^2\ _i\hat{s}_z$, with $i = 1,2$ have the 
following form
\begin{equation}
i| + \rangle = \left(\begin{array}{c}
1 \\ 0
\end{array}\right)_i, \qquad 
i| - \rangle = \left(\begin{array}{c}
0 \\ 1
\end{array}\right)_i.
\label{123}
\end{equation}

A very used operator in a two-fermion system is the total spin 
\begin{equation}
\hat{S} = _1\hat{S} + _2\hat{S}~.
\label{124}
\end{equation}

The spinors of $\hat{s}^2\ \hat{s}_z $
are kets $|\hat{S}, \sigma  \rangle$, which are linear combinations 
of $_i\hat{s}^2\ _i\hat{s}_z$
\begin{eqnarray}
| + +  \rangle = 
\left(\begin{array}{c} 1 \\ 0 \end{array}\right)_1
\left(\begin{array}{c} 1 \\ 0 \end{array}\right)_1, && \qquad 
| + -  \rangle = 
\left(\begin{array}{c} 1 \\ 0 \end{array}\right)_1
\left(\begin{array}{c} 0 \\ 1 \end{array}\right)_2, \nonumber\\
| - +  \rangle = 
\left(\begin{array}{c} 0 \\ 1 \end{array}\right)_2
\left(\begin{array}{c} 1 \\ 0 \end{array}\right)_1, && \qquad 
| - -  \rangle = 
\left(\begin{array}{c} 0 \\ 1 \end{array}\right)_2
\left(\begin{array}{c} 0 \\ 1 \end{array}\right)_2.
\label{125}
\end{eqnarray}

The spinorial functions in (125) are assumed orthonormalized. 
In $E_n$ the ket 
$| ++ \rangle$ has $S_z = 1$ and at the same time it is an eigenfunction
of the operator 
\begin{equation}
\hat{S} = _1\hat{s}^2 + 2(_1\hat{s})(_2\hat{s})+_2\hat{s}^2,
\label{126}
\end{equation}
as one can see from 
\begin{eqnarray}
\label{127}
\hspace*{-30pt}
\hat{S}^2 &=& | + +  \rangle = \textstyle{3 \over 2} | + +  \rangle +
2(_1\hat{s}_x \cdot _2\hat{s}_x + _1\hat{s}_y \cdot _2\hat{s}_y +
  _1\hat{s}_z \cdot _2\hat{s}_z)| + +  \rangle , \hspace*{-30pt}\\
\label{128}\hspace*{-30pt}
\hat{S}^2 &=& | + +  \rangle = 2 | + +  \rangle  =1(1+1) | + +  \rangle .
\hspace*{-30pt}
\end{eqnarray} 

If we introduce the operator  
\begin{equation}
\hat{S}_{-} = _1\hat{s}_{-} + _2\hat{s}_{-},
\label{129}
\end{equation}
one gets
\begin{equation}
[\hat{S}_{-} ,\hat{S}^2] = 0.
\label{130}
\end{equation}
Then $(\hat{S}_{-})^k|1,1\rangle$ can be written in terms of the 
eigenfunctions of the operator $\hat{S}^2$, i.e.
\begin{equation}
\hat{S}_{-}|1,1\rangle = \hat{S}_{-}|+ +\rangle =
\sqrt{2}|+ -\rangle + \sqrt{2}|- +\rangle.
\label{131}
\end{equation}
Thus, $S_z = 0$ in the state $\hat{S}_{-}|1,1\rangle $. 
On the other hand, from the normalization condition, we have
\begin{eqnarray}
|1,0\rangle  = \textstyle{1\over\sqrt{2}}(|+ -\rangle + |- +\rangle)\\
\label{132}
\hat{S}_{-}|1,0\rangle =|- -\rangle + |- -\rangle = \alpha|1, -1\rangle.
\label{133} 
\end{eqnarray}

In addition, the normalization condition gives 
\begin{equation}
|1, -1\rangle  = |-, -\rangle. 
\label{134}
\end{equation}

There is only one other linear-independent combination 
of functions of the type (125), which is different of
$|1, 1\rangle,\ |1, 0\rangle$ and $|1, -1\rangle$, which is
\begin{eqnarray}
\psi_4 = \textstyle{1\over\sqrt{2}}(|+ -\rangle - |- +\rangle), \label{135}\\
\hat{S}_z \psi_4 = 0, \qquad \hat{S}^2 \psi_4.
\label{136}
\end{eqnarray}
Therefore
\begin{equation}
\psi_4 =|0, 0\rangle.
 \label{137}
\end{equation}
$\psi_4$ describes the state of a system of two fermions
having the total spin equal to zero. 
The latter type of state is called {\it singlet\/}. On the other hand,
the state of two fermions of total spin one can be called {\it triplet\/}
having a degree of degeneration $g=3$.

\section*{Total angular momentum}
The total angular momentum is an operator defined as the sum
of the angular and spin momenta, i.e.
\begin{equation}
\hat{J} = \hat{l}+\hat{S},
\label{138}
\end{equation}
where $\hat{l}$ and $\hat{S}$, as we have seen, 
act in different spaces,
though the square of $\hat{l}$ and $\hat{S}$ commute with
$\hat{J}$
\begin{equation}
[\hat{J}_i, \hat{J}_j] = i\varepsilon_{ijk}\hat{J}_k, \qquad
[\hat{J}_i, \hat{l}^2] = 0, \qquad [\hat{J}_i, \hat{S}^2] = 0,
\label{139}
\end{equation}
From (139) one finds that $\hat{l}^2$ and $\hat{S}^2$ have a common
eigenfunction system with $\hat{J}^2$ and $\hat{J}_z$.

Let us determine the spectrum of the projections $\hat{J}_z$ for a fermion.
The state of maximum projection $\hat{J}_z$ can be written 
\begin{eqnarray}
\bar{\psi} &=& \psi_{ll}
\left(\begin{array}{c} 1 \\ 0 \end{array}\right) =
 |l,l,+ \rangle \\
\label{140}
\hat{\jmath}_z\psi &=& (l + \textstyle{1 \over 2})\bar{\psi}, \rightarrow
j= l + \textstyle{1 \over 2}.
\label{141}
\end{eqnarray}

We introduce the operator $\hat{J}_{-}$ defined as 
\begin{equation}
\hat{J}_{-}=\hat{l}_{-}+\hat{S}_{-} = \hat{l}_{-}+
\left(\begin{array}{cc} 0 & 0 \\ 1 & 0 \end{array}\right).
\label{142}
\end{equation}

On account of the normalization $\alpha = \sqrt{(J+M)(J-M+1)}$, one gets
\begin{equation}
\hat{J}_{-}\psi_{ll}\left(\begin{array}{c} 1 \\ 0 \end{array}\right) =
\sqrt{2l}|l,l-1, +\rangle + |l,l-1, -\rangle,
\label{143}
\end{equation}
so that the value of the projection of $\hat{j}_{-}$ in 
$\hat{j}_{-}\bar{\psi}$ will be
\begin{equation}
\hat{\jmath}_z = (l-1) + \textstyle{1 \over 2} = l - \textstyle{1 \over 2}~.
\label{144}
\end{equation}
It follows that $\hat{\jmath}_{-}$ lowers by one unit the action of 
$\hat{J}_z$.

In the general case we have
\begin{equation}
\hat{\jmath}_{-}^k = \hat{l}_{-}^k + k\hat{l}_{-}^{k-1}\hat{S}_{-}~.
\label{145}
\end{equation}
One can see that (145) is obtained from the binomial expansion 
considering that 
$\hat{s}^2_{-}$  and all higher-order powers of $\hat{s}$ are zero.
\begin{equation}
\hat{\jmath}_{-}^k |l,l,+\rangle = \hat{l}_{-}^k |l,l,+\rangle + 
k\hat{l}_{-}^{k-1} |l,l,-\rangle.
\label{146}
\end{equation}

Using 
\[
(\hat{l}_{-})^k\psi_{l,l} =
\textstyle{\sqrt{\frac{k!(2l)!}{(2l-k)!}}\psi_{l,l-k}}
\]
we get
\begin{equation}
\hat{\jmath}_{-}^k | l,l,+\rangle = 
\textstyle{\sqrt{\frac{k!(2l)!}{(2l-k)!}}}| l,l-k,+\rangle +
\textstyle{\sqrt{\frac{(k+1)!(2l)!}{(2l-k+1)!}}} k| l,l-k+1,-\rangle~.
\label{147}
\end{equation}
Now noticing that $m = l-k$
\begin{equation}
\hat{\jmath}_{-}^{l-m} | l,l,+\rangle = 
\textstyle{\sqrt{\frac{(l-m)!(2l)!}{(l+m)!}}}| l,m,+\rangle +
\textstyle{\sqrt{\frac{(l-m-1)!(2l)!}{(2l+m+1)!}}} (l-m)| l,m+1,-\rangle.
\label{148}
\end{equation}

\noindent
The eigenvalues of the projections of the total angular momentum
are given by the sequence of numbers differing by one unit from
$j=l+{1\over 2}$ p\h n\A\ to $j=l-{1\over 2}$.
All these states belong to the same eigenfunction of $\hat{J}$
as $|l,l,+\rangle$ because $[\hat{J}_{-},\hat{J}^2]=0$:
\begin{eqnarray}
\hat{J}^2|l,l,+\rangle &=& 
(\hat{l}^2 + 2\hat{l}\hat{S} + \hat{S}^2)|l,l,+\rangle, \nonumber\\ &=& 
[l(l+1) + 2l\textstyle{1\over 2} + \textstyle{3\over 4}]|l,l,+\rangle
\label{149}
\end{eqnarray}
where $j(j+1) = (l+{1 \over 2})(l + {3 \over 2})$.

In the left hand side of (149) a contribution different of zero gives only 
$j=\hat{l}_z\hat{S}_z$. Thus, the obtained eigenfunctions correspond
to the pair 
$j=l+{1 \over 2}$, $m_j=m+{1 \over 2}$; they are of the form
\begin{equation} 
 |l+{1 \over 2}, m+{1 \over 2} \rangle = 
\sqrt{l+m+1 \over 2l+1}|l, m, + \rangle +
\sqrt{l-m \over 2l+1}|l, m+1, - \rangle.
\label{150}
\end{equation}

The total number of linearly independent states is
\begin{equation}
N=(2l+1)(2s + 1) = 4l+2,
\label{151}
\end{equation}
of which in (150) only (2j+1)=2l+3 have been built. The rest of
$2l-1$ eigenfunctions can be obtained from the 
orthonormalization condition:
\begin{equation}
|l-\textstyle{1 \over 2}, m-\textstyle{1 \over 2} \rangle = 
\sqrt{l-m \over  2l+1} | l, m, + \rangle -
\sqrt{l+m+1 \over  2l+1} | l, m+1, - \rangle.
\label{152}
\end{equation}

If two subsystems are in interaction in such a way that each of the angular
momenta $\hat{j}_i$ is conserved, then the eigenfunctions of the total angular
momentum
\begin{equation}
\hat{J} =\hat{\jmath}_1 + \hat{\jmath}_2,
\label{153}
\end{equation}
can be obtained by a procedure similar to the previous one. 
For fixed eigenvalues of  
$\hat{\jmath}_1$ and  $\hat{\jmath}_2$ there are 
$(2j_1+1)(2j_2+1)$ orthonormalized eigenfunctions of the projection
of the total angular momentum $\hat{J}_z$; the one corresponding to the 
maximum value of the projection $\hat{J}_z$, i.e.  
$M_J = j_1 + j_2$,
can be built in a unique way and therefore $J = j_1 + j_2$ is the maximum value
of the total angular momentum of the system. Applying the 
operator
$\hat{J} = \hat{\jmath}_1 + \hat{\jmath}_2$ repeatingly to the function
\begin{equation}
|j_1 + j_2, j_1 + j_2, j_1 + j_2 \rangle = |j_1 , j_1\rangle \cdot
|j_2 , j_2\rangle,
\label{154}
\end{equation}
one can obtain all the $2(j_1 + j_2) + 1$ eigenfunctions of 
$\hat{J} = j_1 + j_2$ with different $M$s: 
\[
-(j_1 + j_2) \leq M \leq (j_1 + j_2).
\]
For example, the eigenfunction of $M = j_1 + j_2-1$ is
\begin{equation}
|j_1 + j_2, j_1 + j_2-1, j_1 , j_2 \rangle =
\sqrt{j_1 \over j_1 + j_2}|j_1,j_1-1, j_2, j_2 \rangle +
\sqrt{j_2 \over j_1 + j_2}|j_1,j_1, j_2, j_2-1 \rangle .
\label{155}
\end{equation}

Applying iteratively the operator $\hat{J}_{-}$, all the
$2(j_1 + j_2-1) -1$ eigenfunctions of $J = j_1 + j_2-1$ can be obtained.

One can prove that 
\[
|j_1 - j_2| \leq J \leq j_1 + j_2~,
\]
so that
\begin{equation}
\sum_{{\rm min} \ J}^{{\rm max} \ J}(2J +1) = (2J_1 +1)(2J_2 +1)~.
\label{156}
\end{equation}
Thus
\begin{equation}
|J,M,j_1,j_2\rangle = 
\sum_{m_1+m_2 = M} (j_{1} m_{1}j_{2}m_{2}|J M) | j_1, m_1, j_2, m_2 \rangle~,
\label{156}
\end{equation}
where the coefficients $(j_{1} m_{1}j_{2}m_{2}|J M)$ determine the contribution
of the various kets $| j_1, m_1, j_2, m_2 \rangle$ to the eigenfunctions 
of $\hat{J^2}$, $\hat{J_{z}}$ having the eigenvalues $J(J+1)$, $M$. They are 
called Clebsch-Gordan coefficients.

\bigskip
\newpage

\noindent
\underline{References}: 

\noindent

\noindent
1. H.A. Buchdahl, ``Remark concerning the eigenvalues of orbital angular
momentum",

\noindent
Am. J. Phys. {\bf 30}, 829-831 (1962)

\bigskip

\noindent
{\bf 3N. Note}:
 1. The operator corresponding to the Runge-Lenz vector of the
classical Kepler problem is written as 
$$
\hat{\vec{A}}=\frac{{\bf \hat{r}}}{r}+\frac{1}{2}\Bigg[(\hat{l}\times \hat{p})-
(\hat{p}\times \hat{l})\Bigg]~,
$$
where atomic units have been used and the case $Z=1$ (hydrogen atom) was assumed.
This operator commutes with the Hamiltonian of the atomic hydrogen
$\hat{H}=\frac{\hat{p^2}}{2}-\frac{1}{r}$, that is it is an integral of
the atomic quantum motion. Its components have commutators 
of the type $[A_i,A_j]=-2i\epsilon _{ijk}l_{k}\cdot H$; the 
commutators of the Runge-Lenz components with the components of the angular
momentum are of the type
$[l_i,A_j]=i\epsilon _{ijk}A_k$. Thus, they respect the conditions (23).
Proving that can be a useful exercise.

\newpage
\section*{{\huge 3P. Problems}}
\subsection*{Problem 3.1}
Show that any translation operator, for which 
$\psi (y+a) = T_{a}\psi(y)$, can be written as an exponential 
operator. Apply the result for 
$y=\vec{r}$ and for a the finite rotation $\alpha$ around $z$.\\
\noindent
{\bf Solution}\\
The proof can be obtained expanding $\psi(y+a))$ in Taylor series
in the infinitesimal neighborhood around $x$, that is in powers of $a$
$$
\psi(y+a)=\sum _{n=0}^{\infty}\frac{a^{n}}{n!}\frac{d^{n}}{dx^{n}}\psi(x)
$$
We notice that
$$
\sum _{n=0}^{\infty}\frac{a^{n}\frac{d^{n}}{dx^{n}}}{n!}=e^{a\frac{d}{dx}}
$$
and therefore one has $T_{a}=e^{a\frac{d}{dx}}$ in the 1D case. In 3D, 
$y=\vec{r}$ and $a\rightarrow \vec{a}$. The result is 
$T_{\vec{a}}=e^{\vec{a}\cdot 
\vec{\nabla}}
$.

For the finite rotation $\alpha$ around $z$ we has $y=\phi$ and
$a=\alpha$. It follows
$$
T_{\alpha}=R_{\alpha}=e^{\alpha \frac{d}{d\phi}}~.
$$

Another exponential form of the rotation around $z$ is that in terms of the 
angular momentum operator as was already commented in this chapter.
Let $x' = x+dx$ and consider only the first order of the Taylor series 
\begin{eqnarray}
\psi(x',y',z') &=& \psi(x,y,z) + (x'-x)\frac{\partial}{\partial x'}
\psi(x',y',z')\bigg|_{\vec{r'} = \vec{r}}  \nonumber \\
&& + (y'-y)\frac{\partial}{\partial y'} \psi(x',y',z')
\bigg|_{\vec{r'} = \vec{r}} \nonumber \\ 
&& + (z'-z)\frac{\partial}{\partial z'} \psi(x',y',z')
\bigg|_{\vec{r'} = \vec{r}}\, . \nonumber 
\end{eqnarray}

Taking into account 
\begin{eqnarray}
\frac{\partial}{\partial x'_i}\psi(\vec{r}')\bigg|_{\vec{r}'} &=& 
\frac{\partial}{\partial x_i}\psi(\vec{r}), \nonumber\\ 
x' = x -y d \phi, \qquad y' &=& y + xd \phi, \qquad z' = z, \nonumber
\end{eqnarray}
one can reduce the series from three to two dimensions  
\begin{eqnarray}
\psi(\vec{r}') &=& \psi(\vec{r}) + 
  (x-yd\phi-x)\frac{\partial\psi(\vec{r})}{\partial x} 
+ (y+xd\phi -y)\frac{\partial\psi(\vec{r})}{\partial y'}, \nonumber\\
&=& \psi(\vec{r})-y d\phi \frac{\partial\psi(\vec{r})}{\partial x}  
	         +x d\phi x\frac{\partial\psi(\vec{r})}{\partial y},\nonumber\\ 
&=& \left[1 - d\phi\left(-x\frac{\partial}{\partial y} 
            + y\frac{\partial}{\partial x}\right)\right]\psi(\vec{r})~.\nonumber
\end{eqnarray}
Since $i\hat{l}_z = 
 \left(x\frac{\partial}{\partial y} - y\frac{\partial}{\partial x}\right)$
it follows that
$
R = \left[1 - d\phi\left(x\frac{\partial}{\partial y} 
            - y\frac{\partial}{\partial x}\right)\right]~.
$
In the second order one can get 
$\textstyle{1 \over 2!}(i\hat{l}_zd\phi)^2$, and so forth.
Thus, $R$ can be written as an exponential 
\[
R=e^{i\hat{l}_zd\phi}.
\]

\subsection*{Problem 3.2}
Based on the expressions given in (14) show that one can get (15). \\

\bigskip

\noindent
{\bf{Solution}}\\
Let us consider only linear terms in the Taylor expansion 
(infinitesimal rotations)
\[
e^{i\hat{l}_zd\phi} = 1 + i\hat{l}_zd\phi + 
\textstyle{1 \over 2!}(i\hat{l}_zd\phi)^2 + \ldots\, , \nonumber
\]
so that
\begin{eqnarray}
(1 + i\hat{l}_zd\phi)\hat{A}_x(1-i\hat{l}_zd\phi) &=& 
\hat{A}_x -\hat{A}_xd\phi,\nonumber\\
(\hat{A}_x + i\hat{l}_zd\phi\hat{A}_x)(1-i\hat{l}_zd\phi) &=& 
\hat{A}_x -\hat{A}_xd\phi,\nonumber\\
\hat{A}_x -\hat{A}_xi\hat{l}_zd\phi+ i\hat{l}_zd\phi\hat{A}_x +
\hat{l}_zd\phi\hat{A}_x \hat{l}_zd\phi &=&\hat{A}_x -\hat{A}_xd\phi,\nonumber\\
i(\hat{l}_z\hat{A}_x -\hat{A}_x\hat{l}_z)d\phi &=&-\hat{A}_yd\phi.\nonumber
\end{eqnarray} 
We easily arrive at the conclusion
\[
[ \hat{l}_z, \hat{A}_x]  = i\hat{A}_y~. \nonumber
\]
In addition, $[ \hat{l}_z, \hat{A}_y]  = i\hat{A}_x$ can be obtained from
\begin{eqnarray}
(1 + i\hat{l}_zd\phi)\hat{A}_y(1-i\hat{l}_zd\phi) &=& 
\hat{A}_xd\phi -\hat{A}_y,\nonumber\\
(\hat{A}_y + i\hat{l}_zd\phi\hat{A}_y)(1-i\hat{l}_zd\phi) &=& 
\hat{A}_xd\phi -\hat{A}_y,\nonumber\\
\hat{A}_y -\hat{A}_yi\hat{l}_zd\phi+ i\hat{l}_zd\phi\hat{A}_y +
\hat{l}_zd\phi\hat{A}_y \hat{l}_zd\phi &=&\hat{A}_xd\phi -\hat{A}_y,\nonumber\\
i(\hat{l}_z\hat{A}_y -\hat{A}_y\hat{l}_z)d\phi &=&-\hat{A}_xd\phi.\nonumber
\end{eqnarray} 

\subsection*{Problem 3.3}
Determine the operator $\frac{d\hat{\sigma}_{x}}{dt}$ based on the  
Hamiltonian of an electron with spin in a magnetic field of induction $\vec{B}$.

\bigskip

\noindent
{\bf Solution}\\
The Hamiltonian in this case is 
$\hat{H}(\hat{{\bf p}},\hat{{\bf r}},\hat{{\bf \sigma}})=
\hat{H}(\hat{{\bf p}},\hat{{\bf r}})+\hat{{\bf \sigma}}\cdot \vec{{\bf B}}$, 
where the latter term is the Zeeman Hamiltonian of the electron.
Since $\hat{\sigma}_{x}$ commutes with the momenta and the 
coordinates, applying the Heisenberg equation of motion leads to
$$  
\frac{d\hat{\sigma}_{x}}{dt}=\frac{i}{\hbar}[\hat{H},\hat{\sigma}_{x}]=
-\frac{i}{\hbar}\frac{e\hbar}{2m_e}((\hat{\sigma}_{y}B_y+\hat{\sigma}_{z}B_z)
\hat{\sigma}_{x}-\hat{\sigma}_{x}(\hat{\sigma}_{y}B_y+
\hat{\sigma}_{z}B_z))~.
$$
Using $[\sigma _{x},\sigma _y]=i\sigma _z$, one gets :
$$
\frac{d\hat{\sigma}_{x}}{dt}=
\frac{e}{m_e}(\hat{\sigma}_{y}B_{z}-\hat{\sigma}_{z}B_{y})
=\frac{e}{m_e}(\vec{\sigma}\times \vec{B})_{x}~.
$$



%

%


\newpage
\begin{center}
{\huge{4. THE WKB METHOD}}
\end{center}
\setcounter{equation}{0}
\hspace{0.6cm} In order to study more realistic potentials with regard to  
{\em rectangular barriers and wells}, it is necessary to employ approximate
methods allowing to solve the Schr\"odinger equation for more general
classes of potentials and at the same time to give very good approximations
of the exact solutions.  

The aim of the various approximative methods is to offer solutions of acceptable
precision and simplicity that can be used for understanding the behaviour 
of the system in quasianalytic terms.

Within quantum mechanics, one of the oldest and efficient approximate method 
for getting rather good 
Schr\"odinger solutions was developed almost simulataneously
by  
{\em G. Wentzel, H. A. Kramers and L.
Bri\-llouin} in 1926, hence the acronym  
{\em WKB} under which this method is known (or {\em JWKB} as is more correctly 
used by many authors, see note 4N).

It is worth mentioning that the WKB method applies to 1D 
Schr\"odinger equations and that there are serious difficulties 
when trying to generalize it to more dimensions.

In order to solve the Schr\"odinger equation
\begin{equation}
-\frac{\hbar^2}{2m}\frac{d^2\psi}{dy^2}+u(y)\psi=E\psi
\end{equation}
with a potential of the form 
\begin{equation}
u(y)=u_0f\Big(\frac{y}{a}\Big)~,
\end{equation}

\noindent
we first perform the changes of notations and of variable 
\begin{equation}
\xi^2=\frac{\hbar^2}{2mu_0a^2}
\end{equation}
\begin{equation}
\eta=\frac{E}{u_0}
\end{equation}
\begin{equation}
x=\frac{y}{a}~.
\end{equation}
From eq. $(5)$ we get
\begin{equation}
\frac{d}{dx}=
\frac{dy}{dx}\frac{d}{dy}=
a\frac{d}{dy}
\end{equation}
\begin{equation}
\frac{d^2}{dx^2}
=\frac{d}{dx}\Big(a\frac{d}{dy}\Big)
=\Big(a\frac{d}{dx}\Big)\Big(a\frac{d}{dx}
\Big)=a^2\frac{d^2}{dy^2}
\end{equation}
and the Schr\"odinger eq. reads 
\begin{equation}
-\xi^2\frac{d^2\psi}{dx^2}+f(x)\psi=\eta\psi~.
\end{equation}
Multiplying by $-1/\xi^2$ and defining $r(x)=\eta-f(x)$, it is possible to write 
it as folows
\begin{equation}
\frac{d^2\psi}{dx^2}+\frac{1}{\xi^2}r(x)\psi=0~.
\end{equation}
To solve (9), the following form of the solution is proposed
\begin{equation}
\psi(x)=\exp\Bigg[\frac{i}{\xi}\int_a^x{q(x)dx}\Bigg]~.
\end{equation}


Therefore
$$
\frac{d^2{\psi}}{dx^2}
=\frac{d}{dx}\bigg(\frac{d\psi}{x}\bigg)
=\frac{d}{dx}\Bigg\{\frac{i}{\xi}q(x)
\exp\Bigg[{\frac{i}{\xi}\int_a^xq(x)dx\Bigg]}\Bigg\}
$$
$$
\Longrightarrow\frac{d^2\psi}{dx^2}
=\frac{i}{\xi}\Bigg\{\frac{i}{\xi}q^2(x)\exp\Bigg[\frac{i}{\xi}
\int_a^xq(x)dx\Bigg]+\frac{\partial{q(x)}}{\partial{x}}
\exp\Bigg[\frac{i}{\xi}\int_a^xq(x)dx\Bigg]\Bigg\}~.
$$
Factorizing $\psi$, we have
\begin{equation}
\frac{d^2\psi}{dx^2}=\Bigg[-\frac{1}
{\xi^2}q^2(x)+\frac{i}{\xi}\frac{dq(x)}{dx}\Bigg]\psi~.
\end{equation}

\noindent
Discarding for the time being the dependence of $x$, the Schr\"odinger eq. 
can be written
\begin{equation}
\Bigg[-\frac{1}{\xi^2}q^2+\frac{i}{\xi}\frac{\partial{q}}{\partial{x}}
+\frac{1}{\xi^2}r\Bigg]\psi=0
\end{equation}

and since in general $\psi\neq0$, we get:
\begin{equation}
i\xi\frac{dq}{dx}+r-q^2=0~,
\end{equation}
which is a nonlinear differential eq. of the Riccati type whose solutions are
sought in the form of expansions in powers of 
$\xi$ under the assumption that $\xi$ is very small. 

More precisely, the series is taken of the form
\begin{equation}
q(x)=\sum^\infty_{n=0}(-i\xi)^nq_n(x)~.
\end{equation}
Plugging it into the Riccati eq., we get
\begin{equation}
i\xi\sum_{n=0}^\infty(-i\xi)^n\frac{dq_n}{dx}+r(x)-
\sum_{\mu=0}^\infty(-i\xi)^{\mu}q_{\mu}
\sum_{\nu=0}^\infty(-i\xi)^{\nu}q_{\nu}=0~.
\end{equation}
By a rearrangement of the terms one is led to
\begin{equation}
\sum_{n=0}^\infty(-1)^n(i\xi)^{n+1}\frac{dq_n}{dx}+r(x)-
\sum_{\mu=0}^\infty\sum_{\nu=0}^\infty(-i\xi)^{\mu+\nu}q_{\mu}q_{\nu}=0~.
\end{equation}
Double series have the following important property
\begin{displaymath}
\sum_{\mu=0}^\infty\sum_{\nu=0}^\infty{a_{\mu\nu}}=\sum_{n=0}^\infty
\sum_{m=0}^n{a_{m,n-m}}~,
\end{displaymath}
where $\mu=n-m\quad ,\nu=m$~.\\\\
Thus
\begin{equation}
\sum_{n=0}^\infty(-1)^n(i\xi)^{n+1}\frac{dq_n}{dx}+r(x)-
\sum_{n=0}^\infty\sum_{m=0}^n(-i\xi)^{n-m+m}q_{m}q_{n-m}=0~.
\end{equation}

Let us see explicitly the first several terms in each of the series in
eq. (17): 
\begin{equation}
\sum_{n=0}^\infty(-1)^n(i\xi)^{n+1}\frac{dq_n}{dx}=i\xi
\frac{dq_0}{dx}+\xi^2\frac{dq_1}{dx}-i\xi^3\frac{dq_2}{dx}+\dots
\end{equation}
\begin{equation}
\sum_{n=0}^\infty\sum_{m=0}^n(-i\xi)^{n}q_{m}q_{n-m}=q^2_0-i2{\xi}q_0q_1+\dots
\end{equation}
Asking that the first terms in both series contain $i\xi$, one should write them
as
$$
\sum_{n=1}^\infty(-1)^{n-1}(i\xi)^n\frac{dq_{n-1}}{dx}+r(x)-q_0^2-
\sum_{n=1}^\infty\sum_{m=0}^n(-i\xi)^nq_mq{n-m}=0~,
$$
which leads to
\begin{equation}
\sum_{n=1}^\infty\Bigg[-(-i\xi)^n\frac{dq_{n-1}}{dx}-
\sum_{m=0}^n(-i\xi)^nq_mq_{n-m}\Bigg]+\Bigg[r(x)-q_0^2\Bigg]=0~.
\end{equation}

In order that this equation be right the following conditions should be
satisfied 
\begin{equation}
r(x)-q_0^2=0 \quad\Rightarrow\quad q_0=\pm\sqrt{r(x)}
\end{equation}
$$
-(-i\xi)^n\frac{dq_{n-1}}{dx}-\sum_{m=0}^n(-i\xi)^nq_mq_{n-m}=0 \quad 
$$
\begin{equation}
\Rightarrow\quad\quad\frac{dq_{n-1}}{dx}=-\sum_{m=0}^{n}q_mq_{n-m}
\quad\quad{n\geq1}~.
\end{equation}
The latter is a recurrence relatioship, which occurs naturally in the WKB 
method. Recalling that we have defined  
$r(x)=\eta-f(x),\quad\eta=\frac{E}{u_0}\quad\&\quad{f(x)=\frac{u}{u_0}}$, 
by means of eq. $(21)$ we get
\begin{equation}
q_0=\pm\sqrt{\eta-f(x)}=\pm\sqrt{\frac{E}{u_0}-\frac{u}{u_0}}=
\pm\sqrt{\frac{2m(E-u)}{2mu_0}}~.
\end{equation}
This clearly indicates the classical nature of the WKB momentum
of the particle of energy 
$E$ in the potential $u$ and units of $\sqrt{2mu _0}$. Thus
$$
q_0=p(x)=\sqrt{\eta-f(x)}$$ \hspace{1mm} {\bf is not an operator}.
If we approximate till the second order, we get
$$
q(x)=q_0-i{\xi}q_1-\xi^2q_2
$$
and using the WKB recurrence relationship (22) we
calculate $q_1$ and $q_2$
$$
\frac{dq_0}{dx}=-2q_0q_1
\quad \Rightarrow \quad 
q_1=-\frac{1}{2}\frac{\frac{dq_0}{dx}}{q_0}=
-\frac{1}{2}\frac{d}{dx}(\ln\vert{q_0}\vert)
$$
\begin{equation}
\Rightarrow \quad q_1=-\frac{1}{2}\frac{d}{dx}(\ln\vert p(x)\vert)
\end{equation}
\begin{equation}
\frac{dq_1}{dx}=-2q_0q_2-q_1^2 \quad\Rightarrow\quad 
q_2=-\frac{\frac{dq_1}{dx}-q_1^2}{2q_0}~.
\end{equation}

A glance to eq. $(24)$, affords us to consider $q_1$ as the slope, up to a
change of sign, of $\ln\vert q_0\vert$; when $q_0$ is very small, then 
$q_1\ll0\quad\Rightarrow\quad -\xi{q_1}\gg0$ and therefore the series
diverges. To avoid this the following {\bf WKB condition} is imposed
$$
\vert q_0\vert\gg\vert -\xi{q_1}\vert=\xi\vert{q_1}\vert~.$$

It is worth noting that this WKB condition WKB is not fulfilled at those points 
$x_k$ where
$$
q_0(x_k)=p(x_k)=0~.
$$
Since $q_0=p=\sqrt{\frac{2m(E-u)}{2mu_0}}$ the previous equation leads us to
\begin{equation}
E=u(x_k)~.
\end{equation}

In classical mechanics the points $x_k$ that satisfies  
(26) are called {\bf turning points} because the change of the sense
of the motion of a macroscopic particle takes place there.

By means of these arguments, we can say that $q_0$ is a classical solution
of the problem under examination; also that the quantities $q_1$ \& $q_2$ 
are the first and the second quantum corrections, respectively,
in the WKB problem.

To obtain the WKB wavefunctions we shall consider only the classical solution 
and the first quantum correction that we plug in the WKB form of $\psi$
$$
\psi=\exp\Bigg[\frac{i}{\xi}\int_a^x{q(x)dx}\Bigg]=
\exp\Bigg[\frac{i}{\xi}\int_a^x(q_0-i\xi{q_1})dx\Bigg]
$$
$$
\Rightarrow\quad\psi=\exp\Bigg(\frac{i}{\xi}\int_a^xq_0dx\Bigg)\cdot
\exp\Bigg(\int_a^xq_1dx\Bigg)~.
$$

For the second factor, we get
$$
\exp\Bigg(\int_a^xq_1dx\Bigg)=\exp\Bigg[-\frac{1}{2}
\int_a^x\frac{d}{dx}(\ln \vert p(x)\vert)dx\Bigg]=
$$
$$
\quad\quad\quad\quad\quad\quad\quad=\exp\Bigg[-\frac{1}{2}(\ln\vert 
p(x)\vert)\Big{\vert}_a^x\Bigg]=\frac{A}{\sqrt{p(x)}}~, 
$$
where $A$ is a constant, whereas for the first factor we get
$$
\exp\Bigg(\frac{i}{\xi}\int_a^xq_0dx\Bigg)=\exp\Bigg[\pm\frac{i}{\xi}
\int_a^xp(x)dx\Bigg]~.
$$
Thus, we can write $\psi$ in the following form
\begin{equation}
\psi^{\pm}=\frac{1}{\sqrt{p(x)}}\exp\Bigg[\pm\frac{i}{\xi}\int_a^xp(x)dx\Bigg]~.
\end{equation}
The latter are known as the
{\em WKB solutions of the 1D Schr\"odinger equation}.
The general WKB solution in the region in which the WKB condition is satisfied
is written down as
\begin{equation}
\psi=a_+\psi^++a_-\psi^-~.
\end{equation}

\noindent
As already mentioned there is no WKB solution 
at the turning points. This raises the question of the manner in which one has
to do the passing from $\psi(x<x_k)$ to
$\psi(x>x_k)$. The solution of this difficulty 
is achieved by introducing the WKB connection formulas.

\bigskip

\subsection*{The connection formulas}

\hspace{0.6cm} We have already seen that the WKB solutions 
are singular at the classical turning points; however, these solutions
are correct both on the left and right side of these turning points   
$x_k$. A natural question is how do we change $\psi(x<x_k)$ in
$\psi(x>x_k)$ when passing through the turning points.
The explicit answer is given by the connection formulas.

From the theory of differential equations of complex variable it can be proved
that really there are such connection formulas and that they are the following  
$$
\psi_1(x)=
\frac{1}{\left[-r(x)\right]^{\frac{1}{4}}}
\exp\left(-\int_x^{x_k}\sqrt{-r(x)}dx\right)\rightarrow 
$$
\begin{equation}
\rightarrow\frac{2}{\left[r(x)\right]^{\frac{1}{4}}}
\cos\left(\int_{x_k}^x\sqrt{r(x)}dx-\frac{\pi}{4}\right)~,
\end{equation}
where $\psi_1(x)$ has only an attenuated exponential behavior
for $x<x_k$. The first connection formula shows that the function 
$\psi(x)$, which at the left of the turning point behaves 
exponentially decaying, turns at the right of  $x_k$ into a cosinusoide of
phase $\phi=\frac{\pi}{4}$ and double amplitude with regard to the amplitude of
the exponential.

In the case of a more general function $\psi(x)$, such as a function with both
rising and decaying exponential behavior, the connection formula is  
$$
\sin\left(\phi+\frac{\pi}{4}\right)
\frac{1}{\left[-r(x)\right]^{\frac{1}{4}}}
\exp\left(\int_x^{x_k}\sqrt{-r(x)}dx\right)\leftarrow
$$
\begin{equation}
\leftarrow\frac{1}{\left[r(x)\right]^{\frac{1}{4}}}
\cos\left(\int_{x_k}^x\sqrt{r(x)}dx+\phi\right)~,
\end{equation}
under the condition that $\phi$ s\A\ do not take a value that is too
close to
$-\frac{\pi}{4}$. The reason is that if $\phi=-\frac{\pi}{4}$, 
then the sinus function is zero . The latter connection   
formula means that a function whose behavior is of the cosinusoid type at the 
right of a turning point  
changes into a growing exponential with sinusoid-modulated amplitude at the right
of that point.

In order to study the details of the procedure 
of getting the connection formulas we recommend the book
{\em Mathematical Methods of Physics} by 
J. Mathews \& R.L. Walker.\\
\bigskip
\subsection*{Estimation of the WKB error}

\hspace{0.6cm} We have found the solution of the  
Schr\"odinger equation in the regions where the WKB condition is satisfied. 
However, the WKB solutions are divergent at the turning points.  
We thus briefly analyze the error introduced by using the WKB approximation
and tackling the
{\em connection formulas} in a close neighbourhood of the turning points.

Considering $x=x_k$ as a turning point, we have
$q_0(x_k)=p(x_k)=0\quad\Rightarrow\quad 
E=u(x_k)$. At the left of $x_k$, that is on the `half-line'
$x<x_k$, we shall assume  
$E<u(x)$ leading to the WKB solution
$$
\psi(x)=\frac{a}{\left[\frac{u(x)-E}{u_0}\right]^\frac{1}{4}}\exp\left(-\frac{1}{\xi}\int_x^{x_k}\sqrt{\frac{u(x)-E}{u_0}}dx\right)\quad+
$$
\begin{equation}
\quad\quad+\quad\frac{b}{\left[\frac{u(x)-E}{u_0}\right]^\frac{1}{4}}
\exp\left(\frac{1}{\xi}\int_x^{x_k}\sqrt{\frac{u(x)-E}{u_0}}dx\right)~.
\end{equation}
Similarly, at the right of $x_k$ (on the `half-line $x>x_k$) 
we assume $E>u(x)$; therefore the WKB solution in the latter region will be 
$$
\psi(x)=\frac{c}{\left[\frac{E-u(x)}{u_0}\right]^\frac{1}{4}}
\exp\left(\frac{i}{\xi}\int_{x_k}^x\sqrt{\frac{E-u(x)}{u_0}}dx\right)\quad+
$$
\begin{equation}
\quad\quad\quad\quad+
\quad\frac{d}{\left[\frac{E-u(x)}{u_0}\right]^\frac{1}{4}}
\exp\left(-\frac{i}{\xi}\int_{x_k}^x\sqrt{\frac{E-u(x)}{u_0}}dx\right)~.
\end{equation}

If $\psi(x)$ is a real function, 
it will have this property both at the right and the left of
$x_k$. It is usually called the 
{\it ``reality condition''}. It means that if
$a,b\in\Re$, then $c=d^*$.

Our problem consists in connecting the approximations on the two sides of
$x_k$ such that they refer to the same solution. 
This means to find 
$c$ and $d$ if one knows $a$ and $b$, as well as viceversa. 
To achieve this connection,  
we have to use an approximate solution, which should be correct
along a contour connecting the regions on the two sides of
$x_k$, where the WKB solutions are also correct. 
A method proposed by {\em Zwann} and {\em Kemble} is very useful in this case. 
It consists in going out from the real axis
in the neighbourhood of $x_k$ on a contour 
around $x_k$ in the complex plane. It is assumed that on this contour
the WKB solutions are still correct. Here, we shall use this method as a means 
of getting the estimation of the error produced by the WKB method.

The estimation of the error is always an important matter for any approximate
solutions. In the case of the WKB procedure, it is more significant 
because it is an approximation on large intervals of the real axis that can lead
to the accuulation of the errors as well as to some artefacts due to the 
phase shifts that can be introduced in this way.   

Let us define 
{\em the associated WKB functions} as follows
\begin{equation}
W_{\pm}=\frac{1}{\left[\frac{E-u(x)}{u_0}\right]^\frac{1}{4}}
\exp\left(\pm\frac{i}{\xi}\int_{x_k}^x\sqrt{\frac{E-u(x)}{u_0}}dx\right)~,
\end{equation}
that we consider as functions of complex variable.  
We shall use cuts in order to avoid the discontinuities in the zeros
of $r(x)=\frac{E-u(x)}{u_0}$. These functions satisfy the differential equation
that is obtained by differentiating with respect to $x$, leading to
$$
W_{\pm}'=\left(\pm\frac{i}{\xi}\sqrt{r}-\frac{1}{4}\frac{r'}{r}\right)W_{\pm}
$$
\begin{equation}
W_{\pm}''+\left[\frac{r}{\xi^2}+\frac{1}{4}\frac{r''}{r}-\frac{5}{16}
\left(\frac{r'}{r}\right)^2\right]W_{\pm}=0~.
\end{equation}
Let us notice that
\begin{equation}
s(x)=\frac{1}{4}\frac{r''}{r}-\frac{5}{16}\left(\frac{r'}{r}\right)^2~,
\end{equation}
then $W_{\pm}$ are exact solutions of the equation
\begin{equation}
W_{\pm}''+\left[\frac{1}{\xi^2}r(x)+s(x)\right]W_{\pm}=0~,
\end{equation}
although they satisfy only approximately the 
Schr\"odinger equation, which is a regular equation in $x=x_k$, whereas the same
equation for the associate WKB functions is singular at that point. 

We shall now define the functions $\alpha_{\pm}(x)$ satisfying the following two
relationships 
\begin{equation}
\psi(x)=\alpha_+(x)W_+(x)+\alpha_-(x)W_-(x)
\end{equation}
\begin{equation}
\psi'(x)=\alpha_+(x)W_+'(x)+\alpha_-(x)W_-'(x)~,
\end{equation}
where $\psi(x)$ is a solution of the Schr\"odinger equation. 
Solving the previous equations for $\alpha_{\pm}$, we get
$$
\alpha_+=\frac{\psi W_-'-\psi'W_-}{W_+W_-'-W_+'W_-}
\qquad\qquad\alpha_-=-\frac{\psi W_+'-\psi'W_+}{W_+W_-'-W_+'W_-}~,
$$
where the numerator is just the {\em Wronskian} of $W_+$ and $W_-$. It is not
difficult to prove that this takes the value   
$-\frac{2}{\xi}i$, so that $\alpha_{\pm}$ simplifies to the following form 
\begin{equation}
\alpha_+=\frac{\xi}{2}i\left(\psi W_-'-\psi'W_-\right)
\end{equation}
\begin{equation}
\alpha_-=\frac{-\xi}{2}i\left(\psi W_+'-\psi'W_+\right)~.
\end{equation}
Doing the derivative in $x$ in the eqs. $(39)$ and $(40)$, we have
\begin{equation}
\frac{d\alpha_{\pm}}{dx}=\frac{\xi}{2}i\left(\psi'W_{\mp}'+\psi W_{\mp}''
-\psi''W_{\mp}-\psi'W_{\mp}'\right)~.
\end{equation}
In the brackets, the first and the fourth terms are zero; recalling that
$$
\psi''+\frac{1}{\xi^2}r(x)\psi=0\quad\&\quad 
W_{\pm}''+\left[\frac{1}{\xi^2}r(x)+s(x)\right]W_{\pm}=0~,
$$
we can write eq. $(41)$ in the form
$$
\frac{d\alpha_{\pm}}{dx}=\frac{\xi}{2}i\left[-\psi\left(\frac{r}{\xi^2}+s\right)W_{\mp}+\frac{r}{\xi^2}\psi W_{\mp}\right]
$$
\begin{equation}
\frac{d\alpha_{\pm}}{dx}=\mp\frac{\xi}{2}is(x)\psi(x)W_{\mp}(x)~,
\end{equation}
which based on eqs. $(33)$ and $(37)$ becomes
\begin{equation}
\frac{d\alpha_{\pm}}{dx}=
\mp\frac{\xi}{2}i\frac{s(x)}{\left[r(x)\right]^\frac{1}{2}}
\left[\alpha_{\pm}+\alpha_{\mp}\exp\left(\mp\frac{2}{\xi}i
\int_{x_k}^x\sqrt{r(x)}dx\right)\right]~.
\end{equation}

Eqs. $(42)$ and $(43)$ are useful for estimating the WKB error in the 1D case. 

The reason for which $\frac{d\alpha_{\pm}}{dx}$ can be considered as a measure
of the WKB errors is that in the eqs.
$(31)$ and $(32)$ the constants $a$, $b$ and $c$, $d$, 
respectively, give only approximate solutions $\psi$, while the functions
$\alpha_{\pm}$ when introduced in the
eqs. $(37)$ and $(38)$ produce exact $\psi$ solutions. From the geometrical
viewpoint the derivative gives the slope of the tangent 
to these functions and indicates the measure in which 
$\alpha_{\pm}$ deviates from the constants $a$, $b$, $c$ and $d$.

\bigskip

\noindent
\underline{{\bf 4N. Note}}: The original (J)WKB papers are the following:\\

\noindent
G. Wentzel, ``Eine Verallgemeinerung der Wellenmechanik", [``A generalization of
wave mechanics"],

\noindent
Zeitschrift f\"ur Physik {\bf 38}, 518-529 (1926) [received on 18 June 1926]\\

\noindent
L. Brillouin, ``La m\'ecanique ondulatoire de Schr\"odinger: une m\'ethode
g\'en\'erale de resolution par approximations successives",
[``Schr\"odinger's wave mechanics: a general method of solving by succesive
approximations"],

\noindent
Comptes Rendus Acad. Sci. Paris {\bf 183}, 24-26 (1926) [received on 5 July 1926]\\

\noindent
H.A. Kramers, ``Wellenmechanik und halbzahlige Quantisierung",
[``Wave mechanics and half-integer quantization"],

\noindent
Zf. Physik {\bf 39}, 828-840 (1926) [received on 9 Sept. 1926]\\

\noindent
H. Jeffreys, ``On certain approx. solutions of linear diff. eqs. of the
second order",

\noindent
Proc. Lond. Math. Soc. {\bf 23}, 428-436 (1925)

\bigskip

\centerline{{\huge 4P.  Problems}}

{\em Problem 4.1}\\

Employ the WKB method for a particle of energy
$E$ moving in a potential  
$u(x)$ of the form shown in fig.~4.1.

\vskip 2ex
\centerline{
\epsfxsize=280pt
\epsfbox{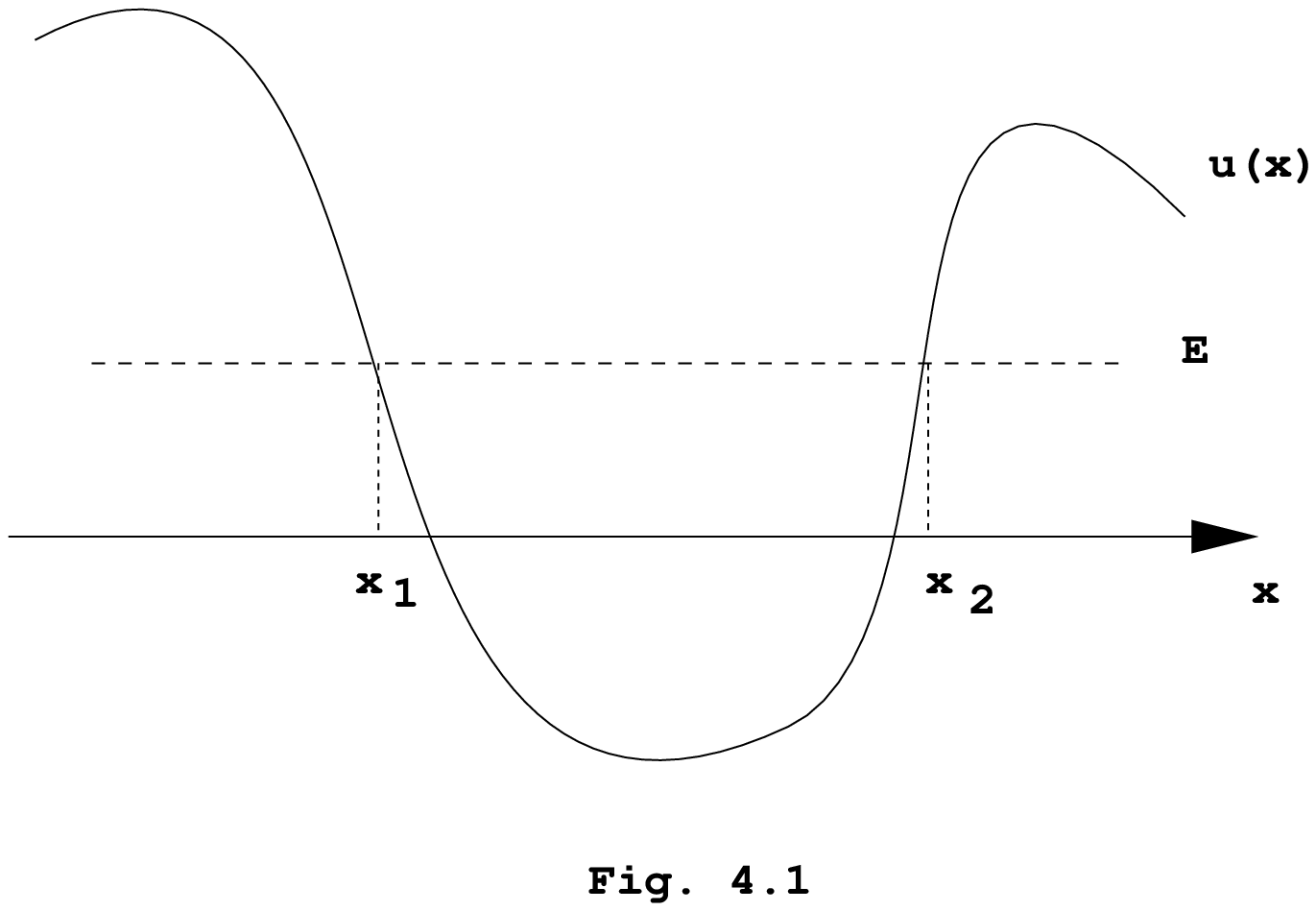}}
\vskip 4ex

{\bf Solution}

The  Schr\"odinger equation is
\begin{equation}
\frac{d^2\psi}{dx^2}+\frac{2m}{\hbar^2}\left[E-u(x)\right]\psi=0~.
\end{equation}

As one can see, we have
$$
r(x)=\frac{2m}{\hbar^2}\left[E-u(x)\right]\qquad
\left\{
\begin{array}{ll}
\mbox{is positive for $a<x<b$}\\
\mbox{is negative for $x<a, x>b$.}
\end{array}
\right.
$$

If $\psi(x)$ corresponds to the region $x<a$, when passing to the interval
$a<x<b$, the connection formula is given by eq. $(29)$ telling us that 
\begin{equation}
\psi(x)\approx\frac{A}{\left[E-u\right]^{\frac{1}{4}}}\cos\left(\int_a^x
\sqrt{\frac{2m}{\hbar^2}(E-u)}dx-\frac{\pi}{4}\right)~,
\end{equation}
where $A$ is an arbitrary constant.

When $\psi(x)$ corresponds to the region $x>b$, when passing to the 
segment $a<x<b$, we have in a similar way 
\begin{equation}
\psi(x)\approx-\frac{B}{\left[E-u\right]^{\frac{1}{4}}}\cos\left(\int_x^b
\sqrt{\frac{2m}{\hbar^2}(E-u)}dx-\frac{\pi}{4}\right)~,
\end{equation}
where $B$ is another arbitrary constant. The reason why the connection formula
is again given by eq. (29) is easily understood examinining
what happens when the 
particle reaches the second classical turning point at $x=b$. This produces
the inversion of the direction of motion. Thus, the particle appears to come
from the right toward the left. In other words, we are in the first case
(from the left to the right), only that as seen in a mirror
placed at the point $x=a$.

These two expressions should be the same, independently of the 
constants $A$ and $B$, so that
$$
\cos\left(\int_a^x\sqrt{\frac{2m}{\hbar^2}(E-u)}dx-\frac{\pi}{4}\right)
=-\cos\left(\int_x^b\sqrt{\frac{2m}{\hbar^2}(E-u)}dx-\frac{\pi}{4}\right)\\
$$
\begin{equation}
\Rightarrow\cos\left(\int_a^x\sqrt{\frac{2m}{\hbar^2}(E-u)}dx-\frac{\pi}{4}
\right)+\cos\left(\int_x^b\sqrt{\frac{2m}{\hbar^2}(E-u)}dx-
\frac{\pi}{4}\right)=0~.
\end{equation}
Recalling that
$$
\cos A+\cos B= 2\cos\left(\frac{A+B}{2}\right)\cos\left(\frac{A-B}{2}\right)~,
$$
eq. $(47)$ can be written
$$
2\cos\left[\frac{1}{2}\left(
\int_a^x\sqrt{\frac{2m}{\hbar^2}(E-u)}dx-\frac{\pi}{4}
+\int_x^b\sqrt{\frac{2m}{\hbar^2}(E-u)}dx-\frac{\pi}{4}\right)\right]\cdot
$$
\begin{equation}
\cdot\cos\left[\frac{1}{2}\left(\int_a^x\sqrt{\frac{2m}{\hbar^2}(E-u)}dx
-\frac{\pi}{4}-\int_x^b\sqrt{\frac{2m}{\hbar^2}(E-u)}dx
+\frac{\pi}{4}\right)\right]=0~,
\end{equation}
which implies that the arguments of the cosinusoids are integer multiples
of  
$\frac{\pi}{2}$. On the other hand, the argument of the second
cosinusoid do not lead to a nontrivial result. Therefore, we pay attention only
to the argument of the first cosinusoid, 
which prove to be essential for getting an important result
$$
\frac{1}{2}\left(\int_a^x\sqrt{\frac{2m}{\hbar^2}(E-u)}dx
-\frac{\pi}{4}+\int_x^b\sqrt{\frac{2m}{\hbar^2}(E-u)}dx
-\frac{\pi}{4}\right)=\frac{n}{2}\pi\quad\mbox{for n odd}
$$
$$
\Rightarrow\quad\quad\int_a^b\sqrt{\frac{2m}{\hbar^2}(E-u)}dx
-\frac{\pi}{2}=n\pi
$$
$$
\Rightarrow\quad\quad\int_a^b\sqrt{\frac{2m}{\hbar^2}(E-u)}dx=
(n+\frac{1}{2})\pi
$$
\begin{equation}
\Rightarrow\quad\quad\int_a^b\sqrt{2m(E-u)}dx=(n+\frac{1}{2})\pi\hbar~.
\end{equation}

\noindent
This result is very similar to the {\em Bohr - Sommerfeld quantization rules}.

We recall that Bohr's postulate says that
the orbital angular momentum of an electron 
moving on an ``allowed atomic orbit" is quantized as $L=n\hbar$, 
$n=1,2,3,\dots$. We also recall that the Wilson - Sommerfeld quantization 
rules  
assert that any coordinate of a system that varies periodically in time
should satisfy the `quantum' condition: $\oint p_qdq=n_q h$, where $q$ is a 
periodic coordinate, $p_q$ is the associated momentum, $n_q$ is an integer, and
$h$ is Planck's constant. One can see that the WKB result is indeed very 
similar.\\

\bigskip

{\em Problem 4.2}\\

Estimate the error of the WKB solution WKB at a point
$x_1\neq x_k$, where $x_k$ is a classical turning point for the differential
equation $y''+xy=0$. {\em The solution of this problem is of importance in the 
study of uniform fields, such as the gravitational and electric fields 
generated by large planes.}\\

\bigskip

{\bf Solution:}\\
For this differential equation we have
$$
\xi=1,\qquad r(x)=x\qquad\&\qquad s(x)=-\frac{5}{16}x^{-2}~.
$$
$r(x)=x$ has a single zero at $x_k=0$, therefore for $x\gg0$: 
\begin{equation} 
W_{\pm}=x^{-\frac{1}{4}}
\exp\left(\pm i\int_0^x\sqrt{x}dx\right)
=x^{-\frac{1}{4}}\exp\left(\pm\frac{2}{3}ix^{\frac{3}{2}}\right)~.
\end{equation}
Derivating $W_{\pm}$ up to the second derivative in $x$, we realize that the 
following differential equation is satisfied
\begin{equation}
W_{\pm}''+(x-\frac{5}{16}x^{-2})W_{\pm}=0~.
\end{equation}

The exact solution $y(x)$ of the latter differential equation can be written as
a linear combination of $W_{\pm}$, as it has been shown in the corresponding 
section where the WKB error has been tackled; recall that 
the following form of the linear combination was proposed therein
$$
y(x)=\alpha_+(x)W_+(x)+\alpha_-(x)W_-(x) 
$$

For large $x$, the general solution of the
differential equation can be written in the WKB approximation as follows 
\begin{equation}
y(x)=Ax^{-\frac{1}{4}}
\cos\left(\frac{2}{3}x^{\frac{3}{2}}
+\delta\right)\qquad \mbox{for} \quad x\rightarrow\infty~.
\end{equation}
Thus, $\alpha_+\rightarrow\frac{A}{2}e^{i\delta}$ and 
$\alpha_-\rightarrow\frac{A}{2}e^{-i\delta}$ for 
$x\rightarrow\infty$. We want to calculate the error due to these WKB solutions. 
A simple measure of this error is the deviation of  
$\alpha_+$ and of $\alpha_-$ relative to the constants $A$. Using the equation
$$
\frac{d\alpha_{\pm}}{dx}
=\mp\frac{\xi}{2}i\frac{s(x)}{\sqrt{r(x)}}
\left[\alpha_{\pm}+\alpha_{\mp}
\exp\left(\mp2i\int_{x_k}^x\sqrt{r(x)}dx\right)\right]
$$
and doiing the corresponding substitutions, one gets
\begin{equation}
\frac{d\alpha_{\pm}}{dx}
=\mp\frac{i}{2}\left(-\frac{5}{16}x^{-2}\right)
x^{-\frac{1}{2}}\left[\frac{A}{2}e^{\pm i\delta}
+\frac{A}{2}e^{\mp i\delta}\exp\left(\mp 2i\frac{2}{3}
x^{\frac{3}{2}}\right)\right]~.
\end{equation}
Taking $\Delta\alpha_{\pm}$ as the changes displayed by  
$\alpha_{\pm}$ when $x$ varies between $x_1$ and $\infty$, 
we can do the required calculation by means of
$$
\frac{\Delta\alpha_{\pm}}{A/2}
=\frac{2}{A}\int_{x_1}^\infty\frac{d\alpha_{\pm}}{dx}dx
=\qquad\qquad\qquad\qquad\qquad\qquad\qquad\qquad\quad
$$
\begin{equation}
=\pm i\frac{5}{32}e^{\pm i\delta}\left[\frac{2}{3}x_1^{-\frac{3}{2}}+e^{\mp 
2i\delta}\int_{x_1}^\infty x^{-\frac{5}{2}}\exp\left(\mp 
i\frac{4}{3}x^\frac{3}{2}\right)dx\right]~. 
\end{equation}
The second term in the parentheses is less important than the first one 
because the complex exponential oscillates between 
$1$ and $-1$ and therefore $x^{-\frac{5}{2}}<x^{-\frac{3}{2}}$. 
Consequently
\begin{equation}
\frac{\Delta\alpha_{\pm}}{A/2}\approx\pm\frac{5}{48}ie^{\pm 
i\delta}x_1^{-\frac{3}{2}}~, \end{equation}
and as we can see the error introduced by the WKB approximation 
is indeed small if we take into account that the complex  
exponential oscillates between $-1$ and $1$, while 
$x_1^{-\frac{3}{2}}$ is also small.\\









\newpage
\newcommand{\bc}{\begin{center}}
\newcommand{\ec}{\end{center}}
\newcommand{\ii}{\'{\i}}
\newcommand{\be}{\begin{equation}}
\newcommand{\ee}{\end{equation}}
\newcommand{\dd}{\dagger}
\newcommand{\ad}{a^{\dd}}
\newcommand{\m}{\mid}

\section*{{\huge 5. THE HARMONIC OSCILLATOR (HO)}}
\section*{The solution of the Schr\"odinger eq. for HO}
\setcounter{equation}{0}
The HO can be considered as a paradigm of Physics.
Its utility is manifest in many areas from classical physics until quantum 
electrodynamics and theories of gravitational collapse.\\
From classical mechanics we know that many complicated potentials can be well 
approximated near their equilibrium positions by HO potentials 
\be
V(x) \sim \frac{1}{2}V^{\prime\prime}(a)(x-a)^2~.
\ee

This is a 1D case. For this case, the classical 
Hamiltonian function of a particle of mass
{\em m}, oscillating at the frequency $\omega$ has the following form:
\be
H=\frac{p^2}{2m}+\frac{1}{2}m\omega^2x^2
\ee
and the quantum Hamiltonian corresponding to the space of configurations
is given by
\be
\hat{H}=\frac{1}{2m}(-i\hbar\frac{d}{dx})^2+\frac{1}{2}m\omega^2x^2
\ee
\be
\hat{H}=-\frac{\hbar^2}{2m}\frac{d^2}{dx^2}+\frac{1}{2}m\omega^2x^2~.
\ee

Since we consider a time-independent potential,
the eigenfunctions $\Psi_n$ and the eigenvalues $E_n$ are obtained by means
of the time-independent Schr\"odinger equation
\be
\hat{H}\Psi_n=E_n\Psi_n~.
\ee

For the HO Hamiltonian, the Schr\"odinger eq. is
\be
\frac{d^2\Psi}{dx^2}+\Bigg[\frac{2mE}{\hbar^2}
-\frac{m^2\omega^2}{\hbar^2}x^2\Bigg]\Psi=0~.
\ee

We cancealed the subindices of $E$ and $\Psi$ because they are not of any
importance here.
Defining
\be
k^2=\frac{2mE}{\hbar^2}
\ee
\be
\lambda=\frac{m\omega}{\hbar}~,
\ee

\noindent
the Schr\"odinger eq. becomes
\be
\frac{d^2\Psi}{dx^2}+[k^2-\lambda^2x^2]\Psi=0~,
\ee

\noindent
which is known as Weber's differential
equation in mathematics. \\
We shall make now the transformation
\be
y=\lambda x^2~.
\ee

In general, by changing the variable from
$x$ to $y$, the differential operators take the form
\be
\frac{d}{dx}=\frac{dy}{dx}\frac{d}{dy}
\ee
\be
\frac{d^2}{dx^2}=\frac{d}{dx}(\frac{dy}{dx}\frac{d}{dy})
=\frac{d^2y}{dx^2}\frac{d}{dy}+(\frac{dy}{dx})^2\frac{d^2}{dy^2}~.
\ee

Applying this obvious rule to the proposed transformation
we obtain the following differential eq. in the $y$ variable 
\be
y\frac{d^2\Psi}{dy^2}+\frac{1}{2}\frac{d\Psi}{dy}+[\frac{k^2}{4\lambda}
-\frac{1}{4}y]\Psi=0~,
\ee

\noindent
and, by definind :
\be
\kappa=\frac{k^2}{2\lambda}=\frac{\bar k^2}{2m\omega}=\frac{E}{\hbar\omega}~,
\ee
we get
\be
y\frac{d^2\Psi}{dy^2}+\frac{1}{2}\frac{d\Psi}{dy}
+[\frac{\kappa}{2}-\frac{1}{4}y]\Psi=0~.
\ee

Let us try to solve this equation by first doing its asymptotic analysis
in the limit
$y\rightarrow\infty$. We first rewrite the previous equation
in the form
\be
\frac{d^2\Psi}{dy^2}+\frac{1}{2y}\frac{d\Psi}{dy}
+[\frac{\kappa}{2y}-\frac{1}{4}]\Psi=0~.
\ee

We notice that in the limit
$y\rightarrow\infty$ the equation behaves as follows
\be
\frac{d^2\Psi_{\infty}}{dy^2}-\frac{1}{4}\Psi_{\infty}=0~.
\ee

This equation has as solution 
\be
\Psi_{\infty}(y)=A\exp{\frac{y}{2}}+B\exp{\frac{-y}{2}}~.
\ee

Taking $A=0$, we eliminate $\exp{\frac{y}{2}}$ since it
diverges in the limit $y\rightarrow\infty$, keeping only the attenuated
exponential. We can now suggest that $\Psi$ has the following form
\be
\Psi(y)=\exp{\frac{-y}{2}}\psi(y)~.
\ee

Plugging it in the differential equation for $y$ ( eq.~$15$) 
one gets:
\be
y\frac{d^2\psi}{dy^2}
+(\frac{1}{2}-y)\frac{d\psi}{dy}+(\frac{\kappa}{2}-\frac{1}{4})\psi=0~.
\ee

The latter is a confluent hypergeometric equation
\footnote{It is also known as Kummer's differential
equation.} :
\be
z\frac{d^2y}{dz^2}+(c-z)\frac{dy}{dz}-ay=0~.
\ee

The general solution of this equation is 
\be
y(z)=A \hspace{.2cm} _1F_1(a;c,z)+
B \hspace{.2cm} z^{1-c} \hspace{.1cm}  _1F_1(a-c+1;2-c,z)~,
\ee

where the confluent hypergeometric equation
is defined by
\be
_1F_1(a;c,z)=\sum_{n=0}^{\infty}\frac{(a)_n x^n}{(c)_n n!}~.
\ee

Comparing now our equation with the standard confluent hypergeometric equation,
one can see that the general solution of the first one is
\be
\psi(y)=A\hspace{.2cm} _1F_1(a;\frac{1}{2},y)+
B \hspace{.2cm} y^{\frac{1}{2}}
\hspace{.2cm} _1F_1(a+\frac{1}{2};\frac{3}{2},y)~,
\ee

where
\be
a=-(\frac{\kappa}{2}-\frac{1}{4})~.
\ee

If we keep these solutions in their present form, 
the normalization condition is not satisfied for the wavefunction
because from the asymptotic behaviour of the confluent hypergeometric 
function \footnote{ The asymptotic behavior for 
$\mid x \mid\rightarrow \infty$ is
\bc
$_1F_1(a;c,z)\rightarrow
\frac{\Gamma(c)}{\Gamma(c-a)}e^{-ia\pi}x^{-a}
+\frac{\Gamma(c)}{\Gamma(a)}e^{x}x^{a-c}~.$
\ec
} it follows
( taking into account ony the dominant exponential behavior ) :
\be
\Psi(y)=e^{\frac{-y}{2}}\psi(y)\rightarrow
\hspace{.3cm}const. \hspace{.2cm} e^{\frac{y}{2}}y^{a-\frac{1}{2}}~.
\ee

The latter approximation leads to a divergence in the normalization integral,
which physically is not acceptable. 
What one does in this case is to impose the termination condition
for the series \footnote{The truncation condition of the confluent hypergeometric
series $_1F_1(a;c,z)$ is $a=-n$, where $n$ is a nonnegative integer
( i.e., zero included).} , that is , the series has only
a finite number of terms and therefore it is a polynomial of $n$ order.\\
{\em We thus notice that asking for a finite normalization constant  
(as already known, a necessary condition for the physical interpretation
in terms of probabilities), 
leads us to the truncation of the series, which simultaneously generates the 
quantization of energy.}\\
In the following we consider the two possible cases

$1)\hspace{.4cm} a=-n \hspace{.3cm}$ and $ B=0$
\be
\frac{\kappa}{2}-\frac{1}{4}=n~.
\ee

The eigenfunctions are given by
\be
\Psi_n(x)=D_n \exp{\frac{-\lambda x^2}{2}}
\hspace{.1cm} _1F_1(-n;\frac{1}{2},\lambda x^2)
\ee

and the energy is:
\be
E_n=\hbar\omega(2n+\frac{1}{2})~.
\ee

$2)\hspace{.4cm} a+\frac{1}{2}=-n \hspace{.3cm}$ and $A=0$
\be
\frac{\kappa}{2}-\frac{1}{4}=n+\frac{1}{2}~.
\ee

The eigenfunctions are now
\be
\Psi_n(x)=D_n \exp{\frac{-\lambda x^2}{2}}
\hspace{.2cm}x \hspace{.2cm}_1F_1(-n;\frac{3}{2},\lambda x^2)~,
\ee

whereas the stationary energies are
\be
E_n=\hbar\omega[(2n+1)+\frac{1}{2}]~. 
\ee

The polynomials obtained by this truncation of the confluent 
hypergeometric series are called Hermite polynomials and in 
hypergeometric notation they are 
\be
H_{2n}(\eta)=(-1)^n \frac{(2n)!}{n!}
\hspace{.2cm} _1F_1(-n;\frac{1}{2},\eta^2)
\ee
\be
H_{2n-1}(\eta)=(-1)^n \frac{2(2n+1)!}{n!}
\hspace{.2cm}\eta \hspace{.2cm} _1F_1(-n;\frac{3}{2},\eta^2)~.
\ee

We can now combine the obtained results 
( because some of them give us the even cases and the others the odd ones ) 
in a single expression for the eigenvalues and eigenfunctions 
\be
\Psi_n (x)=D_n \exp{ \frac{-\lambda x^2}{2}} H_n (\sqrt{\lambda}x)
\ee
\be
E_n =(n+\frac{1}{2})\hbar\omega \hspace{1cm}n=0,1,2~\ldots
\ee

The HO energy spectrum is equidistant, i.e., there is the same energy difference
$\hbar \omega$ \h between any consequitive neighbour levels. Another remark
refers to the minimum value of the energy of the oscillator;
somewhat surprisingly it is not zero. This is considered by many people to be
a pure quantum result because it is zero when $\hbar\rightarrow 0$. 
It is known as {\em the zero point energy} and the fact that it is different
of zero is the main characteristic of all confining potentials.\\

The normalization constant is easy to calculate 
\be
D_n = \Bigg[ \sqrt{\frac{\lambda}{\pi}}\frac{1}{2^n n!}\Bigg]^{\frac{1}{2}}~.
\ee

Thus, one gets the following normalized eigenfunctions
of the 1D operator
\be
\Psi_n (x)= \Bigg[ \sqrt{\frac{\lambda}{\pi}}\frac{1}{2^n n!}\Bigg]^{\frac{1}{2}}
\hspace{.2cm} \exp( \frac{-\lambda x^2}{2})
\hspace{.2cm} H_n( \sqrt{\lambda} x)~.
\ee


\section*{Creation and anihilation operators: $\hat{a}^{\dagger}$ and 
$\hat{a}$}

There is another approach to deal with the HO besides the conventional one
of solving the Schr\"odinger equation. It is the algebraic method, also
known as the method of creation and annihilation (ladder) operators. This is a
very efficient procedure, which can be successfully applied to many 
quantum-mechanical problems, especially when dealing with discrete spectra.\\
Let us define two nonhermitic operators $a$ and $a^{\dd}$ :
\be
a=\sqrt{\frac{m\omega}{2\hbar}}(x+\frac{ip}{m\omega})
\ee
\be
a^{\dd}=\sqrt{\frac{m\omega}{2\hbar}}(x-\frac{ip}{m\omega})~.
\ee

These operators are known as 
\hspace{.1cm} {\em annihilation operator}
\hspace{.1cm}  and \hspace{.1cm} {\em creation operator},
\hspace{.1cm}  respectively  (the reason of this terminology
will be seen in the following,
though one can claim that it comes from quantum field theories).\\
Let us calculate the commutator of these operators
\be
[a,a^{\dd}]=\frac{m\omega}{2\hbar}[x
+\frac{ip}{m\omega},x-\frac{ip}{m\omega}]=\frac{1}{2\hbar}(-i[x,p]+i[p,x])=1~,
\ee

where we have used the commutator
\be
[x,p]=i\hbar~.
\ee

Therefore the annihilation and creation  operators do not commute, 
since we have
\be
[a,a^{\dd}]=1~.
\ee

Let us also introduce the very important {\em number operator} $\hat{N}$:
\be
\hat{N}=\ad a~.
\ee

This operator is hermitic as one can readily prove 
using $(AB)^{\dd}=B^{\dd}A^{\dd}$ :
\be
\hat{N}^{\dd}=(\ad a)^{\dd}=\ad (\ad)^{\dd}=\ad a=\hat{N}~.
\ee

Considering now that
\be
\ad a =\frac{m\omega}{2\hbar}(x^2+\frac{p^2}{m^2\omega^2})+\frac{i}{2\hbar}[x,p]=\frac{\hat{H}}{\hbar\omega}-\frac{1}{2}
\ee

\noindent
we notice that the Hamiltonian can be written in a quite simple form as a 
function of the number operator
\be
\hat{H}=\hbar\omega(\hat{N}+\frac{1}{2})~.
\ee

The number operator bear this name because
its eigenvalues are precisely the subindices of the eigenfunctions on which it
acts
\be
\hat{N}\m n\rangle=n\m n\rangle~,
\ee

\noindent where we have used the notation 
\be
Psi_n = \hspace{.2cm}\m n\rangle~.
\ee

Applying this fact to $(47)$, we get
\be
\hat{H}\m n>=\hbar\omega(n+\frac{1}{2})\m n>~.
\ee

On the other hand, from the Schr\"odinger equation we know that
$\hat{H}\m n>=E\m n>$. In this way, it comes out that the energy eigenvalues
are given by
\be
E_n=\hbar\omega(n+\frac{1}{2})~.
\ee

This result is identical (as it should be) to the result 
$(36)$.\\
We go ahead and show why the 
operators $a$ and $\ad$ bear the names they have. For this, we calculate the
commutators
\be
[\hat{N},a]=[\ad a,a]=\ad[a,a]+[\ad,a]a=-a~,
\ee

which can be obtained from $[a,a]=0$ and $(43)$.
Similarly, let us calculate
\be
[\hat{N},\ad]=[\ad a,\ad]=\ad[a,\ad]+[\ad,\ad]a=\ad~.
\ee
Using these two commutators, we can write
\begin{eqnarray}
\hat{N}(\ad \m n>)&=&([\hat{N},\ad]+\ad\hat{N})\m n>\nonumber\\
&=&(\ad+\ad\hat{N})\m n>\\
&=&\ad(1+n)\m n>=(n+1)\ad\m n>~.\nonumber
\end{eqnarray}

By a similar procedure, one can also obtain
\be
\hat{N}(a\m n>)=([\hat{N},a]+a\hat{N})\m n>=(n-1)a\m n>~.
\ee
The expression $(54)$ implies that one can consider the
ket $\ad \m n>$ as an eigenket of that number operator for which the eigenvalue
is raised by one unit. In physical terms, this means that an energy
quanta has been produced by the action of 
$\ad$ on the ket. This already expains the name of creation
operator. Similar comments with corresponding conclusion can be infered 
for the operator 
$a$, originating the name of annihilation operator (an energy quanta is 
eliminated from the system when this operator is put in action).\\
Moreover, eq. $(54)$ implies the proportionality of the kets $\ad\m n>$ and
$\m n+1>$:
\be
\ad\m n>=c\m n+1>~,
\ee

where $c$ is a constant that should be determined. Considering 
in addition 
\be
(\ad\m n>)^{\dd}=<n\m a=c^*<n+1\m~,
\ee
one can perform the following calculation 
\be
<n\m a( \ad\m n>)=c^*<n+1\m (c\m n+1>)
\ee
\be
<n\m a\ad \m n>=c^*c<n+1\m n+1>
\ee
\be
<n\m a\ad \m n>=\m c\m^2~.
\ee

But from the commutation relation for the operators $a$ and $\ad$ 
\be
[a,\ad]=a\ad-\ad a=a\ad-\hat{N}=1~,
\ee

we have
\be
a\ad=\hat{N}+1~.
\ee

Substituting in $(60)$, we get
\be
<n\m \hat{N}+1\m n>=<n\m n>+<n\m \hat{N}\m n>=n+1=\m c\m^2~.
\ee

Asking conventionally for a positive and real $c$, 
the following value is obtained
\be
c=\sqrt{n+1}~.
\ee

Consequently, we have 
\be
\ad \m n>=\sqrt{n+1}\m n+1>~.
\ee

For the annihilation operator, following the same procedure one can get 
the following relation 
\be
a\m n>=\sqrt{n}\m n-1>~.
\ee

Let us show now that the values of $n$ should be nonnegative integers. 
For this, we employ the positivity requirement for the norm, applying it to the
state vector $a\m n>$. The latter condition tells us that the interior product
of the vector with its adjunct 
($ (a\m n>)^\dd=<n\m \ad$) should always be nonnegative
\be
( <n\m \ad)\cdot(a\m n>)\geq 0~.
\ee

This relationship is nothing else but
\be
<n\m \ad a\m n>=<n\m \hat{N}\m n>=n \geq 0~.
\ee

Thus, $n$ cannot be negativ. It should be an integer since were it not
by applying iteratively the annihilation operator we would be lead to negative
values of $n$, which would be a contradiction to the previous statement.\\
It is possible to express the state $n$ $(\m n \rangle)$ 
directly as a function of the ground state 
$(\m 0\rangle)$ using the creation operator. Let us see how proceeds
this important iteration

\begin{eqnarray}
 \m 1\rangle=\ad \m 0\rangle \\
 \m 2\rangle=[\frac{\ad}{\sqrt{2}}]\m 1\rangle=[\frac{(\ad)^2}{\sqrt{2!}}]
\m 0\rangle \\
 \m 3\rangle=[\frac{\ad}{\sqrt{3}}]\m 2\rangle=
[ \frac{ (\ad)^3}{\sqrt{3!}}]\m 0\rangle 
\end{eqnarray}
\vdots
\begin{eqnarray}
 \m n\rangle=[ \frac{ (\ad)^n}{\sqrt{n!}}]\m 0\rangle ~.
\end{eqnarray}

One can also apply this method to get the eigenfunctions in the 
configuration space. To achieve this, we start with the ground state
\be
a\m 0\rangle=0~.
\ee
In the $x$ representation, we have
\be
\hat{ a}  
\Psi_0(x)=\sqrt{\frac{m\omega}{2\hbar}} (x+\frac{ip}{m\omega}) \Psi_0(x)=0~.
\ee
Recalling the form of the momentum operator in the $x$ representation,
we can obtain a differential equation for the wavefunction 
of the ground state. Moreover, introducing the definition
$x_0=\sqrt{\frac{\hbar}{m\omega}}$, we have
\be
(x+x_0^2\frac{d}{dx})\Psi_0=0~.
\ee
The latter equation can be readily solved, 
and normalizing (its integral from $-\infty$ to $\infty$ 
should be equal to unity), we obtain the wavefunction of the
ground state 
\be
\Psi_0(x)=(\frac{1}{\sqrt{ \sqrt{\pi}x_0}})e^{ -\frac{1}{2}(\frac{x}{x_0})^2}~.
\ee
The rest of the eigenfunctions, which describe the HO excited states, 
can be obtained employing the creation operator. The procedure is the following
\begin{eqnarray}
\Psi_1=\ad \Psi_0 =(\frac{1}{\sqrt{2}x_0})(x-x_0^2\frac{d}{dx})\Psi_0\\
\Psi_2=\frac{1}{\sqrt{2}}(\ad)^2\Psi_0=\frac{1}{\sqrt{2!}}(\frac{1}
{\sqrt{2}x_0})^2(x-x_0^2\frac{d}{dx})^2\Psi_0~.
\end{eqnarray}
By mathematical induction, one can show that
\be
\Psi_n=\frac{1}{\sqrt{ \sqrt{\pi}2^nn!}}\hspace{.2cm}
\frac{1}{x_0^{n+\frac{1}{2}}}
\hspace{.2cm}(x-x_0^2\frac{d}{dx})^n
\hspace{.2cm}e^{-\frac{1}{2}(\frac{x}{x_0})^2}~.
\ee


\section*{Time evolution of the oscillator}

In this section we shall illustrate on the HO example the way of working with
the Heisenberg representation in which the states are fixed in time 
and only the operators evolve. 
Thus, we shall consider the operators as functions of time and obtain explicitly
the time evolution of the HO position and momentum operators, $a$ and $\ad$,
respectively.
The Heisenberg equations of the motion for $p$ and $x$ are 
\begin{eqnarray}
\frac{d\hat{p}}{dt}&=&-\frac{\partial}{\partial\hat{x}}V({\bf \hat{x})}\\
\nonumber\\
\frac{d\hat{x}}{dt}&=&\frac{\hat{p}}{m}~.
\end{eqnarray}

Hence the equations of the motion for $x$ and $p$ in the HO case are
the following 
\begin{eqnarray}
\frac{d\hat{p}}{dt}&=&-m\omega^2\hat{x}\\
\nonumber\\
\frac{d\hat{x}}{dt}&=&\frac{\hat{p}}{m}~.
\end{eqnarray}

These are a pair of coupled equations, 
which are equivalent to a pair of {\em uncoupled} equations for the creation and 
annihilation operators. Explicitly, we have
\begin{eqnarray}
\frac{d a}{dt}&=&\sqrt{\frac{m\omega}{2\hbar}}\frac{d}{dt}(\hat{x}+\frac{i\hat{p}}{m\omega})\\
\nonumber\\
\frac{da}{dt}&=&\sqrt{\frac{m\omega}{2\hbar}}(\frac{d\hat{x}}{dt}+
\frac{i}{m\omega}\frac{d\hat{p}}{dt})~.
\end{eqnarray}
 
Substituting $(82)$ and $(83)$ in $(85)$, we get
\be
\frac{da}{dt}=\sqrt{\frac{m\omega}{2\hbar}}(\frac{\hat{p}}{m}-
i\omega\hat{x})=-i\omega a~.
\ee
Similarly, one can obtain a differential equation
for the creation operator
\be
\frac{d\ad}{dt}=i\omega\ad ~.
\ee
The differential evolution equations for the creation and annihilation
operators can be immediately integrated leading to the explicit evolution
of these operators as follows
\begin{eqnarray}
a(t)&=&a(0)e^{-i\omega t}\\
\ad (t)&=&\ad (0)e^{i\omega t}~.
\end{eqnarray}

It is worth noting based on these results and eqs. $(44)$ and
$(47)$ that both the Hamiltonian and the number operator are not 
time dependent.\\
Using the latter two results, we can obtain the position and momentum operators
as functions of time 
as far as they are expressed in terms of the
creation and annihilation operators 
\begin{eqnarray}
\hat{x}&=&\sqrt{\frac{\hbar}{2m\omega}}(a+\ad)\\
\hat{p}&=&i\sqrt{ \frac{m\hbar\omega}{2}}(\ad-a)~.
\end{eqnarray}

Substituting them, one gets 
\begin{eqnarray}
\hat{x}(t)&=&\hat{x}(0)\cos{\omega t}+\frac{\hat{p}(0)}{m\omega}\sin{\omega t}\\
\nonumber\\
\hat{p}(t)&=&-m\omega\hat{x}(0)\sin{\omega t}+\hat{p}(0)\cos{\omega t}~.
\end{eqnarray}

The time evolution of these operators is the same as for the classical 
equations of the motion.\\
Thus, we have shown here the explicit evolution form of the four HO basic
operators, and also we illustrated the effective way of working in the 
Heisenberg representation.


\section*{The 3D HO}

We commented on the 
importance in physics of the HO at the very beginning of our analysis of the 
quantum HO. 
If we will consider a 3D analog, we would be led to study a Taylor expansion in
three variables\footnote{It is possible to express the Taylor series in the
neighbourhood of ${\bf r_{0}}$ as an exponential 
operator
\bc
$e^{[ (x-x_o)+(y-y_o)+(z-z_o)]
(\frac{\partial}{\partial x}+\frac{\partial}{\partial y}+
\frac{\partial}{\partial z})} \hspace{.1cm} f({\bf r_o})~.$\\
\ec }
retaining the terms up to the second order, 
we get a quadratic form in the most general case. The problem at hand
in this approximation is not as simple as it might look from the examination
of the corresponding potential 

\be
V(x,y,z)=ax^2+by^2+cz^2+dxy+exz+fyz~.
\ee


There are however many systems with spherical symmetry or for which this
symmetry is sufficiently exact. 
\^{I}n acest caz:

\be
V(x,y,z)=K(x^2+y^2+z^2)~,
\ee

\noindent
which is equivalent to saying that the second unmixed partial derivatives 
have all the same value, denoted by
$K$ in our case). We can add that this is a good approximation
in the case in which the values of the mixed second partial derivatves
are small in comparison to the unmixed ones.\\
When these conditions are satisfied and the potential is given 
by $(95)$, we say that the system is a 
{\em 3D spherically symmetric HO}.\\
The Hamiltonian in this case is of the form

\be
\hat{H}=\frac{-\hbar^2}{2m}\bigtriangledown^2 + \frac{m\omega^2}{2}r^2~,
\ee

\noindent
where the Laplace operator is given in spherical coordinates
and $r$ is the spherical radial coordinate.\\
Since the potential is time independent the energy is conserved. In addition,
because of the spherical symmetry the orbital momentum is also conserved.  
having two conserved quantities, we may say that to each of it one can associate
a quantum number. 
Thus, we can assume that the eigenfunctions depend on two quantum numbers
(even though for this case we shall see that another one will occur). Taking care
of these comments, the equation of interest is 

\be
\hat{H}\Psi_{nl}=E_{nl}\Psi_{nl}~.
\ee


The Laplace operator in spherical coordinates reads

\be
\bigtriangledown^2
=\frac{\partial^2}{\partial r^2}+\frac{2}{r}\frac{\partial}{\partial r}
-\frac{\hat{L}^2}{\hbar^2r^2}
\ee

and can be also inferred from the known fact

\be
\hat{L}^2=-\hbar^2[ \frac{1}{\sin{\theta}}
\frac{\partial}{\partial\theta}
( \sin{\theta}\frac{\partial}{\partial\theta})
+\frac{1}{\sin{\theta}^2}\frac{\partial^2}{\partial\varphi^2}]~.
\ee


{\em The eigenfunctions of $\hat{L}^2$ are the spherical harmonics}, i.e.

\be
\hat{L}^2Y_{lm_{l}}(\theta,\varphi)=-\hbar^2l(l+1)Y_{lm_{l}}(\theta,\varphi)
\ee

The fact that the spherical harmonics `wear' the  quantum number
$m_{l}$ introduces it in the total wavefunction
$\Psi_{nlm_{l}}$.\\
In order to achieve the separation of the variables and functions, the following
substitution is proposed

\be
\Psi_{nlm_{l}}(r, \theta,\varphi)=\frac{R_{nl}(r)}{r}
Y_{lm_{l}}(\theta,\varphi)~.
\ee

Once this is plugged in the Schr\"odinger equation, the spatial part is separated
from the angular one; the latter is identified with an operator that is  
proportional to the square of the orbital momentum, for which the eigenfunctions
are the spherical harmonics, whereas for the spatial part the following 
equation is obtained

\be
R_{nl}^{\prime\prime}+(\frac{2mE_{nl}}{\hbar^2}
-\frac{m^2\omega^2}{\hbar^2}r^2-\frac{l(l+1)}{r^2})R_{nl}(r)=0~.
\ee


Using the definitions $(7)$ and $(8)$, the previous equation is precisely of the
form $(9)$, unless the angular momentum term, which is commonly known as the
unghiular, care \h n mod comun se cunoa\c{s}te ca 
{\em angular momentum barrier}

\be
R_{nl}^{\prime\prime}+(k^2-\lambda^2r^2-\frac{l(l+1)}{r^2})R_{nl}=0~.
\ee


\noindent
To solve this equation,
we shall start with its asymptotic analysis. If we shall consider
first $r\rightarrow\infty$, we notice that the orbital momentum term is
negligible, so that in this limit the asymptotic behavior is similar to that
of $(9)$, leading to

\be
R_{nl}(r)\sim\exp{\frac{-\lambda r^2}{2}}\hspace{2cm}\mbox{for}
\hspace{.3cm}\lim\hspace{.1cm}r\rightarrow\infty~.
\ee


If now we pass to the behavior close to zero, we can see that the dominant term 
is that of the orbital momentum, i.e., the differential equation
$(102)$ in this limit turns into

\be
R_{nl}^{\prime\prime}-\frac{l(l+1)}{r^2}R_{nl}=0~.
\ee


This is a differential equation of the Euler type 
\footnote{An equation of the Euler type has the form

\[x^n y^{(n)}(x)+x^{n-1} y^{(n-1)}(x)+\cdots+x y^{\prime}(x)+y(x)=0~.\]

Its solutions are of the type $x^{\alpha}$ that are plugged in the equation
obtaining a polynomial in $\alpha$.} , whose two independent solutions are
\be
R_{nl}(r)\sim \hspace{.2cm}r^{l+1}\hspace{.2cm}\mbox{or}
\hspace{.4cm}r^{-l}\hspace{2cm}\mbox{for}\hspace{.4cm}\lim\hspace{.1cm}r
\rightarrow 0~.
\ee

The previous arguments lead to proposing the substitution
\be
R_{nl}(r)=r^{l+1}\exp{\frac{-\lambda r^2}{2}}\phi(r)~.
\ee

One can also use another substitution
\be
R_{nl}(r)=r^{-l}\exp{\frac{-\lambda r^2}{2}}v(r)~,
\ee
which, however, produces the same solutions as 
$(107)$ (showing this is a helpful exercise).
Substituing $(107)$ in $(103)$, the following differential equation for 
$\phi$ is obtained
\be
\phi^{\prime\prime}+2(\frac{l+1}{r}-\lambda r)\phi^{\prime}
-[ \lambda (2l+3)-k^2]\phi=0~.
\ee

Using now the change of variable $w=\lambda r^2$, one gets
\be
w\phi^{\prime\prime}+(l+\frac{3}{2}-w)\phi^{\prime}-[ \frac{1}{2}(l+
\frac{3}{2})-\frac{\kappa}{2}]\phi=0~,
\ee
where $\kappa =\frac {k^2}{2\lambda}=\frac{E}{\hbar\omega}$ has been introduced. 
We see that we found again a differential equation
of the confluent hypergeometric type
having the solutions (see $(21)$ and $(22)$)
\be
\phi(r)=A\hspace{.2cm}_1F_1[\frac{1}{2}(l+\frac{3}{2}-\kappa);l+\frac{3}{2},
\lambda r^2]+B\hspace{.2cm}r^{-(2l+1)}
\hspace{.3cm}_1F_1[\frac{1}{2}(-l+\frac{1}{2}-\kappa);-l+\frac{1}{2},
\lambda r^2]~.
\ee
 
The second particular solution cannot be normalized because 
diverges strongly in zero. This forces one to take $B=0$, therefore
\be
\phi(r)=A\hspace{.2cm}_1F_1[\frac{1}{2}(l+\frac{3}{2}-\kappa);l+\frac{3}{2},
\lambda r^2]~.
\ee
Using the same arguments as in the 1D HO case,
that is, imposing a regular solution at infinity, leads to the truncation of the
series, which implies the quantization of the energy. 
The truncation is explicitly
\be
\frac{1}{2}(l+\frac{3}{2}-\kappa)=-n~,
\ee
\noindent where introducing $\kappa$ we get the energy spectrum 
\be
E_{nl}=\hbar\omega(2n+l+\frac{3}{2})~.
\ee

One can notice that for the 3D spherically symmetric HO there is a zero point 
energy $\frac{3}{2}\hbar\omega$.\\
The unnormalized eigenfunctions are
\be
\Psi_{nlm}(r,\theta,\varphi)
=r^{l}e^{\frac{-\lambda r^2}{2}}\hspace{.2cm}_1F_1(-n;l
+\frac{3}{2},\lambda r^2)\hspace{.1cm}Y_{lm}(\theta,\varphi)~.
\ee


\section*{{\huge 5P. Problems}}

\subsection*{Problem 5.1}

{\bf Determine the eigenvalues and eigenfunctions of the HO in the momentum
space}.

The quantum HO Hamiltonian reads
\[
\hat{H}=\frac{\hat{p}^2}{2m}+\frac{1}{2}m\omega^2\hat{x}^2~.
\]
In the momentum space, the operators  $\hat{x}$  
and $\hat{p}$ have the following form
\[
\hat{p}\rightarrow\hspace{.2cm}p
\]
\[
\hat{x}\rightarrow\hspace{.2cm}i\hbar\frac{\partial}{\partial p}~.
\]
Thus, the HO quantum Hamiltonian in the momentum representation is
\[
\hat{H}=\frac{p^2}{2m}-\frac{1}{2}m\omega^2\hbar^2\frac{d^2}{dp^2}~.
\]
We have to solve the eigenvalue problem  
(i.e., to get the eigenfunctions and the eigenvalues) given by $(5)$, 
which, with the previous Hamiltonian, turns into the following differential 
equation  
\be
\frac{d^2\Psi(p)}{dp^2}+( \frac{2E}{m\hbar^2\omega^2}-
\frac{p^2}{m^2\hbar^2\omega^2})\Psi(p)=0~.
\ee
One can see that this equation is identical, up to some constants,  
with the differential equation in the space of configurations (eq.~$(6)$ ). 
Just to show another way of solving it, we define two parameters,
which are analogous to those in $(7)$ and $(8)$
\be
k^2=\frac{2E}{m\hbar^2\omega^2} \hspace{1cm}\lambda=\frac{1}{m\hbar\omega}~.
\ee
With these definitions, we get the differential eq. $(9)$ and therefore the 
solution sought for
(after performing the asymptotic analysis) is of the form
\be
\Psi(y)=e^{-\frac{1}{2}y}\phi (y)~,
\ee
where $y=\lambda p^2$ and $\lambda$ is defined in $(117)$.
Substitute $(118)$ in $(116)$ taking care to put $(118)$ in the 
variable $p$. One gets a differential equation in the variable $\phi$ 
\be
\frac{d^2\phi(p)}{dp^2}-2\lambda p\frac{d\phi (p)}{dp}+(k^2-\lambda)\phi (p)=0~.
\ee
We shall now make the change of variable $u=\sqrt{\lambda}p$ 
that finally leads us to the Hermite equation 
\be
\frac{d^2\phi (u)}{du^2}-2u\frac{d\phi (u)}{du}+2n\phi(u)=0~,
\ee
where $n$ is a nonnegative integer and where we have put
\[
\frac{k^2}{\lambda}-1=2n~.
\]
From here and the definitions given in $(117)$  
one can easily conclude that the eigenvalues are given by
\[
E_n=\hbar\omega(n+\frac{1}{2})~.
\]
The solutions for $(120)$ are the Hermite polynomials $\phi(u)=H_n(u)$ and
the unnormalized eigenfunctions are
\[
\Psi(p)=A e^{-\frac{\lambda}{2}p^2}H_n(\sqrt{\lambda}p)~.
\]

\vspace{1mm}

\subsection*{Problem 5.2}

{\bf Prove that the Hermite polynomials can be expressed in the following
integral representation 
\be
H_n(x)=\frac{2^n}{\sqrt{\pi}}\int_{-\infty}^{\infty} (x+iy)^n e^{-y^2}dy~.
\ee
}

This representation of Hermite polynomials is not really usual, though it can
prove useful in many cases. In order to accomplish the proof, we shall expand
expand the integral and next prove that what we've got is identical
to the series expansion of the Hermite polynomials that reads
\be
\sum_{k=0}^{[\frac{n}{2}]} \frac{ (-1)^k n!}{(n-2k)!k!}(2x)^{n-2k}~,
\ee

\noindent
where the symbol $[c]$, indicating where the series terminates,
denotes the greatest integer less or equal to $c$.\\
The first thing we shall do is to expand the 
binomial in the integral by using the well-known binomial theorem 
\[
(x+y)^n = \sum_{m=0}^n \frac{n!}{(n-m)!m!}x^{n-m}y^m~.
\]
Thus
\be
(x+iy)^n= \sum_{m=0}^n \frac{n!}{(n-m)!m!}i^mx^{n-m}y^m~,
\ee

which plugged in the integral leads to
\be
\frac{2^n}{\sqrt{\pi}}\sum_{m=0}^n \frac{n!}{(n-m)!m!}i^m x^{n-m}
\int_{-\infty}^{\infty} y^m e^{-y^2}dy~.
\ee

Inspecting of the integrand we realize that the integral is not zero
when $m$ is even, whereas it is zero when $m$ is odd. 
Using the even notation $m=2k$, we get

\be
\frac{2^n}{\sqrt{\pi}}
\sum_{k=0}^{[\frac{n}{2}]}\frac{n!}{(n-2k)!(2k)!}i^{2k}x^{n-2k}
\hspace{.2cm}2\int_{0}^{\infty} y^{2k}e^{-y^2}dy~.
\ee

Under the change of variable $u=y^2$, the integral turns into a gamma function 
\be
\frac{2^n}{\sqrt{\pi}}\sum_{k=0}^{[\frac{n}{2}]}
\frac{n!}{(n-2k)!(2k)!}i^{2k}x^{n-2k}\int_{0}^{\infty}u^{k-\frac{1}{2}}e^{-u}du~,
\ee

\noindent
more precisely $\Gamma(k+\frac{1}{2})$, which can be expressed in terms of
factorials ( of course for $k$ an integer) 
\[
\Gamma(k+\frac{1}{2})=\frac{(2k)!}{2^{2k}k!}\sqrt{\pi}~.
\]
Plugging this expression in the sum and using 
$i^{2k}=(-1)^k$, one gets

\be
\sum_{k=0}^{[\frac{n}{2}]} \frac{ (-1)^k n!}{(n-2k)!k!}(2x)^{n-2k}~,
\ee

\noindent
which is identical to $(122)$, hence completing the proof.

\vspace{1mm}

\subsection*{Problem 5.3}

{\bf Show that Heisenberg's uncertainty relation  
is satisfied by doing the calculation using the HO eigenfunctions}~.

We have to show that for any $\Psi_n$, we have 
\be
<(\Delta p)^2(\Delta x)^2>\hspace{.3cm}\geq \frac{\hbar^2}{4}~,
\ee
where the  notation $<>$ means the mean value.\\
We shall separately calculate $<(\Delta p)^2>$ and $<(\Delta x)^2>$, 
where each of these expressions is

\[<(\Delta p)^2>=< (p-<p>)^2 >=< p^2 - 2p<p>+<p>^2>=<p^2>-<p>^2~,\]

\[<(\Delta x)^2>=< (x-<x>)^2 >=< x^2 - 2x<x>+<x>^2>=<x^2>-<x>^2~. \]

First of all, we shall prove that both the mean of $x$ as well as of $p$ 
are zero. For the mean of $x$, we have

\[<x>=\int_{-\infty}^{\infty} x[\Psi_n(x)]^2 dx~.\]

This integral is zero because the integrand is odd. 
Thus
\be
<x>=0~.
\ee

The same argument holds for the mean of $p$, 
if we do the calculation in the momentum space, employing the functions
obtained in problem 1. It is sufficient to notice that
the functional form is the same (only the symbol does change). 
Thus
\be
<p>=0~.
\ee

Let us now calculate the mean of $x^2$. We shall use the virial theorem  
\footnote{We recall that the virial theorem in quantum mechanics 
asserts that
\[ 2<T>=<{\bf r}\cdot\bigtriangledown V({\bf r})>~.\]
For a potential of the form $V=\lambda x^n$, the virial theorem gives
\[
2<T>=n<V>~,
\]
where $T$ is the kinetic energy and $V$ is the potential energy.}. We first
notice that
\[
<V>=\frac{1}{2}m\omega^2 <x^2>~.
\]
Therefore, it is possible to relate the mean of $x^2$ directly to the mean of
the potential for this case (implying the usage of the virial theorem).
\be
<x^2>=\frac{2}{m\omega^2}<V>~.
\ee

We also need the total energy
\[
<H>=<T>+<V>~,
\]

for which again one can make use of the virial theorem (for $n=2$)
\be
<H>=2<V>~.
\ee

Thus, we obtain
\be
<x^2>=\frac{<H>}{m\omega^2}=\frac{\hbar\omega(n+\frac{1}{2})}{m\omega^2}
\ee

\be
<x^2>=\frac{\hbar}{m\omega}( n+\frac{1}{2})~.
\ee

Similarly, the mean of $p^2$ can be readily calculated
\be
<p^2>=2m<\frac{p^2}{2m}>=2m<T>=m<H>=m\hbar\omega(n+\frac{1}{2})~.
\ee

Employing $(133)$ and $(135)$, we have
\be
<(\Delta p)^2(\Delta x)^2>=(n+\frac{1}{2})^2\hbar^2~.
\ee

Based on this result, we come to the conclusion that in the HO stationary states
that actually have not been directly used, 
Heisenberg's uncertainty relation is satisfied and it is at the minimum for the
ground state, $n=0$.

\vspace{1mm}

\subsection*{Problem 5.4}

{\bf Obtain the matrix elements of
the operators $a$, $\ad$, $\hat{x}$, and $\hat{p}$}.

Let us first find the matrix elements for the creation and annihilation
operators, which are very helpful for all the other operators.\\
We shall use the relatinships $(65)$ and $(66)$, leading to
\be
<m \m a\m n>=\sqrt{n}<m\m n-1>=\sqrt{n}\delta_{m,n-1}~.
\ee

Similarly for the creation operator we have the result
\be
<m\m \ad \m n>=\sqrt{n+1}<m\m n+1>=\sqrt{n+1}\delta_{m,n+1}~.
\ee

Let us proceed now with the calculation of the matrix elements of the
position operator.
For this, let us express this operator
in terms of creation and annihilation operators.
Using the definitions $(39)$ and $(40)$, one can immediately prove that the 
position operator is given by
\be
\hat{x}=\sqrt{ \frac{\hbar}{2m\omega}}(a+\ad)~.
\ee

Employing this result, the matrix elements of the operator $\hat{x}$ 
can be readily calculated 
\begin{eqnarray}
<m\m \hat{x}\m n>&=&<m\m \sqrt{ \frac{\hbar}{2m\omega}}(a+\ad)\m n>\nonumber\\
&=&\sqrt{ \frac{\hbar}{2m\omega}}[\sqrt{n}\delta_{m,n-1}+
\sqrt{n+1}\delta_{m,n+1}]~.
\end{eqnarray}
Following the same procedure we can calculate the matrix elements
of the momentum operator, just by taking into account that $\hat{p}$ is given 
in terms of the creation and annihilation operators as follows
\be
\hat{p}= i\sqrt{ \frac{m\hbar\omega}{2}}(\ad -a)~.
\ee

This leads us to
\be
<m\m\hat{p}\m n>= i\sqrt{ \frac{m\hbar\omega}{2}}[\sqrt{n+1}\delta_{m,n+1}
-\sqrt{n}\delta_{m,n-1}]~.
\ee

One can realize the ease of the calculations when the matrix elements of the 
creation and annihilation operators are used. 
Finally, we remark on the nondiagonality of the obtained matrix elements.
This is not so much of a surprise because the employed representation 
is that of the number operator and none of the four operators do not commute
with it.

\vspace{1mm}

\subsection*{Problem 5.5}

{\bf Find the mean values of 
$\hat{x}^2$ and $\hat{p}^2$ for the1D HO and use them to calculate the mean
(expectation) values of the kinetic and potential energies. Compare the 
result with the virial theorem.}

First of all, let us obtain the mean value of $\hat{x}^2$. For this, we use
eq. $(139)$ that leads us to
\be
\hat{x}^2 = \frac{\hbar}{2m\omega} (a^2 + (\ad) ^2 +\ad a+a\ad)~.
\ee
Recall that the creation and annihilation operators do not commute.
Based on $(143)$, we can calculate the mean value of $\hat{x}^2$ 
\begin{eqnarray}
<\hat{x}^2>&=&<n\m \hat{x}^2\m n>\nonumber\\
&=&\frac{\hbar}{2m\omega}[ \sqrt{n(n-1)}\delta_{n,n-2}+\sqrt{(n+1)(n+2)}
\delta_{n,n+2}\nonumber\\
&+& \hspace{.2cm}n\hspace{.1cm}\delta_{n,n}\hspace{.2cm}+
\hspace{.2cm}(n+1)\hspace{.1cm}\delta_{n,n}]~,
\end{eqnarray}

which shows that
\be
<\hat{x}^2>=<n\m \hat{x}^2\m n> = \frac{\hbar}{2m\omega}(2n+1)~.
\ee
In order to calculate the mean value of $\hat{p}^2$ we use $(141)$ that helps
us to express this
operator in terms of the creation and annihilation operators 
\be
\hat{p}^2 = -\frac{m\hbar\omega}{2}(a^2+(\ad)^2-a\ad-\ad a)~.
\ee
This leads us to
\be
<\hat{p}^2>=<n\m \hat{p}^2\m n>=\frac{m\hbar\omega}{2}(2n+1)~.
\ee
The latter result practically gives us the mean kinetic energy 
\be
<\hat{T}>=<\frac{\hat{p}^2}{2m}>=\frac{1}{2m}<\hat{p}^2>
=\frac{\hbar\omega}{4}(2n+1)~.
\ee

On the other hand, the mean value of the potential energy
\be
<\hat{V}>=<\frac{1}{2}m\omega^2 \hat{x}^2>=\frac{1}{2}m\omega^2 <\hat{x}^2>
=\frac{\hbar\omega}{4}(2n+1)~,
\ee
where $(145)$ has been used.

We can see that these mean values are equal for any $n$, 
which confirms the quantum virial theorem, telling us that for a quadratic 
(HO) potential,
the mean values of the kinetic and potential energies should be equal and 
therefore be half of the mean value of the total energy.

\vspace{1mm}

\newpage
\def\bi{bigskip}
\def\noi{noindent}
\def\ii{\'{\i}}
\begin{center}{\huge 6. THE HYDROGEN ATOM}
\end{center}
\section*{Introduction} 
In this chapter we shall study the hydrogen atom by solving the  
time-independent
Schr\"odinger equation for the potential due to two charged particles,  
the electron and the proton, and the  
Laplacian operator in spherical coordinates. 
From the mathematical viewpoint, the method of separation of variables 
will be employed, and a physical interpretation of the wavefunction as solution
of the Schr\"odinger equation in this important case will be provided, 
together with the interpretation of the quantum numbers and of the 
probability densities.\\
\setcounter{equation}{0}
The very small spatial scale of the hydrogen atom is a clue that the related 
physical phenomena enter the domain of applicability of the quantum mechanics, 
for which the atomic processes have been a successful area since the early
days of the quantum approaches.
Quantum mechanics, as any other theoretical framework, gives relationships 
between observable quantities.
Since the uncertainty principle leads to a substantial change
in the understanding of observables at the conceptual level, it is important 
to have a clear idea on the notion of atomic observable.
As a matter of fact, the real quantities on which 
quantum mechanics offers explicit answers and connections
are always probabilites.
Instead of saying, for example, that the radius of the electron orbit in the 
fundamental state of the hydrogen atom is always 
$5.3 \times 10^{-11}$ m, quantum mechanics asserts
that this is a truly mean radius (not in the measurable sense). Thus,
if one performs an appropriate 
experiment, one gets, precisely as in the case of
the common arrangement of macroscopic detectors probing macroscopic properties
of the matter, random values around the mean value
$5.3 \times 10^{-11}$ m. In other words, from the viewpoint of the 
experimental errors there is no essential difference 
with regard to the classical physics. The fundamental difference is in the 
procedure of calculating the mean values within the theoretical framework. 

As is known, for performing quantum-mechanical calculations, one needs a 
corresponding wave function 
$\Psi$. Although $\Psi$ has no direct physical interpretation,
the square modulus $\mid \Psi \mid^{2}$ calculated at an arbitrary position
and given moment is proportional to the probability to find the 
particle in the infinitesimal neighbourhood of that point at the given time.
The purpose of quantum mechanics is to determine  
$\Psi$ for a specified particle in the prepared experimental conditions. 

Before proceeding with the rigorous approaches of getting $\Psi$ for the 
hydrogen electron, we will argue on several general requirements regarding the
wave function.
First, the integral of $\mid \Psi 
\mid^{2}$ over all space should be finite if we really want to deal with 
a localizable electron. In addition, if
\begin{equation} 
\int_{-\infty}^{\infty} \mid \Psi \mid^{2} dV = 0~,
\end{equation}
then the particle does not exist. $\mid \Psi \mid^{2}$ cannot be negative or
complex because of simple mathematical reasons.
In general, it is convenient to identify $\mid \Psi \mid^{2}$ with the 
probability P not just the proportionality.
In order that $\mid \Psi \mid^{2}$ be equal to P one imposes
\begin{equation} 
\int_{-\infty}^{\infty}\mid \Psi \mid^{2} dV = 1~,
\end{equation}
because 
\begin{equation} 
\int_{-\infty}^{\infty}{\rm P} dV = 1
\end{equation}
is the mathematical way of saying that the particle exists
at a point in space at any given moment.
A wave function respecting eq. 2 is said to be normalized. 
Besides this, $\Psi$ should be single valued, because 
P has a unique value at a given point and given time.
Another condition is that $\Psi$ and its partial
first derivatives  
$\frac{\partial \Psi}{\partial x}$, $\frac{\partial 
\Psi}{\partial y}$, $\frac{\partial \Psi}{\partial z}$ should be continuous
at any arbitrary point. 

The Schr\"odinger equation is considered as the fundamental equation
of nonrelativistic quantum mechanics in the same sense in which
Newton's force law is the fundamental equation of motion of newtonian  mechanics.
Notice however that we have now a wave equation for a function $\Psi$ which is
not directly measurable.

Once the potential energy is given, one can solve the 
Schr\"odinger equation for
$\Psi$, implying the knowledge of the probability density
$\mid \Psi \mid^{2}$ as a function of $x,y,z,t$.
In many cases of interest, the potential energy does not depend on time. Then, 
the Schr\"odinger equation simplifies considerably.
Notice, for example, that for a 1D free particle the wave function can be 
written
\begin{eqnarray} 
\Psi(x,t) & = &  Ae^{(-i/\hbar)(Et - px)} \nonumber \\
& = & Ae^{-(iE/\hbar)t}e^{(ip/\hbar)x} \nonumber\\
& = & \psi(x) e^{-(iE/\hbar)t}~,
\end{eqnarray}
i.e., $\Psi(x,t)$ is the product of a time-dependent phase  
$e^{-(iE/\hbar)t}$ and a stationary wave function $\psi(x)$. 

In the general case, the stationary Schr\"odinger equation can be solved, under 
the aforementioned requirements, only for certain values of the energy E. 
This is not a mathematical difficulty, but merely a fundamental physical
feature. 
To solve the Schr\"odinger equation for a given system means to get the wave
function $\psi$, as a solution for which certain physical boundary condition
hold and, in addition, as already mentioned, it is continuous together with its
first derivative everywhere in space, is finite, and single valued.
Thus, the quantization of energy occurs as a natural theoretical element in 
wave mechanics, whereas in practice as a universal phenomenon,  
characteristic for all stable microscopic systems.

\section*{Schr\"odinger equation for the hydrogen atom}
In this section, we shall apply the Schr\"odinger equation to the hydrogen  
atom, about which one knows that it is formed of a positive nucleus/proton
of charge +$e$ and an electron of charge -$e$. The latter, being
1836 times smaller in mass than the proton, is by far more dynamic. 

If the interaction between two particles is of the type $u(r)=u ( \mid 
\vec r_{1} - \vec r_{2} \mid)$, the problem of the motion 
is reduced both classically and quantum to the motion of 
a single particle in a field of spherical symmetry. 
Indeed, the Lagrangian
\begin{equation} 
L = \frac{1}{2} m_{1} \dot { \vec r_{1}^{2}} + \frac{1}{2} m_{2} \dot {\vec
r_{2}^{2} } - u ( \mid \vec r_{1} - \vec r_{2} \mid)
\end{equation}
is transformed, using 
\begin{equation} 
\vec r = \vec r_{1} - \vec r_{2}
\end{equation}
and
\begin{equation} 
\vec R = \frac{m_{1} \vec r_{1} + m_{2} \vec r_{2}}{m_{1} + m_{2}}~,
\end{equation}
in the Lagrangian
\begin{equation} 
L = \frac{1}{2} M \dot { \vec R^{2}} + \frac{1}{2} \mu \dot {\vec
r^{2} } - u (r)~,
\end{equation}
where
\begin{equation} 
M = m_{1} + m_{2}
\end{equation}
and
\begin{equation} 
\mu =\frac{m_{1} m_{2}}{m_{1} + m_{2}}~.
\end{equation}

On the other hand, the momentum is introduced through the  
Lagrange formula
\begin{equation} 
\vec P = \frac{\partial L}{\partial \dot { \vec R}} = M \dot { \vec R}
\end{equation}
and
\begin{equation} 
\vec p = \frac{\partial L}{\partial \dot { \vec r}} = m \dot { \vec r}~,
\end{equation}
that allows to write the classical Hamilton function in the form 
\begin{equation} 
H = \frac{P^{2}}{2M} + \frac{p^{2}}{2m} + u(r)~.
\end{equation}

Thus, one can obtain the hamiltonian operator for the corresponding quantum
problem with commutators of the type
\begin{equation} 
[P_{i},P_{k}] = -i \hbar \delta_{ik}
\end{equation}
and
\begin{equation} 
[p_{i},p_{k}] = -i \hbar \delta_{ik}~.
\end{equation}
These commutators implies a Hamiltonian operator of the form 
\begin{equation} 
\hat H = -\frac{\hbar^{2}}{2M}\nabla_{R}^{2} - 
\frac{\hbar^{2}}{2m}\nabla_{r}^{2} + u(r)~,
\end{equation}
which is fundamental for the study of the hydrogen atom
by means of the stationary Schr\"odinger equation
\begin{equation} 
\hat H \psi = E \psi ~.
\end{equation}
This form does not include relativistic effects, i.e., electron velocities
close to the velocity of light in vacuum.


The potential energy $u(r)$ is the electrostatic one 
\begin{equation} 
u = -\frac{e^{2}}{4\pi \epsilon _{0} r}
\end{equation}

There are two possibilities. The first is to express $u$ as a function
of the cartesian coordinates
$x,y,z$, substituing $r$ by $\sqrt{x^{2}+y^{2}+z^{2}}$. The second
is to write the Schr\"odinger equation in spherical polar coordinates 
$r,\theta,\phi$. Because of the obvious spherical symmetry of this case, 
we shall deal with the latter approach, which leads to considerable 
mathematical simplifications.

In spherical coordinates, the Schr\"odinger equation reads
\begin{equation} 
\frac{1}{r^{2}} \frac{\partial}{\partial r}\left(r^{2} \frac{\partial 
\psi}{\partial r}
\right) + \frac{1}{r^{2} \sin\theta} \frac{\partial}{\partial 
\theta} \left(\sin\theta \frac{\partial \psi}{\partial \theta}\right) + 
\frac{1}{r^{2}\sin^{2}\theta} \frac{\partial^{2} \psi}{\partial \phi^{2}} 
+ \frac{2m}{\hbar^{2}}(E - u)\psi = 0
\end{equation}
Substituing (18), and multiplying  the whole equation
by $r^{2}\sin^{2}\theta$, one gets
\begin{equation} 
\sin^{2}\theta \frac{\partial}{\partial r}\left(r^{2}
\frac{\partial \psi}{\partial r}\right) + \sin\theta \frac{\partial}{\partial
\theta}\left(\sin\theta \frac{\partial \psi}{\partial \theta}\right) +
\frac{\partial^{2} \psi}{\partial \phi^{2}} +
\frac{2mr^{2}\sin^{2}\theta}{\hbar^{2}} \left(\frac{e^{2}}{4\pi 
\epsilon_{0}r} + E\right)\psi = 0~. 
\end{equation}
This equation is a partial differential equation for the electron
wavefunction $\psi(r,\theta,\phi)$ `within' the atomic hydrogen. 
Together with the various conditions that the wavefunction
$\psi(r,\theta,\phi)$ should fulfill 
[for example, $\psi(r,\theta,\phi)$ should have a unique value
at any spatial point ($r,\theta,\phi$)],
this equation specifies in a complete manner the behavior of the hydrogen
electron. To see the explicit behavior, we shall solve eq. 20 for
$\psi(r,\theta,\phi)$ and we shall interpret appropriately  the obtained
results.

\section*{Separation of variables in spherical coordinates}
The real usefulness of writing the hydrogen Schr\"odinger equation 
in spherical coordinates consists in the easy way of achieving the separation
procedure in three independent equations, each of them being one-dimensional.  
The separation procedure is to seek the solutions for which the wavefunction
$\psi(r, \theta, \phi)$ has the form of a product of three functions, each
of one of the three spherical variables, namely $R(r)$, 
depending only on $r$; 
$\Theta(\theta)$ depending only on $\theta$, and $\Phi(\phi)$ that depends 
only on $\phi$. This is quite similar to the separation of the 
Laplace equation. Thus
\begin{equation} 
\psi(r, \theta, \phi) = R(r)\Theta(\theta)\Phi(\phi)~.
\end{equation}
The $R(r)$ function describes the differential variation of the electron 
wavefunction 
$\psi$ along the vector radius coming out from the nucleus, 
with $\theta$ and $\phi$ assumed to be constant. The differential variation
of $\psi$ with the polar angle $\theta$ along a meridian of an arbitrary sphere 
centered in the nucleus is described only by the function 
$\Theta(\theta)$ for constant $r$ and $\phi$. Finally, the function 
$\Phi(\phi)$ describes how $\psi$ varies with the azimuthal angle 
$\phi$ along a parallel of an arbitrary sphere centered at the nucleus, 
under the conditions that $r$ and $\theta$ are kept constant.

Using $\psi=R\Theta\Phi$, one can see that
\begin{equation} 
\frac{\partial \psi}{\partial r} = \Theta \Phi \frac{d 
R}{d r}~,  
\end{equation}
\begin{equation} 
\frac{\partial \psi}{\partial \theta} = R\Phi \frac{d
\Theta}{d \theta}~, 
\end{equation} 
\begin{equation} 
\frac{\partial \psi}{\partial \phi} = R\Theta \frac{d
\Phi}{d\phi }~.  
\end{equation}
Obviously, the same type of formulas are maintained for 
the unmixed higher-order derivatives.
Subtituting them in eq. 20, and after deviding by  
$R\Theta \Phi$, we get
\begin{equation} 
\frac{\sin^{2}\theta}{R} \frac{d}{d r}\left(r^{2} \frac{d
R}{d r}\right)+\frac{\sin\theta}{\Theta} \frac{d}{d 
\theta}\left(\sin\theta
\frac{d \Theta}{d \theta}\right)+\frac{1}{\Phi}  
\frac{d^{2} \Phi}{d \phi^{2}} +
\frac{2mr^{2}\sin^{2}\theta}{\hbar^{2}} \left(\frac{e^{2}}{4\pi 
\epsilon_{0}r} + E\right) = 0~.
\end{equation}
The third term of this equation is a function of the angle
$\phi$ only, while the other two terms are functions of $r$ and $\theta$. 
We rewrite now the previous equation in the form
\begin{equation} 
\frac{\sin^{2}\theta}{R} \frac{\partial}{\partial r}\left(r^{2} \frac{\partial
R}{\partial r}\right)+\frac{\sin\theta}{\Theta} \frac{\partial}{\partial
\theta}\left(\sin\theta
\frac{\partial \Theta}{\partial \theta}\right)+
\frac{2mr^{2}\sin^{2}\theta}{\hbar^{2}} \left(\frac{e^{2}}{4\pi 
\epsilon_{0}r} +
E\right) = -\frac{1}{\Phi}\frac{\partial^{2} \Phi}{\partial \phi^{2}}~.
\end{equation}
This equation can be correct only if the two sides are equal to the same
constant, because they are functions of different variables.
It is convenient to denote this (separation) constant by $m_{l}^{2}$.  
The differential equation for the $\Phi$ function is
\begin{equation}  
-\frac{1}{\Phi}\frac{\partial^{2} \Phi}{\partial \phi^{2}} = m_{l}^{2}~.
\end{equation}
If one substitutes $m_{l}^{2}$ in the right hand side of eq. 26 and 
devides the resulting equation by $\sin^{2}\theta$, after regrouping
the terms, the fllowing result is obtained
\begin{equation} 
\frac{1}{R} \frac{d}{d r}\left(r^{2} \frac{d
R}{d r}\right) + 
\frac{2mr^{2}}{\hbar^{2}} \left(\frac{e^{2}}{4\pi \epsilon_{0}r} 
+ E\right) = \frac{m_{l}^{2}}{\sin^{2}\theta} - \frac{1}{\Theta \sin\theta} 
\frac{d}{d\theta}\left(\sin\theta\frac{d 
\Theta}{d \theta}\right)~.
\end{equation}
Once again, we end up with an equation in which different variables occur
in the two sides, thus forcing at equating of both sides to the same constant.  
For reasons that will become clear later on, we shall denote this
constant by $l(l+1)$. The equations for the functions $\Theta(\theta)$ and
$R(r)$ reads
\begin{equation} 
\frac{m_{l}^{2}}{\sin^{2}\theta} - \frac{1}{\Theta 
\sin\theta}\frac{d}{d\theta}\left(sin\theta \frac{d\Theta}{d\theta}\right) 
= l(l +1)
\end{equation}
and
\begin{equation} 
\frac{1}{R}\frac{d}{dr}\left(r^{2}\frac{dR}{dr}\right) + 
\frac{2mr^{2}}{\hbar^{2}}\left(\frac{e^{2}}{4\pi \epsilon_{0}r} + 
E\right) = l(l+1)~.
\end{equation}
The equations 27, 29 and 30 are usually written in the form
\begin{equation} 
\frac{d^{2}\Phi}{d\phi^{2}} + m_{l}^{2}\Phi = 0~,
\end{equation}
\begin{equation} 
\frac{1}{\sin\theta}\frac{d}{d\theta}\left(\sin\theta 
\frac{d\Theta}{d\theta}\right) + 
\left[l(l+1)-\frac{m_{l}^{2}}{\sin^{2}\theta}\right]\Theta = 0~,
\end{equation}
\begin{equation} 
\frac{1}{r^{2}}\frac{d}{dr}\left(r^{2}\frac{dR}{dr}\right) + 
\left[\frac{2m}{\hbar^{2}}\left(\frac{e^{2}}{4\pi \epsilon_{0}r} + 
E\right) - \frac{l(l+1)}{r^{2}}\right]R = 0~.
\end{equation}

Each of these equations is an ordinary differential equation
for a function of a single variable. In this way, the  
Schr\"odinger equation for the hydrogen electron, 
which initially was a partial differential equation for a function  
$\psi$ of three variables, got a simple form of three 1D ordinary differential
equations for unknown functions of one variable.

\section*{Interpreting the separation constants: the quantum numbers}

\subsection*{The solution for the azimuthal part}
Eq. 31 is readily solved leading to the following solution
\begin{equation} 
\Phi(\phi) = A_{\phi}e^{im_{l}\phi}~,
\end{equation}
where $A_{\phi}$ is the integration constant. One of the conditions 
that any wavefunctions should fulfill is to have a unique 
value for any point in space. This applies to
$\Phi$ as a component of the full wavefunction $\psi$.
One should notice 
that\ $\phi$ and $\phi + 2\pi$ are identical in the same meridional plane. 
Therefore, one should have $\Phi(\phi)= \Phi(\phi + 
2\pi)$, i.e., $A_{\phi}e^{im_{l}\phi} = A_{\phi}e^{im_{l}(\phi + 2\pi)}$. This 
can be fulfilled only if $m_{l}$ is zero or a positiv or negative integer
$(\pm 1, \pm 2, \pm 3,...)$. $m_{l}$ is known as the magnetic
quantum number of the atomic electron and is related to the direction of the
projection of
the orbital momentum $L_{z}$. It comes into play whenever the effects of axial
magnetic fields on the electron may show up. 
There is also a deep connection between $m_{l}$ and the orbital
quantum number
$l$, which in turn determines the modulus of the orbital momentum of the 
electron. 

The interpretation of the orbital number $l$ does not miss some problems.
Let us examine eq. 33 that corresponds to the radial wavefunction $R(r)$. 
This equation rules only the radial motion of the electron, i.e., with the 
relative distance with respect to the nucleus along some guiding ellipses. 
However, the total energy of the electron $E$ is also present. 
This energy includes the kineticelctron energy in its orbital motion that is 
not related to the radial motion. This contradiction can be eliminated by the
following argument. The kinetic energy $T$ has two parts:
$T_{radial}$ due to the radial oscillatory motion and
$T_{orbital}$, which is due to the closed orbital motion. 
The potential energy $V$ of the electron is the electrostatic energy.  
Therefore, its total energy is
\begin{equation}  
E = T_{radial} + T_{orbital} - \frac{e^{2}}{4\pi \epsilon_{0}r}~.
\end{equation}
Substituting this expression of $E$ in eq. 33 we get with some regrouping
of the terms
\begin{equation}  
\frac{1}{r^{2}}\frac{d}{dr}\left(r^{2}\frac{dR}{dr}\right) + 
\frac{2m}{\hbar^{2}}\left[T_{radial} + T_{orbital} - 
\frac{\hbar^{2}l(l+1)}{2mr^{2}}\right]R=0~.
\end{equation}
If the last two terms in parentheses compansates between themselves, we get
a differential equation for the pure radial motion. 
Thus, we impose the condition
\begin{equation}  
T_{orbital} = \frac{\hbar^{2}l(l+1)}{2mr^{2}}~.
\end{equation}
However, the orbital kinetic energy of the electron is
\begin{equation}  
T_{orbital} = \frac{1}{2}mv_{orbital}^{2}
\end{equation}
and since the orbital momentum of the electron $L$ is
\begin{equation}  
L = mv_{orbital}r~,
\end{equation}
we can express the orbital kinetic energy in the form 
\begin{equation}  
T_{orbital} = \frac{L^{2}}{2mr^{2}}~.
\end{equation}
Therefore, we have
\begin{equation}  
\frac{L^{2}}{2mr^{2}} = \frac{\hbar^{2}l(l+1)}{2mr^{2}}
\end{equation}
and consequently
\begin{equation}  
L = \sqrt{l(l+1)}\hbar~.
\end{equation}
The interpretation of this result is that since the orbital quantum number 
$l$ is constrained to take the values $l=0,1,2,...,(n-1)$, 
the electron can only have orbital momenta $L$ specified by means of
eq. 42. As in the case of the total energy $E$, the angular momentum is 
conserved and gets quantized. Its natural unit in quantum mechanics is  
$\hbar=h/2\pi=1.054 \times 10^{-34}$ J.s. 

In the macroscopic planetary motion (putting aside the many-body features), 
the orbital quantum number 
is so large that any direct experimental detection is impossible.
For example, an electron  
with $l= 2$ has an angular momentum 
$L=2.6 \times 10^{-34}$ J.s., whereas the terrestrial angular momentum  
is $2.7 \times 10^{40}$ J.s.!

A common notation for the angular momentum states is by means of the letter 
$s$ for 
$l=0$, $p$ for $l=1$, $d$ for $l=2$, and so on.
This alphabetic code comes from the empirical spectroscopic 
classification in terms of the so-called series, which was in use before the 
advent of quantum mechanics.

The combination of the principal quantum number with the latter 
corresponding to the angular momentum is another frequently used notation
in atomic and molecular physics.. 
For example, a state for which $n=2$ and $l=0$ is a state  
$2s$, while a state $n=4$ and $l=2$ is a state $4d$.

On the other hand, for the interpretation of the magnetic quantum number, 
we shall take into account, as we did for the linear momentum, 
that the orbital momentum is a vector operator and therefore one has to specify
its direction, sense, and modulus. 
$L$, being a vector product, is perpendicular on the plane of rotation. The 
geometric
rules of the vectorial products still hold, in particular the rule of the right
hand: its direction and sense are given by the right thumb whenever the 
other four fingers point at the direction of rotation.

But what significance can be associated to
a direction and sense 
in the limited space of the atomic hydrogen ? The answer may be quick if we
think that the rotating electron is nothing but a one-electron loop current that
considered as a magnetic dipole has a corresponding magnetic field.
Consequently, an atomic electron will always interact with an applied magnetic
$B$. The magnetic quantum number $m_{l}$ specifies the spatial
direction of $L$, 
which is determined by the component of $L$ along the direction of the
external magnetic field. This effect is commonly known as the quantization
of the space in a magnetic field. 

If we choose the direction of the magnetic field as the 
$z$ axis, the component of $L$ along this direction is
\begin{equation} 
L_{z} = m_{l}\hbar~.
\end{equation}
The possible values of $m_{l}$ for a given value of $l$, 
go from $+l$ 
to $-l$, passing through zero, so that there are $2l+1$ possible orientations
of the angular momentum $L$ in a magnetic field. When 
$l=0$, $L_{z}$ can be only zero; when$l=1$, $L_{z}$ 
can be $\hbar$, 0, or $-\hbar$; when $l=2$, $L_{z}$ takes only one of the values
$2\hbar$, $\hbar$, 0, $-\hbar$, or $-2\hbar$, and so forth. 
It is worth mentioning that $L$ cannot be put exactly parallel or
anti-parallel to $B$, because $L_{z}$ is always smaller than the 
modulus $\sqrt{l(l+1)}\hbar$ of the total orbital momentum.

The spatial quantization of the orbital momentum for the hydrogen atom is shown 
in fig. 6.1 in a particular case.

\vskip 2ex
\centerline{
\epsfxsize=120pt
\epsfbox{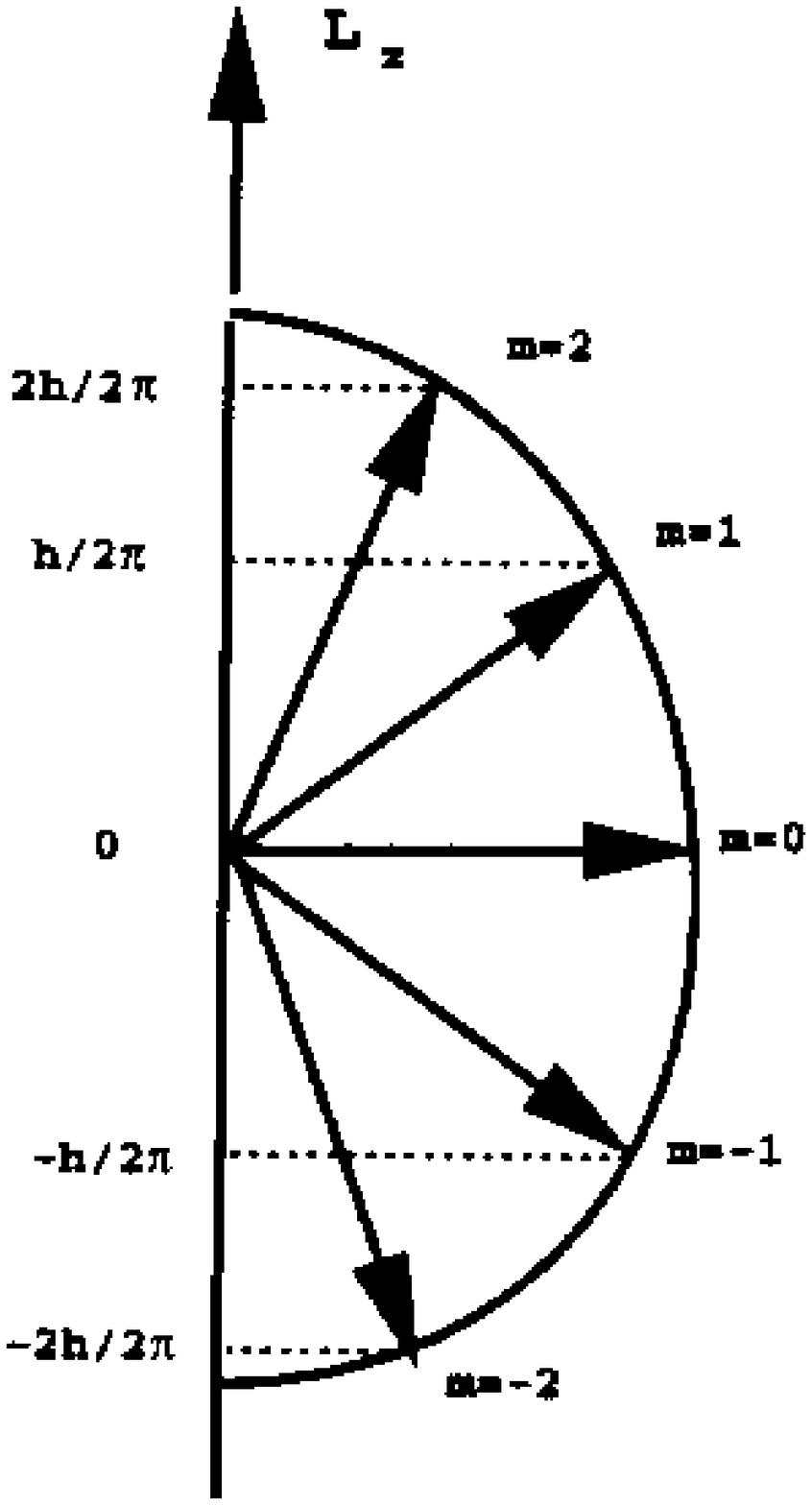}}
\vskip 4ex
\begin{center}
{\small Fig. 6.1:
The spatial quantization of the electron angular momentum for states $l=2$,
$L=\sqrt{6}\hbar$.}
\end{center}

One should consider the atom/electron characterized by a given 
$m_{l}$ as having the orientation of its angular momentum 
$L$ determined relative to the external applied magnetic field.

In the absence of the external magnetic field, the direction of the $z$ axis
is fully arbitrary. Therefore, the component of 
$L$ in any arbitrary chosen direction is $m_{l}\hbar$; the external magnetic 
field 
offers a preferred reference direction from the experimental viewpoint.

Why is quantized only the component $L_{z}$ ?  
The answer is related to the fact that $L$ cannot be put along a direction in an
arbitrary way. Its `vectorial arrow' moves always along a cone centered
on the quantization axis such that its projection $L_{z}$ is 
$m_{l}\hbar$. The reason why such a phenomenon occurs is due to the 
uncertainty principle. If $L$ would be fixed in space, 
in such a way that
$L_{x}$, $L_{y}$ and $L_{z}$ would have well-defined values, the electron 
would have to be confined to a well-defined plane. For example, if $L$ would be
fixed along the $z$ direction, the electron tends to maintain itself in the
plane $xy$  (fig. 6.2a).

\vskip 2ex
\centerline{
\epsfxsize=180pt
\epsfbox{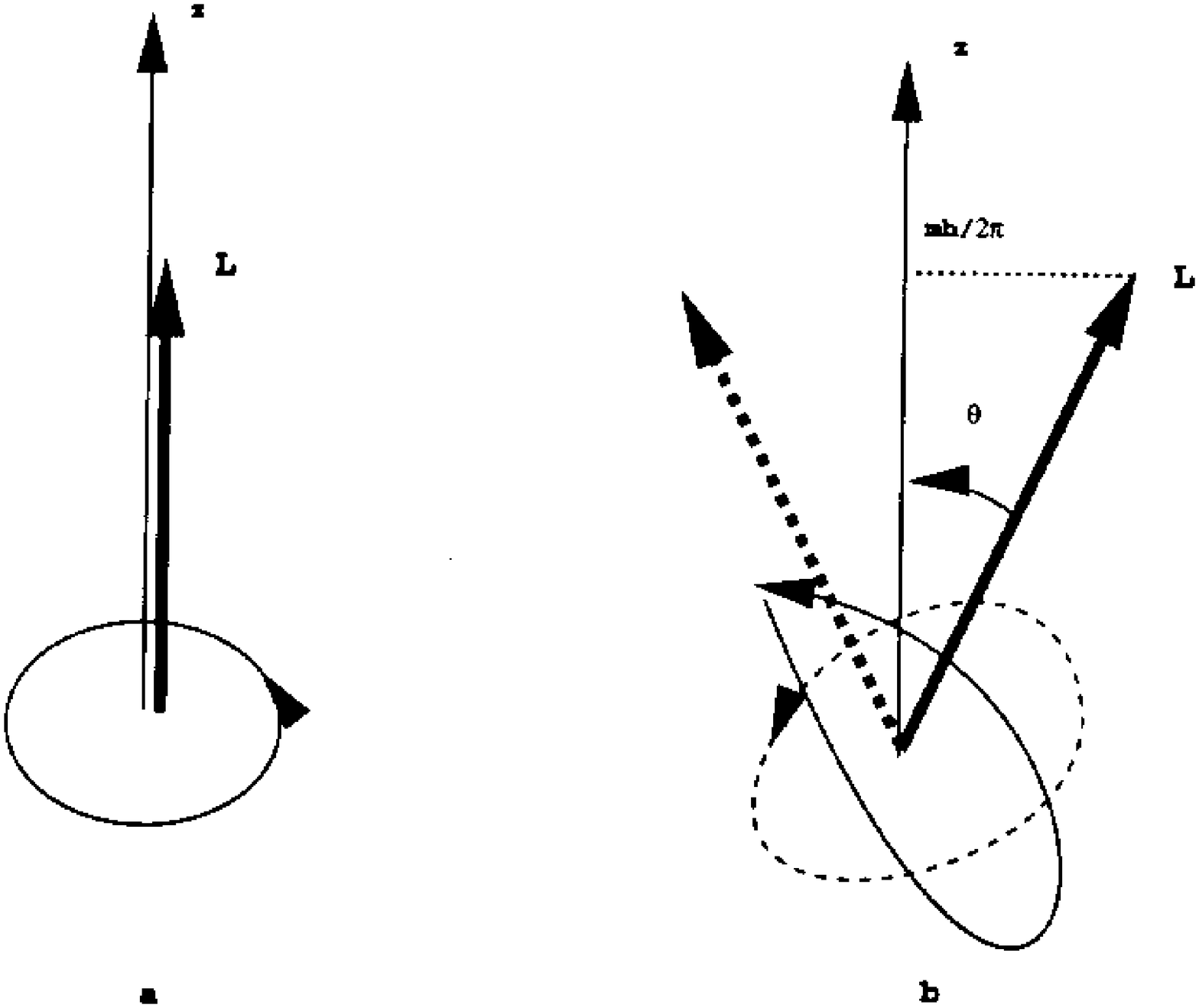}}
\vskip 4ex
\begin{center}
{\small{Fig. 6.2: The uncertainty principle forbids a fixed direction in space of
the angular momentum.}\\
}
\end{center}

This can occur only in the case in which the 
component $p_{z}$ of the
electron momentum is `infinitely' uncertain. This is however impossible if
the electron is part of the hydrogen atom. 
But since in reality just the component
$L_{z}$ of $L$ together with $L^2$ have well-defined values and $\mid 
L \mid > \mid L_{z} \mid$, the electron is not constrained to
a single plane (fig. 6.2b). If this would be the case, an uncertainty would 
exist in the 
coordinate $z$ of the electron. The direction of $L$ changes continuously
(see fig. 6.3), so that the mean values of $L_{x}$ and 
$L_{y}$ are zero, although $L_{z}$ keeps all the time its value $m_{l}\hbar$.

\vskip 2ex
\centerline{
\epsfxsize=180pt
\epsfbox{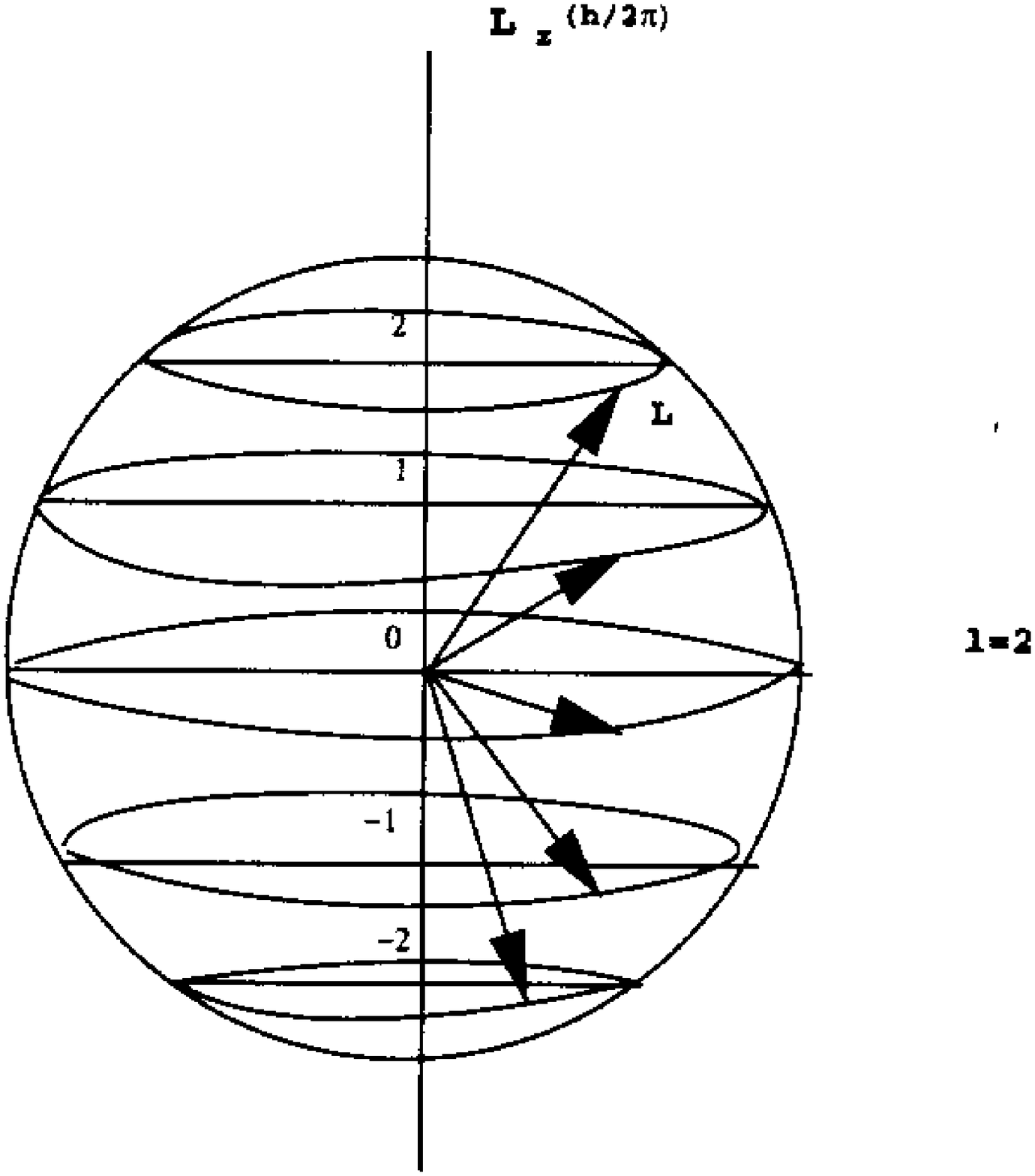}}
\vskip 4ex
\begin{center}
{\small{Fig. 6.3: The angular momentum displays a constant precession
around the $z$ axis.}\\
}
\end{center}

The solution for $\Phi$ should also fulfill the normalization condition 
given by eq. 2. Thus, we have
\begin{equation} 
\int_{0}^{2\pi} \mid \Phi \mid^{2}d\phi = 1
\end{equation}
and substituting $\Phi$, one gets
\begin{equation} 
\int_{0}^{2\pi} A_{\phi}^{2}d\phi = 1~.
\end{equation}
It follows that $A_{\phi}=1/\sqrt{2\pi}$, and thefore the normalized $\Phi$ is
\begin{equation} 
\Phi(\phi) = \frac{1}{\sqrt{2\pi}}e^{im_{l}\phi}~.
\end{equation}

\subsection*{Solution for the polar part}
The solution of the $\Theta(\theta)$ equation is more complicated. It is
expressed in terms of the associated Legendre polynomials
\begin{equation} 
P_{l}^{m_{l}}(x) = 
(-1)^{m_{l}}(1-x^{2})^{m_{l}/2} 
\frac{d^{m_{l}}}{dx^{m_{l}}}P_{l}(x) = 
(-1)^{m_{l}}\frac{(1-x^{2})^{m_{l}/2}}{2^{l}l!}\frac{d^{m_{l} + 
l}}{dx^{{m_{l} + l}}}(x^{2} - 1)^{l}~.
\end{equation}
Their orthogonality relationship is
\begin{equation} 
\int_{-1}^{1} [P_{l}^{m_{l}}(cos\theta)]^{2}dcos\theta = 
\frac{2}{2l+1}\frac{(l+m_{l})!}{(l-m_{l})!}~.
\end{equation}
For the case of quantum mechanics, $\Theta(\theta)$ is given by the normalized
associated Legendre polynomials. Thus, if
\begin{equation} 
\Theta(\theta) = A_{\theta}P_{l}^{m_{l}}(cos\theta)~,
\end{equation}
then the normalization condition is 
\begin{equation} 
\int_{-1}^{1} A_{\theta}^{2}[P_{l}^{m_{l}}(cos\theta)]^{2}dcos\theta = 1~.
\end{equation}
Therefore, the normalization constant for the polar part is given by
\begin{equation} 
A_{\theta} = \sqrt{\frac{2l+1}{2} \frac{(l-m_{l})!}{(l+m_{l})!}}
\end{equation}
and consequently, the function $\Theta(\theta)$ already normalized reads
\begin{equation} 
\Theta(\theta) = 
\sqrt{\frac{2l+1}{2}\frac{(l-m_{l})!}{(l+m_{l})!}} P_{l}^{m_{l}}(cos\theta)~.
\end{equation}

For our purposes here, the most important property of these functions  
is that they exist only when the constant $l$ 
is an integer number greater or at least equal to
$\mid m_{l}\mid$, which is the absolute value of $m_{l}$. This condition
can be written in the form of the set of values available for $m_{l}$  
\begin{equation} 
m_{l} = 0,\pm 1, \pm 2,...,\pm l~. 
\end{equation}

\subsection*{Unification of the azimuthal and polar parts: spherical harmonics}
The solutions of the azimuthal 
and polar parts can be unified within spherical harmonics functions that depend
on both $\phi$ and $\theta$. This simplifies the algebraic manipulations  
of the full wave functions $\psi(r,\theta,\phi)$. 
Spherical harmonics are introduced as follows
\begin{equation} 
Y_{l}^{m_{l}}(\theta,\phi) = (-1)^{m_{l}} \sqrt{\frac{2l+1}{4\pi} 
\frac{(l-m_{l})!}{(l+m_{l})!}} P_{l}^{m_{l}}(cos\theta)e^{im_{l}\phi}~.
\end{equation} 
The supplementary factor $(-1)^{m_{l}}$ does not produce any problem because 
the Schr\"odinger equation is linear and homogeneous. This factor is added
for the sake of convenience in angular momentum studies. 
It is known as the Condon-Shortley phase factor and its effect is to introduce
an alternance of the signs $\pm$ for the spherical harmonics.

\subsection*{Solution for the radial part}
The solution for the radial part $R(r)$ of the wave function  
$\psi$ of the hydrogen atom is somewhat more complicated. It is here where
significant differences with respect to the electrostatic 
Laplace equation do occur. The final result is expressed 
analytically in terms of the associated Laguerre polynomials
(Schr\"odinger 1926). The radial equation 
can be solved in exact way only when E is positive or for one of the 
following negative values $E_{n}$ (in which cases, the electron is in a 
bound stationary state within atomic hydrogen)
\begin{equation} 
E_{n} = 
-\frac{m 
e^{4}}{32\pi^{2}\epsilon_{0}^{2}\hbar^{2}}\left(\frac{1}{n^{2}}\right)~, 
\end{equation}
where $n$ is an integer number called the principal quantum number. It gives 
the quantization of the electron energy in the hydrogen atom.
This discrete atomic spectrum has been first obtained in 1913 by Bohr 
using semi-empirical quantization methods and next by 
Pauli and Schr\"odinger almost simultaneously in 1926.

Another condition that should be satisfied to solve the radial equation
is that $n$ have to be strictly bigger than $l$. Its lowest value is
$l+1$ for a givem $l$. 
Vice versa, the condition on $l$ is
\begin{equation} 
l = 0,1,2,...,(n-1) 
\end{equation}
for given $n$.

The radial equation can be written in the form  
\begin{equation} 
r^{2}\frac{d^{2}R}{dr^{2}} + 2r\frac{dR}{dr} + \left[\frac{2m 
E}{\hbar^{2}}r^{2} + \frac{2me^{2}}{4\pi \epsilon_{0} \hbar^{2}}r - 
l(l+1)\right]R = 0~,
\end{equation}
Dividing by $r^2$ and using the substitution  
$\chi (r) =rR$ to eliminate the first derivative $\frac{dR}{dr}$, one gets the 
standard form of the radial 
Schr\"odinger equation displaying the effective potential 
$U(r)=-{\rm const}/r + l(l+1)/r^2$
(actually, electrostatic potential plus quantized centrifugal barrier). 
These are necessary mathematical steps in order to discuss a new
boundary condition, since the spectrum is obtained 
by means of the $R$ equation.
The difference between a radial Schr\"odinger equation and
a full-line one is that a supplimentary boundary condition should be imposed
at the origin ($r=0$). The coulombian potential belongs to a class of 
potentials that are called weak singular for which 
${\rm lim} _{r\rightarrow 0}=U(r)r^2=0$. In these cases, one tries solutions
of the type $\chi \propto r^{\nu}$, implying
$\nu (\nu -1)=l(l+1)$, so that the solutions are $\nu _1 =l+1$ and 
$\nu _2=-l$, just as in electrostatics.
The negative solution is eliminated for
$l\neq 0$ because it leads to a divergent normalization constant, nor did it
respect the normalization at the delta function for the continuous part of the
spectrum. On the other hand, the particular case
$\nu _2 =0$ is elmininated because the mean kinetic energy is not finite. 
The final conclusion is that $\chi (0)=0$ for any $l$.


Going back to the analysis of the radial equation for $R$, first thing to do is
to write it in nondimensional variables.
This is performed by noticing that the only space and time scales that one can
form on combining the three fundamental constants entering this problem, 
namely
$e^2$, $m_{e}$ and $\hbar$ are the Bohr radius $a_{0}=\hbar ^2/me^2=0.529\cdot 
10 ^{-8}$ cm. and $t_{0}=\hbar ^3/me^4=0.242 10^{-16}$ sec., usually known as
atomic units.
Employing these units, one gets 
\begin{equation} 
\frac{d^{2}R}{dr^{2}} + \frac{2}{r}\frac{dR}{dr} + \left[2 
E + \frac{2}{r} - 
\frac{l(l+1)}{r^2}\right]R = 0~,
\end{equation}
where we are especially interested in the discrete part of the spectrum
($E<0$). The notations 
$n=1/\sqrt{-E}$ and $\rho=2r/n$ leads us to
\begin{equation} 
\frac{d^{2}R}{d\rho ^{2}} + \frac{2}{\rho}\frac{dR}{d\rho} + 
\left[\frac{n}{\rho}-\frac{1}{4} - 
\frac{l(l+1)}{\rho ^2}\right]R = 0~.
\end{equation}
For $\rho \rightarrow \infty$, this equation reduces to  
$\frac{d^{2}R}{d\rho ^{2}}=\frac{R}{4}$, having 
solutions $R\propto e^{\pm\rho /2}$.
Because of the normalization condition only the decaying exponential
is acceptable.
On the other hand, the asymptotics at zero, as we already commented on,
should be $R\propto \rho ^{l}$. Therefore, we can write $R$ as a product of 
three radial functions $R=\rho ^{l}e^{-\rho /2}F(\rho)$, 
of which the first two give the asymptotic behaviors, whereas the third is the 
radial function in the intermediate region. The latter function is of most
interest because its features determine the energy spectrum. 
The equation for $F$ is
\begin{equation} 
\rho\frac{d^{2}F}{d\rho ^{2}} + (2l+2-\rho)\frac{dF}{d\rho} + 
(n-l-1)F = 0~.
\end{equation}
This is a particular case of confluent hypergeometric equation for which the
two `hyper'geometric parameters depend on the pair of quantum numbers
$n,l$. It can be identified as the equation for the associated
Laguerre polynomials $L_{n+l}^{2l+1}(\rho)$.
Thus, the normalized form of $R$ is
\begin{equation} 
R_{nl}(r) = 
-\frac{2}{n^2}\sqrt{\frac{(n-l-1)!}{2n[(n+l)!]^{3}}}
e^{-\rho /2}\rho^{l} L_{n+l}^{2l+1}(\rho)~,
\end{equation}
where the following Laguerre normalization condition has been used 
\begin{equation} 
\int_{0}^{\infty}e^{-\rho}\rho^{2l}[L_{n+l}^{2l+1}(\rho)]^{2}\rho^{2}d\rho = 
\frac{2n[(n+l)!]^{3}}{(n-l-1)!}~.
\end{equation}

We have now the solutions of all the equations 
depending on a single variable
and therefore we can build the wave function for any electronic state
of the hydrogen atom. The full wave function reads
\begin{equation} 
\psi(r,\theta,\phi)={\cal N}_{H}(\alpha r)^{l} 
e^{-\alpha r/2} L_{n+l}^{2l+1}(\alpha r) 
P_{l}^{m_{l}}(cos\theta)e^{im_{l}\phi}~,
\end{equation}
where ${\cal N}_{H}=-\frac{2}{n^2}
\sqrt{\frac{2l+1}{4\pi}\frac{(l-m_{l})!}{(l+m_{l})!} 
\frac{(n-l-1)!}{[(n+l)!]^{3}}}$ and $\alpha=2/na_{0}$.

Using the spherical harmonics, the solution is written as follows  
\begin{equation}  
\psi(r,\theta,\phi)=-\frac{2}{n^2}\sqrt{\frac{(n-l-1)!}{[(n+l)!]^{3}}}
(\alpha r)^{l}
e^{-\alpha r/2} L_{n+l}^{2l+1}(\alpha r)Y_{l}^{m_{l}}(\theta,\phi)~.
\end{equation}

The latter formula may be considered as the final result
for the Schr\"odinger solution of the hydrogen atom  
for any stationary electron state.
Indeed, one can see explicitly both the asmptotic dependence
and the two orthogonal and complete sets of functions, i.e., 
the associated Laguerre polynomials and the spherical harmonics that correspond
to this particular case of linear partial second-order differential equation. 
The parabolic coordinates 
[$\xi=r(1-\cos\theta)$, $\eta =r(1+\cos \theta)$, $\phi=\phi$], 
are another coordinate system in which the
Schr\"odinger hydrogen equation is separable
(E. Schr\"odinger, Ann. Physik {\bf 80}, 437, 1926; 
P.S. Epstein, Phys. Rev. {\bf 28}, 695, 1926; 
I. Waller, Zf. Physik {\bf 38}, 635, 1926). 
The final solution in this case is expressed as the product of factors of
asymptotic nature, azimuthal harmonics, and two sets of associate
Laguerre polynomials in the variables $\xi$ and $\eta$, respectively. 
The energy spectrum  
($-1/n^2$) and the degeneracy ($n^2$) of course do not depend on the 
coordinate system.


\section*{Electronic probability density}
In the Bohr model of the hydrogen atom, the electron rotates around the nucleus
on circular or elliptic trajectories. 
It is possible to think of appropriate experiments
allowing to ``see" that the electron moves within experimental errors at the
predicted radii $r=n^{2}a_{0}$ (where $n$ is the principal quantum number 
labeling the orbit and $a_{0}=0.53$ $\AA$ is the Bohr radius) 
in the equatorial plane $\theta=90^{o}$, 
whereas the azimuthal angle may vary according to the specific 
experimental conditions.

The more rigorous quantum theory changes the conclusions of the Bohr
model in at least two important aspects. 
First, one cannot speak about exact values of
$r,\theta,\phi$, but only of relative probabilities 
to find the electron within an infinitesimal given region of space. 
This feature is a consequence of the wave nature of the electron.
Secondly, the electron does not move around the nucleus in the classical
conventional way because the probability density 
$\mid \psi \mid^{2}$ does not depend on time 
but can vary substantially as a function of the relative position of
the infinitesimal region.   

The hydrogenic electron wave function $\psi$  
is $\psi=R\Theta\Phi$, where $R=R_{nl}(r)$ describes the way  
$\psi$ changes with $r$ when the principal and orbital
quantum numbers have the values
$n$ and $l$, respectively. $\Theta=\Theta_{lm_{l}}(\theta)$ describes in turn
how $\psi$ varies with $\theta$ when the orbital and magnetic quantum numbers
have the values $l$ and $m_{l}$, respectively. Finally, 
$\Phi=\Phi_{m_{l}}(\phi)$ gives the change of $\psi$ with 
$\phi$ when the magnetic quantum number has the value $m_{l}$.  
The probability density $\mid \psi \mid^{2}$ can be written 
\begin{equation}  
\mid \psi \mid^{2} = \mid R \mid^{2} \mid \Theta \mid^{2} \mid \Phi \mid^{2}~.
\end{equation}
Notice that the probability density $\mid \Phi \mid^{2}$, which measures
the possibility to find the electron at a given azimuthal angle
$\phi$, is a constant (does not depend on $\phi$). Therefore, the electronic
probability density is symmetric with respect to the $z$ axis and independent on 
the magnetic substates (at least until an external magnetic field is applied).
Consequently, the electron has an equal probability to be found in any azimuthal
direction.  
The radial part $R$ of the wave function, contrary to $\Phi$, not only
varies with $r$, but it does it differently for any different combination
of quantum numbers $n$ and $l$. Fig. 6.4 shows plots of
$R$ as a function of $r$ for the states $1s$, $2s$,
and $2p$. $R$ is maximum 
at the center of the nucleus ($r=0$) for all the $s$ states, whereas it is zero
at $r=0$ for all the states of nonzero angular momentum.

\vskip 2ex
\centerline{
\epsfxsize=280pt
\epsfbox{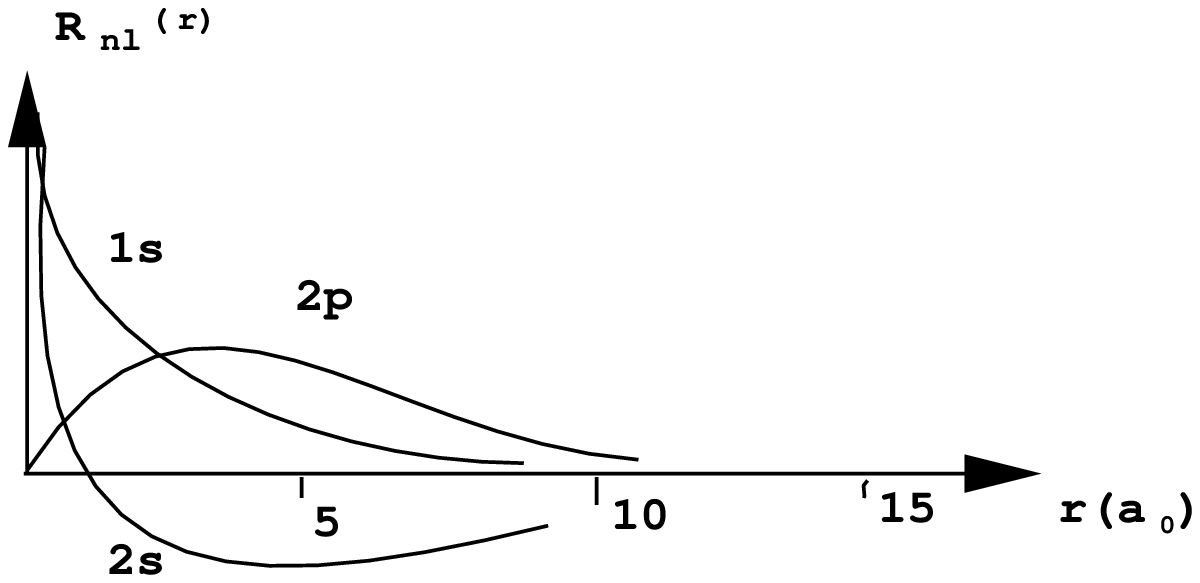}}
\vskip 4ex
\begin{center}
{\small{Fig. 6.4: Approximate plots of the radial functions  
$R_{1s}$, $R_{2s}$, $R_{2p}$; ($a_0=0.53$ \AA ).}\\
}
\end{center}

\vskip 2ex
\centerline{
\epsfxsize=280pt
\epsfbox{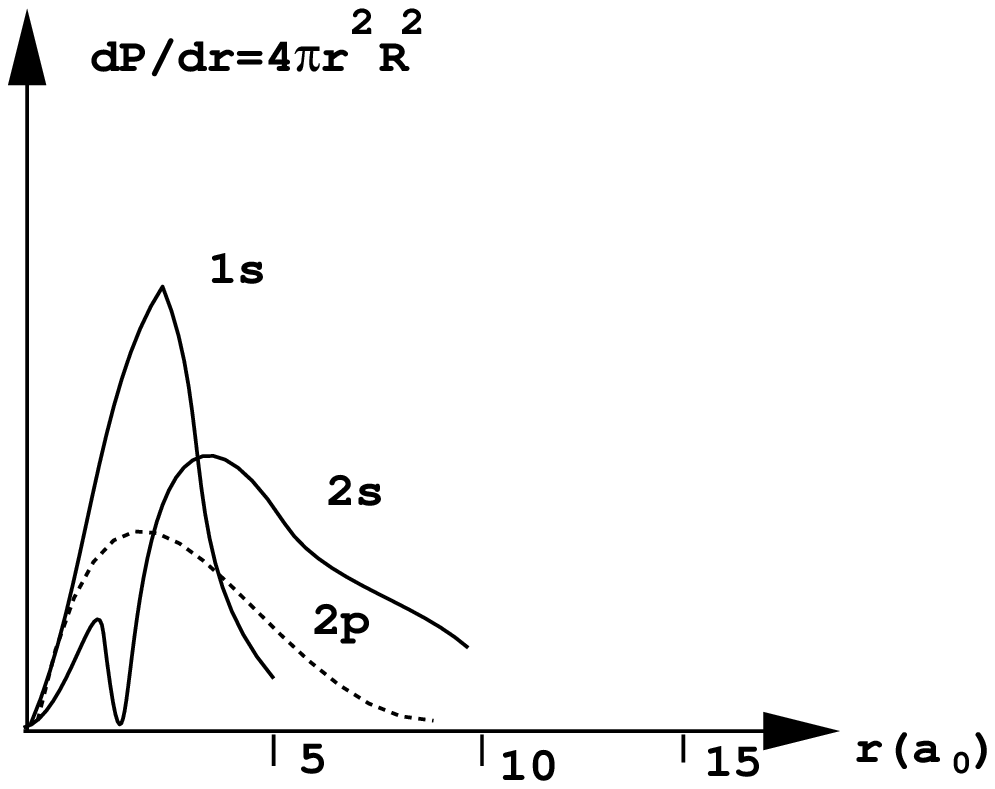}}
\vskip 4ex
\begin{center}
{\small{Fig. 6.5: Probability density of finding the hydrogenic electron 
between $r$ and $r+dr$ with respect to the nucleus for the states
$1s$, $2s$, $2p$.}\\
}
\end{center}

The electronic probability density at the point $r,\theta,\phi$ is
proportional to $\mid \psi \mid^{2}$, but the real probability in the 
infinitesimal volume element $dV$ is 
$\mid \psi \mid^{2}dV$. In spherical coordinates
\begin{equation}  
dV=r^{2}\sin\theta dr d\theta d\phi~,
\end{equation}
and since $\Theta$ and $\Phi$ are normalized functions,  
the real numerical probability $P(r)dr$ to find the electron at a relative 
distance with respect to the nucleus between $r$ and $r+dr$ is
\begin{eqnarray}  
P(r)dr & = & r^{2}\mid R \mid^{2}dr \int_{0}^{\pi} 
\mid\ \Theta \mid^{2} \sin\theta d\theta \int_{0}^{2\pi} 
\mid\ \Phi \mid^{2}d\phi \nonumber\\
& = & r^{2}\mid R \mid^{2}dr
\end{eqnarray}
$P(r)$ is displayed in fig. 6.5 for the same states for which the radial  
functions $R$ appear in fig. 6.4. In principle, the curves are quite different.
We immediately see that $P(r)$ is not maximal in the nucleus
for the states $s$, as happens for $R$. Instead, their maxima are encountered at
a finite distance from the nucleus. The most probable value of
$r$ for a $1s$ electron is exactly $a_{0}$, the Bohr radius. 
However, the mean value of $r$ for a $1s$ electron is $1.5a_{0}$. At first sight
this might look strange, because the energy levels  
are the same both in quantum mechanics and in Bohr's model.
This apparent unmatching is eliminitated if one takes into account
that the electron energy depends on $1/r$ and not on 
$r$, and the mean value of $1/r$ for a $1s$ electron is exactly $1/a_{0}$.

The function $\Theta$ varies with the polar angle $\theta$ for all the quantum
numbers $l$ and $m_{l}$, unless $l=m_{l}=0$, which are
the $s$ states. The probability density $\mid\ \Theta \mid^{2}$ for a  
$s$ state is a constant (1/2). This means that since
$\mid \Phi \mid^{2}$ is also a constant, the electronic probability density 
$\mid \psi \mid^{2}$ has the same value for a given 
$r$ value, not depending on the direction. In other states, 
the electrons present an angular behavior that in many cases 
may be quite complicated.
This can be seen in fig.6.5, where the electronic probability densities
for different atomic states are displayed 
as a function of $r$ and $\theta$. (The plotted term is
$\mid \psi \mid^{2}$ and not $\mid \psi \mid^{2}dV$). Because  
$\mid \psi \mid^{2}$ is independent of $\phi$, a three-dimensional 
representation of $\mid \psi \mid^{2}$ can be obtained by rotating a particular
representation around a vertical axis. This can prove that the probability 
densities for the $s$ states have spherical symmetry, while all the other  
states do not possess it. In this way, one can get more or less pronounced lobes
of characteristic forms depending on state. These lobes are quite important
in chemistry for specifying the atomic interaction in the molecular bulk.
\\
\\
{\bf 6N. Note}:

\noindent
1. In 1933, E. Schr\"odinger has been awarded the Nobel Prize in Physics 
(together with Dirac)
for the ``discovery of new productive forms of atomic theory". 
Schr\"odinger wrote a remarkable series of four papers 
``Quantisierung als Eigenwertproblem"  [``Quantization as an eigenvalue 
problem"] (I-IV, received by Annalen der Physik on
27 January, 23 February, 10 May and 21 June 1926, respectively).

\section*{{\huge 6P. Problems}}
  
{\bf Problem 6.1} - Obtain the formulas for the stable orbits 
and the energy levels of the electron in 
the atomic hydrogen using only 
arguments based on the de Broglie wavelength associated to the
electron and the empirical value $5.3 \cdot 10^{-11}$ m 
for the Bohr radius. 

\noindent
{\bf Solution}: The electron wavelength is given by
$
\lambda = \frac{h}{mv}
$,
whereas if we equate the electric force and the centripetal force 
$
\frac{mv^{2}}{r} = \frac{1}{4\pi \epsilon_{0}}\frac{e^{2}}{r^{2}}
$
we obtain the electron `velocity'
$
v = \frac{e}{\sqrt{4\pi \epsilon_{0} mr}}~.
$
Thus, the wavelength of the electron is 
$
\lambda = \frac{h}{e}\sqrt{\frac{4\pi \epsilon_{0}r}{m}}
$.
If we now use the value $5.3 \times 10^{-11}$m for the radius $r$ of the 
electron orbit, we can see that the wavelength of the electron is
$\lambda=33 \times 10^{-11}$ m. But this is exactly the same value as of the
circumference of the orbit, $2\pi 
r=33 \times 10^{-11}$ m. One may say that the electron orbit in the atomic 
hydrogen corresponds to a wave ``closing into itself" 
(i.e., stationary). 
This fact can be compared to the vibrations of a metallic ring. 
If the wavelengths are multiples of the circumference, the ring
goes on with its vibrations for a long time with very small dissipation
If, on the other hand, the number of wavelengths making a circumference is not  
an integer, the interference of the waves is negative and they dissapear in a
short period of time.  
One may say that the electron will rotate around the nucleus without
radiating its energy for an infinite time as far as its orbit contains an
integer number of  
de Broglie wavelengths. Thus, the stability/stationary condition is 
\begin{eqnarray}
n\lambda = 2\pi r_{n}~,\nonumber
\end{eqnarray}
where $r_{n}$ is the radius of the electron orbit containing $n$ wavelengths.  
Substituting $\lambda$, we have
\begin{eqnarray}
\frac{nh}{e}\sqrt{\frac{4\pi \epsilon_{0}r_{n}}{m}} = 2\pi r_{n}~,\nonumber
\end{eqnarray}
and therefore the stationary electron orbits are 
\begin{eqnarray}
r_{n} = \frac{n^{2}\hbar^{2}\epsilon_{0}}{\pi me^{2}}~.\nonumber
\end{eqnarray}

To get the energy levels, we use $E=T+V$ and substituting the kinetic and 
potential energies leads to
\begin{eqnarray}
E = \frac{1}{2}mv^{2} - \frac{e^{2}}{4\pi \epsilon_{0}r}~,\nonumber
\end{eqnarray}
or equivalently
\begin{eqnarray}
E_{n} = -\frac{e^{2}}{8\pi \epsilon_{0}r_{n}}~.\nonumber
\end{eqnarray}
Plugging the value of $r_{n}$ into the latter equation, we get 
\begin{eqnarray}
E_{n} = 
-\frac{me^{4}}{8\epsilon_{0}^{2}\hbar^{2}} 
\left(\frac{1}{n^{2}}\right)~.\nonumber 
\end{eqnarray}
\\

\noindent
{\bf Problem 6.2} - Uns\"old's theorem tells that for any value
of the orbital number $l$, the probability densities, summed over all possible
substates, from  $m_{l}=-l$ to $m_{l}=+l$ 
give a constant that is independent of the angles $\theta$ and $\phi$, i.e.
\begin{eqnarray}
\sum_{m_{l}=-l}^{+l} \mid \Theta_{lm_{l}} \mid^{2} \mid \Phi_{m_{l}} 
\mid^{2} = ct.\nonumber 
\end{eqnarray}

This theorem shows that any atom or ion with closed (occupied) sublevels
has a spherically-symmetric charge distribution.
Check Uns\"old's theorem for $l=0$, $l=1$, and $l=2$.

\noindent
{\bf Solution}: For $l=0$, $\Theta_{00}=1/\sqrt{2}$ and
$\Phi_{0}=1/\sqrt{2\pi}$, so that
\begin{eqnarray}
\mid \Theta_{0,0} \mid^{2} \mid \Phi_{0} \mid^{2} = \frac{1}{4\pi}~.\nonumber
\end{eqnarray}

For $l=1$, we have
\begin{eqnarray}
\sum_{m_{l}=-1}^{+1} \mid \Theta_{lm_{l}} \mid^{2} 
\mid \Phi_{m_{l}} \mid^{2} = \mid \Theta_{1,-1} \mid^{2} \mid \Phi_{-1} 
\mid^{2} + \mid \Theta_{1,0} \mid^{2} \mid \Phi_{0} \mid^{2} + \mid 
\Theta_{1,1} \mid^{2} \mid \Phi_{1} \mid^{2}~.\nonumber
\end{eqnarray}
On the other hand, the wave functions are given by 
$\Theta_{1,-1}=(\sqrt{3}/2)sin\theta$, 
$\Phi_{-1}=(1/\sqrt{2\pi})e^{-i\phi}$, 
$\Theta_{1,0}=(\sqrt{6}/2)cos\theta$, $\Phi_{0}=1/\sqrt{2\pi}$, 
$\Theta_{1,1}=(\sqrt{3}/2)sin\theta$, $\Phi_{1}=(1/\sqrt{2\pi})e^{i\phi}$~,
which plugged into the previous equation give
\begin{eqnarray}
\sum_{m_{l}=-1}^{+1} \mid \Theta_{lm_{l}} \mid^{2} 
\mid \Phi_{m_{l}} \mid^{2} = \frac{3}{8\pi}sin^{2}\theta + 
\frac{3}{4\pi}cos^{2}\theta + \frac{3}{8\pi}sin^{2}\theta = 
\frac{3}{4\pi}\nonumber 
\end{eqnarray}
and again we've got a constant.

For $l=2$, we have
$$
\sum_{m_{l}=-2}^{+2} \mid \Theta_{lm_{l}} \mid^{2}
\mid \Phi_{m_{l}} \mid^{2} = 
\mid \Theta_{2,-2} \mid^{2} \mid \Phi_{-2} \mid^{2}
\mid \Theta_{2,-1} \mid^{2} \mid \Phi_{-1} \mid^{2}
$$
$$ 
+ \mid \Theta_{2,0} \mid^{2} \mid \Phi_{0} \mid^{2} 
+ \mid \Theta_{2,1} \mid^{2} \mid \Phi_{1} \mid^{2}
+ \mid \Theta_{2,2} \mid^{2} \mid \Phi_{2} \mid^{2}~, 
$$
and the wave functions are
$\Theta_{2,-2}=(\sqrt{15}/4)sin^{2}\theta$, 
$\Phi_{-2}=(1/\sqrt{2\pi})e^{-2i\phi}$, 
$\Theta_{2,-1}=(\sqrt{15}/2)sin\theta cos\theta$,
$\Phi_{-1}=(1/\sqrt{2\pi})e^{-i\phi}$,
$\Theta_{2,0}=(\sqrt{10}/4)(3cos^{2}\theta-1)$,
$\Phi_{0}=1/\sqrt{2\pi}$,
$\Theta_{2,1}=(\sqrt{15}/2)sin\theta cos\theta$,
$\Phi_{1}=(1/\sqrt{2\pi})e^{i\phi}$,
$\Theta_{2,2}=(\sqrt{15}/4)sin^{2}\theta$,
$\Phi_{2}=(1/\sqrt{2\pi})e^{2i\phi}$,
Plugging them into the previous equation give
\begin{eqnarray}
\sum_{m_{l}=-2}^{+2} \mid \Theta_{lm_{l}} \mid^{2}
\mid \Phi_{m_{l}} \mid^{2} = \frac{5}{4\pi}~,\nonumber
\end{eqnarray}
which again fulfills Uns\"old's theorem.
\\

\noindent
{\bf Problem 6.3} - The probability to find an atomic electron
whose radial wave functions is that of the ground state
$R_{10}(r)$ outside a sphere of Bohr radius $a_{0}$ 
centered on the nucleus is
\begin{eqnarray}
\int_{a_{0}}^{\infty} \mid R_{10}(r) \mid^{2}r^{2}dr~.\nonumber 
\end{eqnarray}
Obtain the probability to find the electron
in the ground state at a distance from the nucleus bigger than $a_{0}$.

\noindent
{\bf Solution}: The radial wave function corresponding to the ground state is
\begin{eqnarray}
R_{10}(r) = \frac{2}{a_{0}^{3/2}}e^{-r/a_{0}}~.\nonumber
\end{eqnarray}
Substituting it in the integral, we get
$
\int_{a_{0}}^{\infty} \mid R(r) \mid^{2}r^{2}dr = 
\frac{4}{a_{0}^{3}} \int_{a_{0}}^{\infty} r^{2} e^{-2r/a_{0}}dr ~,\nonumber
$
or
\begin{eqnarray}
\int_{a_{0}}^{\infty} \mid R(r) \mid^{2}r^{2}dr =
\frac{4}{a_{0}^{3}}\left[-\frac{a_{0}}{2}r^{2}e^{-2r/a_{0}}  
-\frac{a_{0}^{2}}{2}re^{-2r/a_{0}}
-\frac{a_{0}^{3}}{4}e^{-2r/a_{0}}\right]_{a_{0}}^{\infty}~.\nonumber
\end{eqnarray}
This leads us to
\begin{eqnarray}
\int_{a_{0}}^{\infty} \mid R(r) \mid^{2}r^{2}dr = \frac{5}{e^{2}}
\approx 68 \% \; !!~,\nonumber
\end{eqnarray}
which is the result asked for in this problem.


\newpage
\begin{center}{\huge 7. QUANTUM SCATTERING}
\end{center}

\section*{\bf Introduction}
\setcounter{equation}{0}
One usually begins the quantum theory of scattering by referring to results
already known from the classical scattering in central fields with some 
simplifying assumptions helping to avoid unnecessary calculations in getting 
basic results. It is generally known that 
studying scatterings in the laboratory provides information on the 
distribution of matter in the target and other details of the interaction
between the incident beam and the target. The hypotheses that we shall assume
correct in the following are

i) The particles are spinless. This, of course, does not mean that spin effects
are not important in quantum scatterings. 

ii) We shall study only elastic scattering for which the internal structure
of the particles is not taken into account.

iii) The target is sufficiently thin to neglect multiple scatterings.

iv) The interactions are described by 
a potential that depends only on the relative distance between the particles
(central potential).

These hypotheses eliminate some quantum effects that are merely
details. They also represent conditions for getting the quantum analogs
of basic classical results. We now define

\begin{equation}
\frac{d\sigma}{d\Omega}\propto \frac{I(\theta,\varphi)}{I_{0}}~,
\end{equation}

\noindent
where $d\Omega$ is the solid angle infinitesimal element, $I_{0}$ is the number
of incident particles per unit transverse area, and 
$I{}d\Omega$ is the number of scattered particles in the solid angle element. 

Employing these well-known concepts, together with the asymptotic 
notion of {\em impact
parameter} $b$ associated to each classical incident particle, one gets in
classical mechanics the following important formula

\begin{equation}
\frac{d\sigma}{d\Omega}=\frac{b}{\sin\theta}\vert 
\frac{db}{d\theta}\vert ~.
\end{equation}

If one wants to study the scattering phenomenology in quantum 
terminology, one should investigate the time evolution of a `scattering'
wave packet.  
Let $F_{i}$ be the flux of incident particles, i.e., the number of particles per
unit of time passing through the unit of transverse surface onto the 
propagation axis. An appropriate detector configuration is usually placed far 
away from 
the effective interaction region, `seeing' a solid angle 
$d\Omega$ of that region. In general, the number of particles 
$dn/dt$ scattered per unit of time in $d\Omega$ in the direction
$(\theta,\varphi)$ is detected.
\vskip 2ex
\centerline{
\epsfxsize=280pt
\epsfbox{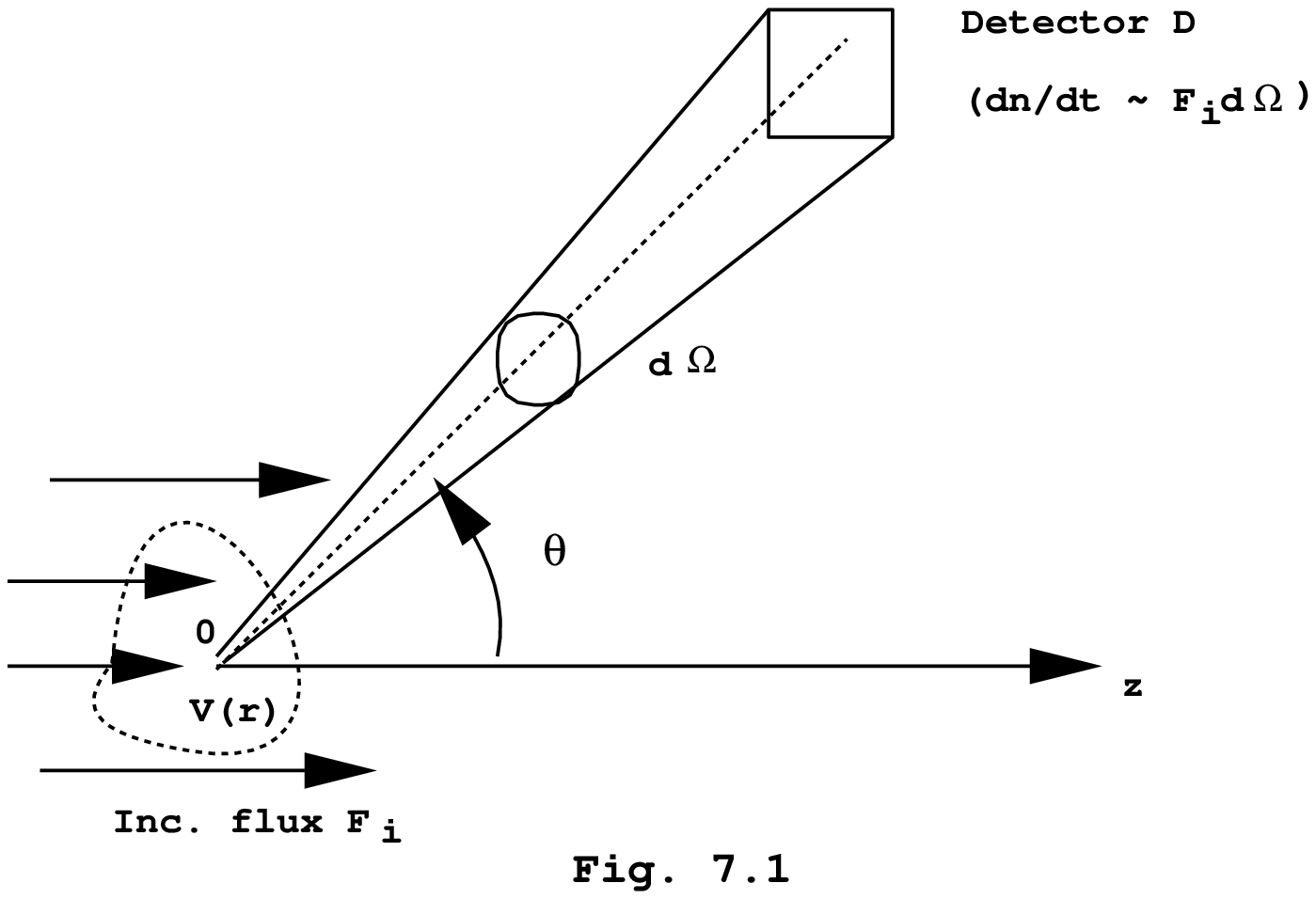}}
\vskip 4ex

\noindent
$dn/dt$ is proportional to $d\Omega$ and 
$F_{i}$. Let us call $\sigma (\theta,\varphi)$  the coefficient of 
proportionality between $dn$ and $F_{i} d\Omega$:
\begin{equation}
 dn=\sigma (\theta,\varphi)F_{i} d\Omega~,
\end{equation}

\noindent
which is by definition the differential cross section.

The number of particles per unit of time reaching the detector 
is equal to the number of particles crossing the surface  
$\sigma (\theta,\varphi) d\Omega$, which is  perpendicular  to the beam axis.
The total section is by definition
\begin{equation}
 \sigma=\int \sigma (\theta,\varphi) d\Omega~. 
\end{equation}

To further simplify the calculation, we choose the z axis along the incident
beam direction. \\
On the negative side of the axis, for large negative $t$, 
the particle is practically free: it is not affected by 
$V({\bf r})$ and its state can be represented by plane waves. Therefore, the
wave function contains terms of the form 
$e^{ikz}$, where $k$ is the constant ocurring in the  
Helmholtz equation. By analogy with optics, the form of the scattered
wave is
\begin{equation}
 f(r)= \frac{e^{ikr}}{r}~. 
\end{equation}

Indeed
\begin{equation}
 (\nabla ^{2} + k^{2})e^{ikr} \neq 0
\end{equation} 

and
\begin{equation} 
(\nabla ^{2} + k^{2}) \frac{e^{ikr}}{r}=0
\end{equation}
for $r>r_{0}$, where $r_{0}$ is any positive number.

We assume that the motion of the particle is described by the  
Hamiltonian
\begin{equation}
H=\frac{{\rm \bf p^2}}{2\mu}+V=H_{0}+V~.
\end{equation}

$V$ is different of zero only in a small neighbourhood close to the origin.
A wave packet at $t=0$ can be written

\begin{equation}
 \psi({\bf{r}},0)=\frac{1}{(2\pi)^\frac{3}{2}}\int 
\varphi({\rm \bf k})\exp[i{\rm \bf k\cdot (r-r_{0})}]
{\rm {\bf d^{3}k}}~,
\end{equation}

\noindent
where $\psi$ is a function that is nonzero in a `width'
$\Delta {\rm \bf k}$
centered on ${\rm \bf k_{0}}$. We also assume
that ${\rm \bf k_{0}}$ is antiparallel
to ${\rm \bf r_{0}}$. In order to see quantitatively
what happens to the wave packet when scatters the target, one can
use the expansion of $\psi({\rm \bf r},0)$
in the eigenfunctions $\psi_{n}({\rm \bf r})$ of $H$, i.e.,
$\psi({\bf{r}},0)=\sum_{n}c_{n}\psi_{n}(\bf{r})$. Thus, the wave packet at
time $t$ is
\begin{equation}
\psi({\bf 
r},t)=\sum_{n}c_{n}\varphi _{n}({\bf r})\exp(-\frac{i}{\hbar}E_{n}t)~.
\end{equation}

This is an eigenfunction of the operator $H_{0}$, not of $H$, 
but we can substitute these eigenfunctions by eigenfunctions of
$H$, which we denote by $\psi_{k}^{(+)}(\bf{r})$.  
The asymptotic form of the latter is
\begin{equation}
\psi _{k}^{(+)}(\bf{r})\simeq e^{i\bf{k\cdot r}} +
f({\rm \bf r})\frac{e^{ikr}}{|r|}~,
\end{equation}

where, as usually
${\rm \bf p}=\hbar {\rm \bf k}$ and $E=\frac{\hbar ^{2}k^{2}}{2m}$.

This corresponds to a plane wave of the incident beam type
and a divergent spherical wave resulting from the interaction between the 
incident beam and the target. One can expand
$\psi ({\rm \bf r},0)$ in plane waves
and $\psi _{k}({\rm \bf r})$
\begin{equation}
\psi({\rm \bf r},0)=\int \varphi ({\rm \bf k})\exp(-i{\rm \bf k\cdot
r_{0}})\psi _{{\rm \bf k}}({\rm \bf r}) d^{3}k~,
\end{equation}
where $ \hbar\omega= \frac{\hbar^{2}k^{2}}{2m}$.
The divergent spherical wave does not contribute to the initial wave packet
because it is an additive part.

\section*{\bf Scattering of a wave packet}

Any wave is dispersed during its propagation. This is why one cannot ignore 
the effect of the divergent wave from this viewpoint. 
One can make use of the following trick

\begin{equation} 
\omega= 
\frac{\hbar}{2m}k^{2}= 
\frac{\hbar}{2m}[{\bf k_{0}+(k-k_{0}})]^{2}= 
\frac{\hbar}{2m}[2{\bf k_{0}\cdot k - k_{0}^{2}+ (k-k_{0})^{2}}]~,
\end{equation}
  
\noindent pentru a neglija ultimul termen \h n paranteze. 
Substituting $\omega$ in $\psi$, we ask that 
\( \frac{\hbar}{2m}({\bf k-k_{0}})^{2}T \ll 1 \),
where $T \simeq \frac{2mr_{0}}{\hbar k_{0}}$. Therefore 

\begin{equation}
\frac{(\Delta k)^{2}r_{0}}{k_{0}} \ll 1~. 
\end{equation}

\noindent This condition tells us that the wave packet does not disperse
significantly even when it moves over amacroscopic distance $r_{0}$.

Choosing the direction of the vector $\bf{k}$ of the incident wave along
one of the three cartesian directions (we use the $z$ one), we can write
in spherical coordinates the following important formula

\( \psi_{k}(r,\theta,\varphi) \simeq e^{ikz} + 
\frac{f(k,\theta,\varphi)e^{ikr}}{r}~. \) 

Since the Hamiltonian $H$, up to now not considered as an operator
(the class of the results presented are the same both at the classical and
quantum level), is invariant under $z$ rotations, 
we can choose boundary conditions of spherical symmetry too. Thus

\( \psi_{k}(r,\theta,\varphi)\simeq e^{ikz}+\frac{f(\theta)e^{ikr}}{r}~.\)

\noindent This type of functions are known as scattering wave functions.
The coefficient $f(\theta)$ of the spherical wave is known as the scattering
amplitude. It is a basic concept in the formal theory of
quantum scatterings.\\

\section*{\bf Probability amplitude in scattering}

We write the Schr\"odinger equation as follows
\begin{equation}
 i\hbar \frac{\partial\psi}{\partial t}= - \frac{\hbar^{2}}{2m} 
\nabla^{2}\psi + V({\bf r},t)\psi~. 
\end{equation}

\noindent
Recall that the expression
\begin{equation}
P({\bf r},t)= \psi^{*}({\bf r},t)\psi ({\bf r},t)=\vert \psi ({\bf 
r},t) \vert ^{2} 
\end{equation}

\noindent can be interpreted, cf. Max Born, as a probability density
under normalization conditions of the type
\begin{equation}
\int \vert \psi ({\rm \bf r},t) \vert ^{2}  d^{3}r = 1~.
\end{equation}

This normalization integral should be time independent. 
This can be noted by writing
\begin{equation}
I= \frac{\partial}{\partial t} \int _{\Omega} P({\rm \bf r},t) d^{3}r=
\int_{\Omega} (\psi^{*}\frac{\partial\psi}{\partial t}
+\frac{\partial\psi^{*}}{\partial t}\psi) d^{3}r~, 
\end{equation}

\noindent and from Schr\"odinger's equation
\begin{equation}
\frac{\partial\psi}{\partial t}= \frac{i\hbar}{2m}
\nabla ^{2}\psi-\frac{i}{\hbar}V({\bf r},t)\psi
\end{equation} 

\noindent one gets
$$
I=\frac{i\hbar}{2m} \int_{\Omega}
[\psi^{*}\nabla^{2}-(\nabla^{2}\psi^{*})\psi]d^{3}r = \frac{i\hbar}{2m} 
\int_{\Omega} \nabla \cdot 
[\psi^{*}\nabla\psi-(\nabla\psi^{*})\psi]d^{3}r=
$$
\begin{equation}
=\frac{i\hbar}{2m} \int_{A}[\psi^{*}\nabla\psi-(\nabla\psi^{*})\psi]_{n}
dA~, 
\end{equation}

\noindent where the Green theorem has been used to evaluate the volume 
integral. 
$dA$ is the infinitesimal surface element on the boundary of the integration 
region and 
$[\quad]_{n}$ denotes the component along the normal direction to the surface
element $dA$.

Defining
\begin{equation}
 {\bf 
S}({\bf r},t)=\frac{\hbar}{2im} [\psi^{*}\nabla\psi-(\nabla\psi^{*})\psi]~, 
\end{equation}

\noindent
we get
\begin{equation}
 I= \frac{\partial}{\partial t} \int_{\Omega} P({\bf r},t) d^{3}r= - 
\int _{\Omega} \nabla\cdot {\bf S} d^{3}r = -\int_{A} S_{n} dA~,
\end{equation}

\noindent for well-bahaved wave packets (not funny asymptotically)
so that the normalization integral converges. The surface integral is  
zero when $\Omega$ covers the whole space. One can prove (see
P. Dennery \& A. Krzywicki, {\it Mathematical methods for physicists}) that
the surface integral is zero. Therefore, the normalization integral is 
constant in time and the initial condition holds. 
From the same equation for ${\bf S}$, we get
\begin{equation}
 \frac{\partial P({\bf r},t)}{\partial t} + \nabla \cdot {\bf S}({\bf 
r},t)= 0~, 
\end{equation}

\noindent which is the continuity equation for the density flux  
$P$ and the current density ${\bf S}$ in the absence of any type of sources
or sinks.
If we interpret $\frac{\hbar}{im}\nabla$ as a sort of velocity `operator'
(as for time, it is difficult to speak rigorously about a velocity operator
in quantum mechanics!), then
\begin{equation}
{\bf S}({\bf r}, t)= Re(\psi ^{*}\frac{\hbar}{im}\nabla\psi)~.
\end{equation}

To calculate the quantum current density for a scattering wave function is a
tricky and inspiring (not illustrative) exercise! The final result is
$j_{r}=\frac{\hbar k}{mr^2}|f(\theta)|^2$, where the direction $\theta =0$
should not be included.

\section*{\bf Green's function in scattering theory}

Another way of writing the Schr\"odinger equation at hand is
$(-\frac{\hbar^{2}}{2m} \nabla^{2} + V)\psi = E\psi $, or
$(\nabla^{2} + k^{2})\psi = U\psi $, where 
$ k^{2}=\frac{2mE}{\hbar^{2}}$, \c{s}i $U=\frac{2mV}{\hbar^{2}}$.

It follows that it is more convenient to put
this equation in an integral form.
This can be done if we consider $U\psi$ in the right hand side of the equation
as a inhomogeneity. This allows to build the solution by means of Green's 
function (integral kernel), 
which, by definition, is the solution of 
\begin{equation}
\label{eq:e1}
(\nabla^{2}+k^{2})G(\bf{r,r'}) =  
\delta(\bf{r-r'})~. 
\end{equation} 

\noindent One can write now the Schr\"odinger solution as the sum of the 
homogeneous equation and the inhomogeneous one of Green's type
\begin{equation}
\psi(\bf{r})=\lambda(\bf{r})-\int
G(\bf{r,r'})U(\bf{r'})\psi(\bf{r'})d^{3}r'~.
\end{equation}

We seek now a $G$ function in the form of a product of linear independent 
functions, for example, plane waves
\begin{equation}
G({\bf r,r'}=\int A({\bf q})e^{i{\bf q\cdot (r-r')}}dq~.
\end{equation}

\noindent Using eq. \ref{eq:e1}, we have
\begin{equation}
\int A({\bf q})(k^{2}-q^{2})e^{i{\bf q\cdot(r-r')}}dq=
\delta{\bf(r-r')}~,
\end{equation}
 
\noindent which turns in an identity if
\begin{equation}
A({\bf q})= (2\pi)^{-3}(k^{2}-q^{2})^{-1}~.
\end{equation}

\noindent Thus

\begin{equation}
G({\bf r,r'})=\frac{1}{(2\pi)^{3}} \int 
\frac{e^{iqR}}{k^{2}-q^{2}}d^{3}q~,
\end{equation}
where $R=\vert {\bf r-r'} \vert$.
After performing a calculation of complex variable 
\footnote{See problem 7.1.}, we get
\begin{equation}
G(r)= - \frac{1}{4\pi} \frac{e^{ikr}}{r}~.
\end{equation}

This function is not determined univoquely since the Green function can be 
any solution of the eq. \ref{eq:e1}. 
The right particular solution is chosen by imposing boundary conditions 
on the eigenfunctions $\psi_{k}({\bf r})$.

The Green function obtained in this way is
\begin{equation}
G({\bf r,r'})= -\left( \frac{e^{ik \vert {\bf r-r'} \vert}}{4\pi
\vert{\bf r-r'}\vert }\right)~.
\end{equation}

Thus, we finally get the integral equation for the scattering wave function 
\begin{equation}
\psi (k,{\bf r})= \varphi (k,{\bf r}) - \frac{m}{2 \pi \hbar^{2}} \int
\frac{e^{ik \vert {\bf r-r'} \vert}}{ {\bf r-r'} } U({\bf r'}) \psi (k,{\bf 
r})d{\bf r}~,
\end{equation}
where $\varphi$ is a solution of the Helmholtz equation. Noticing that
$ \vert {\bf r-r'} \vert = R $, then
\begin{equation}
(\nabla^{2}+k^{2})\psi=(\nabla^{2}+k^{2})[\varphi + \int G({\bf r,r'}) 
U({\bf r'}) \psi({\bf r'}) d^{3}r']
\end{equation}

\noindent and assuming that we can change the order of operations 
and put the $\nabla$ operator inside the integral, we get
\begin{equation}
(\nabla^{2}+k^{2})\psi= \int (\nabla^{2}+k^{2}) G ({\bf r,r'}) U({\bf
r'}) \psi({\bf r'}) d^{3}r'= U({\bf r}) \psi ({\bf r})~,
\end{equation}
which shows us that $G(R)= \frac{1}{4\pi} \frac{e^{ikR}}{R}$ is indeed a 
solution.

\section*{\bf Optical theorem}
 
The total cross section is given by
\begin{equation}
\sigma_{tot}(k)= \int \frac{d\sigma}{d\Omega} d\Omega~.
\end{equation}

Let us express now $f(\theta)$ as a function of the phase shift
$S_{l}(k)=e^{2i\delta_{l}(k)}$ in the form
\begin{equation}
f(\theta)=  \frac{1}{k} \sum_{l=0}^{\infty} (2l+1) e^{i\delta_{i}(k)} 
\sin \delta_{l}(k) P_{l}(\cos \theta)~.
\end{equation}

\noindent
Then
$$
\sigma_{tot} = \int [\frac{1}{k} \sum_{l=0}^{\infty} (2l+1)
e^{i\delta_{l}(k)}\sin \delta_{l}(k) P_{l}(\cos \theta)]
$$
\begin{equation}
[\int 
[\frac{1}{k} \sum_{l'=0}^{\infty} (2l'+1)e^{i\delta_{l'}(k)}\sin 
\delta_{l'}(k) P_{l'}(\cos \theta)]~.
\end{equation}
Using now $\int P_{l}(\cos\theta)P_{l'}(\cos\theta)= \frac{4\pi}{2l+1}
\delta_{ll'}$, we get
\begin{equation}
\sigma_{tot}= \frac{4\pi}{k^{2}} \sum_{l=0}^{\infty} (2l+1)\sin
\delta_{l}(k)^{2}~.
\end{equation}
Of interest is the relationship
$$
{\rm Im} f(0)=\frac{1}{k} \sum_{l=0}^{\infty} (2l+1)
{\rm Im}[e^{i\delta_{l}(k)}\sin \delta_{l}(k)]P_{l}(1) =
\frac{1}{k} \sum _{l=0}^{\infty} (2l+1) \sin
\delta_{l}(k)^{2}=
$$
\begin{equation}
\frac{k}{4\pi} \sigma_{tot}~,
\end{equation}
which is known as the {\em optical theorem}. 
Its physical significance
is related to the fact that
the interference of the incident wave with the dispersed wave 
at zero/forward angle produces the ``getting out" of the particle from the
incident wave, allowing in this way the conservation of the probability.

\section*{\bf Born approximation} 

Let us consider the situation of Fig. 7.2:

\vskip 2ex
\centerline{
\epsfxsize=120pt
\epsfbox{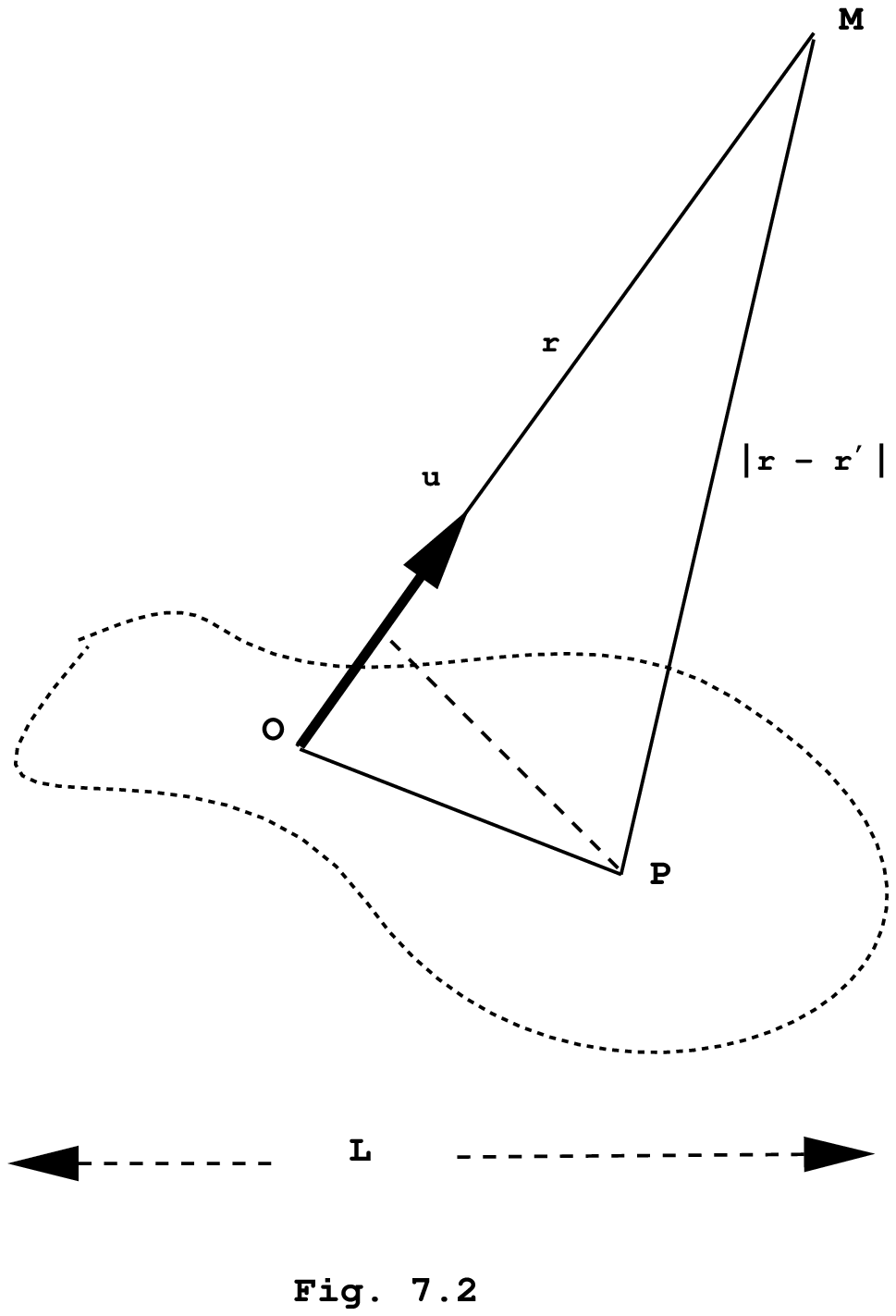}}
\vskip 2ex
The observation point M is far away from P, which is in the range of the 
potential $U$. The geometrical conditions are $r\gg L$, $r'\ll l$.
The segment MP that corresponds to $\vert {\bf r-r'} \vert$ is in the 
aforementioned geometrical conditions
approxiamtely equal to the projection of MP onto MO
\begin{equation}
\vert {\bf r-r'} \vert \simeq r-{\bf u \cdot r'}~,
\end{equation}
\noindent where ${\bf u}$ is a unit vector (versor) in the 
${\bf r}$ direction. Then, for large $r$ 
\begin{equation}
G=- \frac{1}{4\pi} \frac{e^{ik \vert {\bf r-r'} \vert}}{\vert {\bf 
r-r'}\vert} \simeq_{r \rightarrow \infty}  -\frac{1}{4 \pi} 
\frac{e^{ikr}}{r} e^{-ik {\bf u \cdot r}}~.
\end{equation}

\noindent We now substitute $G$ in the integral expression
for the scattering wave function
\begin{equation}
\psi({\bf r})= e^{ikz} - \frac{1}{4\pi} \frac 
{e^{ikr}}{r} 
\int e^{-ik {\bf u \cdot r}}U({\bf r'})\psi ({\bf r'}) 
d^{3}r'~.
\end{equation}

\noindent The latter is already not a function of the distance $r=OM$, but only
of $\theta$ and $\psi$. Thus 
\begin{equation}
f(\theta, \psi)= - \frac{1}{4\pi} \int e^{-ik {\bf u\ 
\cdot r}} U({\bf r'}) \psi ({\bf r'}) d^{3}r'~.
\end{equation}
We define now the incident wave vector ${\bf k_{i}}$
as a vector of modulus $k$ directed along the polar axis 
of the beam. Then 
$ e^{ikz}=e^{i {\bf k_{i} \cdot r}}$. 
Similarly, ${\bf k_{d}}$, of modulus
$k$ and of direction fixed by $\theta$ and $\varphi$, is called the shifted  
wave vector in the direction $(\theta, \varphi)$:
$ {\bf k_{d}}= k{\bf u}$.

The {\em momentum transfer} in the direction $(\theta, 
\varphi)$ is introduced as the vectorial difference ${\bf K}= {\bf k_{d}-k_{i}}$.

\vskip 1ex
\centerline{
\epsfxsize=80pt
\epsfbox{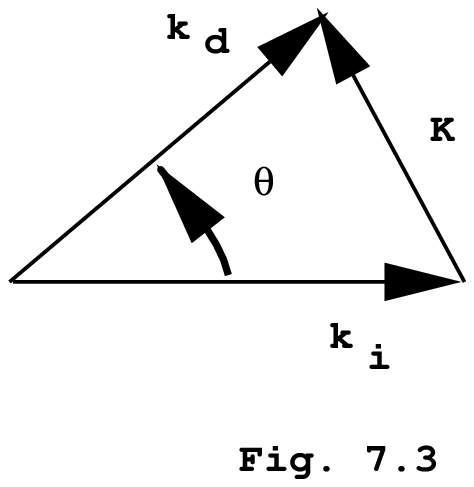}}
\vskip 2ex

Hence we can write the integral equation in the form
\begin{equation}
\label{eq:e3}
\psi ({\bf r})= e^{i{\bf k_{i}\cdot r}} + \int
G({\bf r,r'}) U({\bf r'}) \psi({\bf r'}) d^{3}r'
\end{equation}

One can try to solve this equation iteratively. Putting 
${\bf r} \rightarrow {\bf r'}; {\bf r'} \rightarrow {\bf r''}$, we can write
\begin{equation}
\psi ({\bf r'})= e^{i{\bf k_{i}\cdot r'}} + \int G({\bf r',r''}) U({\bf
r''}) \psi({\bf r''}) d^{3}r''~.
\end{equation}

Substituting in \ref{eq:e3}, we get 
$$
\psi({\bf r})= e^{i{\bf k}_{i}\cdot r} + \int G({\bf r,r'})U({\bf
r'})e^{i{\bf k_{i} \cdot r'}}d^{3}r'
+ 
$$
\begin{equation} \label{eq:e4}
\int \int G({\bf r,r'})U({\bf
r'})G({\bf r',r''})U({\bf r''}) \psi({\bf r''})d^{3}r'' d^{3}r'~.
\end{equation}

The first two terms in the right hand side are known  
and it is only the third one that includes the unknown function 
$\psi({\bf r})$. We can repeat the procedure: substituting ${\bf r}$ 
by ${\bf r''}$, and ${\bf r'}$ by ${\bf r'''}$, we get 
$\psi ({\bf r''})$~, 
that we can reintroduce in the eq. \ref{eq:e4}
$$
\psi({\bf r}) = e^{i {\bf k_{i} \cdot r}} + \int G({\bf r,r'})U({\bf 
r'}) e^{i {\bf k_{i} \cdot r'}}
+
$$
$$
\int \int G({\bf r,r'})U({\bf r'}) G({\bf r',r''})U({\bf r''})e^{i {\bf
k_{i} \cdot r''}}d^{3}r'd^{3}r''+$$
\begin{equation}
 \int \int \int  G({\bf r,r'})U({\bf r'}) G({\bf r',r''})U({\bf
r''})e^{i {\bf k_{i}\cdot r''}} G({\bf r'',r'''})U({\bf r'''}) \psi ({\bf 
r'''})~.
\end{equation}

\noindent The first three terms are now known and the unknown function
$\psi({\bf r})$ has been sent to the fourth term. In this way, 
by succesive iterations we can build the stationary dispersed
wave function. Notice that each term of the series expansion 
has one more power in the potential with respect to the previous one. We can go
on until we get a negligible expression in the right hand side, 
obtaining $\psi({\bf r})$ as a function of only known quantities.

Substituting the expression of $\psi({\bf r})$ in $f(\theta, \varphi)$, we get 
the expansion in Born series of the scattering amplitude. 
In first order in $U$, one should replace
$\psi({\bf r'})$ by $e^{i{\bf k_{i}\cdot r'}}$ in the right hand side to get
$$
f^{(B)}(\theta, \varphi)= \frac{-1}{4\pi}  \int e^{i{\bf k_{i}\cdot
r'}} U({\bf r'}) e^{-ik {\bf u\cdot r'}} d^{3}r'=
\frac{-1}{4\pi} \int e^{-i{\bf (k_{d}-k_{i})\cdot r'}} U({\bf r'})
d^{3}r'=
$$
\begin{equation}
\frac{-1}{4\pi} \int e^{-i{\bf K \cdot r'}}U({\bf r'})d^{3}r'
\end{equation}

\noindent
${\bf K}$ is the momentum transfer vector. Thus, the differential cross
section is simply related to the potential,
$V({\bf r})= \frac{\hbar^{2}}{2m} U({\bf r})$. Since 
$\sigma (\theta,\varphi)= \vert f(\theta, \varphi) \vert^{2}$, the result is
\begin{equation}
\sigma^{(B)} (\theta,\varphi)=\frac{m^{2}}{4\pi^{2}\hbar^{4}} \vert
\int e^{-i{\bf K \cdot r}} V({\bf r})d^{3}r \vert^{2}
\end{equation}

\noindent
The direction and modulus of ${\bf K}$ depends on the modulus
$k$ of ${\bf k_{i}}$ and ${\bf k_{d}}$ as well as on the scattering direction
$(\theta,\varphi)$. For given $\theta$ and $\varphi$, 
it is a function of $k$, the energy of the incident beam.
Analogously, for a given energy,
$\sigma^{(B)}$ is a function of 
$\theta$ and $\varphi$. Born's approximation allows one to get information
on the potential $V({\bf r})$ from the dependence of the differential cross 
section on the scattering direction and the incident energy.\\


\noindent \underline{{\bf 7N. Note}} - The following paper of Born 
was practically the first dealing with quantum scattering:

\noindent
M. Born, ``Quantenmechanik der Stossvorg\"ange" [``Quantum mechanics of
scattering processes "],
Zf. f. Physik {\bf 37}, 863-867 (1926)

\bigskip

\section*{{\huge 7P. Problems}}

{\bf Problem 7.1}

\noindent{\bf Calculus of complex variable for the scattering Green function}

We recall that we already obtained the result

\( G({\bf r,r'})=\frac{1}{(2\pi)^{3}} \int
\frac{e^{iqR}}{k^{2}-q^{2}}d^{3}q~, \)
cu $R=\vert {\bf r-r'} \vert$.
Since $d^{3}q=q^{2} \sin\theta dq d\theta d\phi$, we get
after integrating in angular variables 

\( G({\bf r,r'})= \frac{i}{4\pi^{2}R}\int_{-\infty} ^{\infty}
\frac{(e^{-iqR}-e^{iqR})}{k^{2}-q^{2}} q dq~. \)

\noindent Putting
$C=\frac{i}{4\pi^{2}R}$, we separate the integral in two parts 

\( C(\int _{-\infty} ^{\infty} \frac{e^{-iqR}}{k^{2}-q^{2}} q dq -
\int _{-\infty} ^{\infty}\frac{e^{iqR}}{k^{2}-q^{2}} q dq)~. \)

\noindent Let us make now $q \rightarrow -q$ in the first integral

\(  \int _{-\infty} ^{\infty} \frac{e^{-i(-q)R}}{k^{2}-(-q)^{2}} (-q)
d(-q)= \int _{\infty} ^{-\infty} \frac{e^{iqR}}{k^{2}-q^{2}} q dq
= -\int _{-\infty} ^{\infty} \frac{e^{iqR}}{k^{2}-q^{2}} q dq~, \)

\noindent so that

\( G({\bf r,r'})= -2C ( \int_{-\infty} ^{\infty} 
\frac{qe^{iqR}}{k^{2}-q^{2}}
dq)~.\)

\noindent Substituting $C$,
leads to

\( G({\bf r,r'})= \frac{-i}{2\pi^{2}R}\int_{-\infty} ^{\infty}
\frac{qe^{iqR}}{k^{2}-q^{2}}dq \)

In this form, the integral can be calculated by means
of the theorem of residues of its poles. 
Notice the presence of simple poles at $q=_{-}^{+}k$.

\vskip 2ex
\centerline{
\epsfxsize=280pt
\epsfbox{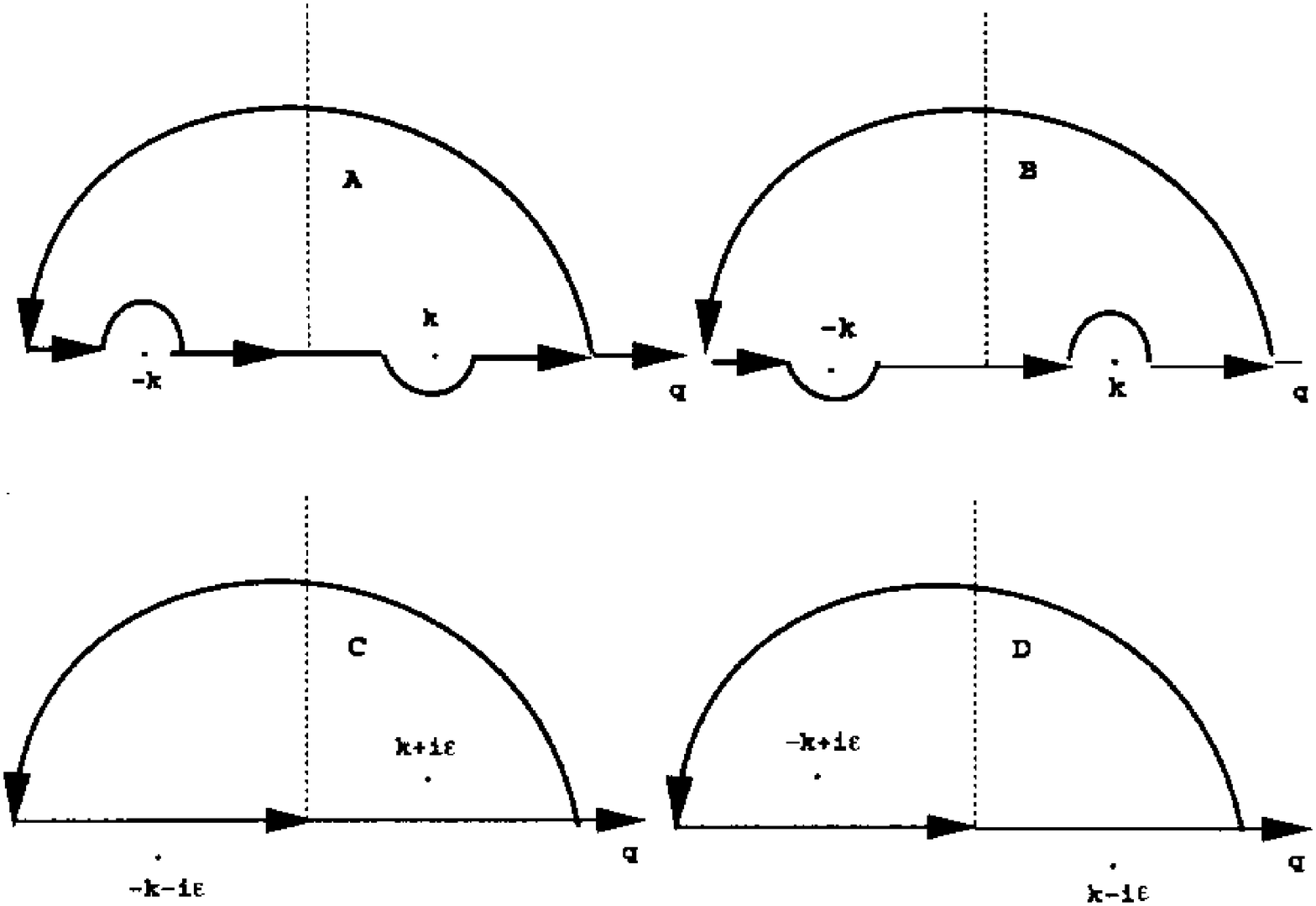}}
\vskip 4ex
\begin{center}
{\small{Fig. 7.4: Contour rules around the poles for $G_{+}$ and 
$G_{-}$}\\
}
\end{center}

We use the contour of fig. 7.4 encircling the poles 
as shown, because in this way we get the physically correct effect from the
theorem of residues

\( G(r)= - \frac{1}{4\pi} \frac{e^{ikr}}{r}\quad({\rm Im} k > 0) \) ,

\(G(r)= - \frac{1}{4\pi} \frac{e^{-ikr}}{r}\quad ({\rm Im} k < 0)~. \)

The solution of interest is the first one, because it provides divergent
waves, whereas the latter solution holds for convergent waves (propagating 
towards the target). Moreover, the linear combination

\( \frac{1}{2} \lim_{\epsilon\rightarrow 0} [G_{k+i\epsilon} +
G_{k-i\epsilon}] = - {\frac{1}{4\pi}} \frac{\cos kr}{r} \)

\noindent corresponds to stationary waves.

\noindent
The formal calculation of the integral can be performed by taking
$k^{2}-q^{2}\rightarrow k^{2}+i\epsilon-q^{2}$, so that:
\(\int _{-\infty} ^{\infty}\frac{qe^{iqR}}{k^{2}-q^{2}}dq \rightarrow
\int _{-\infty} ^{\infty}\frac{qe^{iqR}}{(k^{2}+i\epsilon)-q^{2}}dq~. \)
This is possible for $R>0$. This is why the contour for the calculation 
will be placed in the upper half plane. Thus, the poles of the integrand
are located at
$q=\pm\sqrt{k^{2}+i\epsilon} \simeq
\pm(k+\frac{i\epsilon}{2k})$.
The procedure of taking the limit $\epsilon \rightarrow 0$ should be applied 
{\em after} calculating the integral.\\

{\bf Problem 7.2}

\noindent {\bf Asymptotic form of the radial function}

As we have already seen in the chapter {\it Hydrogen atom} 
the radial part of the Schr\"odinger equation can be written

\( ( \frac {d^2}{dr^{2}} + \frac{2}{r} \frac{d}{dr} ) 
R_{nlm}(r)-\frac{2m}{\hbar^{2}}[V(r)+\frac{l(l+1) 
\hbar^{2}}{2mr^{2}}]R_{nlm}(r)+\frac{2mE}{\hbar^{2}}R_{nlm}(r)=0~. \)

\noindent $n,l,m$ are the spherical quantum numbers. For the sake of convenience
of writing we shall discard them hereafter. 
$R$ is the radial wave function (i.e., depends only on
$r$). We assume that the potential goes to zero stronger than
$1/r$, and that $\lim_{r \rightarrow 0} r^{2}V(r)=0$.

Using $u(r)=rR$, since
$(\frac{d^{2}}{dr^{2}} + \frac{2}{r} \frac{d}{dr})\frac{u}{r} =
\frac{1}{r} \frac{d^{2}}{dr^{2}}u$, we have

\( \frac{d^{2}}{dr^{2}}u + 
\frac{2m}{\hbar^{2}}[E-V(r)-\frac{l(l+1)\hbar^{2}}{2mr^{2}}] u=0~. \)

\noindent Notice that the potential displays a 
supplementary term 

\( V(r)\rightarrow V(r)+\frac{l(l+1)\hbar^{2}}{2mr^{2}}~,\)

\noindent which corresponds to a repulsive centrifugal barrier.
For a free particle $V(r)=0$, and the equation becomes 

\( [\frac{d^{2}}{dr^{2}} + \frac{2}{r} 
\frac{d}{dr})-\frac{l(l+1)}{r^{2}}]R + k^{2}R=0~. \)

\noindent Introducing the variable $\rho=kr$, we get

\( \frac{d^{2}R}{d\rho^{2}} + \frac{2}{\rho} \frac{dR}{d\rho} - 
\frac{l(l+1)}{\rho^{2}}R + R=0~. \)

\noindent
The solutions are the so-called spherical Bessel functions. 
The regular solution is

\( j_{l}(\rho)=(-\rho)^{l} (\frac{1}{\rho} \frac{d}{d\rho})^{l} (\frac{\sin 
\rho}{\rho})~, \)

\noindent while the irregular one 

\( n_{l}(\rho)= - (-\rho)^{l} (\frac{1}{\rho} \frac{d}{d\rho})^{l} 
(\frac{\cos \rho}{\rho})~. \)

For large $\rho$, the functions of interest are the spherical Hankel functions

\( h_{l}^{(1)}(\rho)=j_{l}(\rho)+ in_{l}(\rho) \)
\c{s}i
\( h_{l}^{(2)}(\rho)=[ h_{l}^{(1)}(\rho)]^{*}~. \)

The behaviour for $\rho \gg l$ is of special interest
\begin{equation}
\label{eq:P1}
j_{l}(\rho) \simeq \frac{1}{\rho} \sin
(\rho-\frac{l\pi}{2})
\end{equation}  
\begin{equation}
\label{eq:P2}
 n_{l}(\rho) \simeq - \frac{1}{\rho} \cos(\rho-\frac{l\pi}{2})~. 
\end{equation}

\noindent Then

\( h_{l}^{(1)} \simeq  -\frac{i}{\rho} e^{i(\rho - l\pi/2)}~. \)

The solution regular at the origin is 
$R_{l}(r)=j_{l}(kr)~.$

The asymptotic form is (using eq. \ref{eq:P1})

\( R_{l}(r) \simeq \frac{1}{2ikr}[e^{-ikr-l\pi/2}-e^{ikr-l\pi/2}]~.\)

\bigskip

{\bf Problem 7.3}

\noindent {\bf Born approximation for Yukawa potentials}

Let us consider the potential of the form 

\begin{equation} 
V({\bf r})= V_{0} \frac{e^{-\alpha r}}{r}~, 
\end{equation}

\noindent where $V_{0}$ and $\alpha$ are real constants and $\alpha$ is 
positive. The potential is either attractive or repulsive 
depending on the sign of $V_{0}$; 
the larger  $\vert V_{0} \vert$, the stronger the potential.
We assume that $\vert V_{0} \vert$ is sufficiently small that Born's 
approximation holds. According to a previous formula, 
the scattering amplitude is given by\\

\( f^{(B)}(\theta, \varphi)= - \frac{1}{4\pi} \frac{2mV_{0}}{\hbar^{2}} 
\int e^{-i {\bf K \cdot r}} \frac{e^{-\alpha r}}{r} d^{3}r~. \)

Since this potential depends only on $r$, the angular integrals are
trivial leading to the form

\( f^{(B)}(\theta, \varphi)=  \frac{1}{4\pi} \frac{2mV_{0}}{\hbar^{2}} 
\frac{4\pi}{\vert {\bf K} \vert} \int_{0}^{\infty} \sin \vert {\bf K} 
\vert r \frac{e^{-\alpha r}}{r} r dr~. \)

Thus, we obtain

\( f^{(B)}(\theta, \varphi)=  -\frac{2mV_{0}}{\hbar^{2}} \frac 
{1}{\alpha^{2} + \vert {\bf K}\vert^{2}}~.\)


From the figure we can notice that
$\vert {\bf K} \vert = 2k \sin \frac{\theta}{2}$. Therefore

\( \sigma^{(B)}(\theta)=\frac{4m^{2}V_{0}^{2}}{\hbar^{4}} 
\frac{1}{[\alpha^{2} + 4k^{2} \sin \frac{\theta}{2}^{2}]^{2}}~. \)

The total cross section is obtained by integrating

\( \sigma^{(B)} = \int \sigma^{(B)}(\theta) d\Omega= 
\frac{4m^{2}V_{0}^{2}}{\hbar^{4}} \frac{4\pi}{\alpha^{2}(\alpha^{2}+4k^{2})}
~. \)


\newpage
\def\bi{bigskip}
\def\noi{noindent}
\begin{center}
{\huge 8. PARTIAL WAVES}
\end{center}

\section*{Introduction}
\setcounter{equation}{0}
The partial waves method is quite general and applies to 
particles interacting  
in very small spatial regions with another one, which is usually known 
as scattering center because of its physical characteristics.
(for example, because it can be considered as fixed). 
Beyond the interaction region, the interaction between the two particles
is usually negligible. Under this circumstances, it is possible to describe
the scattered particle by means of the Hamiltonian
\begin{equation}
H=H_0+V~,
\end{equation}

\noindent
where $H_0$ corresponds to the free particle Hamiltonian.
Our problem is to solve the equation
\begin{equation}
(H_0+V) \mid \psi \rangle = E \mid \psi \rangle~.
\end{equation}

Obviously, the spectrum will be continuous 
since we study the case of elastic scattering. The solution will be
\begin{equation}
\mid \psi \rangle = \frac {1}{E-H_0} V\mid \psi \rangle + \mid \phi \rangle~.
\end{equation}

It is easy to see that for $V=0$  one can obtain the solution
$\mid \phi \rangle $, i.e., the solution corresponding to the free particle.
It is worth noting that in a certain sense the operator $\frac{1}{E-H_0}$
is anomalous, because it has a continuum of poles on the real axis
at positions coinciding with the eigenvalues of $H_0$. To get out of this 
trouble, it is common to produce a small shift in the imaginary direction
($\pm i\epsilon$) of the cut on the real axis
\begin{equation}
\mid \psi^{\pm} \rangle = \frac {1}{E-H_0 \pm i\varepsilon} V\mid \psi^{\pm} 
\rangle + \mid \phi \rangle
\end{equation}

This equation is known as the Lippmann-Schwinger equation.
Finally, the shift of the poles is performed in the positive sense of the 
imaginary axis because in this case the causality principle holds
(cf. 
Feynman). Let us consider the x representation
\begin{equation}
\langle {\bf {x}}\mid \psi^{\pm} \rangle =\langle {\bf {x}}\mid \phi \rangle + 
\int d^{3} x^{'}\left \langle {\bf {x}} \vert \frac {1}{E-H_0 \pm i\varepsilon 
}\vert {\bf {x^{'}}} \right \rangle \langle {\bf {x^{'}}} \mid V\mid 
\psi^{\pm}\rangle~.
\end{equation}

The first term on the right hand side  
corresponds to a free particle, while the second one is interpreted as a 
spherical wave getting out from the scattering center. 
The kernel of the previous integral can be considered as a Green
function (also called propagator in quantum mechanics). It is a simple matter 
to calculate it
\begin{equation}
G_{\pm}({\bf {x}},{\bf {x^{'}}})=\frac{\hbar^{2}}{2m}\left \langle {\bf {x}} \vert 
\frac {1}{E-H_0 \pm i\varepsilon}\vert {\bf {x^{'}}} \right \rangle = 
-\frac{1}{4\pi} \frac{e^{\pm ik\mid {\bf {x}}-{\bf {x^{'}}}\mid}}{\mid {\bf 
{x}}-{\bf {x^{'}}}\mid}~,
\end{equation}

\noindent where  $E={\hbar^{2}}{k^2}/2m$.
Writing the wave function as a plane wave plus a divergent spherical one
(up to a constant factor),
\begin{equation}
\langle {\bf {x}}\mid \psi^{+} \rangle =e^{{\bf {k}}\cdot {\bf {x}}} + 
\frac{e^{ikr}}{r} f({\bf {k}},{\bf {k^{'}}})~.
\end{equation}
\noindent the quantity
$f({\bf {k}},{\bf {k^{'}}})$ is known as the scattering amplitude
and is explicitly
\begin{equation}
f({\bf {k}},{\bf {k^{'}}})=-\frac{1}{4\pi} {(2\pi )^3}\frac{2m}{\hbar^2}\langle 
{\bf {k^{'}}}\mid V \mid \psi^{+} \rangle~. 
\end{equation}

Let us now define an operator $T$ such that
\begin{equation}
T\mid \phi \rangle = V\mid \psi^{+} \rangle
\end{equation}

If we multiply the Lippmann-Schwinger equation by $V$ and make use of the 
previous definition, we get
\begin{equation}
T\mid \phi \rangle = V\mid \phi \rangle + V\frac{1}{E-H_0+
i\varepsilon}T\mid \phi 
\rangle ~. 
\end{equation}
Iterating this equation (as in perturbation theory) 
we can get the Born approximation and its higher-order corrections.

\section*{Partial waves method}

Let us now consider the case of a central potential. In this case,
using the definition~(9), it is found that the operator $T$ commutes with
$\vec {L}^{2} $ and $\vec {L}$; it is said that  
$T$ is a scalar operator. To simplify the calculations it is convenient to
use spherical coordinates, 
because of the symmetry of the problem that turns the $T$ operator
diagonal. Let us see now a more explicit form of the scattering amplitude 
\begin{equation}
f({\bf {k}},{\bf {k^{'}}})={\rm const.}\sum_{lml^{'}m^{'}} 
\int dE\int 
dE^{'}\langle {\bf {k^{'}}}\mid E^{'} l^{'} m^{'} \rangle \langle E^{'} 
l^{'} m^{'}\mid T\mid Elm\rangle \langle Elm\mid \bf {k} \rangle~,
\end{equation}
where ${\rm const.}=-\frac{1}{4\pi}\frac{2m}{\hbar^2} {(2\pi)^3}$.
After some calculation, one gets
\begin{equation}
f({\bf {k}},{\bf {k^{'}}})=-\frac{4\pi^2}{k}\sum_{l}\sum_{m} T_{l} (E) 
Y^{m}_{l} ({\bf {k^{'}}})Y^{m^{*}}_{l}(\bf {k})~.
\end{equation}

Choosing the coordinate system such that the vector $\bf {k}$ have the same 
direction with the z axis, one infers that  
only the spherical harmonics of $m=0$ will contribute to the scattering 
amplitude. If we define by $\theta$ the angle between ${\bf {k}}$ and
${\bf {k^{'}}}$, we will get
\begin{equation}
Y^{0}_{l} ({\bf {k^{'}}})=\sqrt {\frac{2l+1}{4\pi}} P_{l}(cos\theta)~.
\end{equation}

Employing the following definition 
\begin{equation}
f_{l}(k)\equiv-\frac{\pi T_{l} (E)}{k}~,
\end{equation}
\noindent
eq.~(12) can be written as follows 
\begin{equation}
f({\bf {k}},{\bf {k^{'}}})=f(\theta)=\sum^{\infty}_{l=0} 
(2l+1)f_{l}(k)P_{l}(cos\theta)~.
\end{equation}

For $f_{l}(k)$ a simple interpretation can be provided, which is based on the 
expansion of a plane wave in spherical waves.
Thus, we can write 
the function $\langle {\bf {x}}\mid \psi^{+} \rangle$ for large values of $r$ 
in the following form
$$
\langle {\bf {x}}\mid \psi^{+} \rangle = \frac{1}{{(2\pi )^{3/2}}}\left[ 
{e^{ikz}}+f(\theta ) \frac{{e^{ikr}}}{r}\right] =
$$
$$
\frac{1}{{(2\pi)^{3/2}}}\left[ \sum_{l} (2l+1)P_{l}(\cos\theta 
)\left(\frac{{e^{ikr}}-{e^{i(kr-l\pi )}}}{2ikr} \right) 
+\sum_{l}(2l+1)f_{l}(k)P_{l}(\cos\theta )\frac{{e^{ikr}}}{r}\right]
$$
\begin{equation}
=\frac{1}{{(2\pi )^{3/2}}}\sum_{l} 
(2l+1)\frac {P_{l}(\cos\theta )}{2ik}\left[ \left[ 
1+2ikf_{l}(k)\right]\frac{{e^{ikr}}}{r}-\frac{{e^{i(kr-l\pi 
)}}}{r} \right]~.
\end{equation}

This expression can be interpreted as follows. 
The two exponential terms correspond to spherical waves: the first to a divergent
wave, and the latter to a convergent one. Moreover, the scattering effect
is conveniently displayed in the coefficient of the divergent wave, which is 
unity when there are no scattering centers.

\section*{Phase shifts}

We consider now a surface enclosing the scattering center.
Assuming that there is no creation and annihilation of particles, one has 
\begin{equation}
\int {\bf {j}}\cdot d{\bf {S}}=0~,
\end{equation}

\noindent
where the integration region is the aforementioned surface,
and ${\bf {j}}$ is the probability current density. 
Moreover, because of the conservation of the orbital momentum,
the latter equation should hold for each partial wave. The theoretical 
formulation of the
problem does not change if one assumes the wave packet as a flux of 
noninteracting particles propagating through a region of central potential 
for which the angular momentum of each particle is conserved, so that the 
`particle' content of the wave packet really does not change. Thus, one 
may think even intuitevely that only phase factor effects can be introduced
under these circumstances. Thus, if one defines
\begin{equation}
S_{l}(k)\equiv 1+ 2ikf_{l}(k)
\end{equation}

\noindent we should have
\begin{equation}
\mid S_{l}(k)\mid =1~.
\end{equation}

These results can be interpreted using the conservation of probabilities.
They are natural and expected because we assumed that there is no 
creation and annihilation of particles. 
Therefore, the effects of the scattering center 
is reduced to adding a phase factor in the components of the divergent wave.
Taking into account the unitarity of the phase factor, we can write it in the 
form
\begin{equation}
S_{l}=e^{2i\delta_{l}}~,
\end{equation}

\noindent 
where $\delta_{l}$ is a real function of $k$. 
Taking into account the definition (18), we can write
\begin{equation}
f_{l}=\frac{{e^{2i\delta_{l}}}-1}{2ik}=\frac{{e^{i\delta_{l}}}\sin 
(\delta_{l})}{k}=\frac{1}{k\cot (\delta_{l})-ik}~. 
\end{equation}

The total cross section has the following form 
$$
\sigma_{total}=\int \mid f(\theta){\mid ^2}d\Omega =
$$
$$
\frac{1}{{k^2}}{\int _{0} ^{2\pi}}d\phi {\int _{-1} ^{1}}d(\cos (\theta 
))\sum_{l} \sum_{{l^{'}}}(2l+1)(2{l^{'}}+1){e^{i\delta_{l}}}\sin 
(\delta_{l}){e^{i\delta_{{l^{'}}}}} \sin (\delta_{{l^{'}}})P_{l}P_{{l^{'}}}
$$
\begin{equation}
=\frac{4\pi }{{k^2}}\sum_{l} (2l+1)\sin {^2}(\delta_{{l^{'}}})~.
\end{equation}

\section*{Getting the phase shifts}
Let us consider now a potential V that is zero for $r>R$, 
where the parameter $R$ is known as the range of the potential. Thus, the region 
$r>R$  corresponds to a spherical unperturbed/free wave.
On the other hand, the general form of the expansion 
of a plane wave in spherical ones is
\begin{equation}
\langle {\bf {x}}\mid \psi^{+} \rangle =\frac{1}{{(2\pi )^{3/2}}}\sum_{l} 
{i^{l}} (2l+1)A_{l}(r)P_{l}(\cos \theta ) \quad (r>R)~,
\end{equation}

\noindent
where the coefficient $A_{l}$ is by definition 
\begin{equation}
A_{l}={c_{l} ^{(1)}}{h_{l} ^{(1)}}(kr)+{c_{l} ^{(2)}}{h_{l} ^{(2)}}(kr)~.
\end{equation}

${h_{l} ^{(1)}}$ and ${h_{l} ^{(2)}}$ are the spherical 
Hankel functions whose asymptotic forms are the following
$$ 
{h_{l} ^{(1)}} \sim \frac{{e^{i(kr-l\pi /2)}}}{ikr}
$$
$$
{h_{l} ^{(2)}} \sim - \frac{{e^{-i(kr-l\pi /2)}}}{ikr}~.
$$

Inspecting the following asymptotic form of the expression (23)
\begin{equation}
\frac{1}{{(2\pi )^{3/2}}}\sum_{l}(2l+1)P_{l}\left[ \frac{{e^{ikr}}}{2ikr}-
\frac{{e^{-i(kr-l\pi)}}}{2ikr} \right]~,
\end{equation}

one can see that
\begin{equation}
{c_{l} ^{(1)}}=\frac{1}{2} e^{2i\delta_{l}} \qquad {c_{l} ^{(2)}}=\frac{1}{2}~.
\end{equation}

This allows to write the radial wave function for $r>R$ in the form
\begin{equation}
A_{l}=e^{2i\delta _{l}}\left[ \cos \delta _{l} j_{l} (kr)
- \sin \delta _{l}n_{l} 
(kr)\right]~. \end{equation}

Using the latter equation, we can get the logarithmic derivative in 
$r=R$, i.e., at the boundary of the potential range 
\begin{equation}
\beta _{l}\equiv \left( \frac{r}{A_{l}}\frac{dA_{l}}{dr}\right)_{r=R}=
kR\left[
\frac{{j_{l}^{'}}\cos \delta _{l}-{n_{l}^{'}}(kR)\sin \delta _{l}}{j_{l}\cos
\delta _{l}-{n_{l}}(kR)\sin \delta_{l}}\right]~.
\end{equation}

\noindent
$j_{l}^{'}$ is the derivative of $j_{l}$ with respect to $r$ evaluated at
$r=R$. Another important result that can be obtained from the knowledge of the
previous one is the phase shift
\begin{equation}
\tan \delta _{l}=\frac{kR{j_{l}^{'}}(kR)-\beta _{l}
j_{l}(kR)}{kR{n_{l}^{'}}(kR)-
\beta_{l} n_{l}(kR)}~.
\end{equation}

To get the complete solution of the problem in this case, it is necessary to 
make the calculations for $r<R$, i.e., within the range of the potential.
For a central potential, the 3D Schr\"odinger equation reads
\begin{equation}
\frac{{d^{2}}u_{l}}{d{r^{2}}}+\left( {k^{2}}-\frac{2m}{{\hbar ^{2}}} 
V-\frac{l(l+1)}{{r^{2}}} \right) u_{l}=0~,
\end{equation}

\noindent
where $u_{l}=rA_{l}(r)$ is constrained by the boundary condition 
$u_{l}\mid _{r=0} \quad =0$. Thus, one can calculate the logarithmic derivative,
which, taking into account the continuity of the log-derivative
(equivalent to the continuity condition of the derivative at a 
discontinuity point) leads to
\begin{equation}
\beta_{l} \mid_{in}=\beta_{l}\mid_{out}~.
\end{equation}

\section*{An example: scattering on a hard sphere}
Let us now consider an important illustrative case, that of the hard
sphere potential
\begin{equation}
V=\left\{ 
\begin{array} {ll}
\infty & \mbox{ $r<R$} \\
0      & \mbox {$r>R~.$} 
\end{array}
 \right.
\end{equation}

It is known that a particle cannot penetrate
into a region where the potential is infinite. Therefore, the wave function 
should be zero at $r=R$. Since we deal with an impenetrable sphere we also have
\begin{equation}
A_{l}(r)\mid_{r=R} =0~.
\end{equation}

Thus, from eq.~(27), we get
\begin{equation}
\tan \delta_{l} = \frac{j_{l} (kR)}{n_{l} (kR)}~.
\end{equation}

One can see that the phase shift calculation is an easy one for any $l$.
In the $l=0$ case (s wave scattering), we have
$$
\delta_{l} = -kR
$$

\noindent
and from eq.~(27)

\begin{equation}
A_{l=0}(r)\sim \frac{\sin kr}{kr}\cos\delta_{0}+\frac{\cos 
kr}{kr}\sin\delta_{0}=\frac{1}{kr}\sin (kr+\delta_{0})~.
\end{equation}

\noindent
We immediately see that there is an additional phase contribution with regard 
to the motion of the free particle. It is also clear that in more general cases
the various waves will have different phase shifts leading to 
a transient distortion of the scattered wave packet.
At small energies, i.e., $kR<<1$,
the spherical Bessel functions (entering the formulas for the spherical 
Hankel functions) are the following
\begin{equation}
j_{l} (kr)\sim \frac{(kr)^{l}}{(2l+1)!!}
\end{equation}
\begin{equation}
n_{l} (kr)\sim -\frac{(2l-1)!!}{(kr)^{l+1}}~,
\end{equation}

\noindent
leading to
\begin{equation}
\tan\delta_{l} = \frac{-(kR)^{2l+1}}{(2l+1)[(2l-1)!!]^{2}}~.
\end{equation}

From this formula, one can see that a substantial contribution to the phase 
shift is given by the $l=0$ waves. Moreover, since 
$\delta_{0}=-kR$ the cross section is obtained as follows
\begin{equation}
\sigma_{total}=\int\frac{d\sigma}{d\Omega}d\Omega=4\pi R^{2}~.
\end{equation}

One can see that the total scattering cross section is four times bigger than 
the classical one and coincides with the total area of the impenetrable sphere.
For large values of the incident energy, one can work in the hypothesis that
all values of $l$ up to a maximum value
$l_{max}\sim kR$ contribute to the total cross section
\begin{equation}
\sigma_{total}
=\frac{4\pi}{k^{2}}{\sum_{l=0} ^{l\sim kR}}(2l+1){\sin}^{2}\delta_{l}~.
\end{equation}

In this way, from eq.~(34), we have
\begin{equation}
{\sin}^{2}\delta_{l}=\frac{\tan^{2}\delta_{l}}{1+\tan^{2}\delta_{l}}=
\frac{[j_{l} (kR)]^{2}}{[j_{l} (kR)]^{2}+[n_{l} (kR)]^{2}}\sim\sin^{2}\left( 
kR-\frac{l\pi}{2}\right)~,
\end{equation}
where the expressions
$$
j_{l} (kr)\sim\frac{1}{kr}\sin\left( kr-\frac{l\pi}{2}\right)
$$
$$
n_{l} (kr)\sim -\frac{1}{kr}\cos\left( kr-\frac{l\pi}{2}\right)~.
$$
have been used.

Inspection of $\delta_{l}$ shows a negative jump of $\frac{\pi}{2}$
whenever $l$ is augmented by a unity. Thus, it is clear that  
${\sin}^{2}\delta_{l}+{\sin}^{2}\delta_{l+1}=1$ holds.
Approximating ${\sin}^{2}\delta_{l}$ by its mean value 
$\frac{1}{2}$ over a period and using the sum of odd numbers, one gets 
\begin{equation}
\sigma_{total}=\frac{4\pi}{k^{2}}(kR)^{2}\frac{1}{2}=2\pi R^{2}~.
\end{equation}

Once again the quantum-mechanical result, although quite similar to the 
corresponding classical result is nevertheless different.
What might be the origin of the factor of two that makes the difference ? 
To get an explanation, we first separate eq.~(15) in two parts
\begin{equation}
f(\theta )=\frac{1}{2ik}{\sum_{l=0} ^{l=kR}}(2l+1){e^{2i\delta_{l}}}P_{l}
\cos (\theta )+\frac{i}{2k}{\sum_{l=0} ^{l=
kR}}(2l+1) P_{l}\cos (\theta )=f_{\mbox{refl}}+f_{\mbox{shadow}}~.
\end{equation}

Calculation of $\int |f_{\mbox{ refl}}|^{2}d\Omega$ gives
\begin{equation}
\int |f_{\mbox{ refl}}|^{2}d\Omega=\frac{2\pi}{4k^2}{\sum_{l=0} 
^{l_{max}}}{{\int_{-1}}^{1}}(2l+1)^{2}[P_{l}\cos (\theta )]^{2} d(\cos \theta 
)=\frac{\pi l^{2}_{max}}{k^{2}}=\pi R^{2}~.
\end{equation}

Analysing now $f_{\mbox{shadow}}$ at small angles, we get
\begin{equation}
f_{\mbox{shadow}}\sim\frac{i}{2k}\sum (2l+1)J_{0}(l\theta )\sim ik{\int_{0} 
^{R}}bJ_{0}(kb\theta )db=\frac{iRJ_{1}(kR\theta )}{\theta}~.
\end{equation}

This formula is rather well known in optics. It corresponds to the  
Fraunhofer diffraction. Employing the change of variable 
$z=kR\theta$ one can calculate the   
integral $\int |f_{\mbox{ shadow }}|^{2}d\Omega$
\begin{equation}
\int |f_{\mbox{shadow}}|^{2}d\Omega \sim 2\pi R^{2}{\int_{0} 
^{\infty}}\frac{[J_{1}(z)]^{2}}{z} dz\sim\pi R^{2}~.
\end{equation}

Finally, neglecting the interference between $f_{\mbox{refl}}$ 
and $f_{\mbox{ shadow }}$ 
(since the phase oscillates between $2\delta_{l+1}=2\delta_{l}-\pi$), one gets 
the result (42). The label `shadow' for one of the terms is easily explained
if one thinks of the wavy behaviour of the scattered particle
(from the physical viewpoint there is no difference
between a wave packet and a particle in this case).
Its origin can be traced back to the backward-scattered components of the 
wave packet leading to a phase shift with respect to the incident waves
and destructive interference.

\section*{Coulomb scattering}
In this section we briefly consider the Coulomb scattering 
in the quantum-mechanical approach. 
For this case, the Schr\"odinger equation is
\begin{equation}
\left( -\frac{\hbar ^{2}}{2m}\nabla ^{2} - \frac{Z_{1}Z_{2}e^{2}}{r}\right)\psi 
({\bf {r}})=E\psi ({\bf {r}}), \qquad E>0~,
\end{equation}

\noindent
where $m$ is the reduced mass of the system,
$E>0$ since we deal with the simple scattering case
where no kind of bound states are allowed to form.
The previous equation is equivalent to the following expression 
(for adequate values of the constants $k$ and $\gamma$)
\begin{equation}
\left( \nabla ^{2} +{k^{2}} +\frac{2\gamma k}{r}\right)\psi ({\bf {r}})=0~.
\end{equation}

If we do not consider the centrifugal barrier,
i.e., we look only to the $s$ waves, we really deal with a pure coulombian 
interaction, for which one can propose a solution of the following form
\begin{equation}
\psi ({\bf {r}})={e^{i{\bf {k\cdot r}}}}\chi (u)~,
\end{equation}

\noindent
where
$$
u=ikr(1-\cos\theta )=ik(r-z)=ikw~,
$$
$$
{\bf {k\cdot r}}=kz~.
$$

\noindent
$\psi ({\bf {r}})$ is the complete solution of the
Schr\"odinger equation with an asymptotic `physical' behaviour to which  
a plane wave 
${e^{i{\bf {k\cdot r}}}}$ and a spherical wave are expected to contribute 
${r^{-1}e^{ikr}}$ are expected to contribute. Defining new variables
\begin{displaymath}
z=z \qquad w=r-z \qquad \lambda =\phi~,
\end{displaymath}

\noindent
and by employing of previous relationships, eq. (48) takes the form
\begin{equation}
\left[ u \frac{d^{2}}{du^{2}}+(1-u)\frac{d}{du}-i\gamma\right]\chi (u)=0~.
\end{equation}

To solve this equation, one should first study its asymptotic behaviour.
Since we have already tackled this issue, we merely present the asymptotic
normalized wave function that is the final result of all previous 
calculations 
\begin{equation}
\psi_{\bf k} ({\bf {r}})=\frac{1}{(2\pi )^{3/2}}\left( {e^{i[{\bf {k\cdot 
r}}-\gamma ln(kr-{\bf {k\cdot r}})]}}+
\frac{f_{c}(k,\theta){e^{i[kr+\gamma 
ln2kr]}}}{r}\right)~.
\end{equation}

As one can see, this wave function displays terms that turns it quite 
different from the form in eq. (7). This is due to the fact that the Coulomb
potential is of infinite range. 
Performing the exact calculation for the Coulomb scattering amplitude 
is not an easy matter. 
Here we give only the final result for the `normalized' wave function 
\begin{equation}
\psi_{\bf k} ({\bf {r}})=\frac{1}{(2\pi )^{3/2}}\left( {e^{i[{\bf {k\cdot r}}-
\gamma ln(kr-{\bf {k\cdot 
r}})]}}+\frac{g_{1}^{*}(\gamma )}{g_{1}(\gamma )}\frac{\gamma}{2k\sin 
(\theta /2) ^{2}}\frac{e^{i[kr+\gamma ln2kr]}}{r}\right)~,
\end{equation}

\noindent where $g_{1}(\gamma )=\frac{1}{\Gamma (1-i\gamma )}$.

In addition, we reduce the partial wave analysis to a clear cut presentation
of the results, of which some have already been mentioned.
First of all, we write the wave function $\psi ({\bf {r}})$ in (49) 
as follows
\begin{equation}
\psi ({\bf {r}})={e^{i{\bf {k\cdot r}}}}\chi (u)=A{e^{i{\bf {k\cdot 
r}}}}\int_{C}{e^{ut}}{t^{i\gamma -1}}(1-t)^{-i\gamma}dt~,
\end{equation}

\noindent
where $A$ is a `normalization' constant, while all the integral part
is the inverse  
Laplace transform of the direct transform of eq.~(50). A convenient form
of the latter equation is
\begin{equation}
\psi ({\bf {r}})=A\int_{C}{e^{i{\bf {k\cdot r}}}(1-t)}{e^{ikrt}}(1-t)
d(t,\gamma )dt
\end{equation}

\noindent where
\begin{equation}
d(t,\gamma )={t^{i\gamma -1}}(1-t)^{-i\gamma -1}~.
\end{equation}

Within the partial wave analysis we proceed by writing  
\begin{equation}
\psi ({\bf {r}})={\sum_{l=0} ^{\infty}}(2l+1){i^{l}}P_{l}(\cos\theta )A_{l}(kr)~,
\end{equation}

\noindent
where
\begin{equation}
A_{l}(kr)=A\int_{C}{e^{ikrt}}j_{l}[kr(1-t)](1-t)d(t,\gamma )~.
\end{equation}

Applying the relationships between the 
spherical Bessel functions and the Hankel functions, we get
\begin{equation}
A_{l}(kr)=A_{l}^{(1)}(kr)+A_{l}^{(2)}(kr)~.
\end{equation}

We shall not sketch here how these coefficients are obtained 
(this is quite messy). They are
\begin{equation}
A_{l}^{(1)}(kr)=0
\end{equation}
\begin{equation}
A_{l}^{(2)}(kr)\sim -\frac{Ae^{\pi\gamma /2}}{2ikr}[2\pi ig_{1}(\gamma)]
\left( 
e^{-i[kr-(l\pi /2)+\gamma \ln 2kr]}-{e^{2i\eta_{l} (k)}}
e^{i[kr-(l\pi /2)+\gamma 
\ln 2kr]}\right)
\end{equation}

\noindent
where
\begin{equation}
{e^{2i\eta_{l} (k)}}=\frac{\Gamma (1+l-i\gamma )}{\Gamma (1+l+i\gamma )}~.
\end{equation}

\section*{Calculation of the Coulomb scattering amplitude}

If we perform the Laplace transform of eq.~(50), we get
\begin{equation}
\chi (u)=A\int_{C} e^{ut}t^{i\gamma-1}(1-t)^{-i\gamma}dt~.
\end{equation}

The contour $C$ goes from $-\infty $ to $\infty$ on the real axis 
and closes through the upper half-plane. 
There are two poles in this case at
$t=0$ and $t=1$. By the change of variable $s=ut$, we get 
\begin{equation}
\chi (u)=A\int_{C_{1}}e^{s}s^{i\gamma -1}(u-s)^{-i\gamma}~.
\end{equation}

$\chi (u)$ should be regular in zero. Indeed, we get
\begin{equation}
\chi (0)=(-1)^{-i\gamma}A\int_{C_{1}}\frac{e^{s}}{s}ds~.
=(-1)^{-i\gamma}A2\pi i
\end{equation}

Performing now the limit $u\to\infty$, let's do an infinitesimal shift to
avoid the location of the poles on the contour. Moreover, by the change of
variable $\frac{s}{u}=
-\frac{(s_{0}\pm i\varepsilon)}{i\kappa}$, 
we see that this expression goes to zero when $u\to\infty$. Thus, 
we can expand 
$(u-s)$ in power series of $\frac{s}{u}$ for the pole with $s=0$. 
This expansion is not the right one in $s=1$, because in this case
$s=-s_{0}+i(\kappa\pm\varepsilon)$. It comes out that $\frac{s}{u}=
1-\frac{(s_{0}\pm 
i\varepsilon )}{\kappa}$ tends to $1$ when $\kappa\to\infty$. If instead
we do the change of variable 
$s^{'}=s-u$, we get rid of this difficulty 
\begin{equation}
\chi (u)=A\int_{\rm C_{2}}\left([e^{s}s^{i\gamma
-1}(u-s)^{-i\gamma}]ds+[e^{s^{'}+u}(-s^{'})^{i\gamma}(u+s^{'})^{i\gamma-1}]
ds^{'}\right)~.
\end{equation}

Expanding the power series, it is easy to calculate the previous integrals,
but one should take the limit 
$\frac{s}{u}\to 0$ in the result in order to get the correct asymptotic
forms for the Coulomb scattering
$$
\chi (u)\sim 2\pi iA\left[u^{-i\gamma}g_{1}(\gamma )-(-u)^{i\gamma 
-1}e^{u}g_{2}(\gamma )\right]
$$
$$
2\pi g_{1}(\gamma )=i\int_{\rm C_{2}}e^{s}s^{i\gamma -1}ds
$$
\begin{equation}
2\pi g_{2}(\gamma )=i\int_{\rm C_{2}}e^{s}s^{-i\gamma}ds~.
\end{equation}

After all this chain of variable changes, we get back to the original 
$s$ one to obtain
$$
(u^{*})^{i\gamma}=(-i)^{i\gamma}[k(r-z)]^{i\gamma}
=e^{\gamma\pi /2}e^{i\gamma \ln k(r-z)}
$$
\begin{equation}
(u)^{-i\gamma}=(i)^{-i\gamma}[k(r-z)]^{-i\gamma}=e^{\gamma\pi /2}e^{-i\gamma 
\ln k(r-z)}~.
\end{equation}

The calculation of $\chi$, once effected, is equivalent with having
$\psi_{\bf k} ({\bf {r}})$ starting from (49).

\section*{Eikonal approximation}

We shall briefly expound on the eikonal approximation whose philosophy
is the same to that used when one wants to pass from the wave optics to the 
geometrical optics. Therefore, it is the right approximation
when the potential varies slowly over distances comparable to
to the wavelength of the scattered wave packet, i.e., for the case
$E>>|V|$. Thus, this approximation may be considered as a quasiclassical
one. First, we propose that the quasiclassical wave function  
has the known form
\begin{equation}
\psi\sim e^{iS({\bf r})/\hbar}~,
\end{equation}

\noindent
where $S$ satisfies the Hamilton-Jacobi equation, having the solution
\begin{equation}
\frac{S}{\hbar}=\int_{-\infty}^{z}\left[ k^{2}-\frac{2m}{\hbar ^{2}}V\left( \sqrt 
{b^{2}+z'^{2}}\right)\right]^{1/2}dz'+ {\mbox{ constant}}~.
\end{equation}

The additive constant is chosen in such a way to fulfill
\begin{equation}
\frac{S}{\hbar}\to kz\qquad {\rm for} \qquad V\to 0~.
\end{equation}

The term multiplying the potential can be interpreted as a change of phase of 
of the wave packet, having the following explicit form
\begin{equation}
\Delta (b)\equiv \frac{-m}{2k\hbar^{2}}\int_{-\infty}^{\infty} V\left( 
\sqrt{b^{2}+z^{2}}\right)dz~.
\end{equation}

Within the method of partial waves, the eikonal approximation has the following 
application. We know it is correct at high energies, 
where many partial waves do contribute
to the scattering. Thus, we can consider $l$ as a continuous variable
and by analogy to classical mechanics we let $l=bk$.
Moreover, as we already mentioned $l_{max}=kR$, which plugged into
eq.~(15) leads to
\begin{equation}
f(\theta )=-ik\int bJ_{0}(kb\theta )[e^{2i\Delta (b)}-1]db~.
\end{equation}

\bigskip

\section*{{\huge 8P. Problems}}

{\em Problem 8.1}

\bigskip

\noindent
Obtain the phase shift and the differential cross section at small angles
for a scattering centre of potential $U(r)=\frac{\alpha}{r^2}$. It should be 
taken into account that for low-angle scattering the main contribution is given
by the partial waves of large $l$.

\bigskip

\noindent
{\bf Solution}:

Solving the equation 
$$
R_{l}^{''}+\Bigg[k^2-\frac{l(l+1)}{r^2}-\frac{2m\alpha}{\hbar ^2 r^2}\Bigg]=0
$$
with the boundary conditions $R_{l}(0)=0$, $R_{l}(\infty)=N$, where $N$ is a
finite number, we get
$$
R_{l}(r)=A\sqrt{r}I_{\lambda}(kr)~,
$$
where $\lambda=\Bigg[(l+\frac{1}{2})^2+\frac{2m\alpha}{\hbar ^2}\Bigg]^{1/2}$ 
and $I$ is the first modified Bessel function.

To determine $\delta _{l}$, one should use the asymptotic expression
of $I_{\lambda}$:
$$
I_{\lambda}(kr)\propto \left(\frac{2}{\pi kr}\right)^{1/2}\sin (kr-
\frac{\lambda\pi}{2}+\frac{\pi}{4})~.
$$
Therefore
$$
\delta _{l}=-\frac{\pi}{2}\left(\lambda-l-\frac{1}{2}\right)=
-\frac{\pi}{2}\left(\Bigg[(l+\frac{1}{2})^2
+\frac{2m\alpha}{\hbar ^2}\Bigg]^{1/2}-\left(l+\frac{1}{2}\right)\right)~.
$$
The condition of large $l$ leads us to 
$$
\delta _{l}=-\frac{\pi m \alpha}{(2l+1)\hbar ^2}~,
$$
whence one can see that $|\delta _{l}|\ll 1$ for large $l$.

From the general expression of the scattering amplitude
$$
f(\theta)=\frac{1}{2ik}\sum _{l=0}^{\infty}(2l+1)P_{l}(\cos \theta)
(e^{2i\delta _{l}}-1)~,
$$
at small angles one gets $e^{2i\delta _{l}}\approx 1+2i\delta _{l}$, so that
$$
\sum _{l=0}^{\infty}P_{l}(\cos \theta)=\frac{1}{2\sin \frac{\theta}{2}}~.
$$
Thus
$$
f(\theta)=-\frac{\pi \alpha m}{k \hbar ^2}\frac{1}{2 \sin \frac{\theta}{2}}~.
$$
The final result is
$$
\frac{d\sigma}{d\theta}=\frac{\pi ^3 \alpha ^2 m}{2\hbar ^2 E}{\mbox ctg}
\frac{\theta}{2}~.
$$



\end{document}